\documentclass[12pt,preprint]{aastex}
\usepackage{graphicx}
\begin{document}
\title{HOW BAD/GOOD ARE THE EXTERNAL FORWARD SHOCK AFTERGLOW MODELS OF GAMMA-RAY BURSTS?}

\author{Xiang-Gao Wang\altaffilmark{1,2,3}, Bing Zhang\altaffilmark{2}, En-Wei Liang\altaffilmark{1,3},
He Gao\altaffilmark{4}, Liang Li\altaffilmark{5,6}, Can-Min Deng\altaffilmark{1,3},
Song-Mei Qin\altaffilmark{7,2}, Qing-Wen Tang\altaffilmark{8},
D. Alexander Kann\altaffilmark{9}, Felix Ryde\altaffilmark{10}, Pawan Kumar\altaffilmark{11}}

\altaffiltext{1}{GXU-NAOC Center for Astrophysics and Space Sciences, Department of Physics, Guangxi University, Nanning 530004, China; wangxg@gxu.edu.cn; lew@gxu.edu.cn}
\altaffiltext{2}{Department of Physics and Astronomy, University of Nevada Las Vegas, NV 89154, USA; zhang@physics.unlv.edu}
\altaffiltext{3}{Guangxi Key Laboratory for the Relativistic Astrophysics, Nanning 530004, China}
\altaffiltext{4}{Department of Astronomy and Astrophysics, Pennsylvania State University, 525 Davey
Laboratory, University Park, PA 16802}
\altaffiltext{5}{Department of Physics, Stockholm University, SE-106 91 Stockholm, Sweden}
\altaffiltext{6}{Erasmus Mundus Joint Doctorate in Relativistic Astrophysics}
\altaffiltext{7}{Mathematics and Physics Section, Guangxi University of Chinese Medicine, Nanning 53001,China}
\altaffiltext{8}{School of Astronomy and Space Science, Nanjing University, Nanjing, Jiangsu 210093, China}
\altaffiltext{9}{Th$\ddot{\rm u}$ringer Landessternwarte Tautenburg, Sternwarte 5, 07778 Tautenburg, Germany}
\altaffiltext{10}{Department of Astronomy, University of Texas at Austin, Austin, TX 78712, USA}
\altaffiltext{11}{Department of Physics, KTH Royal Institute of Technology, SE-106 91 Stockholm, Sweden}

\begin{abstract}

The external forward shock models have been the standard
paradigm to interpret the broad-band afterglow data of gamma-ray bursts (GRBs).
One prediction of the models is that some afterglow temporal breaks at different
energy bands should be {\em achromatic}, namely, the break times should be the
same in different frequencies. Multi-wavelength observations in the {\em Swift}
era have revealed chromatic afterglow behaviors at least in some GRBs, casting
doubts on the external forward shock origin of GRB afterglows.
In this paper, using a large sample of GRBs with both X-ray and optical afterglow
data, we perform a systematic study to address the question:
how bad/good are the external forward shock models?
Our sample includes 85 GRBs up to March 2014 with well-monitored X-ray and optical
lightcurves. Based on how well the data abide by the external forward shock
models, we categorize them into five grades and three samples.
The first two grades (Grade I and II)
include 45/85 GRBs. They show evidence of, or are consistent with having, an achromatic
break. The temporal/spectral behaviors in each afterglow segment are consistent with
the predictions (the ``closure relations'') of the forward shock models. These GRBs
are included in the Gold sample. The next two grades (Grade III and IV) include 37/85
GRBs. They are also consistent with having an achromatic break, even though one or more
afterglow segments do not comply with the closure relations. These GRBs are included in
the Silver sample. Finally, Grade V (3/85) shows direct evidence of chromatic
behaviors, suggesting that the external shock models are inconsistent with the data.
These are included in the Bad sample.
We further perform statistical analyses of various observational properties
(temporal index $\alpha$, spectral index $\beta$, break time $t_b$) and model parameters
(energy injection index $q$, electron spectral index $p$, jet opening angle $\theta_j$,
radiative efficiency $\eta_\gamma$, etc) of the GRBs in the Gold Sample, and
derive constraints on the magnetization parameter $\epsilon_B$ in the forward shock.
Overall, we conclude that the simplest external forward shock models can account for
the multi-wavelength afterglow data of at least half of the GRBs. When more advanced modeling
(e.g., long-lasting reverse shock, structured jets, arbitrary circumburst medium density profile)
is invoked, up to $>90 \%$ of the
afterglows may be interpreted within the framework of the external shock models.

\end{abstract}

\section{Introduction\label{sec:intro}}

Gamma-ray bursts (GRBs) are the most luminous explosions in the universe.
They signify the birth of a stellar-mass black hole or a rapidly
rotating magnetized neutron star during core collapses of massive
stars or mergers of compact objects \citep[][for a recent review]{kumarzhang15}.

Multi-wavelength GRB afterglows were predicted \citep{meszarosrees97}
before their first discoveries \citep{costa97,vanparadijs97,frail97}. This was based on a generic external forward
shock model. Regardless of the physical nature of progenitor and central
engine, a relativistic jet is launched, which is decelerated by a
circumburst medium by a pair of external (forward and reverse) shocks.
The reverse shock is likely short-lived. The forward shock, on the
other hand, continues to plough into the medium as the jet is
decelerated. Synchrotron radiation of electrons accelerated from the
external forward shock powers broad-band electromagnetic radiation
with a decreasing amplitude. This is the broad-band afterglow of GRBs \citep{meszarosrees97,sari98,meszaros98,rhoads99,sari99,chevalier00}.

Before 2004, the observations of the broad band late time afterglow
emission of GRBs generally show broken power-law lightcurves and
instantaneous spectra. Detailed studies  \citep[e.g.,][]{wijers97,waxman97,wijers99,harrison99,huangyf99,huangyf00,panaitescu01,panaitescu02,yost03,wu04} suggested that
these late-time data are generally
consistent with the predictions of the external forward shock models.

The launch of the {\em Swift} satellite in 2004 \citep {gehrels04}
allowed systematic observations of the multi-wavelength GRB afterglow
at early epochs. These data, especially the early X-ray afterglow
data, presented surprises to modelers. The overall X-ray lightcurves
include five distinct temporal components \citep{zhang06}:
I: an early time steep decay phase connected to the prompt emission
\citep{tagliaferri05,barthelmy05,zhang07c};
II: a shallow decay (or plateau) phase, which may signify continuous
energy injection of energy into the blastwave \citep{zhang06,nousek06,liang07}; III: a normal decay phase
consistent with the forward shock emission of a constant-energy fireball;
IV: a late steep decay phase likely due to a jet break origin \citep[e.g.,][]{liang08,racusin09};
and V: erratic X-ray flares, likely powered by late central engine
activities \citep{ioka05,burrows05,fan05,zhang06,liang06,lazzati07,chincarini07,maxham09,margutti10}. The components I and V are are believed to be of an internal
origin (in contrast to the external shock origin). The other three
components (II, III and IV) may be interpreted within the framework of
the external shock models.

The optical afterglow light curves also show interesting temporal
behaviors \citep{liang06,nardini06,kann06,kann10,kann11,panaitescu08,panaitescu11,lil12,liang13,wang13,yi13}. In the similar spirit as \cite{zhang06}, \cite{lil12} attempted to summarize a ``synthetic'' light curveof
optical emission. They found more components with distinct physical
origins: Ia: prompt optical flares; Ib: an early optical flare of an
external reverse shock origin; II: an early shallow-decay segment; III:
the standard afterglow component (the normal decay component, sometimes with an
early onset hump); IV: the post-jet-break phase; V: late optical flares;
VI: late rebrightening humps; and VII: late supernova (SN) bumps.
The components II, III and IV can
find their counterparts in the canonical X-ray light curve
(components II, III, and IV in \citealt{zhang06}).
Some flares in the optical band have counterparts in X-rays, but some
others do not \citep{swenson13}. Some components (e.g., the reverse
shock component Ib and the supernova component VII) are unique for the
optical band only.

There are two types of temporal breaks in the external shock models.
One type corresponds to the crossing of a characteristc frequency in the
observational band \citep{sari98}. Such spectrally-related breaks
occur at different epochs in different energy bands, and therefore
are {\em chromatic}. A testable feature of such a break is that the
spectral indices before and after the temporal break should be
distinctly different. The second type of breaks are related to the
hydrodynamic or geometric properties of the system. Since both effects
affect the global behavior of the blastwave, these breaks should be
{\em achromatic}, i.e. the temporal breaks in different energy bands
should occur around the same observational time.

Most observed breaks in the GRB lightcurves are likely of a hydrodynamic
or geometric origin. Observationally, essentially all the temporal breaks
observed in the X-ray lightcurves are consistent with having no spectral
changes across the break times \citep{liang07,liang08}. Theoretically,
the spectral breaks, especially the cooling break, are predicted to be
very smooth, and are barely observable from the data (\citealt{uhm14a},
see also \citealt{granot02,vanerten09}).
As a result, one expects that the temporal breaks seen by {\em Swift}
should be strictly achromatic based on the external
forward shock models.

Broad-band afterglow data of GRBs are rapidly accumulating.
Shortly after {\em Swift} detected early X-ray afterglow lightcurves
of GRBs, some authors noticed that the basic requirement of
achromaticity of GRB afterglows is violated at least in some GRBs \citep[e.g.,][]{panaitescu06,fan06,huangky07,liang07,liang08}. In particular, while a significant break
is seen in the X-ray lightcurves of some GRBs, the optical lightcurve
does not show evidence of a break at the corresponding time \citep[e.g.,][]{troja07,molinari07}.
Such a puzzling effect led theorists to suggest various non-forward-shock
models of the X-ray afterglow: the long-lasting reverse shock model
\citep{genet07,uhm07}, the dust scattering model
(Shao \& Dai 2007), and the long-lasting central engine model \citep{ghisellini07,kumar08a,kumar08b}.

Indeed, if most GRB afterglows are chromatic, one must throw away the
standard forward shock paradigm, and probably attribute other factors,
in partular, the long-lasting central engine, to account for the X-ray
afterglow. This would have profound implications for our understanding
of the GRB central engine and emission physics. Yet, there seem to
exist some GRBs (e.g., the latest bright GRB 130427A) whose multi-wavelength
data are consistent with the simplest forward shock afterglow model \citep[e.g.,][]{maselli14,perley14}.

It is therefore natural to ask the following question: in general
how bad or how good are the external forward shock models in interpreting
the GRB afterglow data?

This paper aims at addressing this question through a systematic data analysis
and theoretical modeling of a large sample of multi-wavelength afterglows.
We study a sample of 85 {\em Swift} GRBs up to March 2014, which all have
high-quality X-ray and optical light curve data to allow us to study the compliance
of the data to the external forward shock models.
The sample selection and data analyses are described in \S2.
The theoretical external forward shock model, in particular, the so-called
{\em closure relations} are presented in \S3. In \S4, we grade the afterglows
based on how well they abide by the forward shock models, and categorize
them into five grades and three samples.
A statistical analysis of various observational and theoretical
parameters for the Gold sample is presented in \S5.
Our results are summarized in \S6 with some discussion.
We notice that \cite{lil15} recently carried out a similar analysis,
with the focus on the consistency of the data with afterglow models
in individual temporal segments of X-ray and optical lightcurves,
without analyzing the global achromatic/chromatic behaviors of the afterglows.

Throughout the paper, the subscripts ``O'' and ``X'' denote the
optical and X-ray band, respectively, and the subsripts ``1'' and ``2''
denote the pre- and post-break segments, respectively.
In addition, two spectral regimes are defined:
``I'' for $\nu> {\rm max} (\nu_m, \nu_c)$, and ``II'' for $\nu_m<\nu<\nu_c$,
where $\nu_m$ and $\nu_c$ are the minimum injection frequency and cooling frequency
for synchrotron radiation, respectively.

\section{Sample and Data\label{sec:data}}

We systematically investigate all the {\em Swift} GRBs that have X-ray and optical
afterglow data, over a span of almost 10 years from the launch of {\em Swift} to
March 2014. A sample of $\sim$260
optical light curves are compiled from published papers or GCN
Circulars, and a sample of $\sim$900 X-ray light curves are obtained from
the {\em Swift} XRT data archive. Well-sampled light curves in both X-ray and
optical bands are available for 99 GRBs. Fifteen GRBs do not have well constrained spectral indices either in optical or in X-ray bands to allow us to perform some theoretical constraints (see details below). Fourteen of them are removed from the sample. GRB 070420 is the only GRB without adequate spectral information that is included in our sample. This is because it has a clear chromatic feature, which allows us to group it into the Bad sample even if the spectral information is not available (see details in \S \ref{sec:graderesutl}). The remaining 84 GRBs are included in our final sample, whose information is presented in Table \ref{table:sample}. For the optical data, the correction due to Galactic extinction is taken
into account using the reddening map presented by Schlegel et al. (1998).
Due to large uncertainties, we do not make corrections to the extinction
in the GRB host galaxies.

In order to quantify the rich temporal features of GRB lightcurves,
we fit the lightcurves with a model of multiple components. The basic
component of our model is either a single power-law (SPL) function
\begin{equation}
 F_1 = F_{01} t^{-\alpha}
\end{equation}
or a smooth broken power-law (BPL) function
\begin{equation}
F_2 =  F_{02} \left[\left(\frac{t}{t_{\rm
b}}\right)^{\alpha_1\omega} +\left(\frac{t}{t_{\rm
b}}\right)^{\alpha_2\omega}\right]^{-1/\omega}, \label{F2}
\end{equation}
where $\alpha$, $\alpha_1$, $\alpha_2$ are the temporal slopes,
$t_{\rm b}$ is the break time, and $\omega$ measures the sharpness
of the break. In some afterglow models, a double broken power-law
light curve is expected. For example, it is theoretically expected
that the afterglow light curve may have a shallow segment early on
due to energy injection, then transits to a normal decay segment
when energy injection is over, and finally steepens due to a jet
break (e.g., in the canonical X-ray afterglow lightcurve, Zhang et
al. 2006). We therefore also consider a smooth triple-power-law
(TPL) function to fit some lightcurves. In these cases, we extend equation
(\ref{F2}) (with $t_b$ defined as $t_{b,1}$)
to the following function \citep{liang08}
\begin{equation}\label{STPL}
F_3=(F_{2}^{-\omega_2}+F_4^{-\omega_2})^{-1/\omega_2}
\end{equation}
where $\omega_2$ is the sharpness factor of the second break at
$t_{b,2}$, and
\begin{equation}\label{PL}
F_4=F_2(t_{\rm b,2})\left(\frac{t}{t_{b,2}}\right)^{-\alpha_3}.
\end{equation}

 We perform best fits to the data using a subroutine called
MPFIT\footnote{http://www.physics.wisc.edu/~craigm/idl/fitting.html.}.
The sharpness parameter $\omega$ is usually adopted as 3 or 1 in
our fitting. The parameter $t_{b}$ is not significantly affected by
the choice of $\omega$, but the pre- and post-break slopes (i.e. $\alpha_{1}$
and $\alpha_{2}$) somewhat depend on the value of $\omega$ \citep[][]{liang07}.
The larger the value of $\omega$, the sharper the break. The breaks in most X-ray
and optical light curves at later times (e.g. the energy injection breaks and
the jet breaks) can be well fit with $\omega=3$, which is consistent with the fitting
results using other empirical models \citep[e.g.][]{willingale07}. Some very smooth
breaks (e.g., the onset breaks in the early optical lightcurve curves) require $\omega$
being around 1 \cite[][]{liang07,lil12}, and we adopt this value when it is needed.

One focus of our analysis is to study the ``chromaticity'' of the
lightcurves in the X-ray and optical bands. In principle there are
two approaches to do this. The first approach is to blindly search
for $t_b$ using the best fits to the optical and X-ray data,
respectively, and compare how different the two $t_b$ values are.
Such an approach usually gives different break times in the two
bands \citep[][]{liang07,liang08,lil12,lil15}.
The second approach is to start with the achromatic
assumption and investigate how bad the data violate such an assumption.
By doing so, we reduce one free parameter, and impose a same $t_b$ in
both bands in the model. We believe that this second approach is more
reasonable to address the question ``how {\em bad} the external forward shock
models are'', so we adopt the second approach with the assist of the
first approach.
The detailed procedure of our light curve fitting is as follows:

\begin{itemize}
  \item For each GRB, we first fit the optical and X-ray afterglow light
curves separately, and get the respective fitting parameters, such as
$t_{\rm O,b}$, $t_{\rm X,b}$, and the $\omega$ values of each break).
A minimum number of components (SPL, BPL, or TPL)
are introduced based on eye inspection
of the global features in the lightcurve. If the reduced $\chi^2$ is
much larger than 1, we continue to add more components and re-do the
fits, until the reduced $\chi^2$ becomes close to 1 (usually less than 1.5).
The reduced $\chi^2$ values for some lightcurves are much smaller than 1,
indicating that some model parameters are poorly constrained. For these
cases, we fix some parameters and redo the fits until the reduced $\chi^2$
becomes close to 1. Some GRBs have erratic fluctuations in the lightcurves
with small error bars, so that the reduced $\chi^2$ is much larger than 1.
For these cases, we do not add additional
components to fit the lightcurves, so that their $\chi^2$ values
remain much larger than 1.
  \item Next, we jointly fit both optical and X-ray lightcurves by introducing
a same $t_b$. We search for a possible achromatic break time in the range
[$t_{\rm O,b}$, $t_{\rm X,b}$]. We still fit the optical and X-ray lightcures at a test break time $t_{b}$ separately in this step. The individual $\chi^2$ of the optical or X-ray band could not represent the goodness of the jointly fit. To evaluate the goodness of the fits for optical and X-ray lightcurves at $t_{b}$, we introduce a weighted reduced $\chi^2_{total}$, which is essentially the average reduced $\chi^2$ in both bands. Taking GRB 050922C as an
example: a best join fit is achieved at $t_b=17.3$ ks, where the reduced
$\chi^2_{X}$ values are 175/157 and $175/148 \simeq 186/157$ for the
optical and X-ray bands, respectively, so that $\chi^2_{total}$ can be expressed as 361/314.
For all the GRBs, we search for the common $t_b$ with the best $\chi^2_{total}$. We accept the fits with the $\chi^2_{total}\leq 3$, and regard it as not inconsistent with being achromatic\footnote{ The adoption of a separation line at $\chi^2_{total}$ around 3 is somewhat arbitrary, but the value is determined based on close inspection of the fitting results of individual bursts. Our results indicate that most GRB afterglow light curves are well fit with the BPL or SPL light curves models, with a typical value $\chi^2_{total}=1.21\pm0.50$. However, some GRBs (e.g., GRB 050730, 060904B, 080319C, 100901A, 120326A) show a relatively large $\chi^2_{total}$, which are around or even slightly larger than 3. Inspecting their light curves, the relatively large $\chi^2_{total}$ is caused by complicated features in the light curves (such as small flares and fluctuations), especially in the optical band (e.g., GRB 060904B).
However, the PL and BPL fits in any case catch the general features of these light curves.
Since we are interested in the achromatic/chromatic properties rather than the flaring features of the light curves, a relatively loose criterion ($\chi^2_{total} \sim 3$) is reasonable. }. Usually the parameters of this best join fits
does not correspond to the best reduced $\chi^2$ in each band.
  \item If both the optical and X-ray lightcurves decay as a SPL, we
do not need to search for a common break time. The weighted reduced
$\chi^2_{total}$ is calculated based on the above algorithm for the SPL fits
in each band.
  \item If one band decays as a BPL, while the other band does not have enough
data to search for a break time and decays as a SPL (e.g., the Grade II or IV
in Section 3), we impose $t_b$ identified in the first band as the common $t_b$,
and perform the $\chi^2_{total}$ analysis as described above.
\end{itemize}

The fitted results are presented in Figure \ref{gradeI}-\ref{gradeV}. The
parameters of the PL or BPL fits of all the lightcurves are presented in
Table \ref{table:sample}. Some lightcurves have additional features (e.g.,
steep decay phase, flares, rebrightening features) in one band. We do not
report them in Table \ref{table:sample}. Our analysis below discards these
extra components since they likely arise from additional emission components
(e.g., in the internal dissipation regions such as internal shocks and
internal magnetic dissipation sites) other than the external shock.

\section{External Shock Models: Closure Relations and Light Curve Types}

\subsection{Closure Relations}

The standard external shock models of GRB afterglows have clear theoretical
predictions that can be verified or falsified by the observational data.
These models attribute the multi-wavelength afterglow emission to synchrotron
radiation of electrons accelerated in the shock front as the fireball jet
interacts with the circumburst medium. The models largely do not depend on
the details of the central engine activities, so that the afterglow behaviors
only depend on a limited number of parameters.
In the convention of $F_\nu \propto t^{-\alpha}
\nu^{-\beta}$, where $\alpha$ and $\beta$ are the temporal and spectral indices of
the afterglows that can be measured directly from observations, the models predict
certain relationships between $\alpha$ and $\beta$ values, which are called the
``closure relations'' of the models \citep[e.g.,][]{zhangmeszaros04,zhang06,gao13}.
Technically there are many sub-models (e.g., ISM vs. wind,
adiabatic vs. radiative, whether or not there is energy injection), physical regimes
(reverse shock crossing phase, self-similar phase, post-jet-break phase, Newtonian
phase), and spectral regimes (different orders among the observed frequency ($\nu$)
and several characteristic frequencies ($\nu_m$, $\nu_c$, the self-absorption
frequency $\nu_a$). We refer to a comprehensive review of \cite{gao13} and
references therein for the details of various models.

For the time frame of our interest (hours to weeks after the trigger), the reverse
shock crossing phase is usually over, and the blastwave is still in the relativistic
phase. This greatly reduces the number of relevant models. In Table \ref{Tab:alpha-beta},
we summarize the $\alpha$ and $\beta$ predictions of various models studied in this
paper following \cite{zhang06}  and  Gao et al. (2013). This includes the ISM and
wind models for adiabatic blastwaves\footnote{ In general, the circumburst medium can be described by an arbitrary profile $n \propto r^{-k}$. The ISM model corresponds to $k=0$, and the wind model corresponds to $k=2$. In our closure relations, we only consider these two cases, since they are naturally expected from the ISM and a pre-explosion stellar wind. For other $k$ values, it is not straightforward to imagine a physical mechanism to produce such profiles over a large distance scale of interest. We therefore do not include the arbitrary $k$ models in the standard afterglow models, but discuss them as possible modified afterglow models.}, for both pre- and post-jet break temporal phases,
with and without continuous energy injection, and for two spectral regimes
(I: $\nu> \nu_c$ and II: $\nu_m < \nu < \nu_c$) in the slow cooling ($\nu_m < \nu_c$)
regime. By doing so, we have assumed that $\nu_a< {\rm min}(\nu_m, \nu_c)$, and
${\rm min}(\nu_{\rm X}, \nu_{\rm O}) > \nu_{m}$, which is usually satisfied for optical
and X-ray afterglow emission for typical GRB parameters.

The energy injection model invokes either a long-lasting central engine \citep{dai98,zhangmeszaros01},
or a Lorentz-factor-stratified ejecta \citep{rees98,sari00,uhm12}.
The two scenarios are equivalent with each other in terms of lightcurve
behaviors given a relationship between the central engine parameter
$q$ and the stratification parameter $k$ (Zhang et al. 2006). We adopt the
description of a long-lasting central engine with a power-law luminosity history
$L(t)=L_{0}(\frac{t}{t_{0}})^{-q}$ (Zhang \& M\'esz\'aros 2001), so that the injected
energy is $E_{inj}=\frac{L_{0}t_{0}^{q}}{1-q}t^{1-q}$. The prescription applies when
$q<1$. The relevant closure relations are presented in Table \ref{Tab:alpha-beta}.

Many observations suggest that GRB outflows are collimated. Assuming a conical
jet with opening angle $\theta_j$, a steepening in the afterglow light curve is predicted
when $1/\Gamma > \theta_j$ ($\Gamma$ is the bulk Lorentz factor of the blastwave).
The main reason of this steepening is the so-called ``edge effect'' \citep[e.g.,][]{panaitescu98}\footnote{Sideways expansion has been discussed as another factor of steepening the lightcurves
\citep{rhoads99,sari99}. However, later numerical simulations suggest that this effect
is not important \citep[e.g.,][]{zhangw09}. We do not consider this effect in this paper.}:
The $1/\Gamma$ cone is no longer filled with emission beyond the jet break time (when $1/\Gamma >
\theta_j$). There is a reduction factor in flux $\theta_{j}^{2}/(1/\Gamma)^{2}=\Gamma^{2}\theta_{j}^{2}$.
The  relevant closure relations are also presented in Table \ref{Tab:alpha-beta}.

It is possible that in some GRBs the energy injection phase lasts longer than the jet
break time, so that a jet break with energy injection both pre- and post-break phases
can be observed. The relevant closure relations of such models were derived in
\cite{gao13} and are also presented in Table \ref{Tab:alpha-beta}.

\subsection{Type of Afterglow Lightcurves}

\begin{figure*}
\includegraphics[angle=0,scale=1.2]{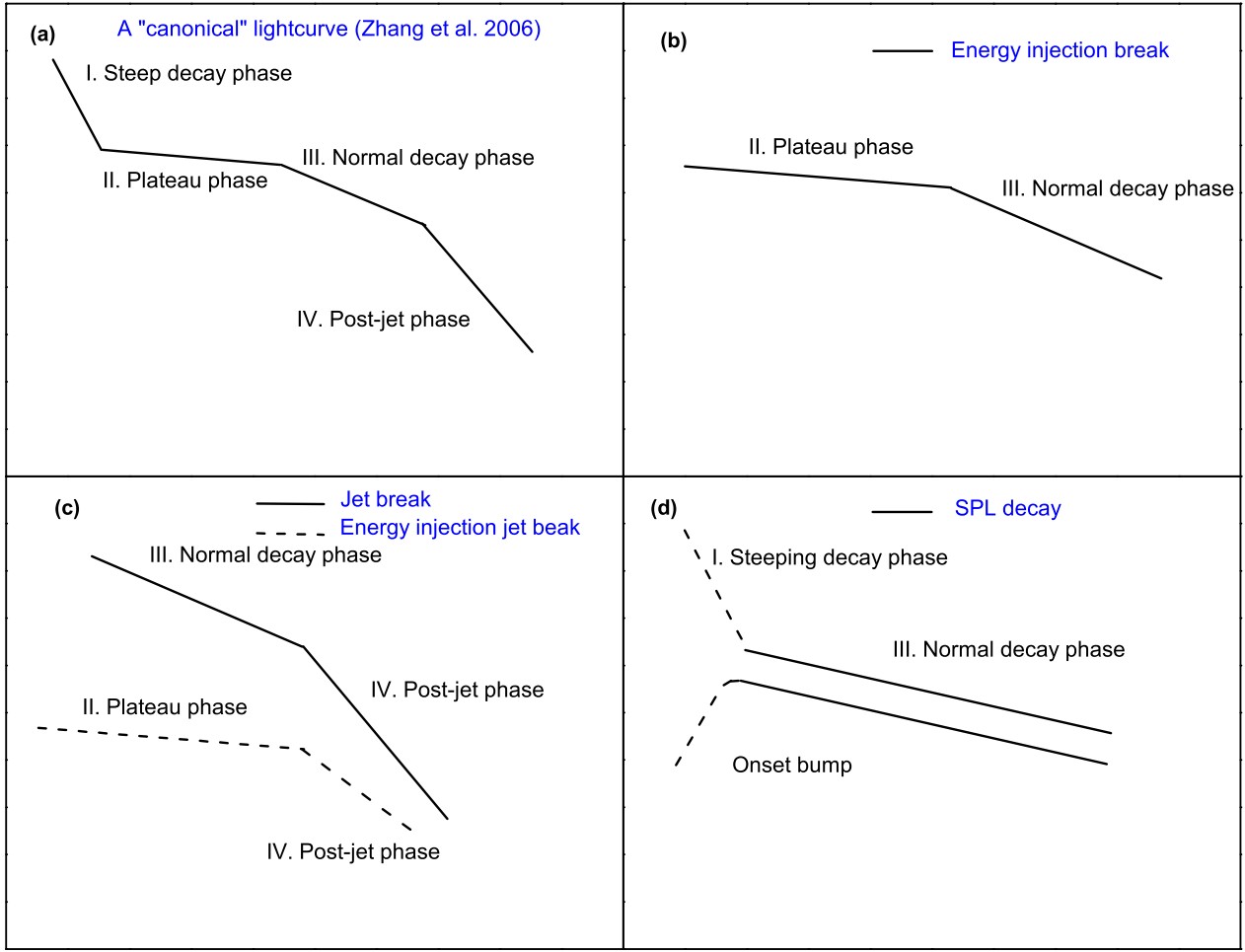}

\caption{Typical light curve behaviors: (a) The canonical X-ray light curve, reproduced from
Zhang et al (2006), with four characteristic temporal segments marked;
(b) The energy injection break case, with a transition
from the shallow decay phase (segment II) to the normal decay phase (segment III);
(c) The jet break cases without (solid) or with (dashed) energy injection. For the former,
it is a transition from the normal decay phase (segment III) to the post-jet-break
phase (segment IV); for the latter, both segments have a shallower decay slope;
(d) the single power-law (SPL) case. The steep decay phase in the X-ray light curve and the
early rising phase in the optical light curve are not included in the analysis.}\label{fit}
\end{figure*}

For the time domain we are interested in and for the optical and X-ray bands, there are
four types of lightcurves (Fig.\ref{fit}):

(1) Broken power-law lightcurves with an energy injection break:
In reference of the canonical X-ray light curve (Zhang et al. 2006),
as reproduced in Fig.\ref{fit}(a), the energy injection break connects
the shallow decay phase (segment II) to the normal decay phase (segment III),
and a typical light curve is shown in Figure \ref{fit}(b). Before and after
the break, the adiabatic deceleration $\alpha(\beta)$ relations
with and without energy injection (as listed in Table \ref{Tab:alpha-beta},
\citealt{zhang06,gao13}) are used to check whether the data are
consistent with model predictions.

(2) Broken power-law lightcurves with a jet break:
This corresponds to transition from segment III to IV in the canonical
lightcurve, and a typical light curve is shown in Figure \ref{fit}(c)
upper curve. Lightcurves of such a category should satisfy the constant-energy,
isotropic closure relations before the break, and the edge-effect
post-jet-break closure relations after the break, with no energy injection
effect both before and after the break (Table \ref{Tab:alpha-beta}).
The post-break decay index is required to be steeper than 1.5 for this model.

(3) Broken power-law lightcurves with a jet break with energy injection:
This model allows the energy injection extend to a duration longer
than the jet break. The temporal break is still defined by the edge effect
of a canonical jet, but the decay slopes before and after the break
are shallower than the previous case (lower curve in Fig.\ref{fit}(c)),
so that a $q$ parameter is introduced for both pre- and post-break phases.

(4) Single power-law decay: For some GRBs, a SPL function is adequate to
describe the afterglow data (Figure \ref{fit}(d)) after the deceleration
phase. In the X-ray band, there might be a steeper decay phase before
this SPL phase, which is due to the tail emission from the prompt emission
\citep{tagliaferri05,barthelmy05,zhang06}.
We ignore the steep decay phase and treat it as a SPL decay (upper curve
of Fig.\ref{fit}(d)). Similarly, in the optical band, some GRBs show an
early rising phase, which is a signature of the onset of afterglow at
the deceleration radius (peak of the lightcurve, lower curve of
Fig.\ref{fit}(d)). We treat these lightcurves also as SPL decay ones.

For all the types, sometimes there are X-ray flares overlapping on
the power-law decay segments. We do not include the flares in our
data fitting, since they originate from a different emission component
due to late central engine activities \citep[e.g.,][]{zhang06,maxham09}.

One important task is to perform a self-consistency check between the
optical and X-ray bands. If a GRB is consistent with the external
forward shock model, we demand that the GRB satisfies the following
criteria:
\begin{itemize}
 \item The X-ray and optical lightcurves are consistent with having
an achromatic break if any;
 \item Both the X-ray and optical lightcurves should satisfy closure
relations of a same circumburst medium type (ISM or wind) in both
pre- and post-break temporal segments;
 \item Either both bands belong to the same spectral regime, or the
two bands are separated by a cooling break $\nu_c$, with the X-ray
band above the break and the optical band below the break (with
allowance of a grey zone, see more discussion below);
 \item The inferred electron spectral index $p$ from both bands
and from both pre- and post-break segments should be consistent
with each other within error;
 \item For energy injection models, the energy injection parameter
$q$ values derived from the X-ray and optical bands should be consistent
with each other.
\end{itemize}

Technically, we check the consistency between
the closure relations for individual temporal segment
in individual energy band. To ensure a same $p$ value derived for
different bands, we also check the consistency between the data
and models in the
$\bigtriangleup \beta_{\rm X,O}- \bigtriangleup \alpha_{\rm X,O}$ plane.
Here $\bigtriangleup \alpha_{\rm X,O}=\alpha_{\rm X}-\alpha_{\rm O}$
is the difference between the decay indices in the X-ray and optical
bands, respectively, in a same temporal segment, and $\bigtriangleup
\beta_{\rm X,O}=\beta_{X}-\beta_{O}$ is the difference between the
spectral indices in the X-ray and optical bands, respectively.
Based on the closure relations (Table \ref{Tab:alpha-beta}), one can
derive the $\bigtriangleup \beta_{\rm X,O}- \bigtriangleup \alpha_{\rm X,O}$
relations of all the models (Table \ref{Tab:delta-alpha-betahigh2} and
\ref{Tab:delta-alpha-betalow2}). One can see that even though
$\alpha$ and $\beta$ values can be very different in different models, the
$\Delta \alpha_{\rm X,O}$ and $\Delta \beta_{\rm X,O}$ values have several
well-predicted values. In particular, for the SPL, and jet break models,
both pre- and post-break values are well-defined constants. For the energy
injection breaks, the post-break segment does not depend on the free parameter
$q$. As a result, if one focuses on the second component only, all the models
can be expressed as several representative coordinate values in the
$\bigtriangleup \beta_{\rm X,O}- \bigtriangleup \alpha_{\rm X,O}$ plane.
Considering the possible grey zones (see below for details),
these points define several straight lines in the $\bigtriangleup \beta_{\rm X,O}-
\bigtriangleup \alpha_{\rm X,O}$ space (Fig.\ref{deltaapha-beta} for details).
If the observed data intersect with these model lines
(within error), one can regard them as being consistent with the
model predictions.

Take the energy injection break as an example, our analysis uses the
following procedure: (a) Use the observed spectral indices $\beta_{\rm O}$
and $\beta_{\rm X}$ to predict the post-break temporal
indices $\alpha_{\rm O,2}$ and $\alpha_{\rm X,2}$ in two possible spectral
regimes. Then compare these theoretical predictions with the observational
values. If theoretical values are consistent with the
fitting results within error then go to next step. Otherwise, it
indicates that this GRB does not fall into this light curve type;
(b) Use the identified spectral regime to calculate the electron spectral
index $p$ from the spectral index $\beta$, i.e. $p = 2\beta+1$ for
$\nu_m < \nu < \nu_c$, or $p = 2\beta$ for $\nu > \nu_c$. Compare
the $p$ values derived from the optical and X-ray data, respectively.
If $p_{\rm O} = p_{\rm X}$ within error, then move to the next step. Otherwise,
this GRB does not fall into such a light curve type;
(c) Use the inferred $p$ value and spectral regimes to infer the energy injection
parameter $q$ using the temporal index before the break ($\alpha_{\rm O,1}$
and $\alpha_{\rm X,1}$). Compare the derived $q$ values from optical and X-ray
bands, respectively. If $q_{\rm O} = q_{\rm X}$ within error, then move to the next
step. Otherwise, this GRB does not fall into such a light curve type; (d) Using the
$\bigtriangleup \beta_{\rm X,O}- \bigtriangleup \alpha_{\rm X,O}$ relation to double
check the data, if the data fall into the predicted region in the $\bigtriangleup
\beta_{\rm X,O}- \bigtriangleup \alpha_{\rm X,O}$ plane, then this burst
can be fully interpreted by such a model. Otherwise, the burst does not fall
into this category.

The simplest analytical model (Sari et al. 1998) predicts
$\beta=p/2$ for Regime I ($\nu>\nu_{c}$) and
$\beta=(p-1)/2$ for Regime II ($\nu_{m}<\nu<\nu_{c}$).
Detailed numerical calculations \citep{uhm14a} showed that the
transition between the two regimes may take several orders of magnitude
in observer time. As a result, some ``grey zones'', with
$(p-1)/2 < \beta < p/2$ are allowed by the model.
Therefore the parameter space between the two closure relation lines
defined by the two spectral regimes in the $\alpha-\beta$ plane
is allowed by the theory. Data points falling into this grey zone
should be regarded as consistent with the model.
There are three possibilities: (1) the optical band is in Regime
II, while the X-ray band is in the grey zone; (2) the X-ray band
is in Regime I, while the optical band is in the grey zone; and (3)
both bands are in the grey zone.

For the cases that both the optical and X-ray bands are in the same
spectral regime, we demand that three spectral indices be the same
within error, i.e. $\beta_{\rm O}=\beta_{\rm OX}=\beta_{\rm X}$, where
$\beta_{\rm OX}$ is the spectral index between the optical and
X-ray band in the joint spectral energy distribution
(SED)\footnote{In order to obtain $\beta_{\rm OX}$, we roughly fit
the SED from optical to X-ray bands. For the optical band, we chose
the R-band where extinction correction is negligible. For the X-rays,
we use the Swift XRT data and adopt a typical band 1.5-2 keV, where
the absorption effect is negligible.}. If the
two bands are in different spectral regimes, we demand
$\beta_{\rm O}<\beta_{\rm OX}\leq\beta_{\rm X}$ or
$\beta_{\rm O}\leq\beta_{\rm OX}<\beta_{\rm X}$.

\section{Confronting Data with Models \label{sec:Confron}}

\subsection{Grading criteria and sample definitions}

With the above preparation, everything is in place for us to systematically confront the broad-band data with the external forward shock afterglow models. Based on how badly the data violate the models,
we define the following five grades (see also Table \ref{table:grades}):

\begin{itemize}
  \item Grade I: Both X-ray and optical bands have SPL lightcurves
or BPL lightcurves with an acceptable achromatic break. Both bands
satisfy closure relations and are
self-consistent (same medium type, $p$ and $q$ values).
These are the best examples where the GRB afterglow data abide by the
external shock model predictions;

  \item Grade II: Some GRBs have a clear break at $t_{b}$ in one
band (e.g., X-rays), but do not have a break in another
band (e.g., optical). The missing break is likely due to incomplete
observational coverage before or after the break. The data are
consistent with the hypothesis of an achromatic break, and both
bands satisfy closure relations self-consistently. These GRBs
are almost as good as Grade I in terms of abiding by the external
shock models;

  \item Grade III: Both X-ray and optical bands have SPL lightcurves
or BPL lightcurves with an acceptable achromatic break. However, at least one
temporal segment in one band does not satisfy the closure relations
in a self-consistent manner with respect to other segments/band.

  \item Grade IV: This is the Grade II equivalent for Grade III. One band
does not have a break, but the data are consistent with the hypothesis
of having an achromatic break. At least one
temporal segment in one band does not satisfy the closure relations
in a self-consistent manner with respect to other segments/band.

  \item Grade V: Clear evidence of chromatic breaks and violation of
closure relations. These GRBs cannot be interpreted within the
one-component external shock models\footnote{Some of these GRBs may be still
interpreted within two-component external shock models with each component
dominating one band \citep[e.g.,][]{depasquale09}. However, the demanded
parameters for the two components are rather contrived.}.

\end{itemize}

With these five grades, we define three samples:
\begin{itemize}
 \item {\em Gold Sample}: The GRBs in Grade I and II are defined as the Gold sample
GRBs, since no observed information violates any predictions of the external
shock models;
 \item {\em Silver Sample}: The GRBs in Grade III and IV are included in this sample.
Even though at least one segment/band does not satisfy the closure relations
self-consistently, the basic requirement of achromaticity is not violated.
We note that the closure relations are the predictions of the simplest analytical
external forward shock models. More complicated models invoking, e.g., a structured
jet \citep{zhangmeszaros02,rossi02,kumar03,granot03} or a circumburst density medium with an arbitrary $k$ value (at least for a certain distance range), predict light curve behaviors
that may not fully abide by the simple closure relations. Furthermore, if the
GRB engine is long-lived and a long-lasting reverse shock outshines the forward
shock, a variety of rich light curve behaviors can be generated, which do not follow
the simple closure relations \citep[e.g.,][]{uhm12,uhm14b}.
So it is possible that the GRBs in the silver sample are still consistent with
the external shock models;
 \item {\em Bad Sample}: The GRBs in Grade V violate the basic achromaticity
principle of the external shock models and do not abide by the closure relations,
and therefore cannot be interpreted within the framework of the external shock
models.
\end{itemize}

\subsection{Grading results \label{sec:graderesutl}}

\begin{figure*}
\includegraphics[angle=0,scale=0.35,width=0.325\textwidth,height=0.30\textheight]{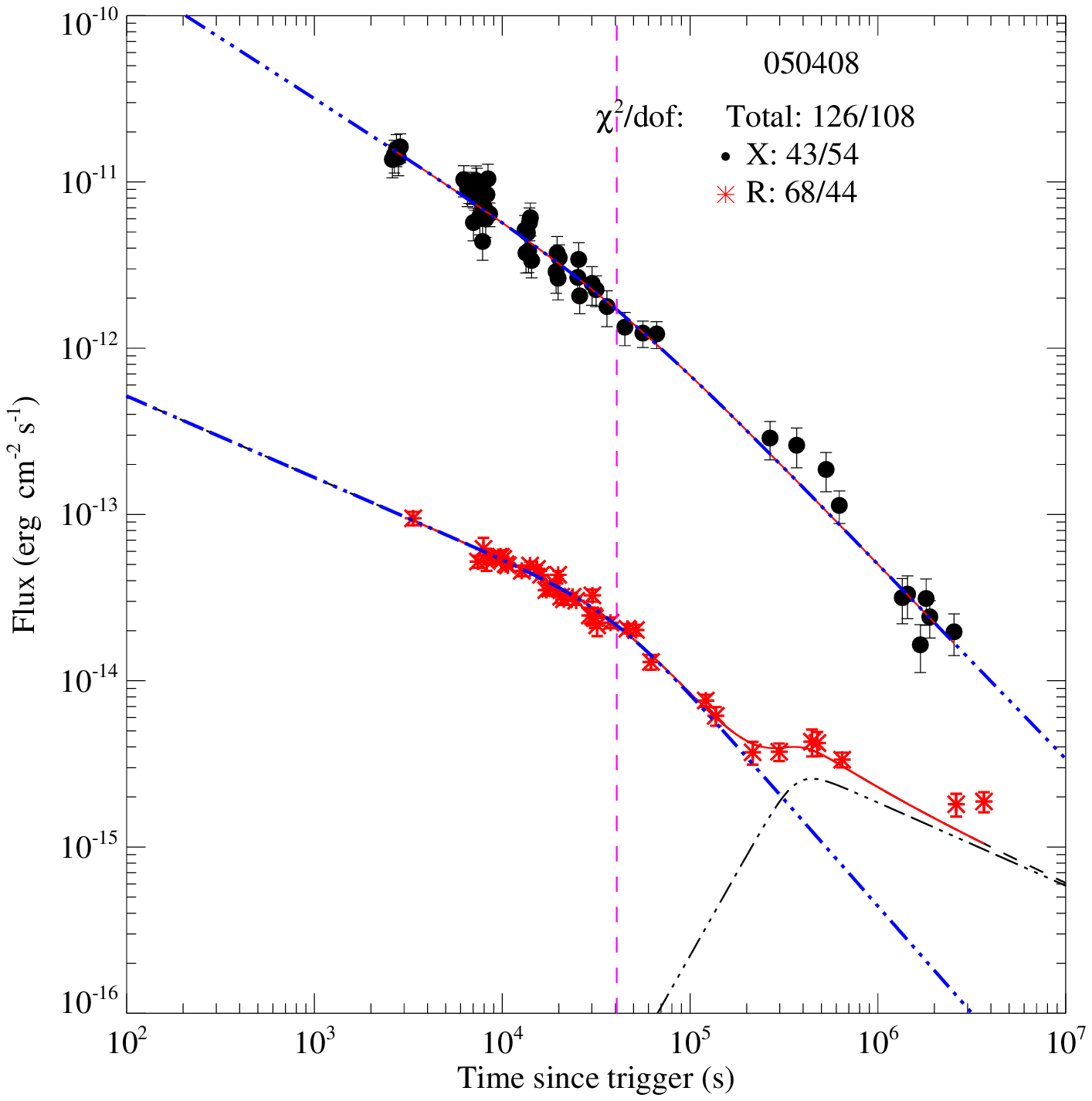}
\includegraphics[angle=0,scale=0.35,width=0.325\textwidth,height=0.30\textheight]{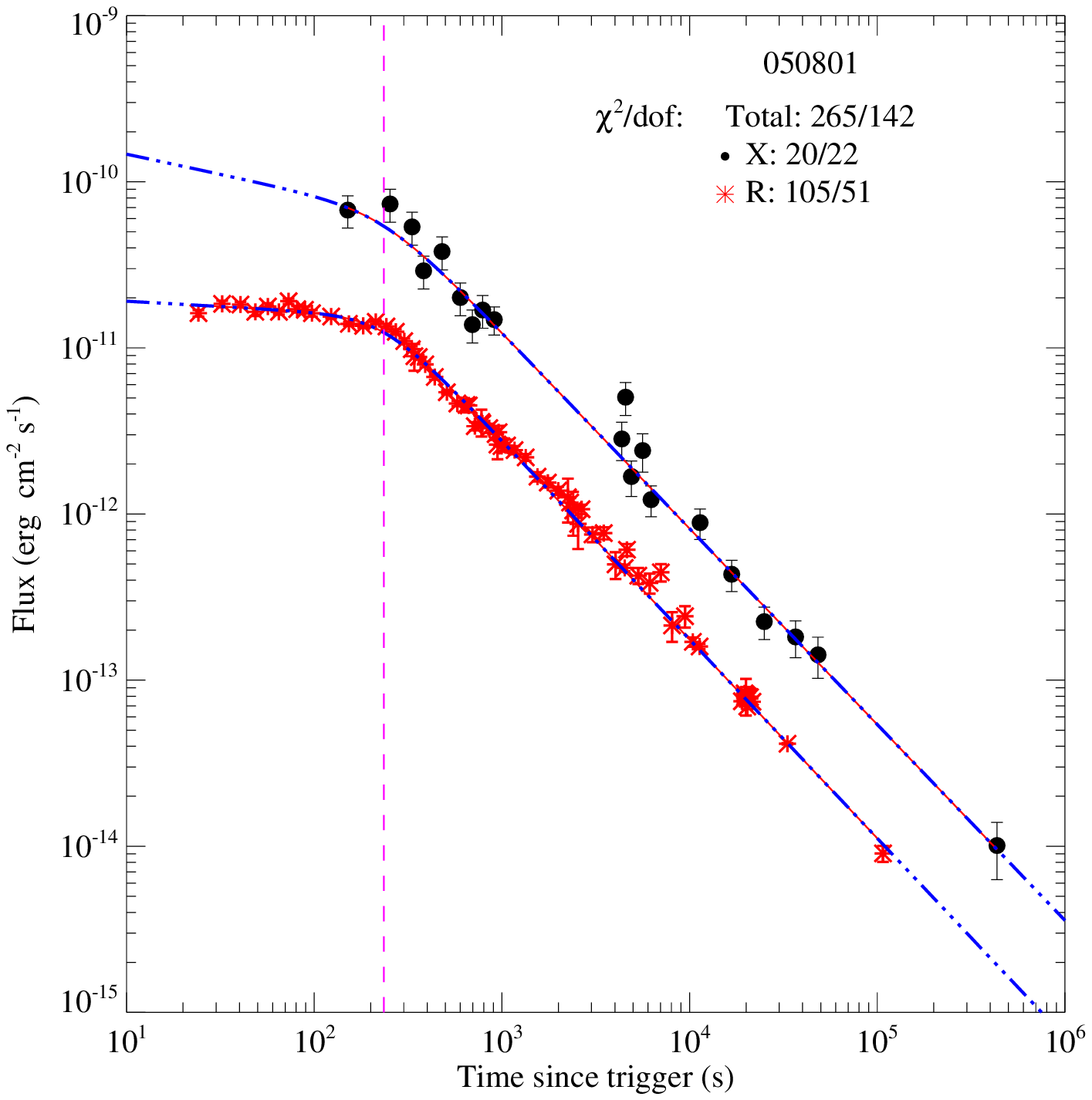}
\includegraphics[angle=0,scale=0.35,width=0.325\textwidth,height=0.30\textheight]{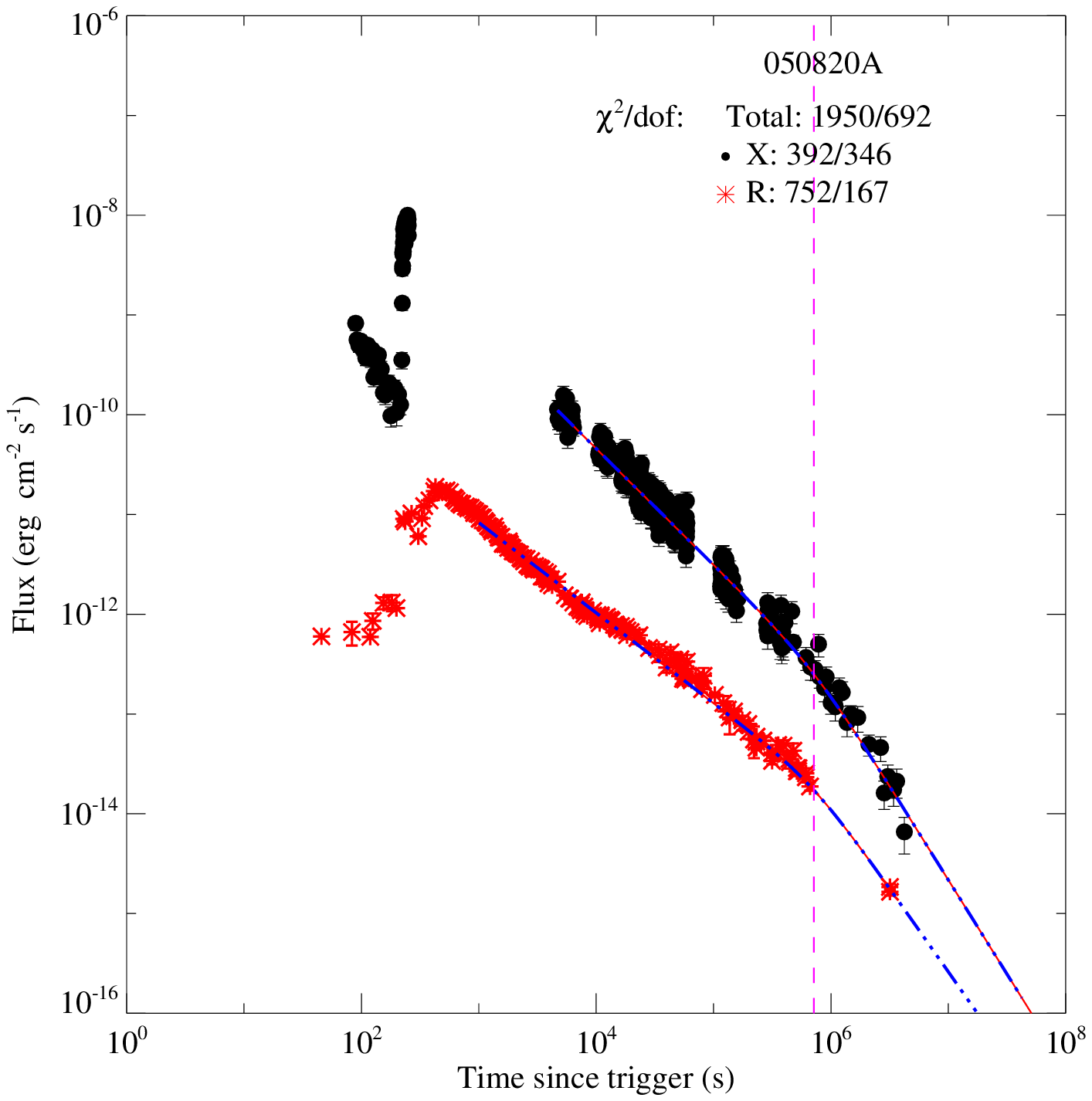}
\includegraphics[angle=0,scale=0.35,width=0.325\textwidth,height=0.30\textheight]{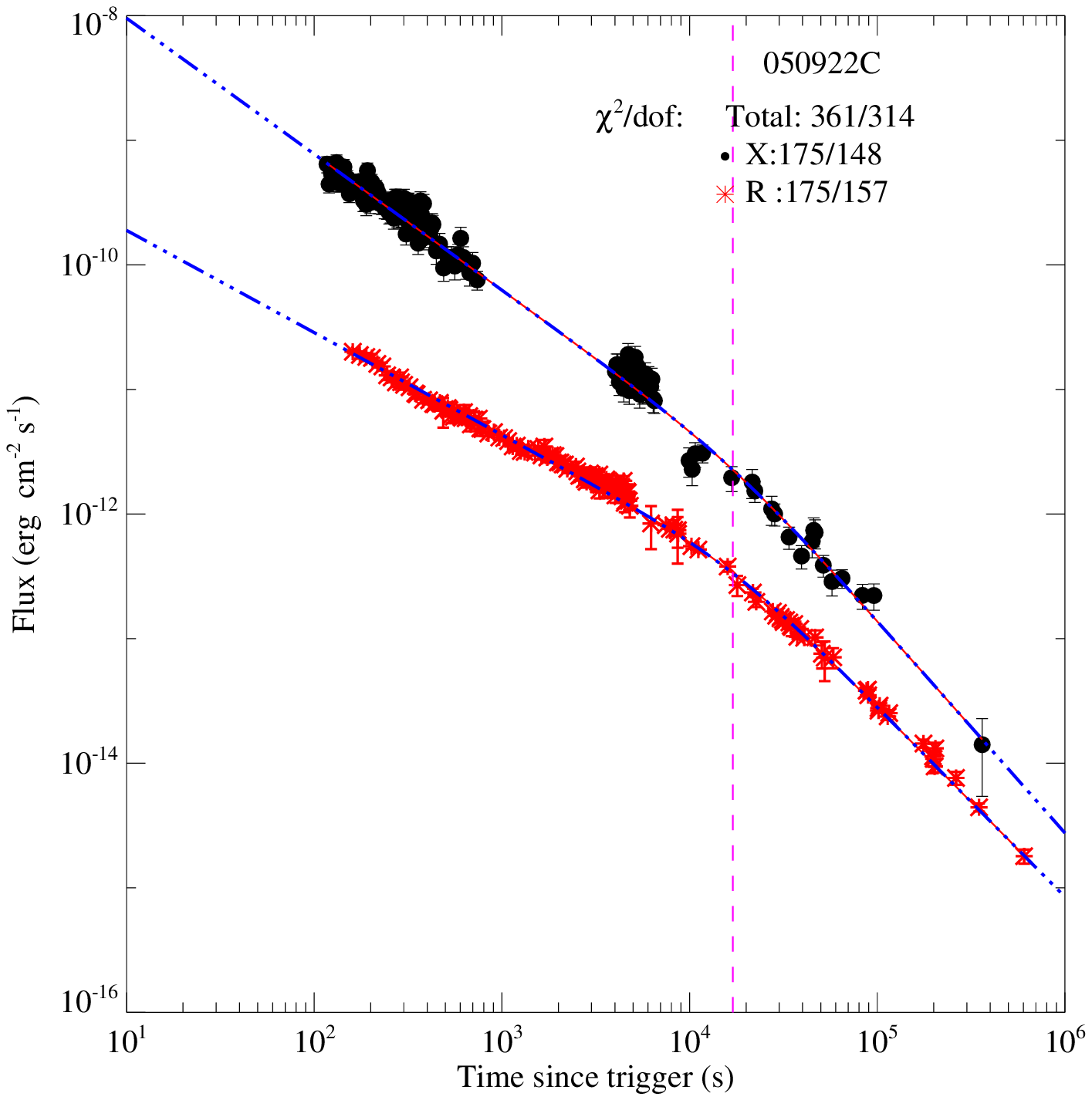}
\includegraphics[angle=0,scale=0.35,width=0.325\textwidth,height=0.30\textheight]{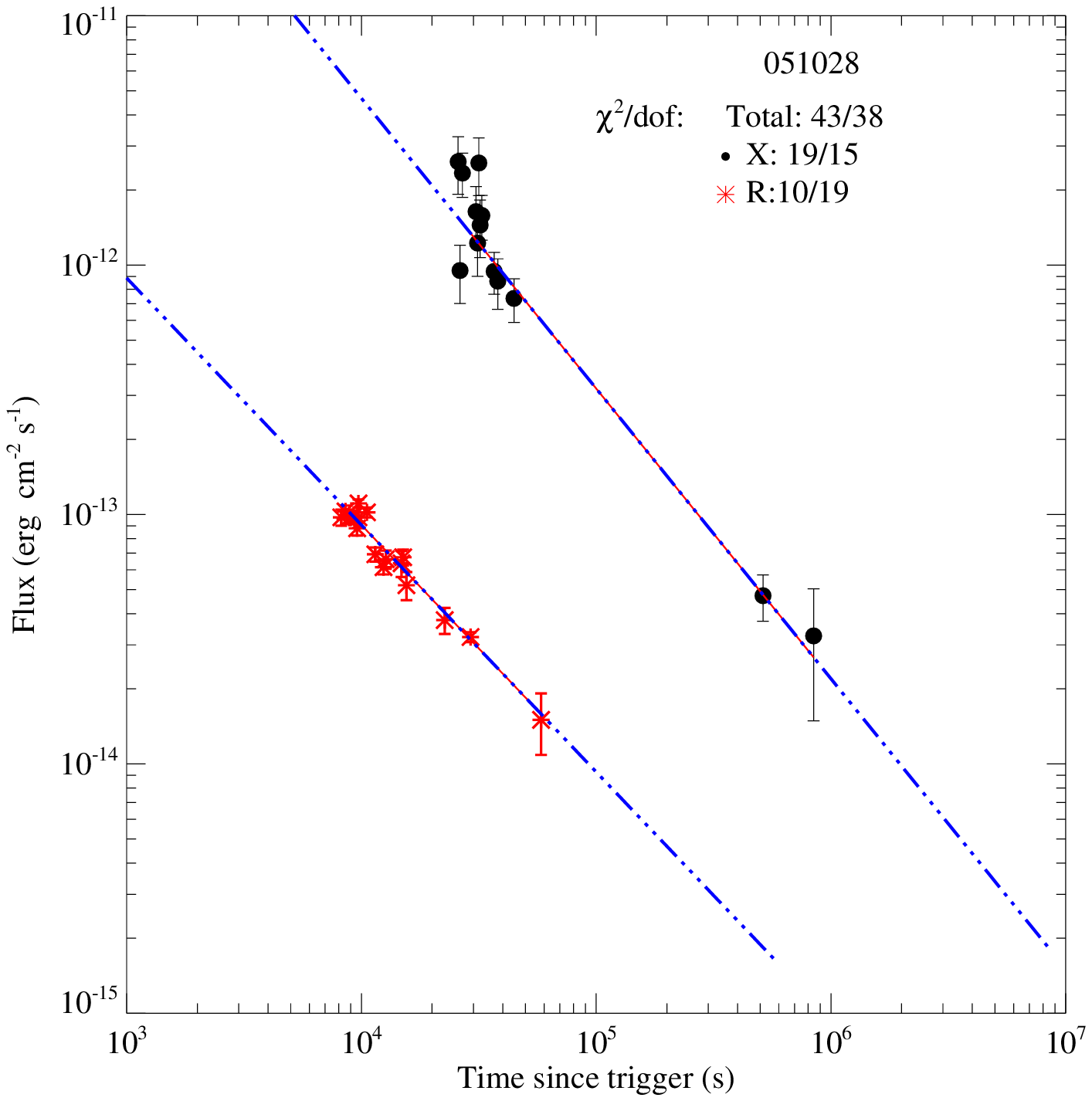}
\includegraphics[angle=0,scale=0.35,width=0.325\textwidth,height=0.30\textheight]{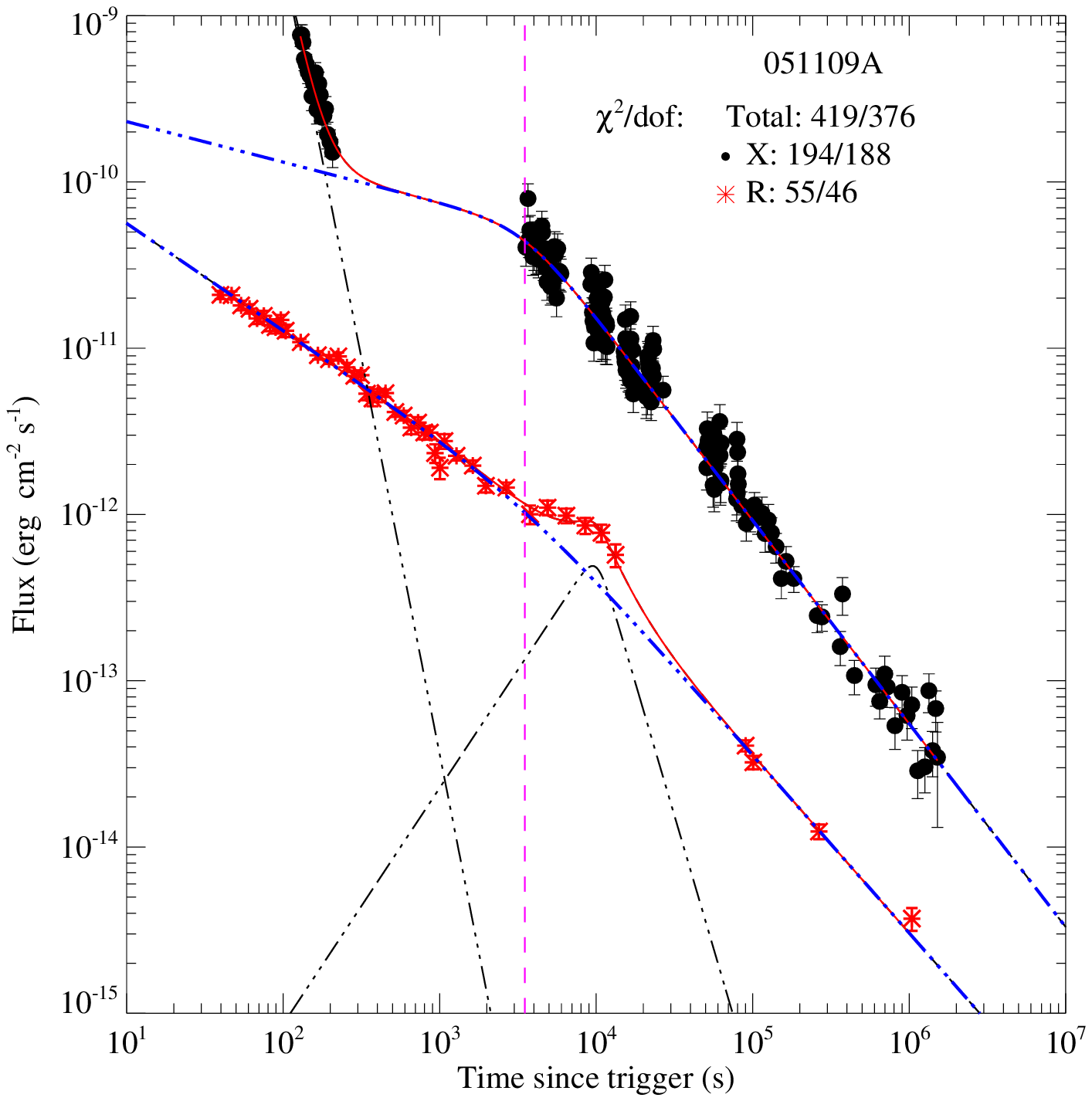}
\includegraphics[angle=0,scale=0.35,width=0.325\textwidth,height=0.30\textheight]{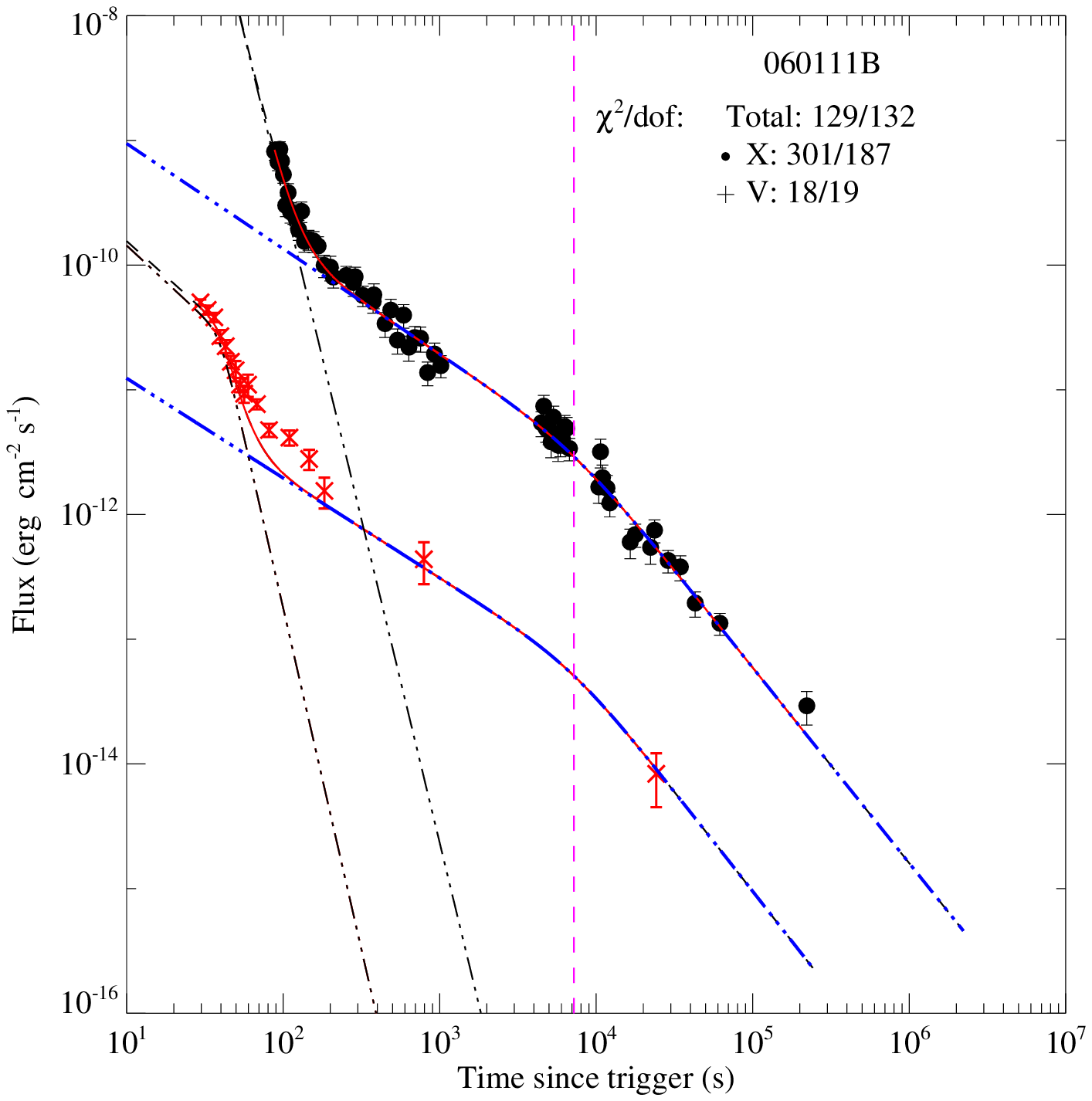}
\includegraphics[angle=0,scale=0.35,width=0.325\textwidth,height=0.30\textheight]{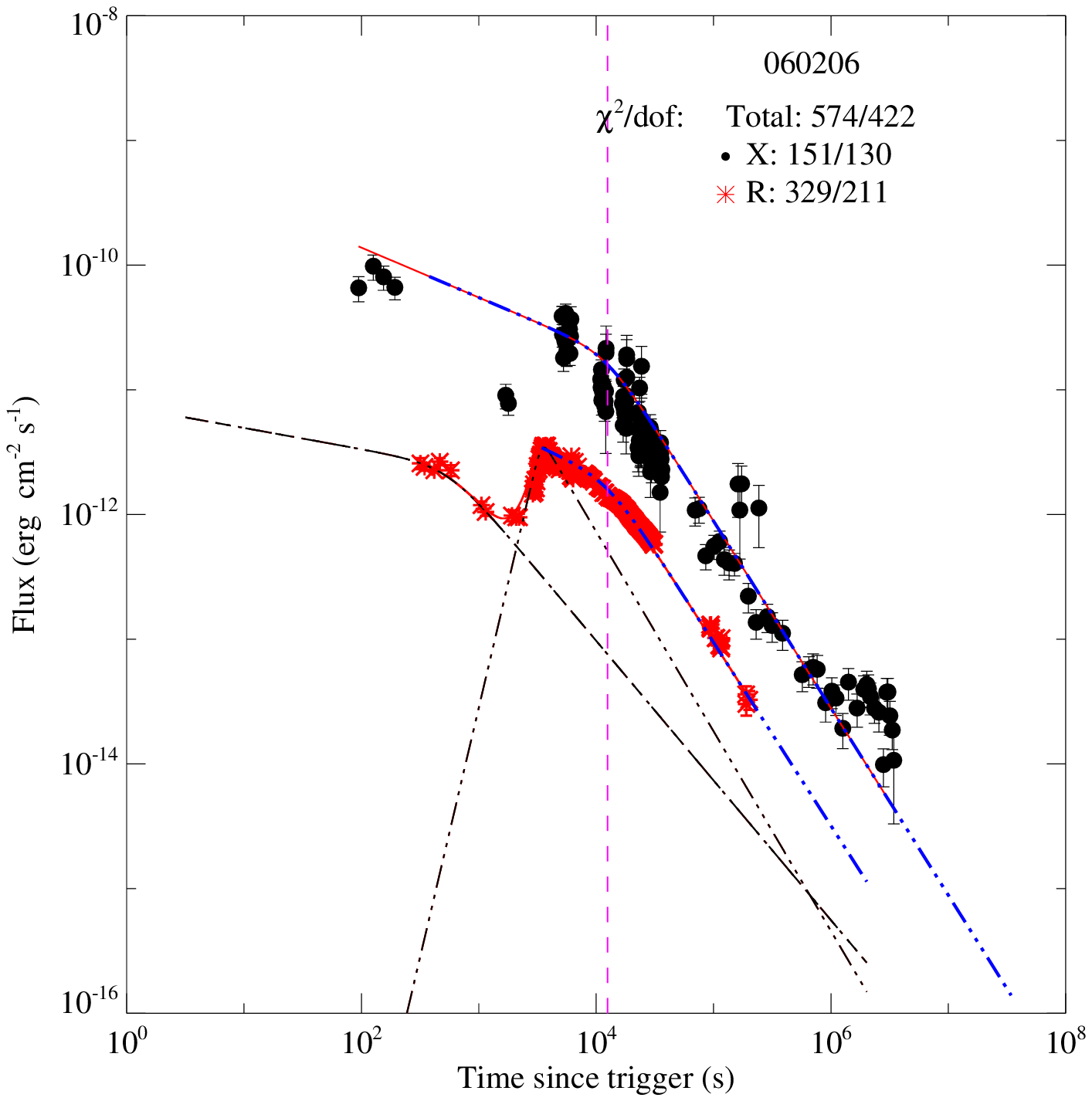}
\includegraphics[angle=0,scale=0.35,width=0.325\textwidth,height=0.30\textheight]{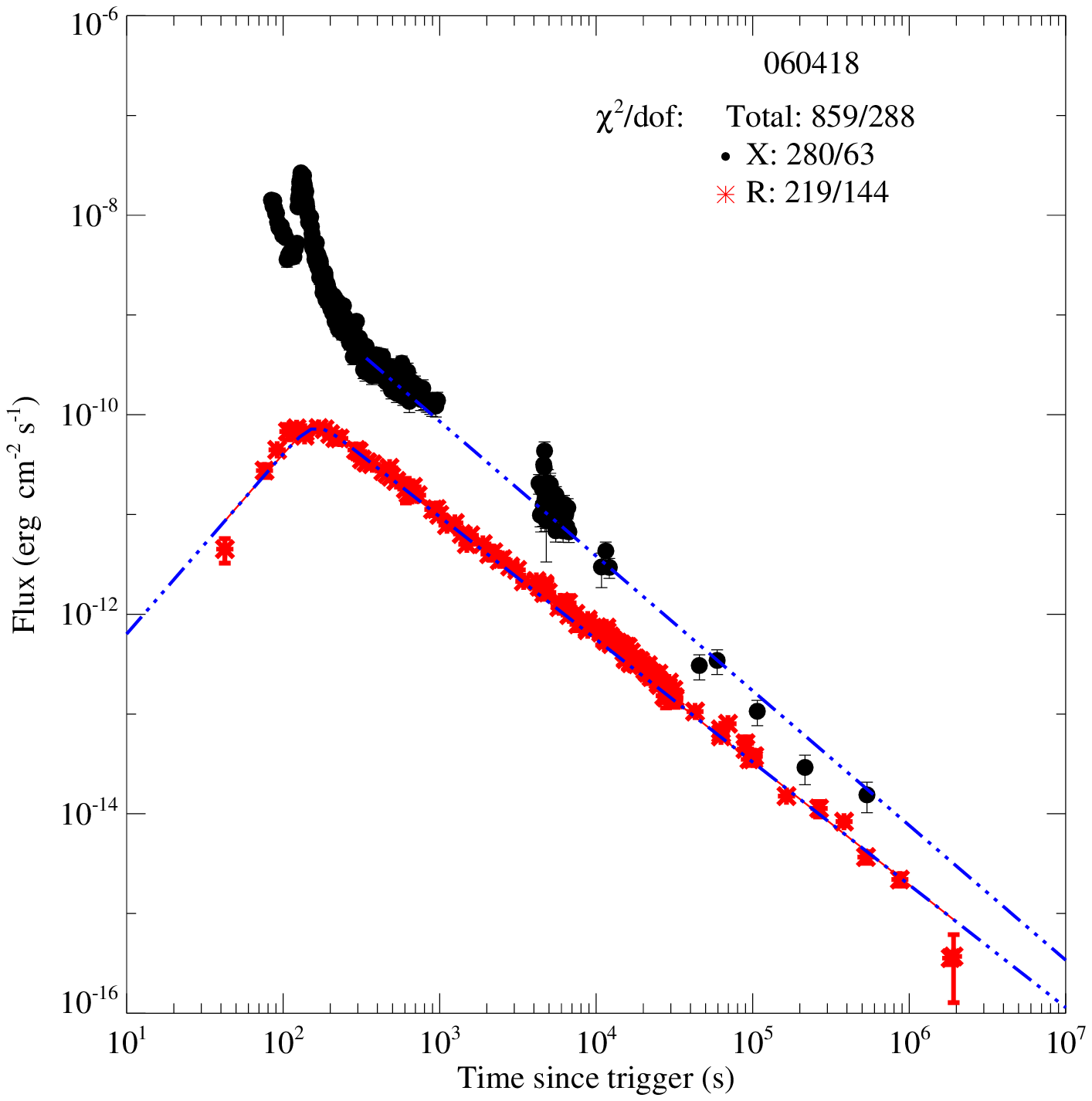}
\caption{The X-ray and optical light curves, as well as the fitting
results (blue dotted-dashed lines) for the GRBs in the Grade I sample. If an achromatic
break exists, the achromatic break time is shown by a purple vertical dashed
line.} \label{gradeI}
\end{figure*}

\begin{figure*}

\includegraphics[angle=0,scale=0.35,width=0.325\textwidth,height=0.30\textheight]{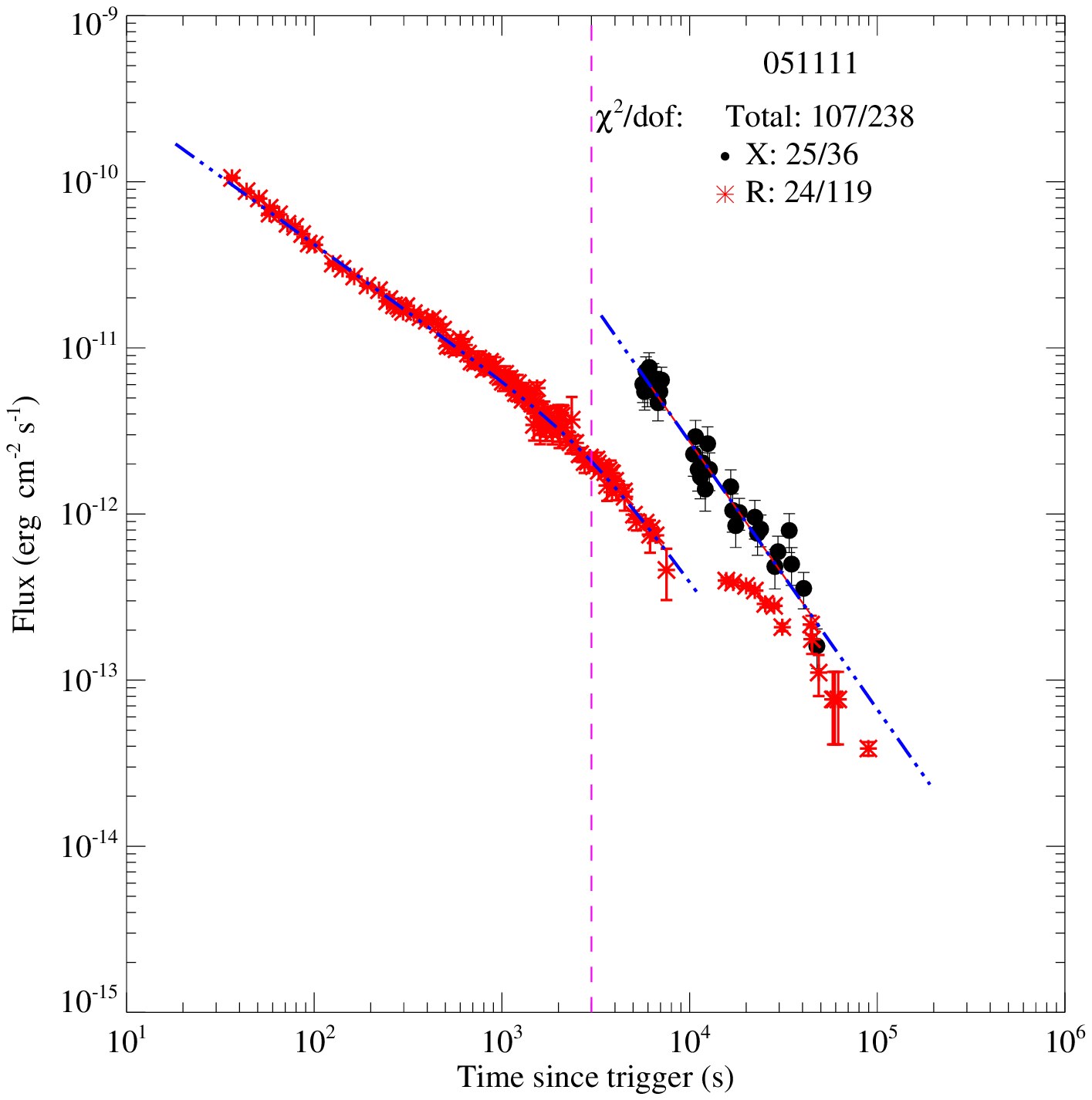}
\includegraphics[angle=0,scale=0.35,width=0.325\textwidth,height=0.30\textheight]{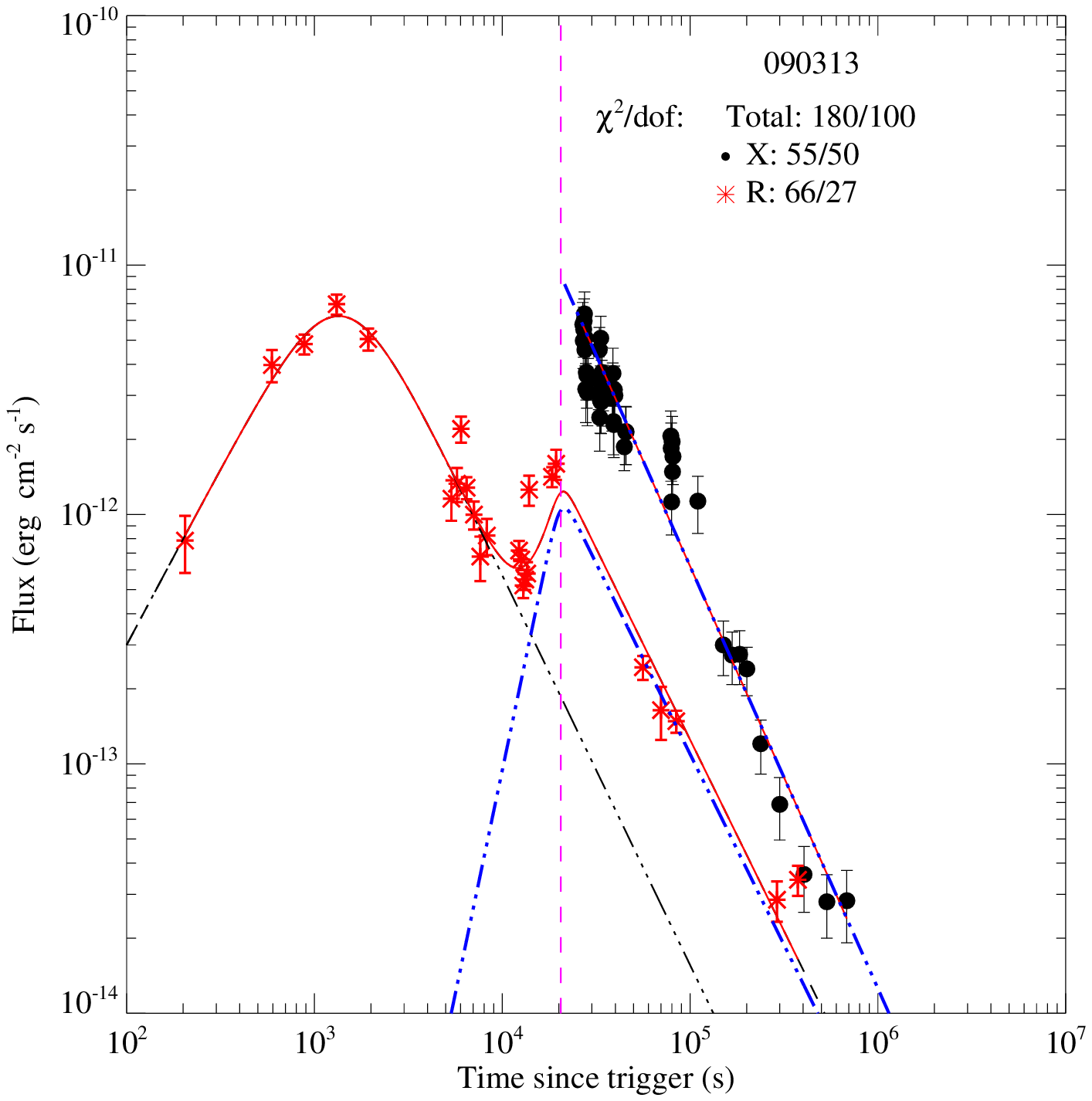}

\caption{Same as Figure \ref{gradeI}, but for the Grade II sample}
\label{gradeII}
\end{figure*}

\begin{figure*}

\includegraphics[angle=0,scale=0.35,width=0.325\textwidth,height=0.30\textheight]{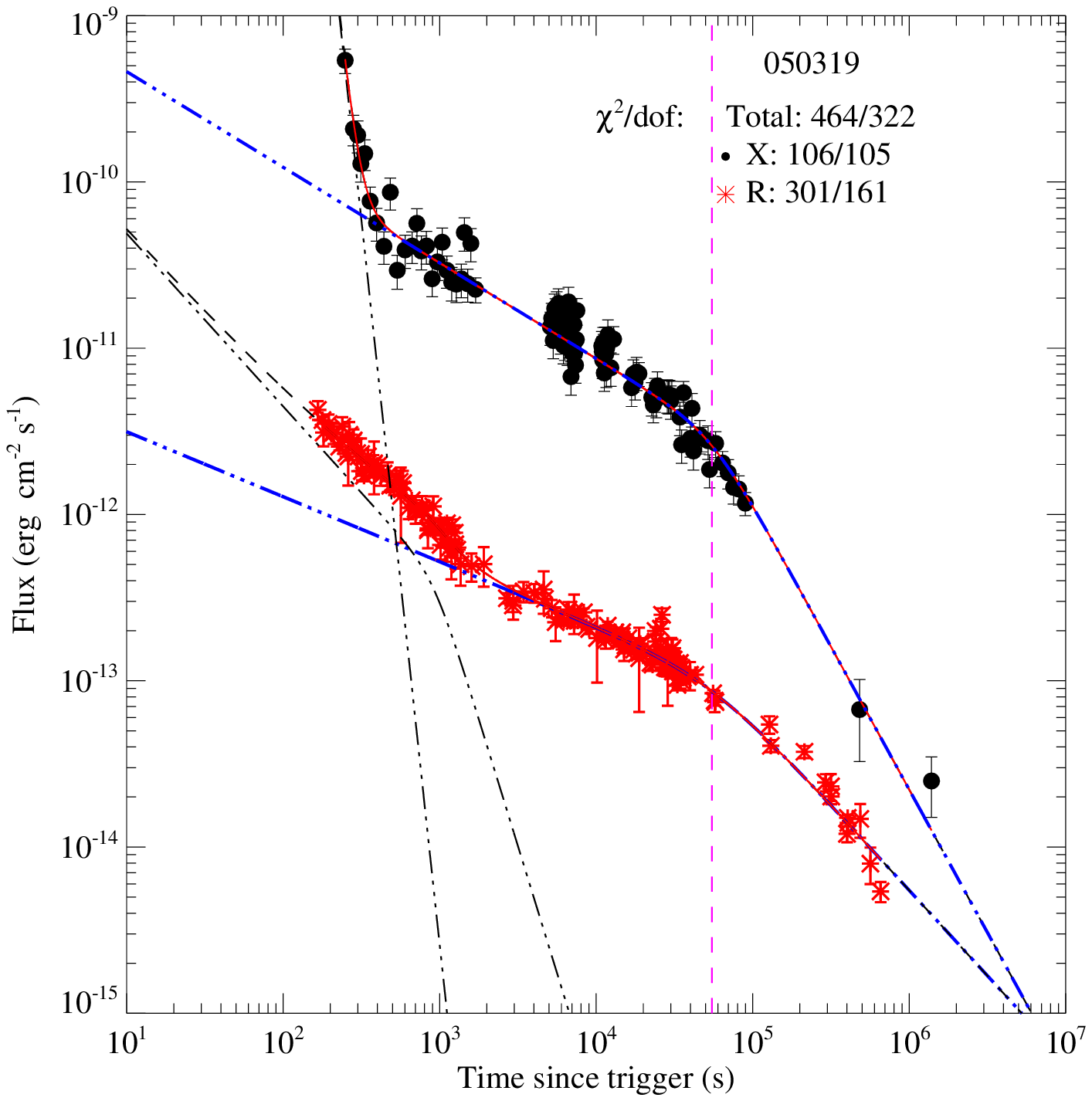}
\includegraphics[angle=0,scale=0.35,width=0.325\textwidth,height=0.30\textheight]{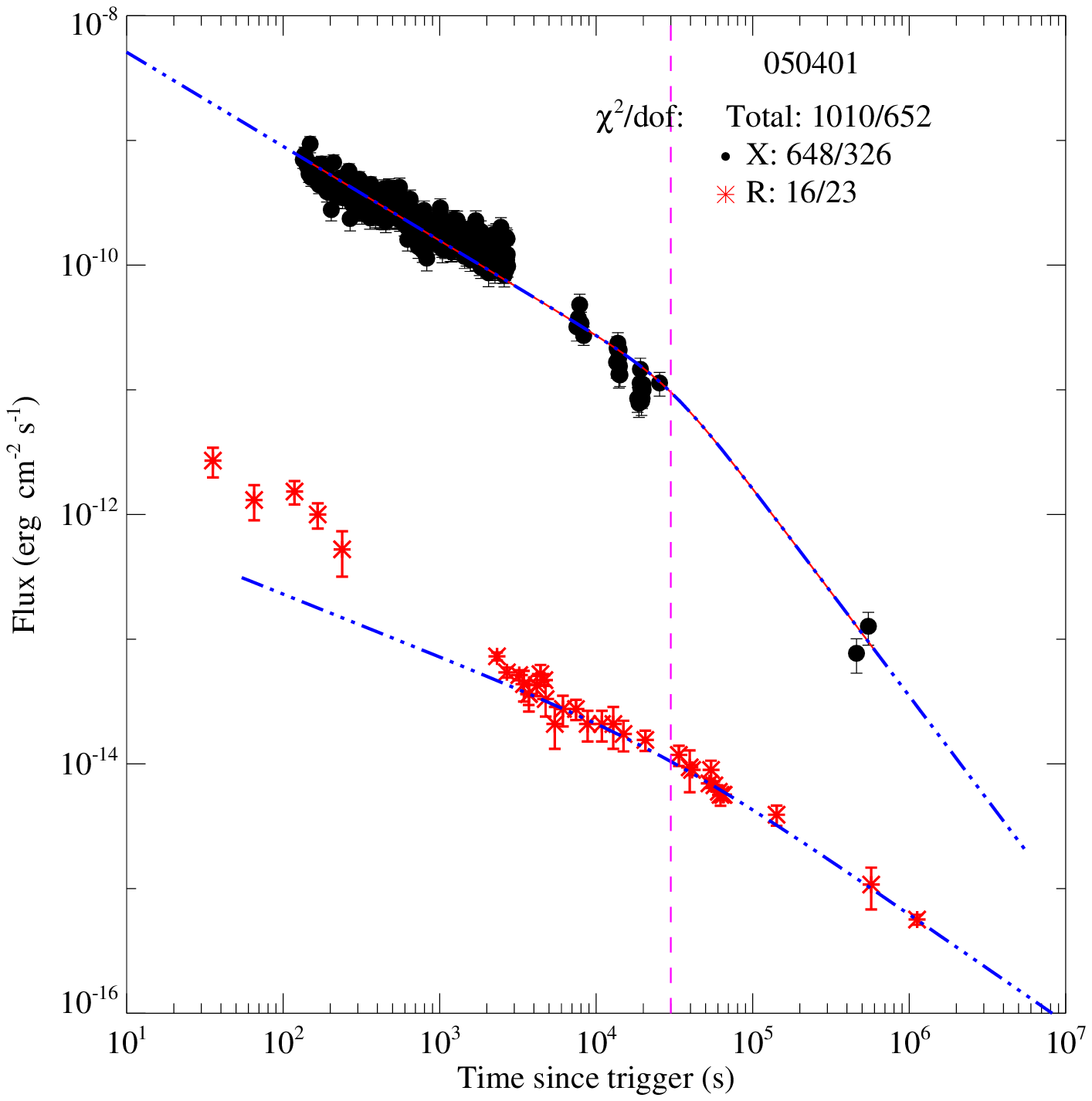}
\includegraphics[angle=0,scale=0.35,width=0.325\textwidth,height=0.30\textheight]{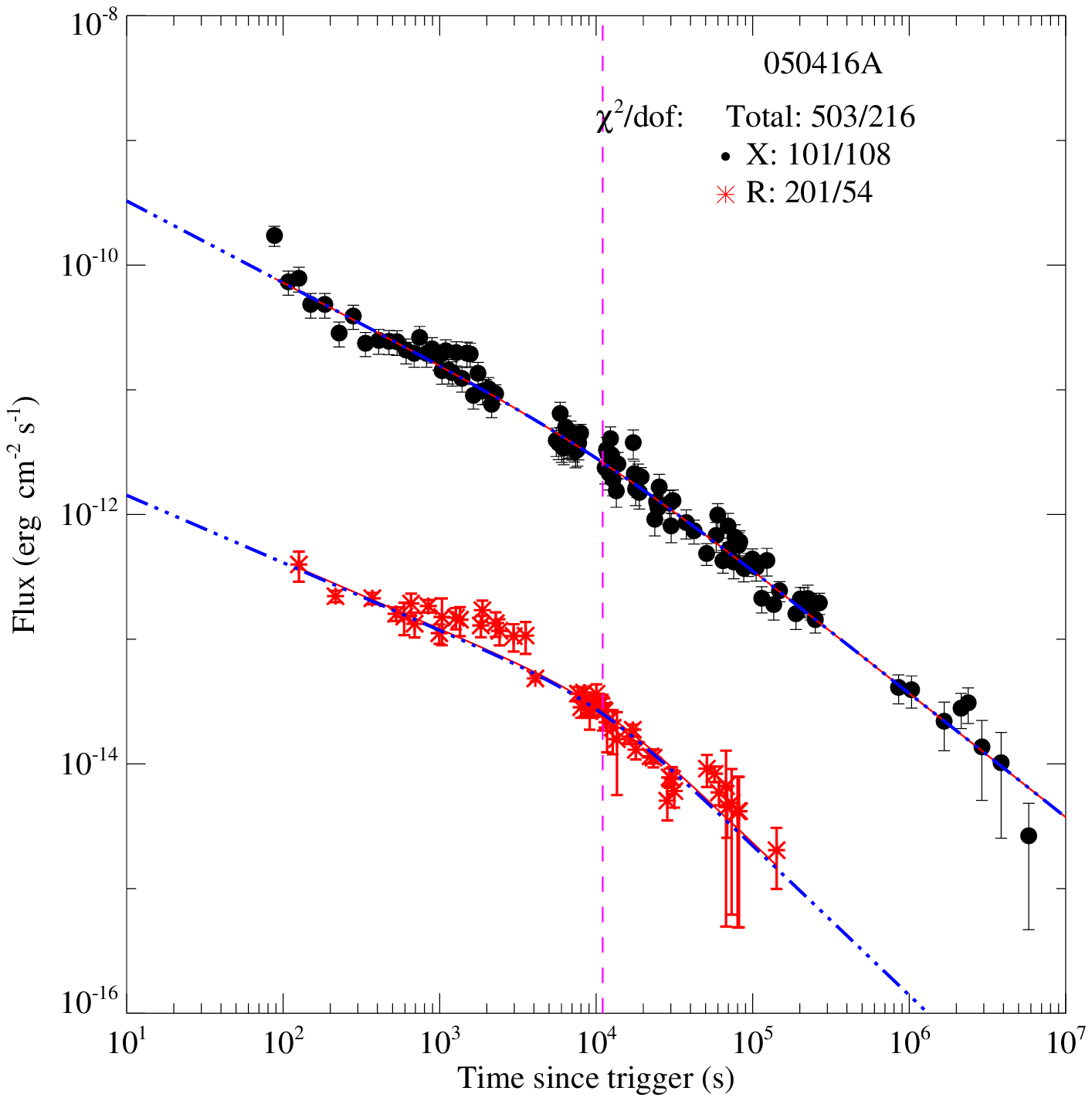}
\includegraphics[angle=0,scale=0.35,width=0.325\textwidth,height=0.30\textheight]{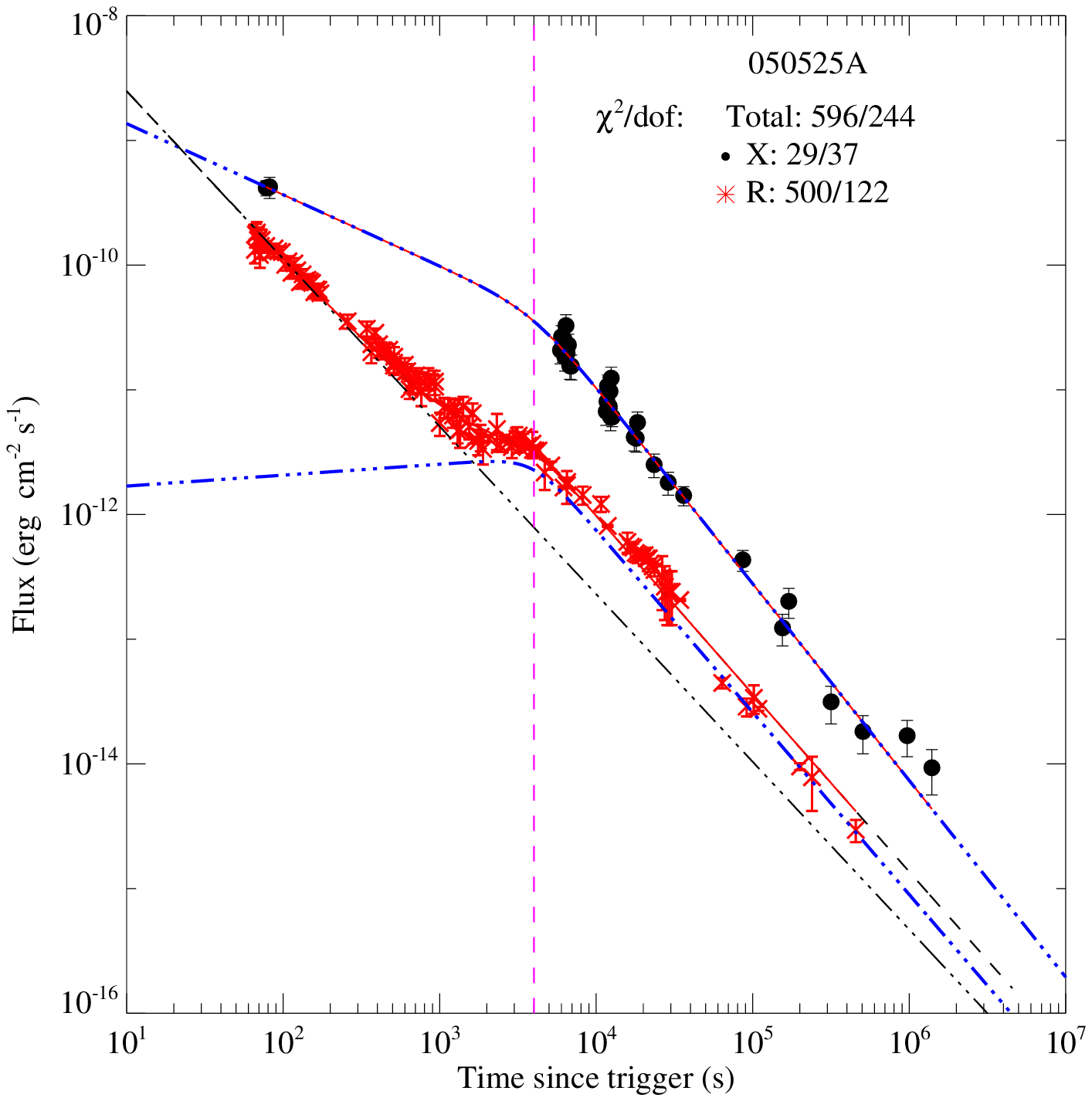}
\includegraphics[angle=0,scale=0.35,width=0.325\textwidth,height=0.30\textheight]{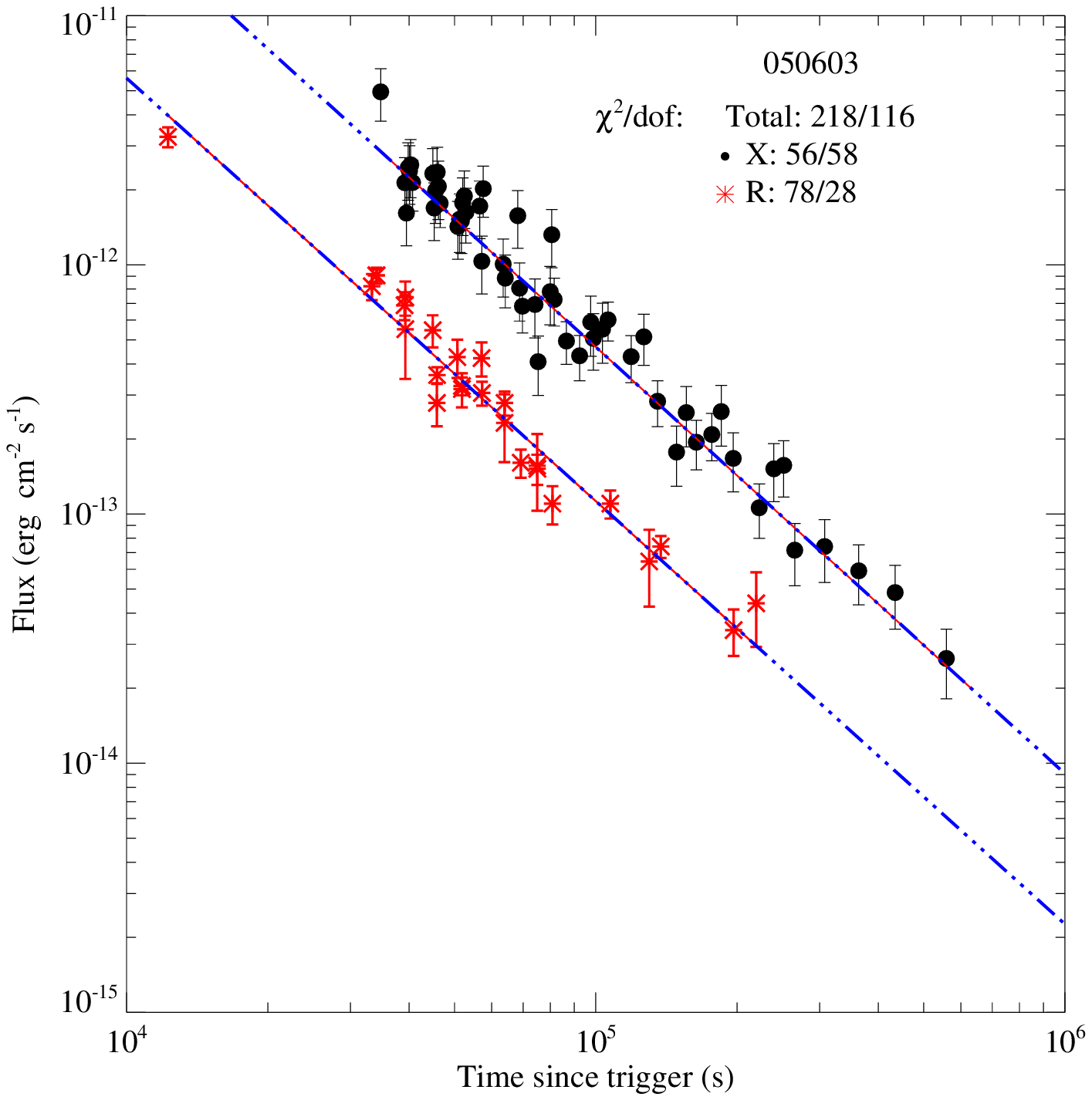}
\includegraphics[angle=0,scale=0.35,width=0.325\textwidth,height=0.30\textheight]{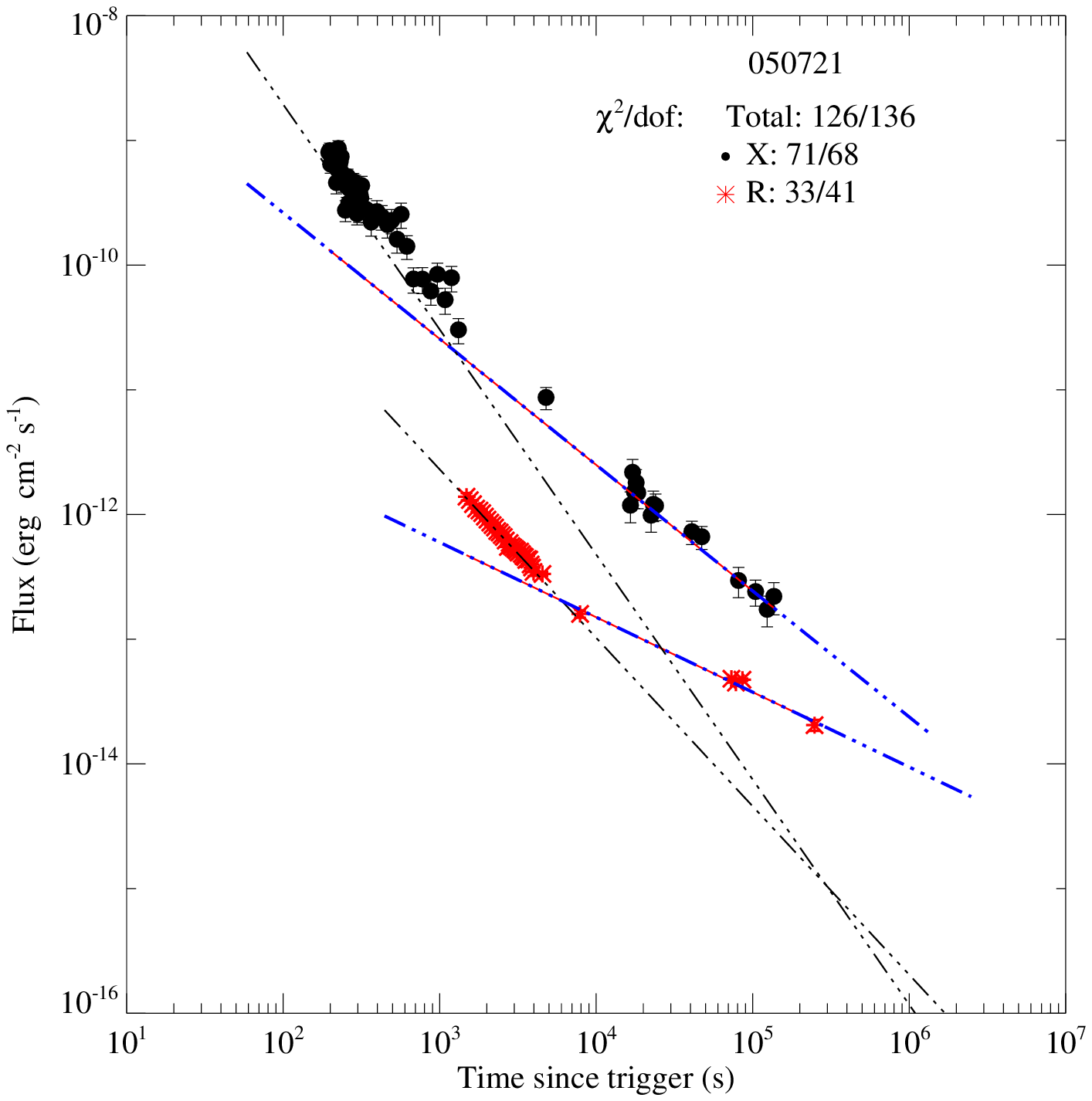}
\includegraphics[angle=0,scale=0.35,width=0.325\textwidth,height=0.30\textheight]{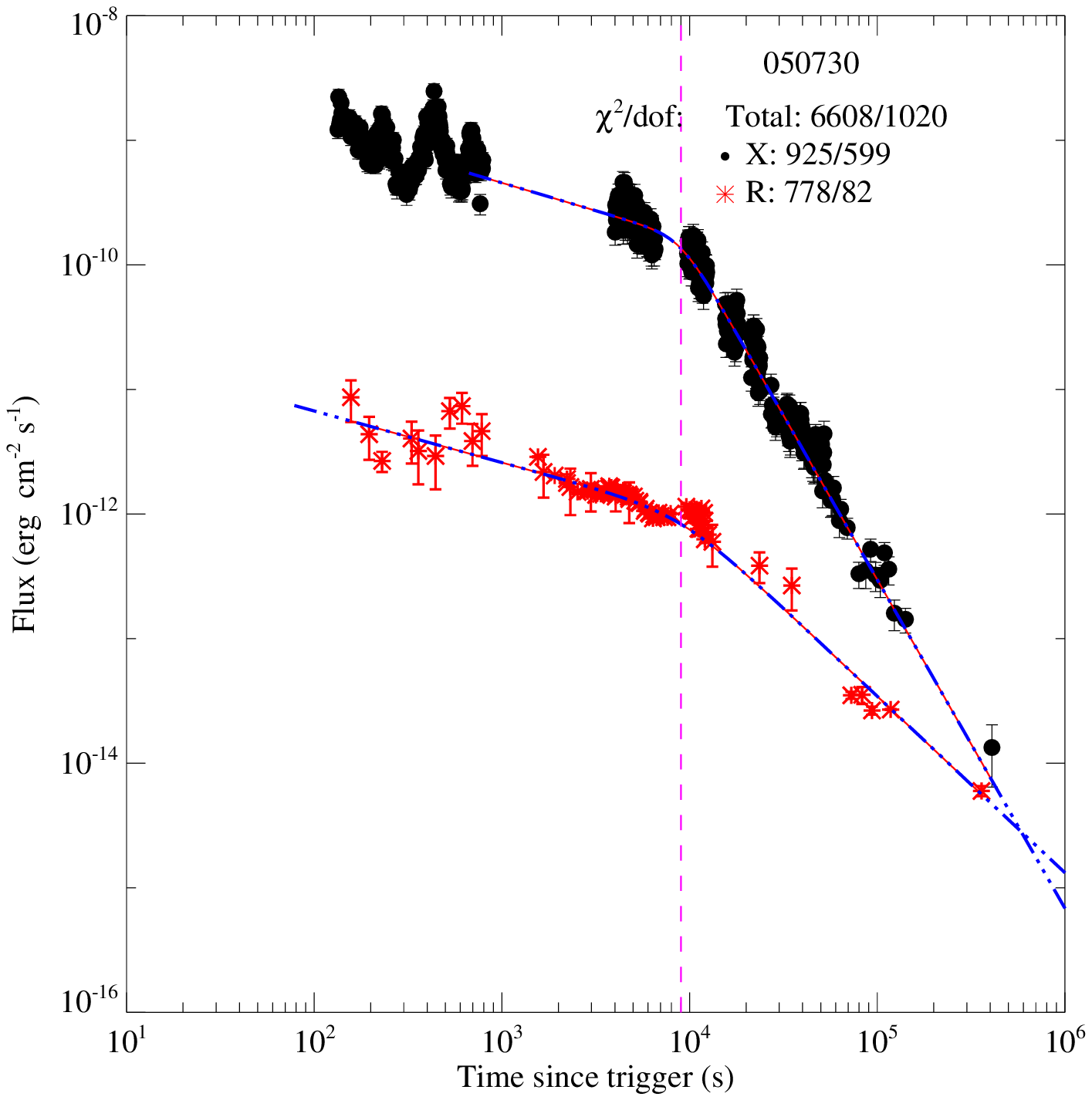}
\includegraphics[angle=0,scale=0.35,width=0.325\textwidth,height=0.30\textheight]{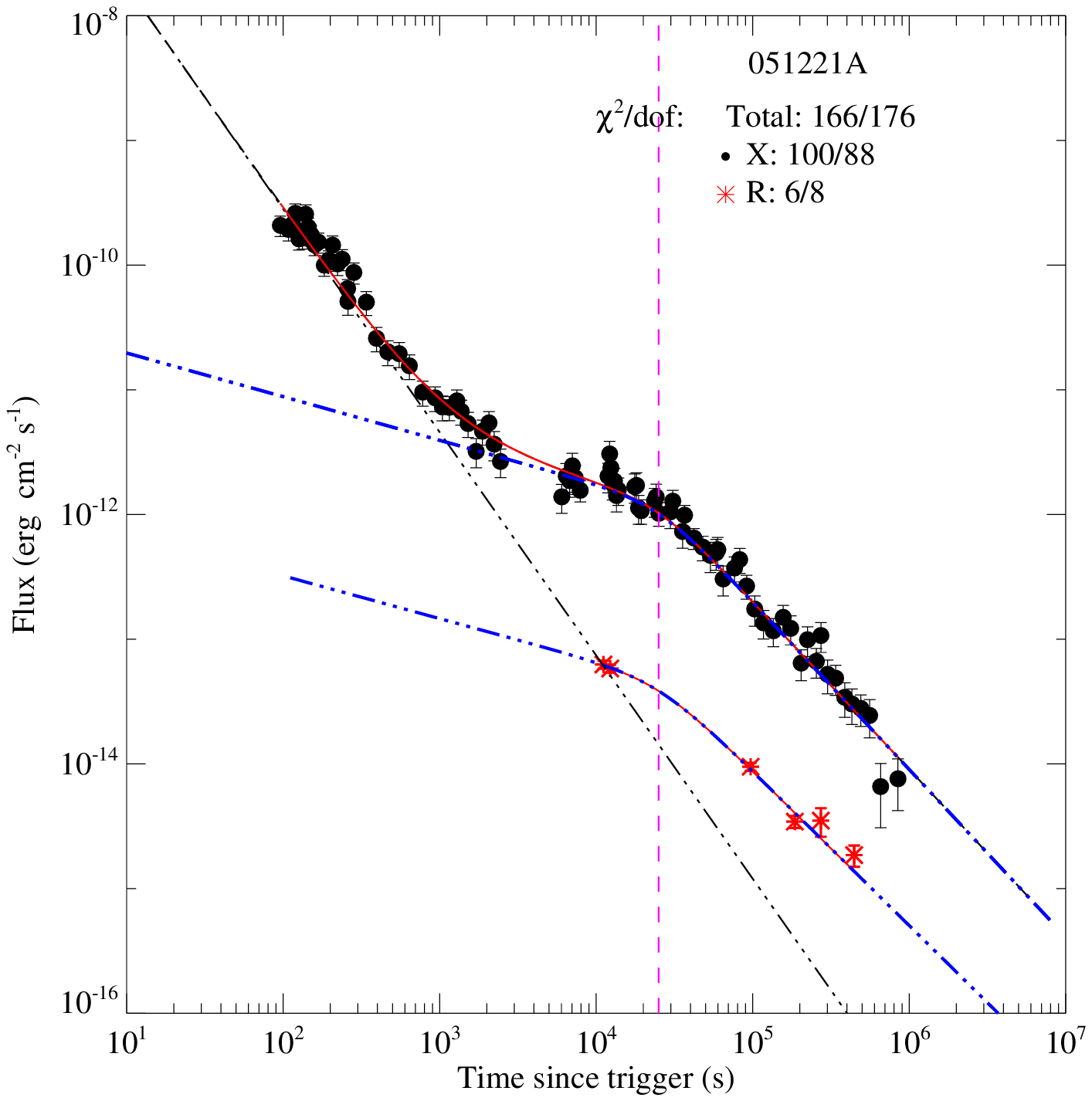}
\includegraphics[angle=0,scale=0.35,width=0.325\textwidth,height=0.30\textheight]{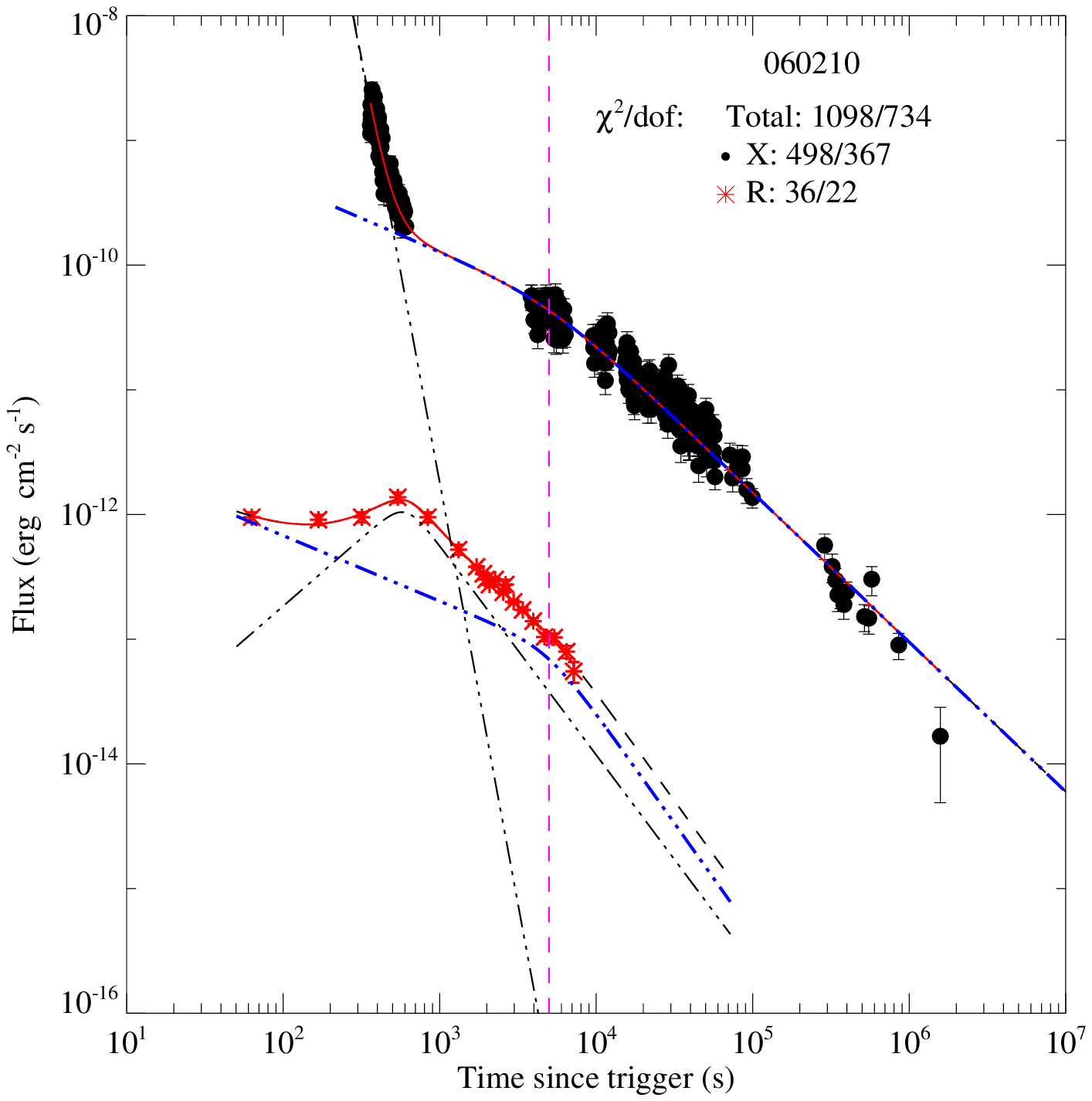}
\caption{Same as Figure \ref{gradeI}, but for the Grade III sample}
\label{gradeIII}
\end{figure*}

\begin{figure*}
\includegraphics[angle=0,scale=0.35,width=0.325\textwidth,height=0.30\textheight]{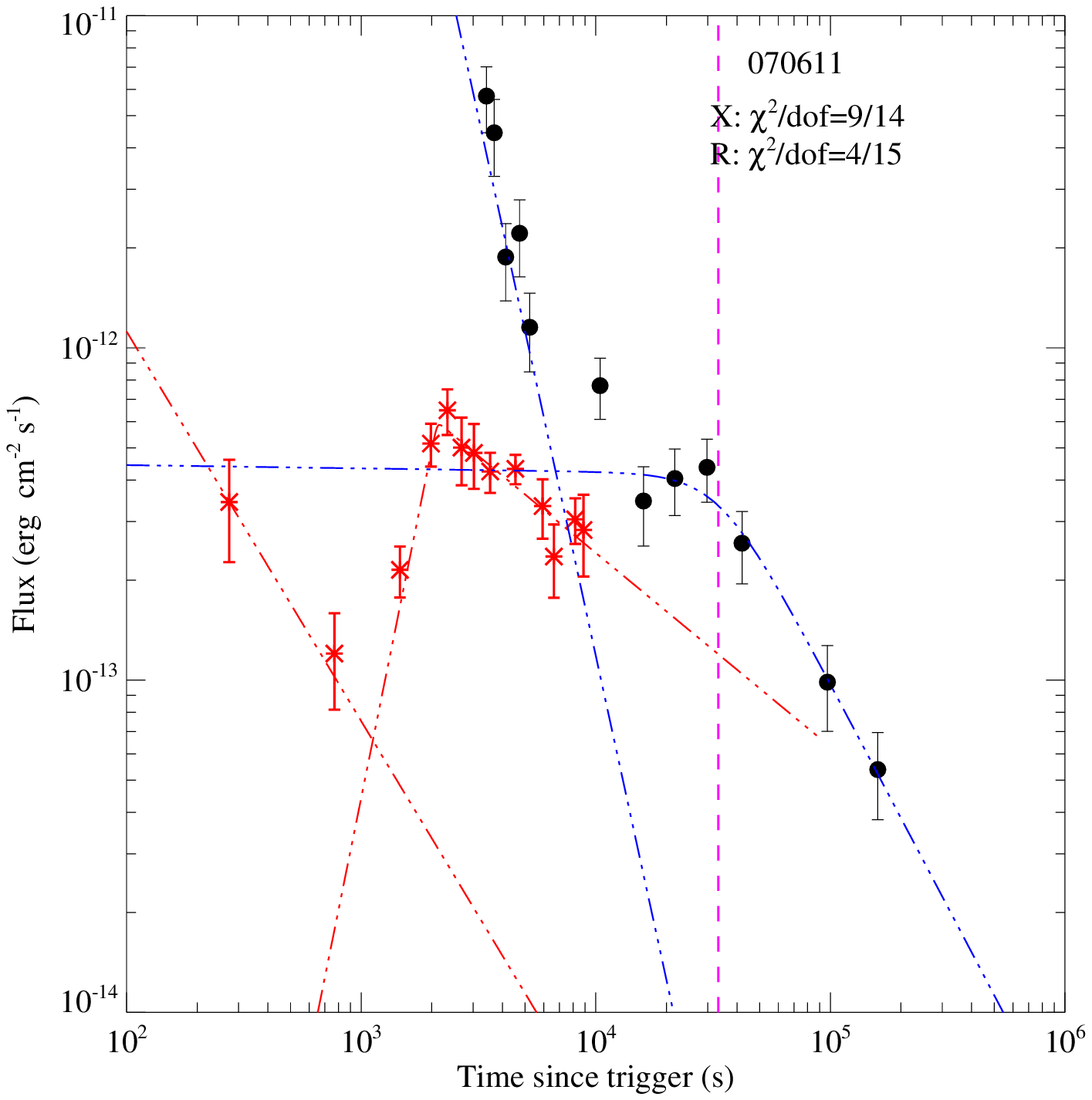}
\includegraphics[angle=0,scale=0.35,width=0.325\textwidth,height=0.30\textheight]{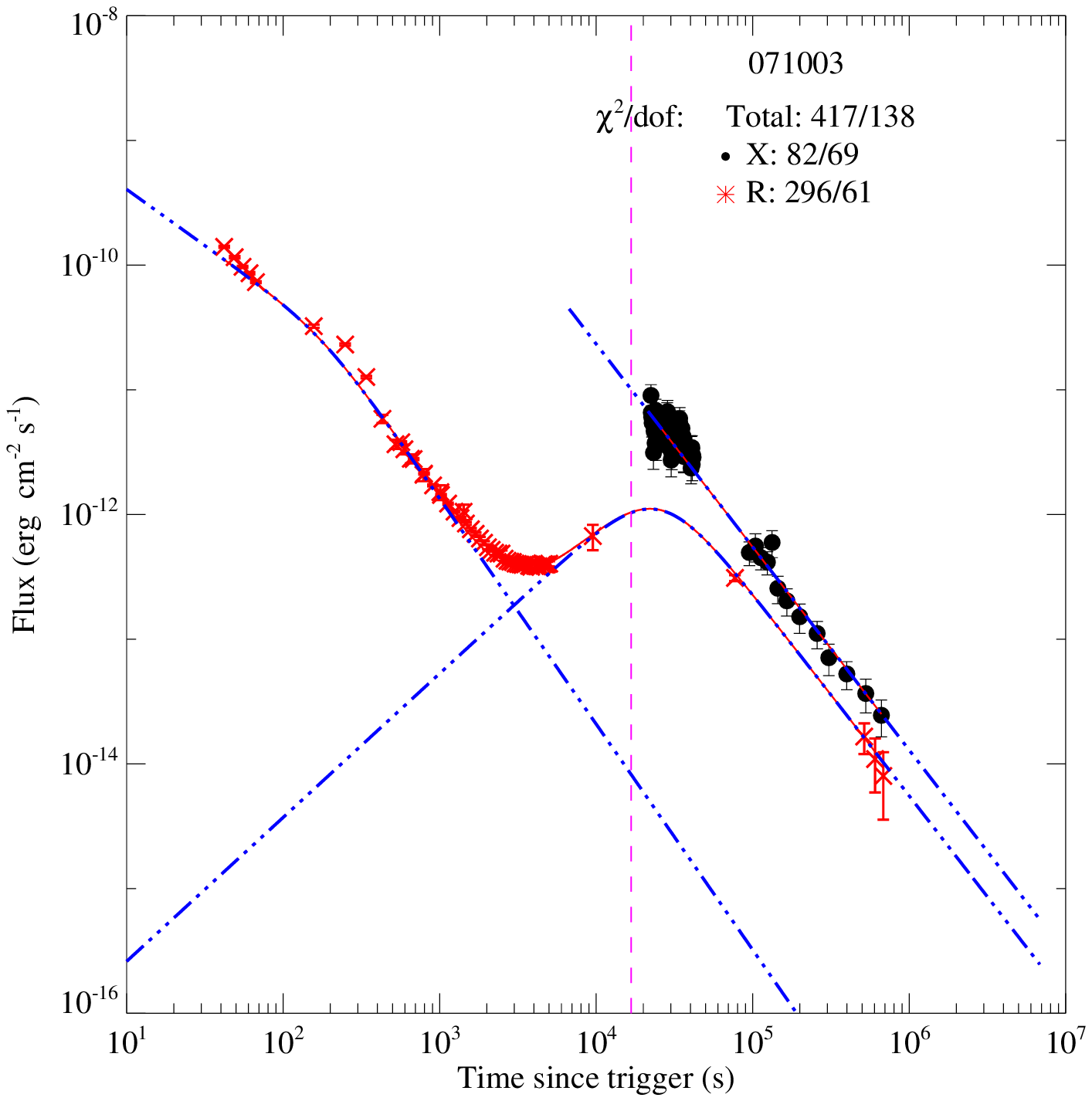}
\includegraphics[angle=0,scale=0.35,width=0.325\textwidth,height=0.30\textheight]{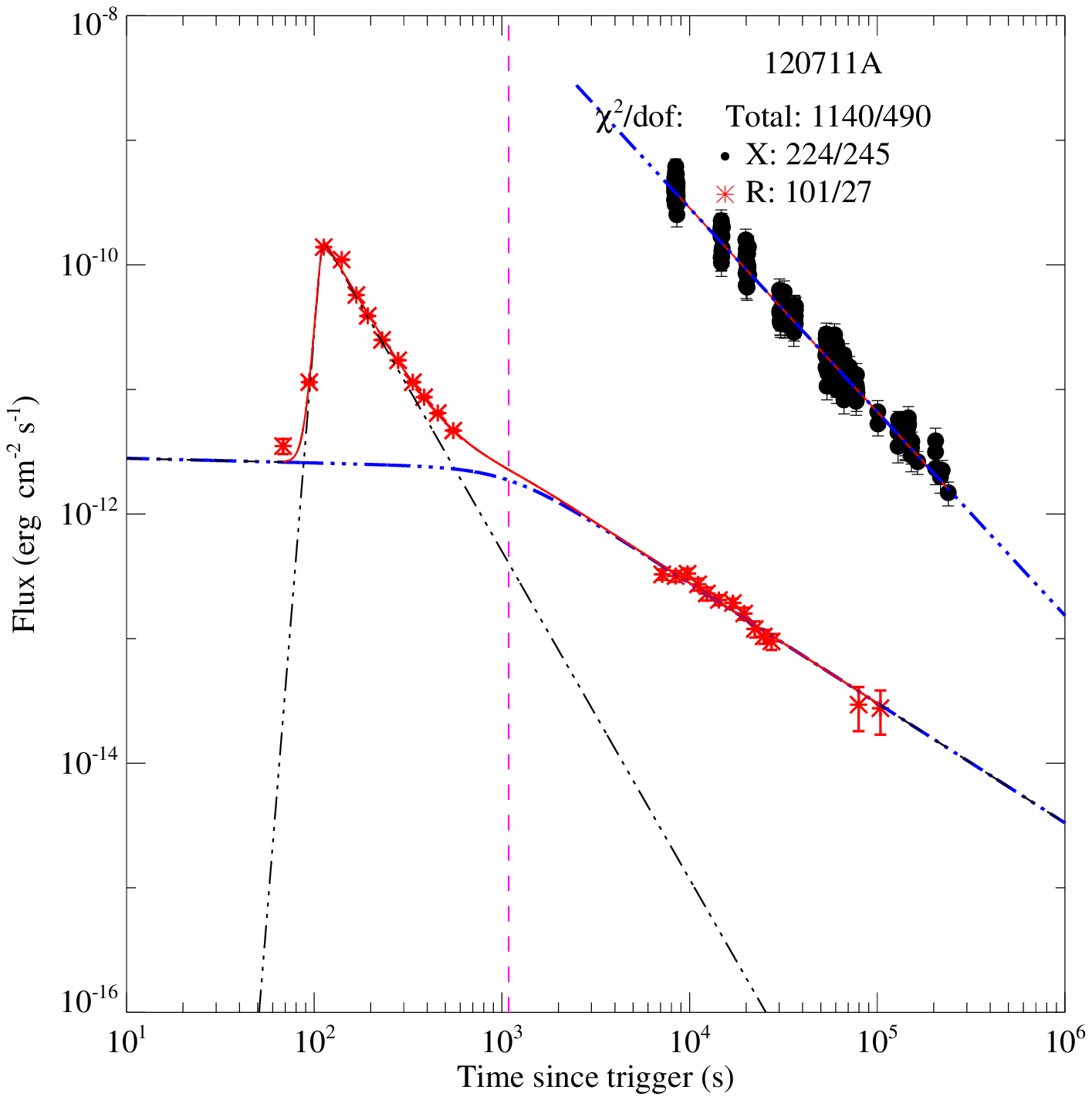}

\caption{Same as Figure \ref{gradeI}, but for the Grade IV sample}
\label{gradeIV}
\end{figure*}

\begin{figure*}
\includegraphics[angle=0,scale=0.35,width=0.325\textwidth,height=0.30\textheight]{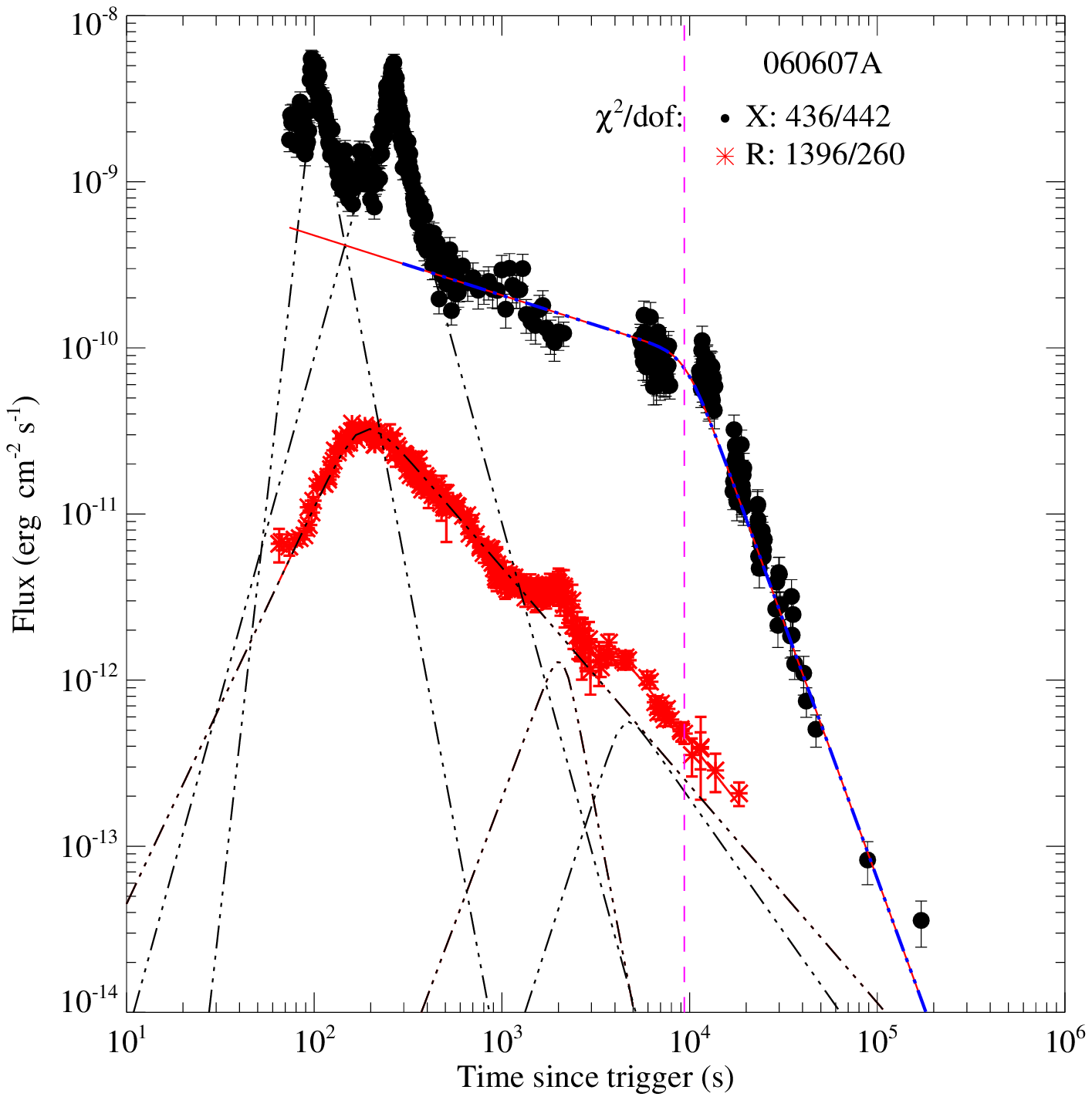}
\includegraphics[angle=0,scale=0.35,width=0.325\textwidth,height=0.30\textheight]{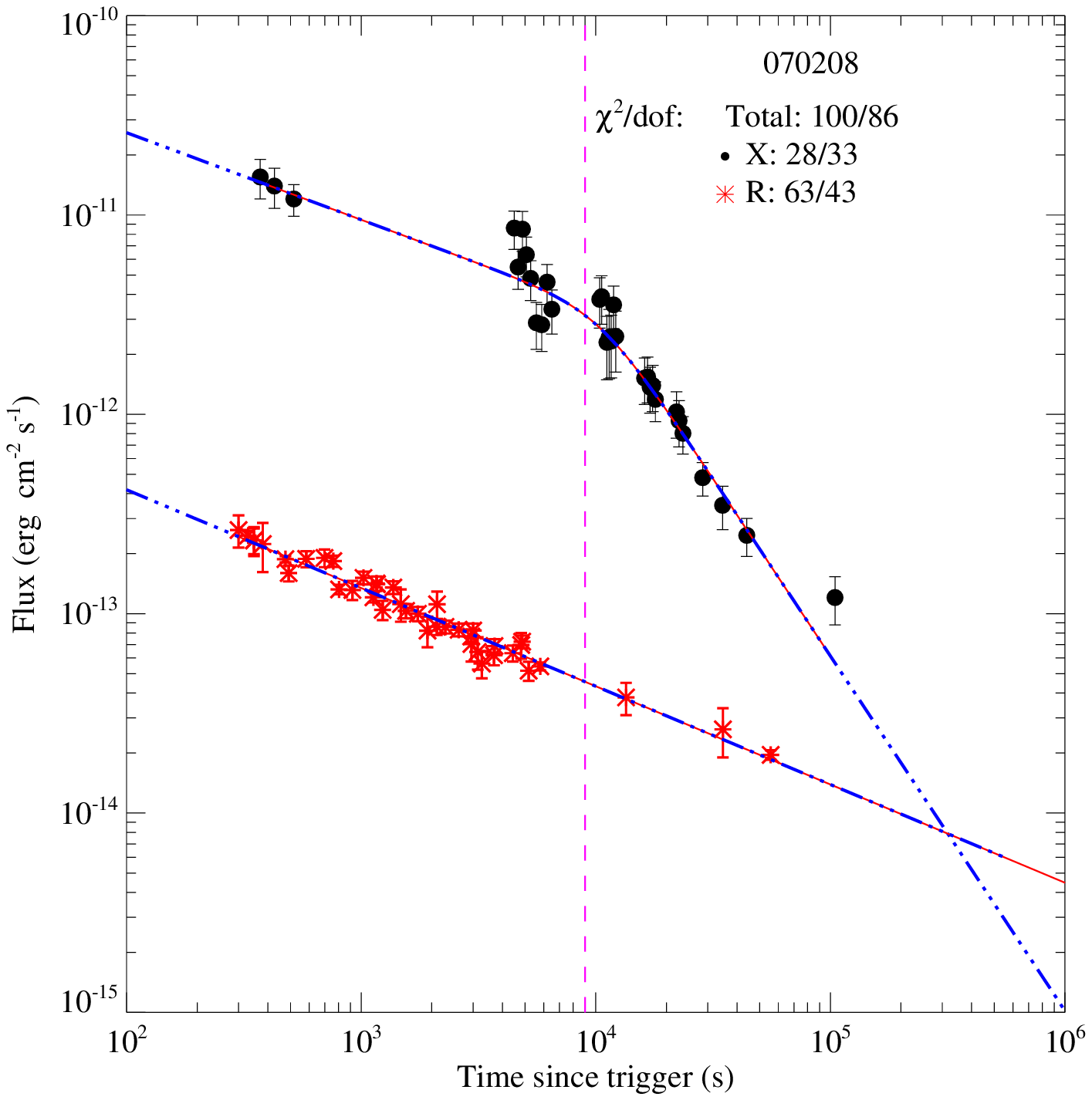}
\includegraphics[angle=0,scale=0.35,width=0.325\textwidth,height=0.30\textheight]{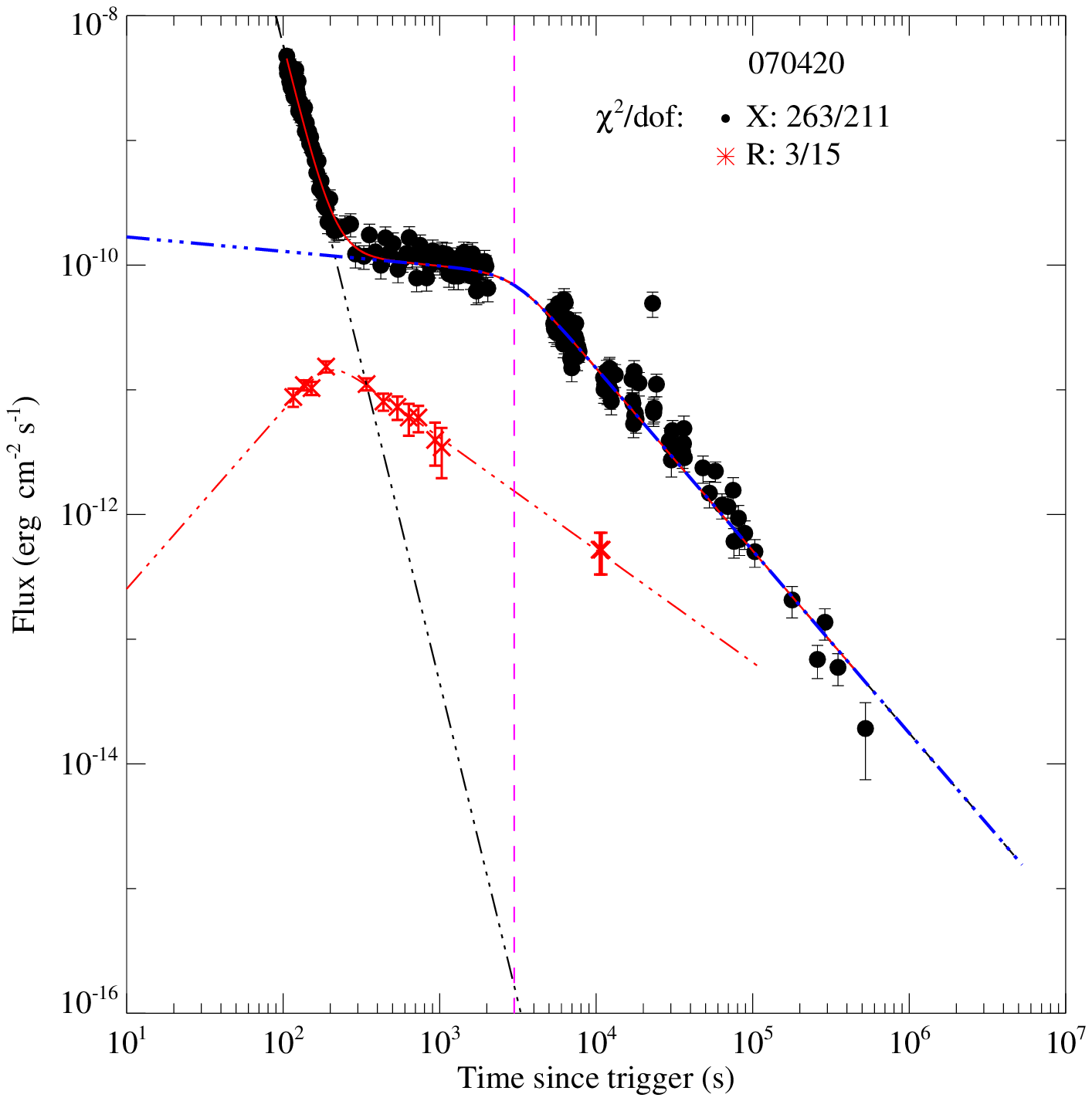}
\caption{Same as Figure \ref{gradeI}, but for the Grade V sample}
\label{gradeV}
\end{figure*}

The 85 well-sampled GRBs in our sample are graded based on the above-defined
grading criteria. The GRBs in the five grades are presented in Figures
\ref{gradeI} - \ref{gradeV}, respectively. The relevant data of different
grades are presented in Table \ref{table:sample} and \ref{table:parameter}.

\begin{itemize}
  \item Grade I: As can be seen from Table \ref{table:sample} and \ref{table:parameter},
and Figure \ref{gradeI}, within errors 43/85 GRBs satisfy the Grade I
criteria. Out of 43 GRBs, 13, 8 and 22 GRBs are constrained to have an energy
injection break, jet break and SPL decay,
respectively.

  \item Grade II: within error 2/85 GRBs fall into this grade (Fig.\ref{gradeII}).
  \item Grade III: there are 34/85 GRBs falling into this grade (Fig.\ref{gradeIII}).
Among the sample, 15/34 and 19/34 GRBs have SPL and BPL lightcurves,
respectively.  GRBs 060906, 080319B and 100219A have
two beaks at different times, respectively.
  \item Grade IV: there are 3/85 GRBs falling into this grade (Fig.\ref{gradeIV}).
  \item Grade V: there are 3/85 GRBs falling into this grade (Fig.\ref{gradeV}).
Two of them (GRBs 060607A and 070208) show clear chromatic
breaks with good temporal coverage in both bands
at the break times.  One GRB (GRB 070420) shows a chromatic behavior
based on the available data and simple model fitting, even though no observational
data are available in the optical band at the break time of the X-ray band, so that
the existence of a break in the optical band (even though very contrived in shape)
at the same epoch is not completely ruled out.

\end{itemize}

Consequently, we get three samples:
\begin{itemize}
  \item Gold sample: This sample has 45/85 GRBs, including 13/49, 8/49 and 24/49 GRBs satisfying the energy injection, jet break, jet break with energy injection,
and SPL decay models, respectively. Among them, 27/49 and 18/49 are consistent with the ISM
and wind models, respectively; 17/49, 4/49 and 24/49 GRBs are consistent with being in a same
spectral regime, different spectral regimes (X-ray band in regime I and optical band in
regime II), and grey zone, respectively.
Among the 17 GRBs with the same spectral regime, 15 and 2 GRBs are consistent with being in the
ISM II and wind II spectral regimes, respectively. For the 4 GRBs with different spectral regime,
all of them are consistent with having an ISM medium.
  \item Silver sample: This sample has 37/85 GRBs, which may (or may not) be interpreted
within the more complicated numerical external shock models.
  \item Bad sample: Only 3/85 GRBs definitely violate the basic achromaticity principle
of the external shock models and therefore belong to the bad sample.
 \end{itemize}

\begin{figure*}

\includegraphics[angle=0,scale=0.35]{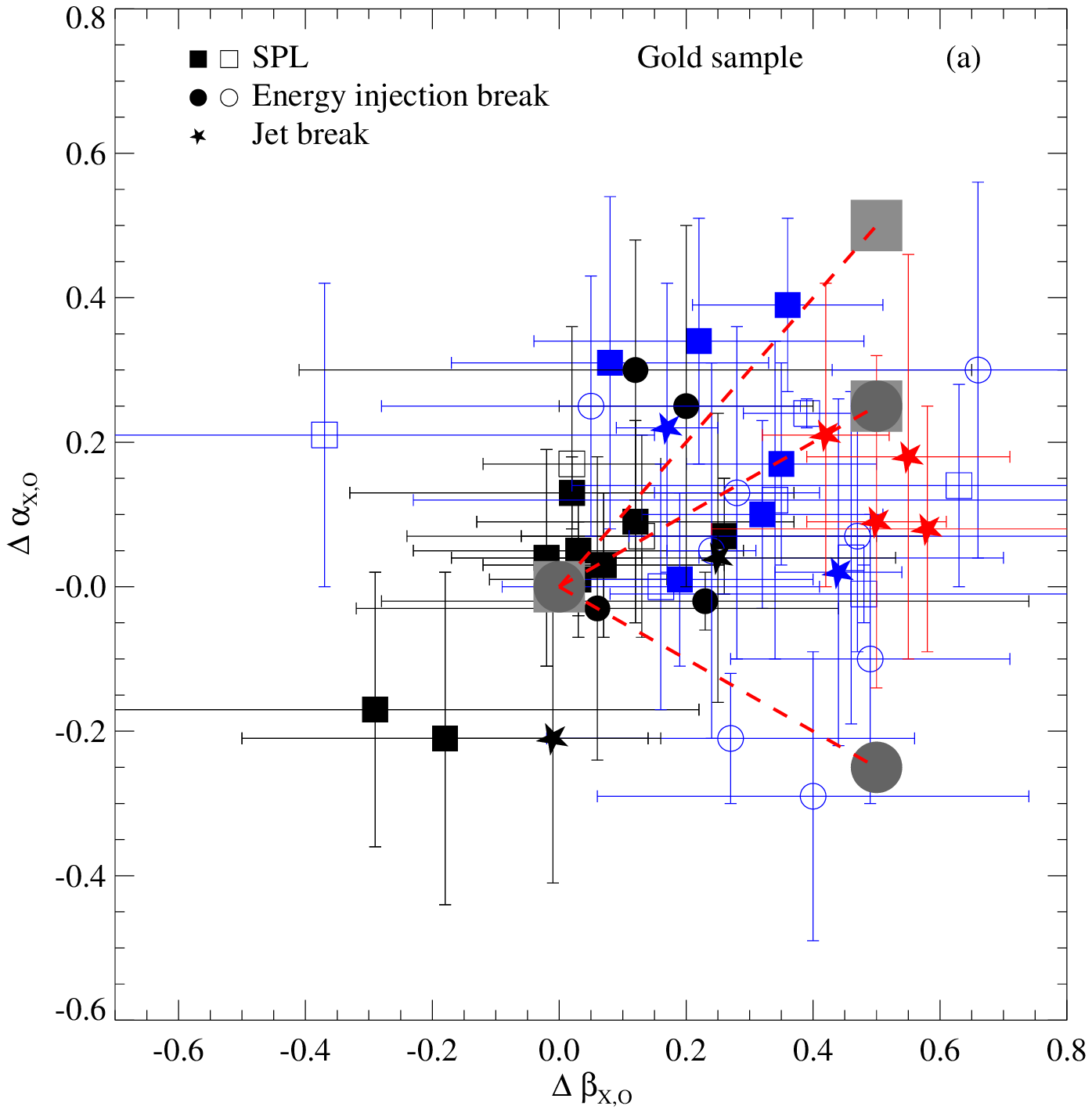}
\includegraphics[angle=0,scale=0.35]{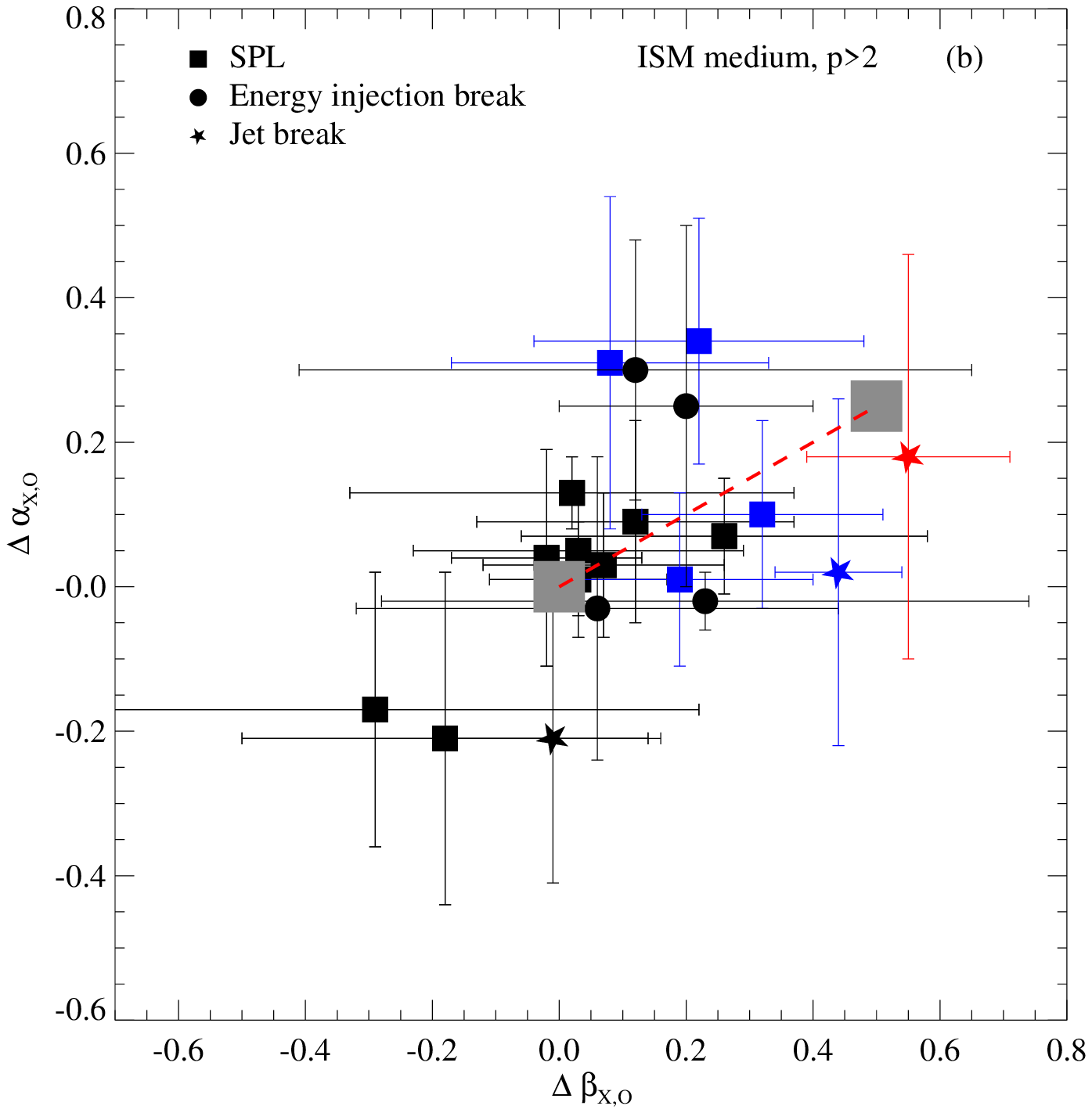}
\includegraphics[angle=0,scale=0.35]{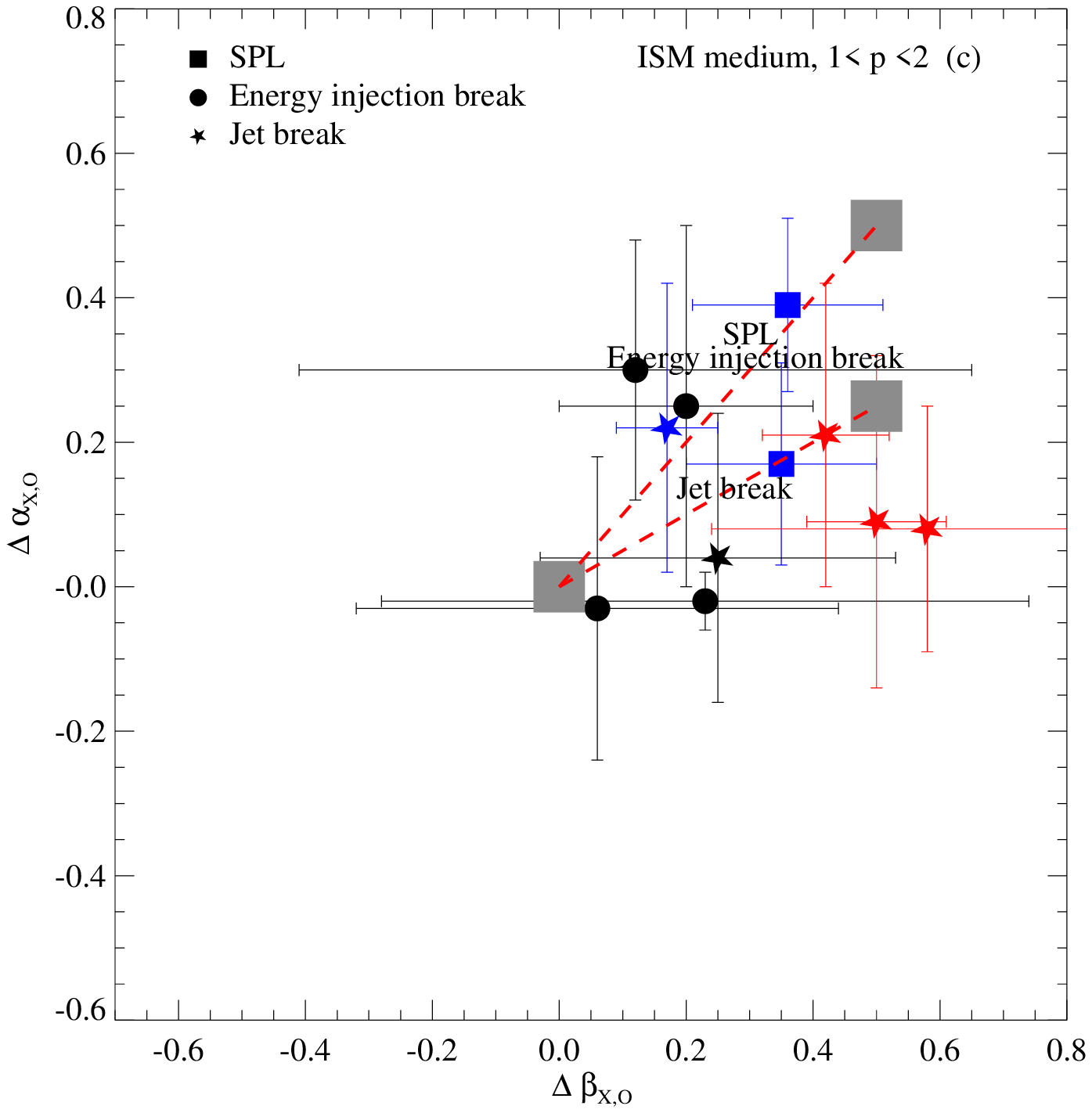}
\includegraphics[angle=0,scale=0.35]{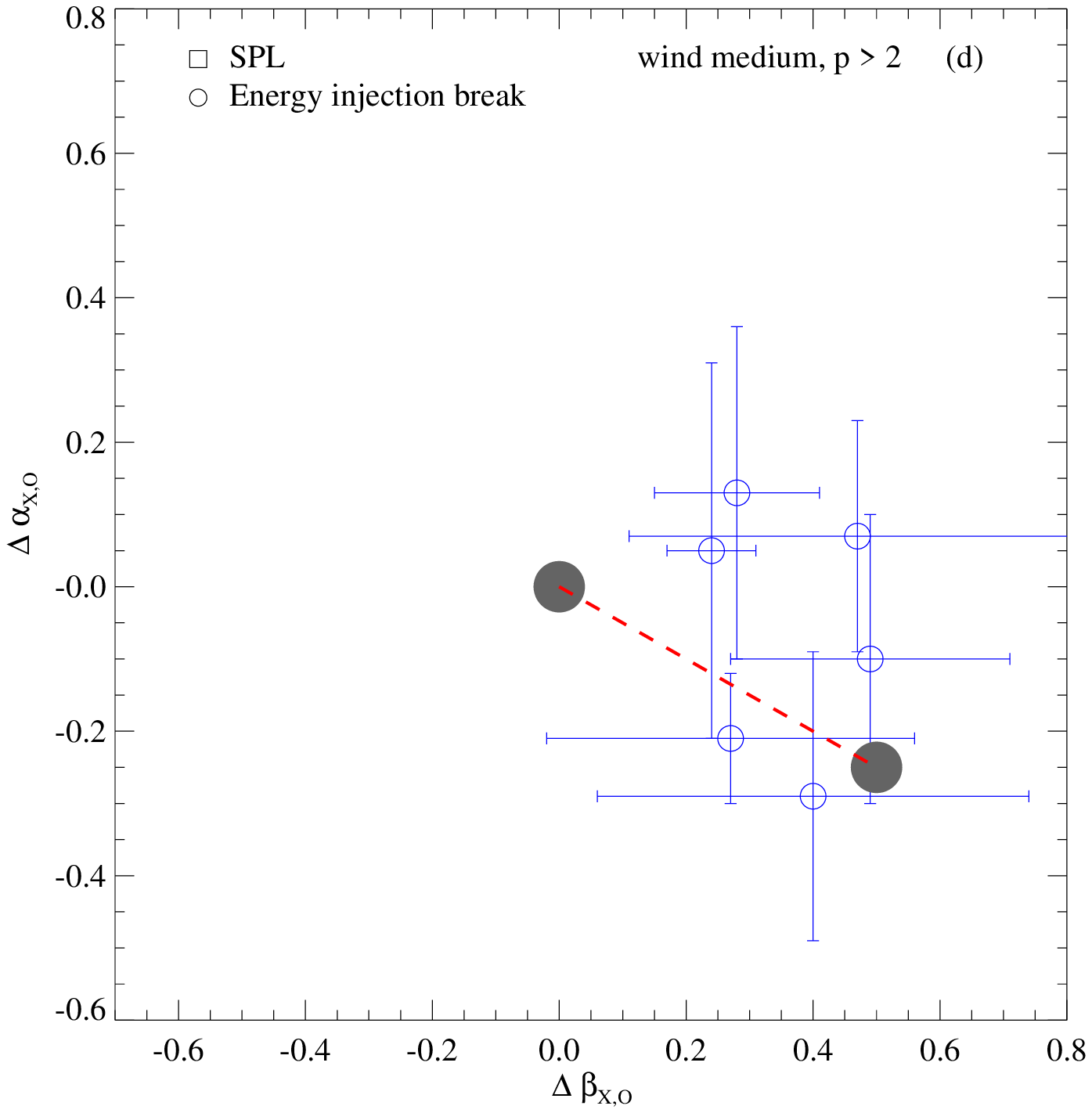}
\includegraphics[angle=0,scale=0.35]{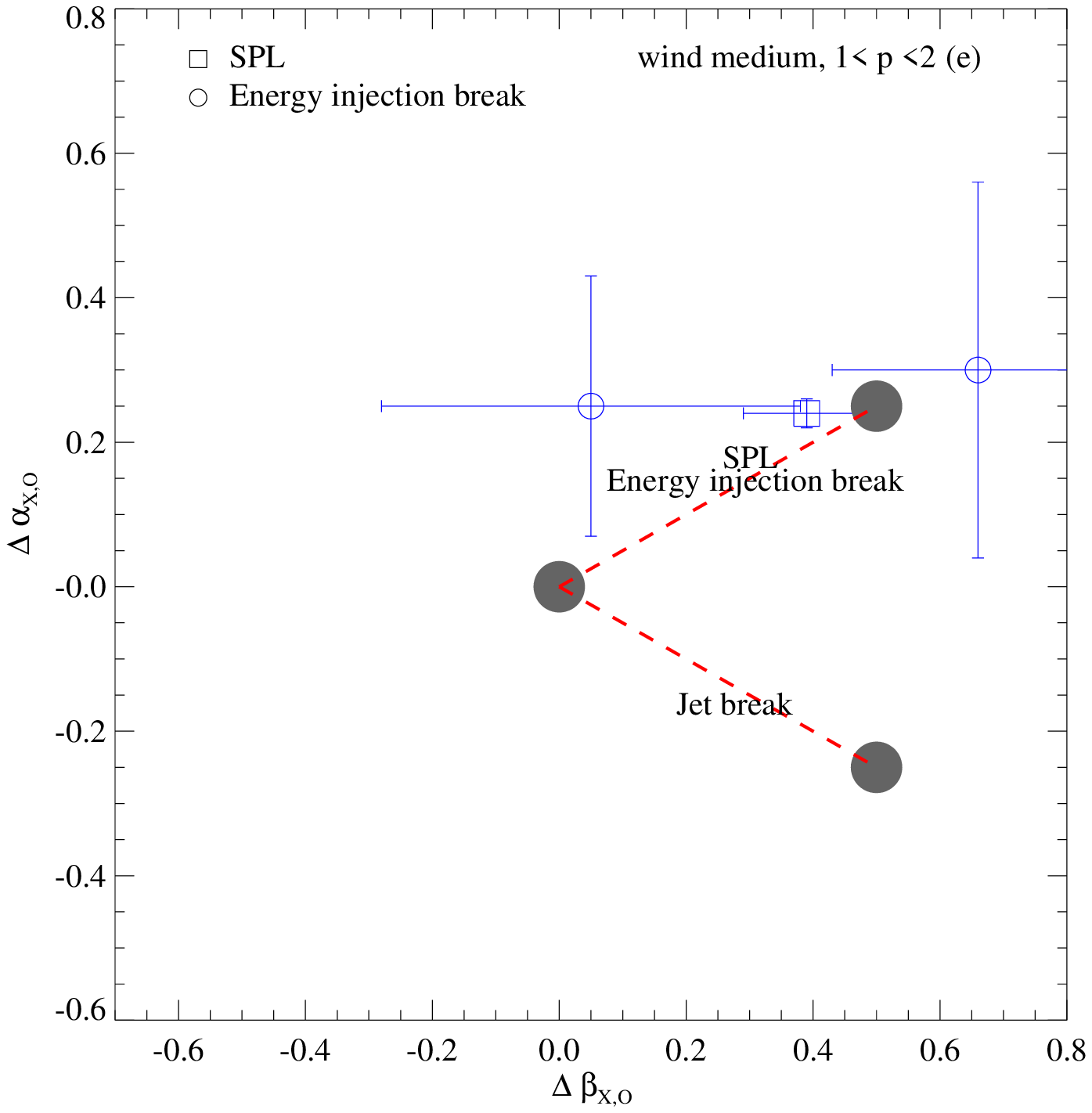}
\includegraphics[angle=0,scale=0.35]{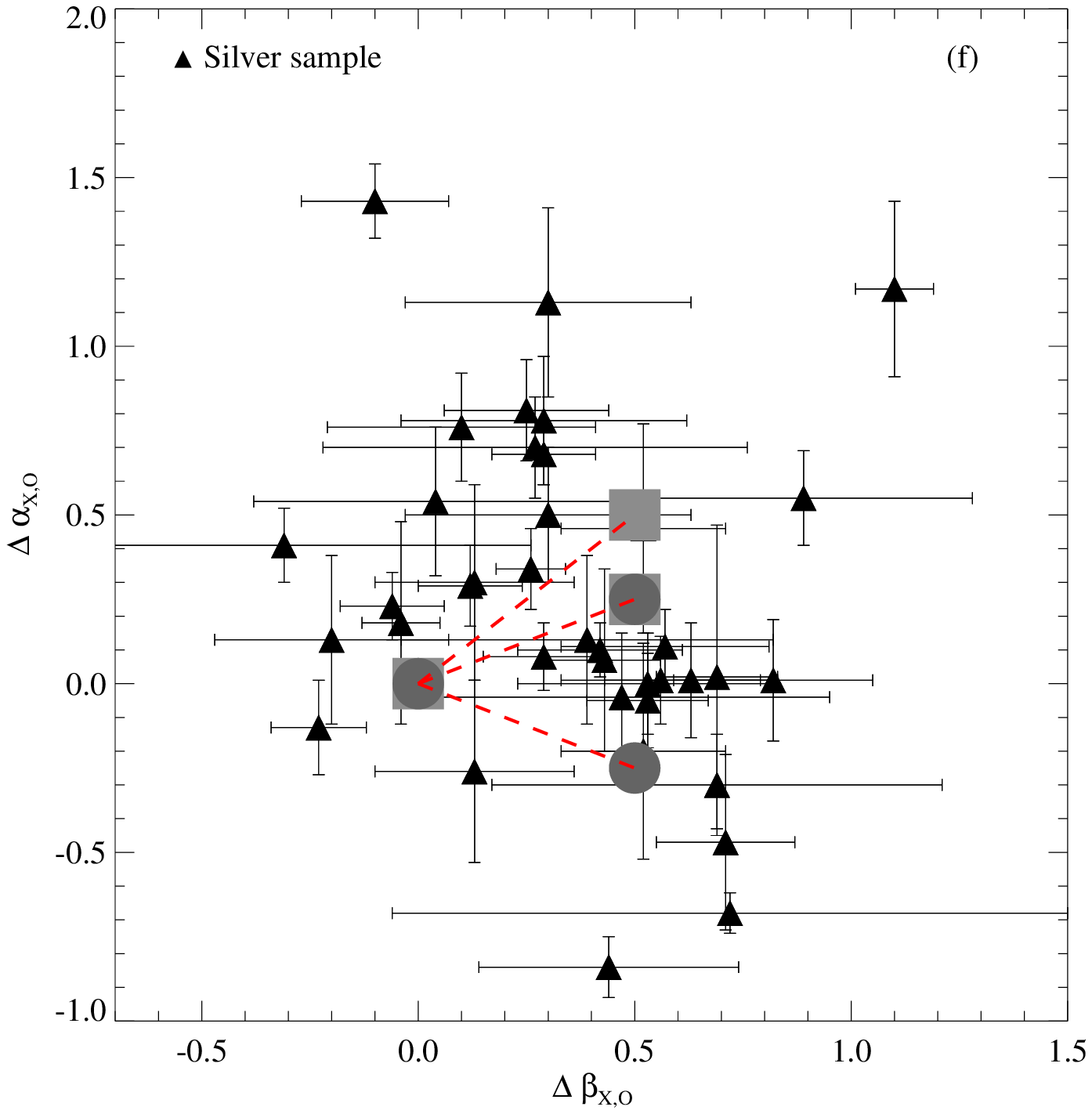}

\caption{The confrontation of the data with model predictions in the
$\Delta \beta_{\rm X,O} - \Delta \alpha_{\rm X,O}$ plane.
The ($\Delta \beta_{X,O}$, $\Delta \alpha_{X,O}$) model predictions for the
ISM and wind medium in different spectral regimes are denoted as large grey
symbols: squares for ISM and circles for wind, respectively. The red dashed
lines define the ($\Delta \beta_{X,O}$, $\Delta \alpha_{X,O}$) range for the grey zones. The SPL, energy injection, and jet break samples
are presented by square, circle, and star, respectively. The ISM and wind samples are
marked with filled and open symbols, respectively. The X-ray and optical bands in the same spectral regime, different spectral regimes, and grey zone are marked with black, red, and blue, respectively. (a): the entire Gold sample GRBs; (b)and (c): the ISM models with $p>2$ and $1<p<2$, respectively; (d) and (e): ththe wind models with $p>2$ and $1<p<2$, respectively; (f): the silver sample GRBs (denoted in triangles).}
\label{deltaapha-beta}
\end{figure*}

Figure \ref{deltaapha-beta}a shows $\bigtriangleup \beta_{\rm X,O}- \bigtriangleup
\alpha_{\rm X,O}$ distributions for Gold sample. For the energy injection sample, we
only used the post-break segment to remove the $q$-dependence.
These are the GRBs that also satisfy
the closure relations in all temporal segments. We do not show the closure relation
$\alpha-\beta$ plots since the energy injection models have an extra $q$-dependence on
the $\alpha$ values. To show the details of how each burst may fall into the model
predictions of each model (grey zone included), in Figure \ref{deltaapha-beta}(b-e)
we show the $\bigtriangleup \beta_{\rm X,O}- \bigtriangleup
\alpha_{\rm X,O}$ distributions of those Gold-Sample GRBs that satisfy the ISM and
wind medium models with $p>2$ and $1<p<2$, respectively.
The Silver sample GRBs are collected in
Figure \ref{deltaapha-beta}f). About half of them  fall outside the predicted region
(red box) defined by the models. Even though some fall into the box, they do not satisfy the
closure relations in all the temporal segments in all energy bands.

\section{Statistics of the External Shock Afterglow Model Parameters\label{sec:normal}}

Since the Gold sample (Grade I and II) GRBs comply with the external shock models well,
they serve as an excellent sample to study external shock model parameters. The derived
external shock parameters of the Gold sample GRBs are presented in Table \ref{table:parameter}.
We present some statistical properties of these model parameters in this section.

\subsection{Temporal indices $\alpha$}

\begin{figure*}
\includegraphics[angle=0,scale=1.00]{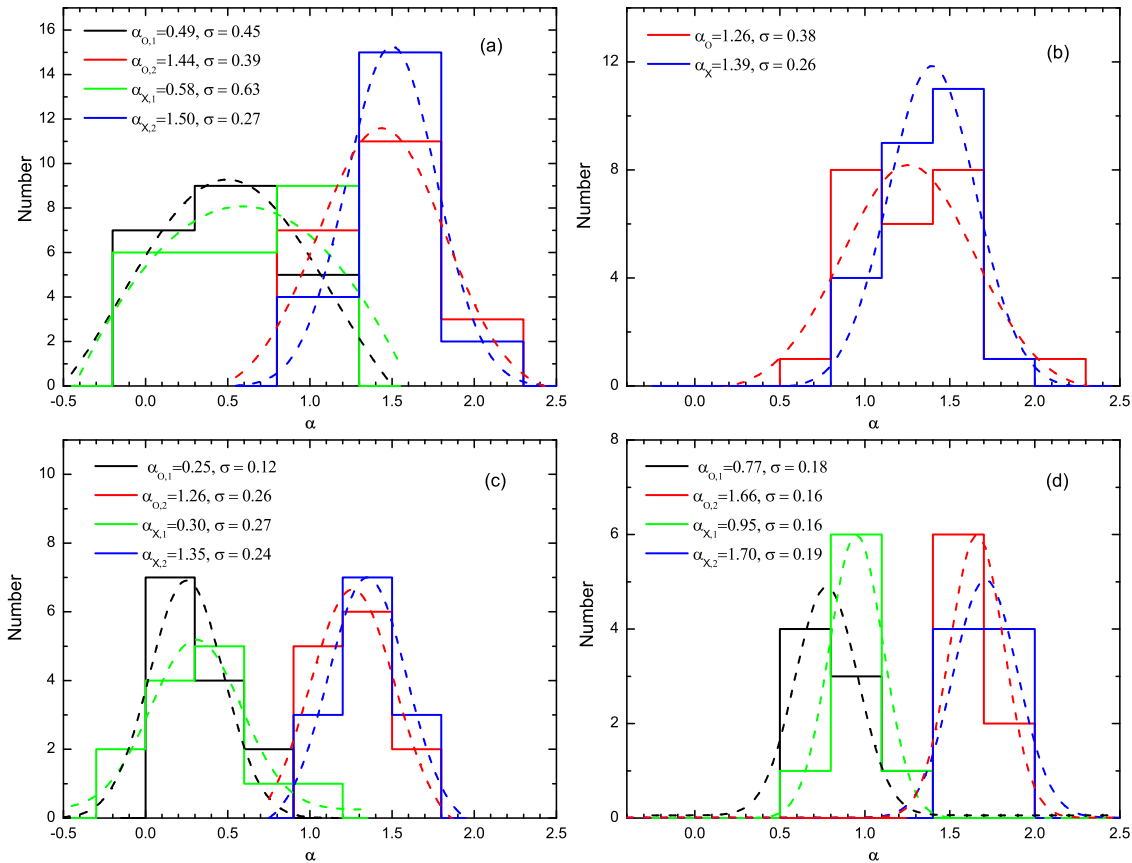}
\caption{The distributions of the temporal decay index $\alpha$
for various sub-samples in the Gold sample and their best Gaussian fits (dashed curves): (a) all the BPL GRBs in the Gold sample, with best
fits $\alpha_{\rm O,1}=0.49\pm0.45$, $\alpha_{\rm O,2}=1.44\pm0.39$, $\alpha_{\rm X,1}=0.58\pm0.63$,
and $\alpha_{\rm X,2}=1.50\pm0.27$; (b) all the SPL GRBs in the Gold sample, with best
fits $\alpha_{\rm O}=1.26\pm0.38$, $\alpha_{\rm X}=1.39\pm0.26$; (c) the energy injection sample,
with $\alpha_{\rm O,1}=0.25\pm0.12$, $\alpha_{\rm O,2}=1.26\pm0.26$, $\alpha_{\rm X,1}=0.30\pm0.27$,
and $\alpha_{\rm X,2}=1.35\pm0.24$; (d) the jet break sample, with $\alpha_{\rm O,1}=0.77\pm0.18$,
$\alpha_{\rm O,2}=1.66\pm0.16$, $\alpha_{\rm X,1}=0.95\pm0.16$ and $\alpha_{\rm X,2}=1.70\pm0.19$.}\label{alpha}
\end{figure*}

Figure \ref{alpha} shows the distributions of the temporal indices $\alpha$ in different
energy bands and different temporal segments. They are all well fitted with
Gaussian distributions for each band/temporal segment.
For the GRBs having a BPL lightcurve, the typical $\alpha$ values are
$\alpha_{\rm O,1}=0.49\pm0.45$, $\alpha_{\rm O,2}=1.44\pm0.39$, $\alpha_{\rm X,1}=0.58\pm0.63$
and $\alpha_{\rm X,2}=1.50\pm0.27$, respectively (Fig.\ref{alpha}a).
For the GRBs with a SPL lightcurve, one has
$\alpha_{\rm O}=1.26\pm0.38$, and $\alpha_{\rm X}=1.39\pm0.26$  (Fig.\ref{alpha}b).
For the BPL sample, we also separate it into the energy injection sample and the jet break
sample and perform the statistics.
For the energy injection breaks, one has $\alpha_{\rm O,1}=0.25\pm0.12$,
$\alpha_{\rm O,2}=1.26\pm0.26$, $\alpha_{\rm X,1}=0.30\pm0.27$ and $\alpha_{\rm X,2}=1.35\pm0.24$, respectively
(Fig.\ref{alpha}c). For the jet breaks, one has $\alpha_{\rm O,1}=0.77\pm0.18$,
$\alpha_{\rm O,2}=1.66\pm0.16$, $\alpha_{\rm X,1}=0.95\pm0.16$ and $\alpha_{\rm X,2}=1.70\pm0.19$, respectively
(Fig.\ref{alpha}d). Both the pre-break and the post-break $\alpha$ values in the energy injection
sample are systematically shallower than those in the jet break sample.
 On average, the X-ray lightcurves are steeper than the optical lightcurves,
consistent with the expectations of the theoretical models (i.e. the X-ray band is more likely
above $\nu_c$ while the optical band is more likely below $\nu_c$).

Another self-consistency check is to compare the observed change of decay slope, $\Delta \alpha =
\alpha_2 - \alpha_1$, with the model predictions. From the closure relations (Table \ref{Tab:alpha-beta}), one can derive

\begin{figure*}
\includegraphics[angle=0,scale=0.90]{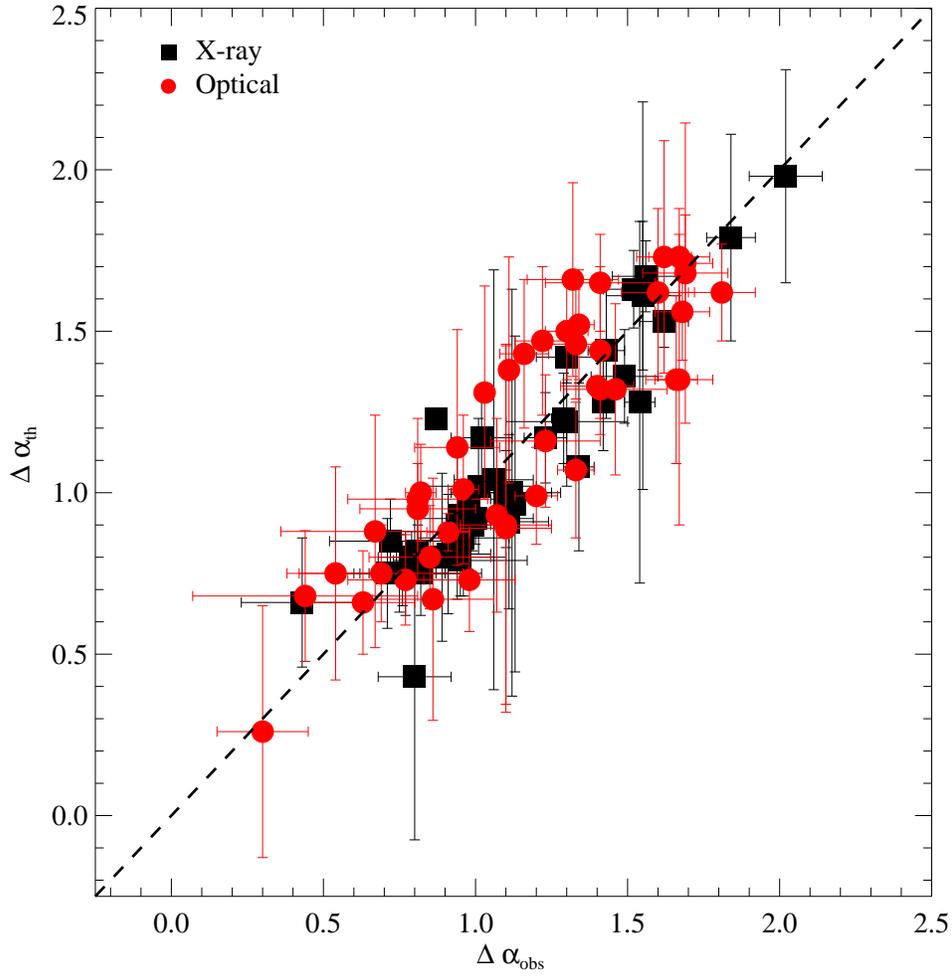}
\caption{The observed temporal break change $\Delta\alpha=\alpha_2 -\alpha_1$ compared against the theoretical value predicted from the closure relations.
The square (black) and circle (red) data points correspond to X-ray and optical bands, respectively.}
\label{a1a2gold}
\end{figure*}

\begin{itemize}
\item For energy injection breaks:

\begin{equation}
\label{deltaalpha1} \Delta \alpha=\cases{ \frac{(1-q)(2+\beta)}{2},
            & ISM II ($p>2$)\cr
 \frac{(1-q)(19+2\beta)}{16},
            & ISM II ($1<p<2$) ,\cr
\frac{(1-q)(1+\beta)}{2},
            & ISM I ($p>2$), wind II ($p>2$), wind I ($p>2$) \cr
\frac{(1-q)(7+\beta)}{8},
            & ISM I ($1<p<2$)\cr
\frac{(1-q)(5+2\beta)}{8},
            & wind II ($1<p<2$)\cr
\frac{(1-q)(3+\beta)}{4},
            & wind I ($1<p<2$)\cr
            }
\end{equation}

\item For jet breaks:

\begin{equation}
\label{deltaalpha2} \Delta \alpha=\cases{ \frac{3}{4},
            & ISM I and II ($p>2$ and $1<p<2$) \cr
 \frac{1}{2},
            & wind I and II ($p>2$ and $1<p<2$) \cr
            }
\end{equation}

\item For jet breaks with energy injection:

\begin{equation}
\label{deltaalpha3} \Delta \alpha=\cases{ \frac{(q+2)}{4},
            & ISM II ($p>2$), ISM I ($p>2$ and $1<p<2$)\cr
 \frac{(3q+6)}{16},
            & ISM II ($1<p<2$),\cr
 \frac{q}{2},
            & wind II and wind I ($p>2$ and $1<p<2$) \cr
            }
\end{equation}

\end{itemize}
Figure \ref{a1a2gold} shows a comparison between the observed
$\Delta\alpha_{\rm obs}$ and the theoretically predicted $\Delta\alpha_{\rm th}$
for each GRB derived from the measured $\beta$ and $q$ values using the
corresponding closure relations. One can see that the two are consistent with each other.

\begin{figure*}
\includegraphics[angle=0,scale=1.00]{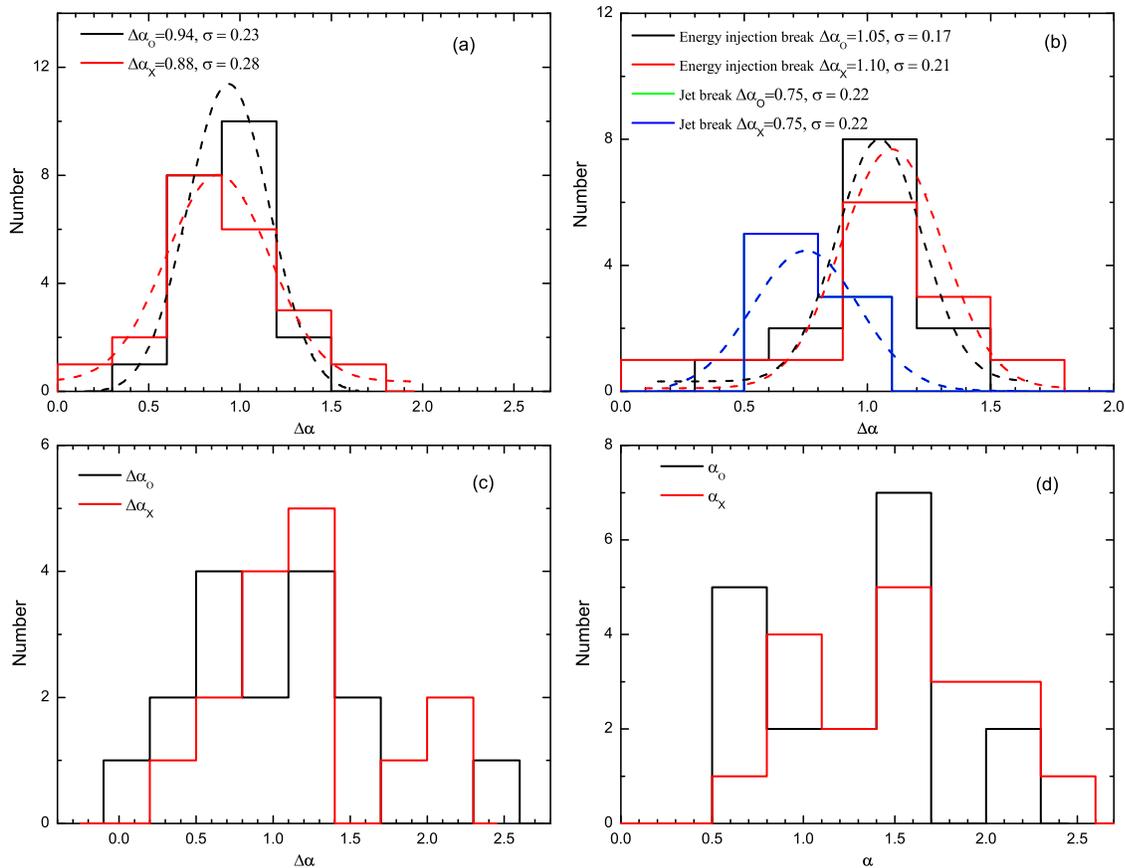}
\caption{The distributions of the temporal break change $\Delta\alpha=\alpha_2 -\alpha_1$ for various
sub-samples in the Gold sample, and the $\Delta\alpha$ and $\alpha$ distributions of the Silver sample.
Best Gaussian fits are marked with dashed curves:
(a) the $\Delta\alpha$ distributions of the entire Gold sample, with $\Delta\alpha_{\rm O}=0.94\pm0.23$
and $\Delta\alpha_{\rm X}=0.88\pm0.28$; (b) the $\Delta\alpha$ distributions of different sub-samples in
the Gold sample, with $\Delta\alpha_{\rm O}=1.05\pm0.17$ and $\Delta\alpha_{\rm X}=1.10\pm0.21$ for the
energy injection breaks, and $\Delta\alpha_{\rm O}=0.75\pm0.22$ and $\Delta\alpha_{\rm X}=0.75\pm0.22$
for the jet breaks; (c) and (d) the $\Delta\alpha$ distributions for the BPL and SPL GRBs in the Silver sample, respectively. The data
are dispersed.}
\label{deltaalpha}
\end{figure*}

Figure \ref{deltaalpha} displays the observed $\Delta\alpha$ distributions of
various samples. For the Gold sample, the $\Delta \alpha$ distributions of
optical and X-ray data are consistent with each other, i.e.
$\Delta\alpha_{\rm O}=0.94\pm0.23$ and $\Delta\alpha_{\rm X}=0.88\pm0.28$ (Fig.\ref{deltaalpha}a).
Furthermore, the $\Delta\alpha$ values of both bands in sub-groups (energy injection breaks and
jet breaks) are also consistent with each other: $\Delta\alpha_{\rm O}=1.05\pm 0.17$
and $\Delta \alpha_{\rm X}=1.10 \pm 0.21$ for the energy injection breaks, and
$\Delta\alpha_{\rm O}=0.75\pm 0.22$ and $\Delta \alpha_{\rm X}=0.75 \pm 0.22$ for
the jet breaks (Fig.\ref{deltaalpha}b). The Silver sample,
on the other hand, shows a poorer statistical behavior (Fig.\ref{deltaalpha}c and d).

\subsection{Spectral indices $\beta$}

\begin{figure*}
\includegraphics[angle=0,scale=1.10]{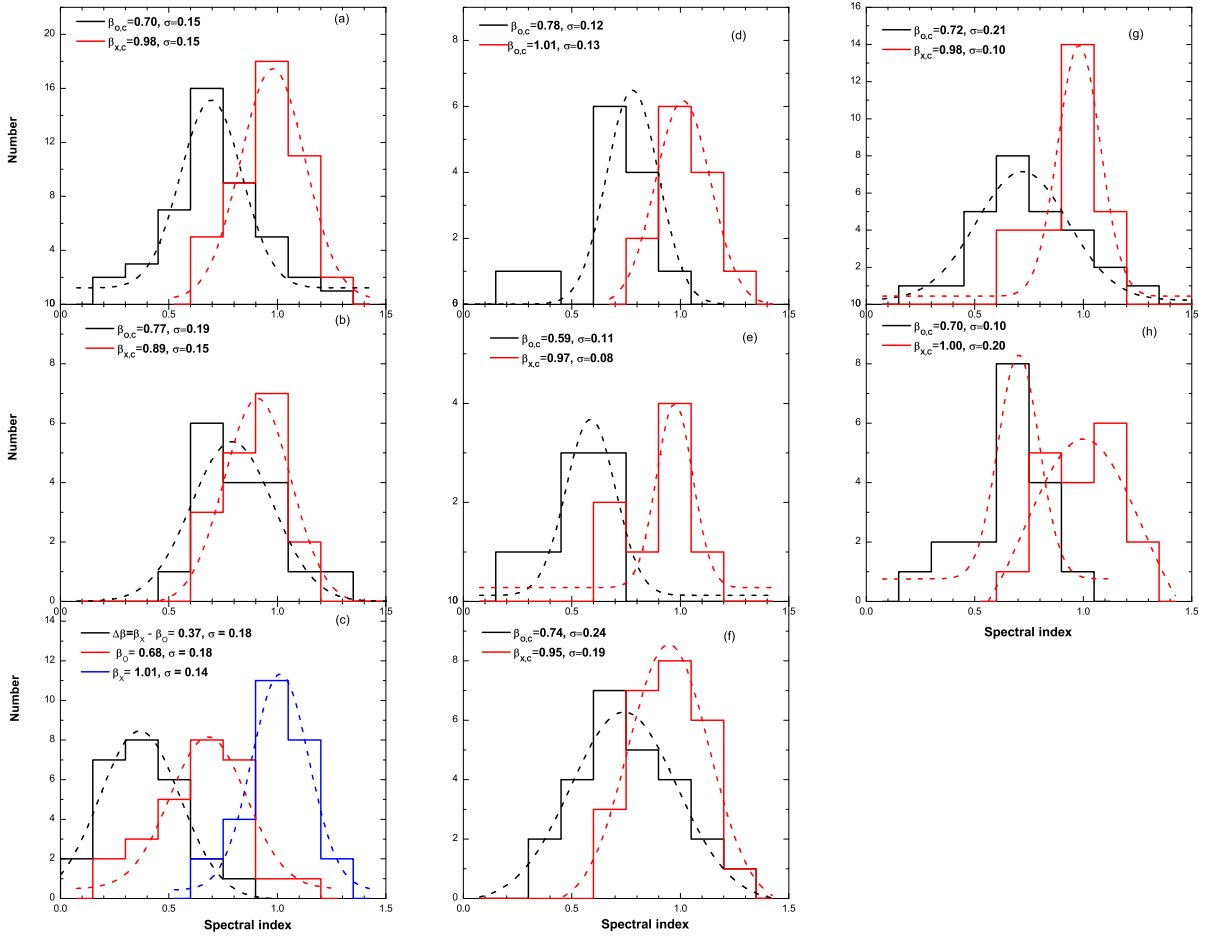}
\caption{The distributions of the spectral indices $\beta$ of
various sub-samples in the Gold sample and their best Gaussian fits (dashed curves):
(a) the distributions for the entire Gold sample, with $\beta_{\rm O}=0.70\pm0.15$,
and $\beta_{\rm X}=0.98\pm0.15$; (b) the $\beta$ distributions of the GRBs with
optical and X-ray bands constrained in the same spectral regime, with
$\beta_{\rm O}=0.77\pm0.19$, $\beta_{\rm X}=0.89\pm0.15$; (c) the $\beta$
distributions for the GRBs with optical and X-ray bands
constrained in different spectral regimes, with
$\beta_{\rm O}=0.68\pm0.18$, $\beta_{\rm X}=1.01\pm0.14$ and
$\bigtriangleup\beta=\beta_{\rm X}-\beta_{\rm O}=0.37\pm0.18$, which is
consistent with the theoretical value 0.5 as expected for a cooling break;
(d) the $\beta$ distributions for energy injection, with $\beta_{\rm O}=0.78\pm0.12$,
and $\beta_{\rm X}=1.01\pm0.13$; (e) the $\beta$ distributions for the jet break sample,
with $\beta_{\rm O}=0.59\pm0.11$, and $\beta_{\rm X}=0.97\pm0.08$; (f) the $\beta$
distributions for the SPL decay sample, with $\beta_{\rm O}=0.74\pm0.24$, and
$\beta_{\rm X}=0.95\pm0.19$; (g) the $\beta$ distributions for the GRBs with an ISM
medium, with $\beta_{\rm O}=0.72\pm0.21$, and $\beta_{\rm X}=0.98\pm0.10$; (h) the
$\beta$ distributions for the GRBs with a wind medium, with $\beta_{\rm O}=0.70\pm0.10$,
and $\beta_{\rm X}=1.00\pm0.20$.}\label{beta}
\end{figure*}

Figure \ref{beta} shows the spectral index
distributions for the Gold sample.
In general, the distributions can be fitted with gaussian functions.
For the global sample, one has $\beta_{\rm O}=0.70\pm0.15$, and
$\beta_{\rm X}=0.98\pm0.15$ (Fig.\ref{beta}a). In the Gold sample, 17/45
GRBs have both the optical and X-ray bands in the same spectral regime.
One has $\beta_{\rm O}=0.77\pm0.19$, and $\beta_{\rm X}=0.89\pm0.15$,
which are consistent with each other (Fig.\ref{beta}b).
The rest 28/45 GRBs are identified to have X-ray and optical bands
separated by a cooling break. The results show
$\beta_{\rm O}=0.68\pm0.18$, $\beta_{\rm X}=1.01\pm0.14$, with
$\Delta \beta=\beta_{\rm X}-\beta_{\rm O}=0.37\pm 0.18$, which is
consistent with the theoretically expected value $0< \Delta \beta \leq0.5$
(Fig.\ref{beta}c).

We investigate the $\beta$ distributions in different types of lightcurves.
For the energy injection sample, one has $\beta_{\rm O}=0.78 \pm 0.12$, and
$\beta_{\rm X} = 1.01 \pm 0.13$ (Fig.\ref{beta}d); for the jet break sample,
one has $\beta_{\rm O}=0.59 \pm 0.11$, and
$\beta_{\rm X} = 0.97 \pm 0.08$ (Fig.\ref{beta}e); and for the SPL sample,
one has $\beta_{\rm O}=0.74 \pm 0.24$, and
$\beta_{\rm X} = 0.95 \pm 0.19$ (Fig.\ref{beta}f).

We also investigate the $\beta$ distributions in different ambient medium types.
For the ISM model (27/45 GRBs), one has $\beta_{\rm O}=0.72 \pm 0.21$, and
$\beta_{\rm X} = 0.98 \pm 0.10$ (Fig.\ref{beta}g); and for the wind model (18/45 GRBs),
one has $\beta_{\rm O}=0.70 \pm 0.10$, and
$\beta_{\rm X} = 1.00 \pm 0.20$ (Fig.\ref{beta}h). The ISM model is more favored
than the wind model, which is consistent with the previous results
\citep[e.g.,][]{panaitescu02,yost03,zhang06,schulze11}.

\subsection{Electron spectral index $p$}

\begin{figure*}
\includegraphics[angle=0,scale=1.0]{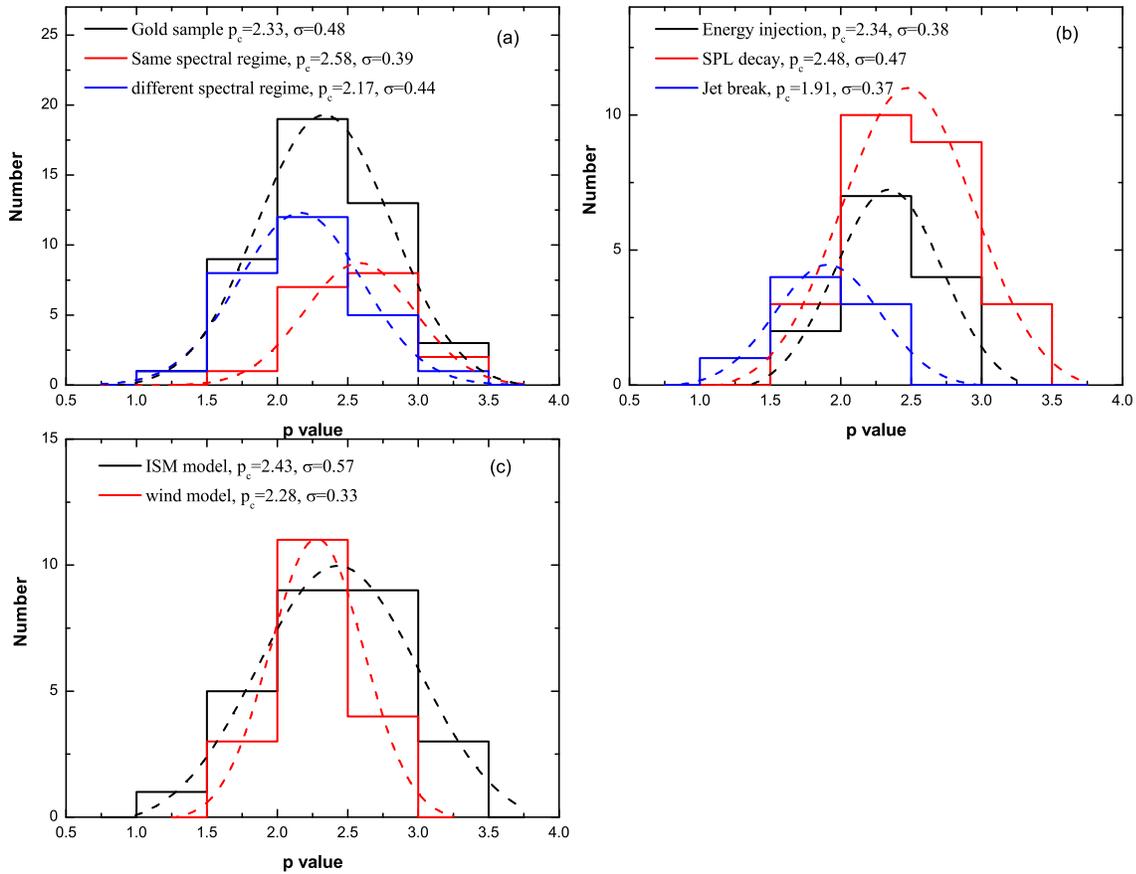}

\caption{The distribution of the inferred electron spectral index $p$ from various
sub-samples in the Gold sample, and their best Gaussian
fits (dashed lines): (a) the entire Gold sample with $p=2.33\pm0.48$, and the sub-samples
with the optical and X-ray bands constrained in the same ($p=2.58\pm0.39$) and different
($p=2.17\pm0.44$) spectral regimes, respectively; (b) the sub-samples with different
light curve behaviors: energy injection ($p=2.34\pm0.38$); jet break ($p=1.91\pm0.37$);
and SPL decay ($p=2.48\pm0.47$); (c) the sub-samples with different medium types:
ISM ($p=2.43\pm0.57$) and wind ($p=2.28\pm0.33$).}\label{pvalue}
\end{figure*}

Figure \ref{pvalue} shows the distributions of the electron spectral
index $p$ of the Gold sample. It has a Gaussian distribution with
$p=2.33\pm0.48$ (Fig.\ref{pvalue}a), which is very consistent with
the typical value of $p$ for relativistic shocks due to 1st-order
Fermi acceleration \citep[e.g.,][]{achterberg01,ellison02}. It also
has a wide distribution, which is consistent with previous studies
\citep[e.g.,][]{shen06,liang07,liang08,curran10}.

The $p$ distribution in different sub-samples are also generally
consistent with each other. Within the Gold sample, those GRBs with
optical and X-ray bands in the same spectral regime have $p=2.58 \pm 0.39$,
whereas those with optical and X-ray bands in different spectral regimes
have $p=2.17 \pm 0.44$ (Fig.\ref{pvalue}a). For the three light curve sub-samples,
one has $p=2.34 \pm 0.38$ for the energy injection break sample, $p=1.91 \pm
0.37$ for the jet break sample, and $p=2.48 \pm 0.47$ for the SPL sample,
respectively (Fig.\ref{pvalue}b).
For the two medium type models, one has $p=2.43 \pm 0.57$ for the ISM
model, and $p=2.28 \pm 0.33$ for the wind model, respectively (Fig.\ref{pvalue}c).

\subsection{Break time $t_{b}$}

\begin{figure*}
\includegraphics[angle=0,scale=0.8]{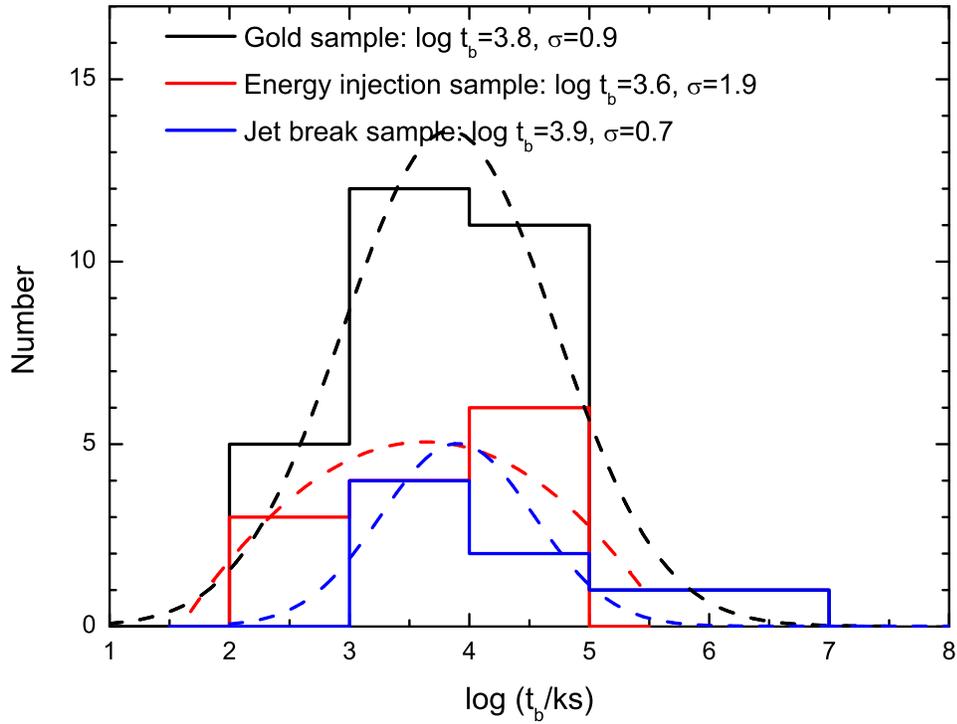}

\caption{The distributions of the inferred $t_{\rm b}$ for the entire Gold sample, the energy
injection sample, and the jet break sample. The dash lines are the best Gaussian
fits: the entire Gold sample ($\rm log (t_{\rm b} \rm /ks) =(3.8\pm0.9$)), the energy injection break sample ($\rm log (t_{\rm b} \rm /ks) =(3.6\pm1.9$)) and the jet break sample ($\rm log (t_{\rm b} \rm /ks) =(3.9\pm0.7$)).}\label{tbdist}
\end{figure*}

Figure \ref{tbdist} shows the distributions of the observed achromatic break times, $t_b$.
The global distribution in the Gold sample gives $\log (t_b \rm /ks) = (3.8 \pm 0.9)$.
Separating the energy injection sample and jet break sample, one has $\log (t_b \rm /ks)
= (3.6 \pm 1.9)$ for the energy injection break sample, $\log (t_b \rm /ks) = (3.9 \pm 0.7)$
for the jet break sample. The energy injection ends (which depends on central engine) is
on average earlier than the jet break time (which depends on geometry of the jet).
The distribution of the energy injection break time is wider
than the jet break time distribution.

\subsection{Energy injection parameter $q$}

\begin{figure*}
\includegraphics[angle=0,scale=0.8]{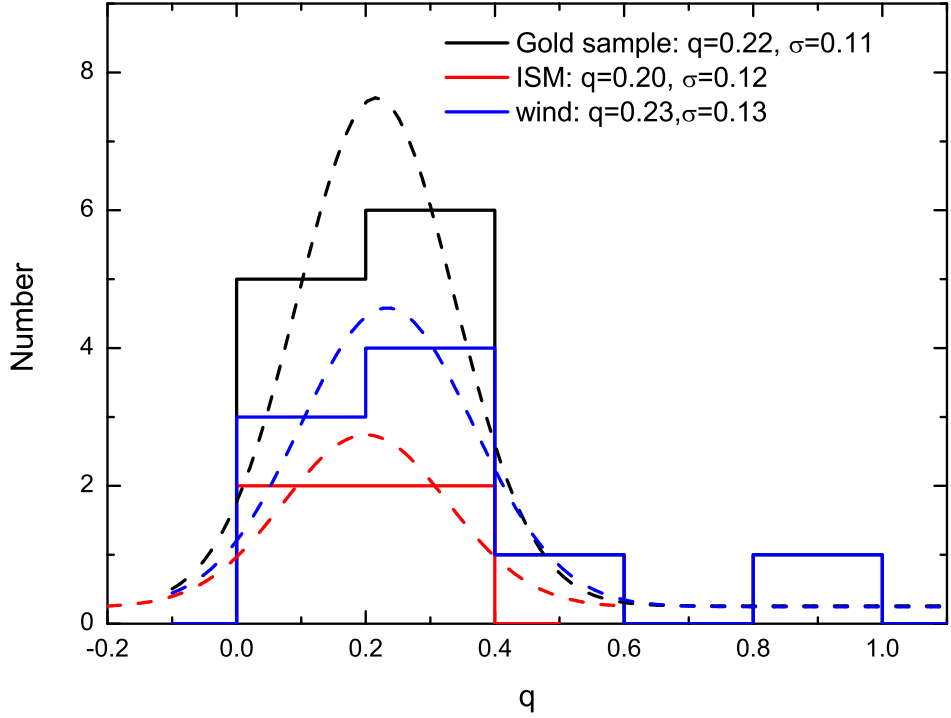}

\caption{The distributions of the inferred energy injection parameter $q$ and their best Gaussian fits (dashed lines): the entire Gold sample ($q=0.22\pm0.11$), the ISM sample ($q=0.20\pm0.12$) and the wind sample ($q=0.23\pm0.13$).}\label{qvalue}
\end{figure*}

Within the Gold sample, 13/45 GRBs show an energy injection type break, i.e.
either an energy injection break or a jet break with energy injection. Among them,
4/13 and 9/13 GRBs satisfy the ISM and wind model, respectively. The
distributions of the energy injection parameter $q$ of various samples are
shown in Figure \ref{qvalue}.
The global sample has $q=0.22\pm0.11$. The ISM and wind models have
$q=0.20\pm0.12$ and $q=0.23\pm0.13$, respectively, which are consistent
with each other.

\subsection{Shock magnetic field equipartition factor $\epsilon_{B}$}

Among the derived shock parameters, the magnetic field equipartition factor
$\epsilon_B$  is of special interest. If the shock simply compresses the
upstream magnetic field, then the expected $\epsilon_B$ is low, of the order
of $10^{-6}-10^{-7}$. If, however, various plasma instabilities are playing
a role to amplify the magnetic fields \citep[e.g.,][]{medvedev99,nishikawa09},
one would expect a relatively large $\epsilon_B$ as high as 0.1.
Early afterglow modeling \citep[e.g.,][]{wijers99,panaitescu01,panaitescu02,yost03}
derived a relatively large $\epsilon_B$, with a typical value $\sim 0.01$.
On the other hand, modeling of GeV emission in several Fermi/LAT-detected
GRBs led to the suggestion that $\epsilon_B$ should be relatively low at
least for some GRBs \citep{kumar09,kumar10}. \cite{santana14} derived
$\epsilon_B$ for a large sample of GRBs, and derived a medium value of
$10^{-5}$.

With our Gold sample, we can constrain $\epsilon_B$ independently.
Even though in most GRBs, $\epsilon_B$ cannot be constrained due to
the degeneracy of the data, one can still place interesting upper limits
to $\epsilon_B$ based on the medium type and spectral regime of the GRBs.
For example, in the ISM model $\nu_c$ decreases with time. For a regime II
($\nu_m < \nu < \nu_c$) GRB, the last data point in the light curve would set
a lower limit on $\nu_c$ at that epoch, and hence, an upper limit on
$\epsilon_B$. Similarly, in the wind model $\nu_c$ increases with time.
For a regime II GRB, the first data point in the light curve would set
a lower limit on $\nu_c$ at that point, and hence, an upper limit on
$\epsilon_B$.

In the Appendix, we present expressions of
$\nu_m$, $\nu_c$, $F_{\nu,\rm{max}}$ and the kinetic energy of the afterglow,
$E_{\rm k,iso}$. For the $p>2$ cases, we adopt the formalism in previous
works \citep{zhang07,gao13,lv14}. New expressions for the $1<p<2$ regime
are also presented following \cite{gao13}. We then derive the expressions
of $\epsilon_B$ in various models, Eqs. (\ref{ISMeb1}), (\ref{ISMeb2}),
(\ref{windeb1}), and (\ref{windeb2}), which are used to constrain $\epsilon_B$.

\begin{figure*}
\includegraphics[angle=0,scale=0.80]{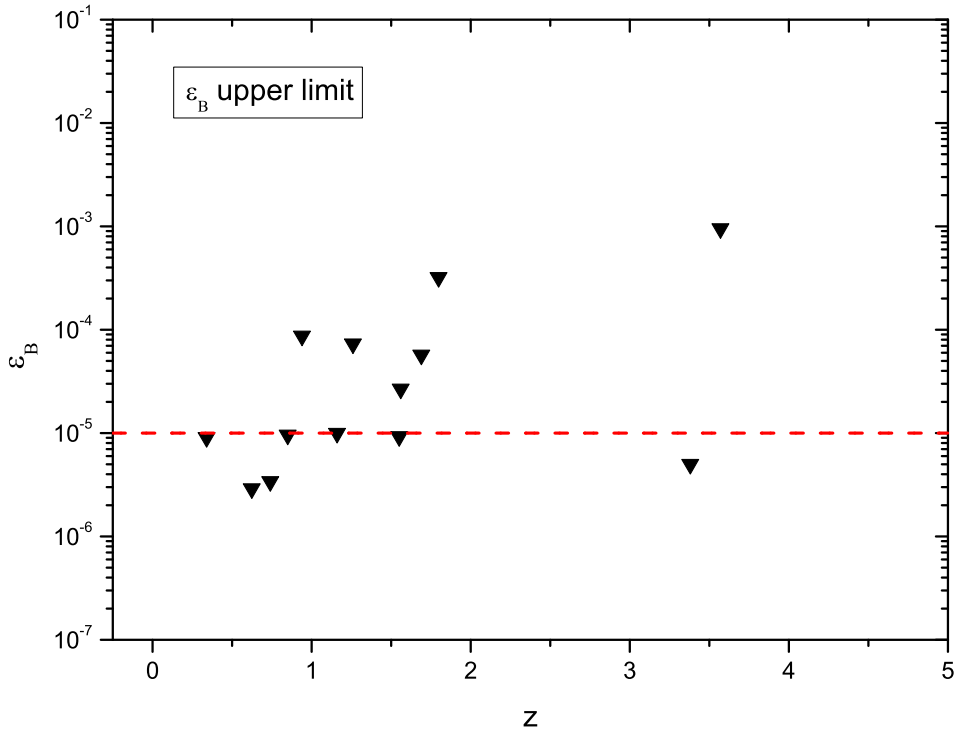}
\caption{The derived upper limits of $\epsilon_B$ of GRBs that are in spectral regime II ($\nu_m < \nu < \nu_c$). Other parameters are fixed as $\epsilon_e = 0.1$, $n=1$, and $A_\ast=1$. The dashed line denotes $\epsilon_B = 10^{-5}$. Most GRBs are consistent with this value, with a few having upper limits even below this value.}\label{epsilonB}
\end{figure*}

The derived upper limits of $\epsilon_B$ are presented in Figure \ref{epsilonB}, with other
parameters fixed as $\epsilon_e = 0.1$, $n=1$ or $A_\ast=1$. One can see that in general
these upper limits point towards a relatively low $\epsilon_B$ value. In some cases for the
ISM model, the upper limits are even lower than $10^{-5}$. These results are consistent with
the findings of \cite{kumar09,kumar10}, \cite{santana14}, and \cite{duran14}.

\subsection{Energetics}

\begin{figure*}
\includegraphics[angle=0,scale=1.0]{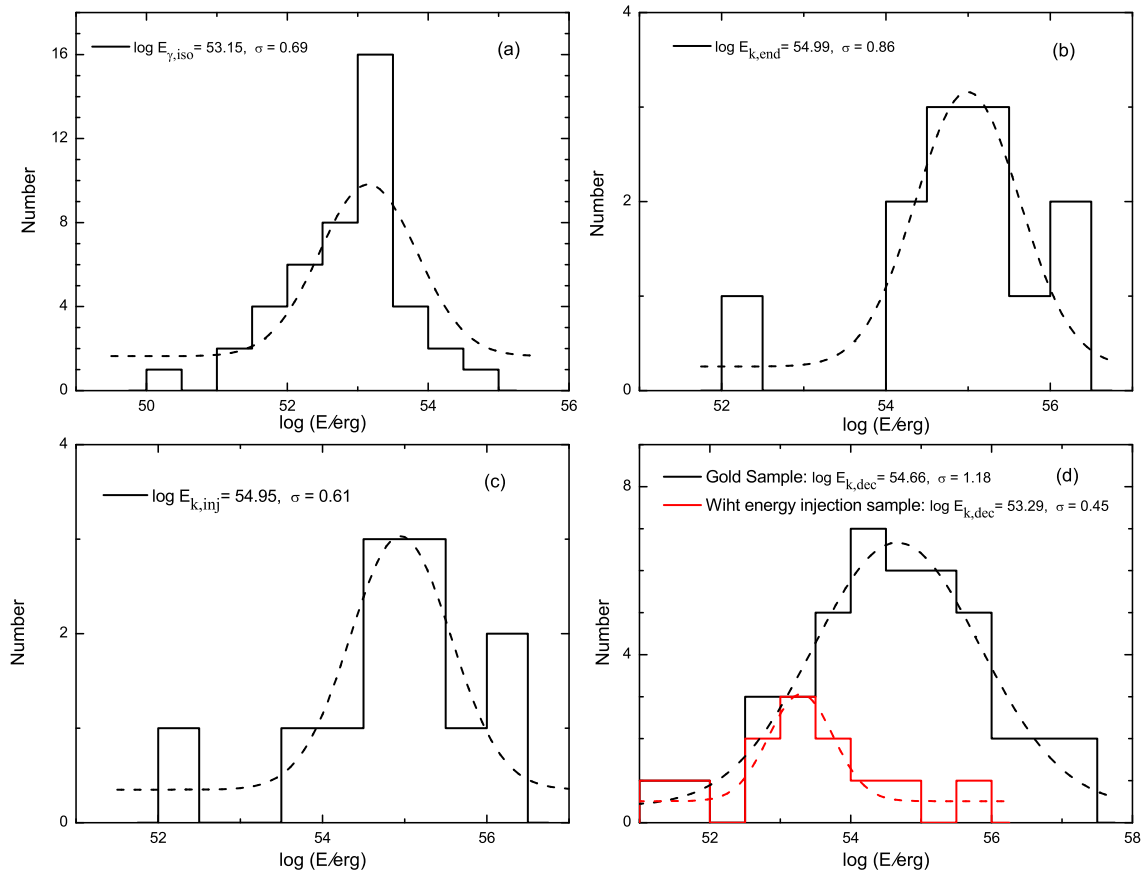}
\caption{The distributions of various energy components derived from the Gold sample and their best Gaussian fits (dashed lines): (a) isotropic $\gamma$-ray energy, $E_{\rm \gamma,iso}$, with a typical value $\log (E_{\rm \gamma,iso} \rm /erg)=(53.15\pm0.69)$; (b) the total isotropic kinetic energy at the end of energy injection for the energy injection sample, with a typical value $\log (E_{\rm K,end} \rm /erg )=(54.99\pm0.86)$; (c) the distribution of the isotropic injected energy in the energy injection sample, with $\log (E_{\rm K,inj} \rm /erg)=(54.95\pm0.61)$; (d) the isotropic kinetic energy at the deceleration time for the energy injection sample ($\log (E_{\rm K,dec} \rm /erg)=(53.29\pm0.45)$), and for the entire Gold sample ($\log (E_{\rm K,dec} \rm /erg)=(54.66\pm1.18)$). The following parameters are adopted in the kinetic energy calculations: $\epsilon_e = 0.1$, $n=1$ or $A_\ast=1$, $Y=1$, and $\epsilon_B$ = $10^{-5}$.}
\label{Edistri}
\end{figure*}

The isotropic $\gamma$-ray energy $E_{\rm \gamma,iso}$ is calculated as
\begin{equation}\label{Eiso}
E_{\rm \gamma,iso}=\frac{4 \pi D^{2}_{\rm L}S_{\gamma}k}{1+z},
\end{equation}
where $S_{\rm \gamma}$ is the gamma-ray fluence in the BAT band,
$D_{\rm L}$ is the luminosity distance of the source at redshift
$z$, and the parameter $k$ is a factor to correct the observed $\gamma$-ray
energy in a given band pass to a broad band (e.g., $1-10^4$ keV in the rest
frame) with the observed GRB spectra \citep{bloom01}. It is well
known that a typical GRB spectrum is well fitted with the so-called
Band function \citep{band93}. If the Band parameters are measured
for a burst, these parameters are used to derive the $k$ parameter.
However, owing to the narrowness of the {\em Swift}/BAT
band, the spectra of many {\em Swift} GRBs in our sample are adequately
fitted with a single power-law, $N\propto E^{-\Gamma}$, so that the Band
parameters are not well constrained. For these GRBs, we use an empirical
relation between $E_p$ and BAT-band photon index $\Gamma$
\citep{zhang07b,sakamoto09,virgili12} to estimate $E_p$.
Taking typical values of the photon indices
$\alpha=-1.1$ and $\beta=-2.2$ \citep{preece00,kaneko06},
we can derive $E_{\rm \gamma,iso}$ values of the GRBs with redshift measurements
in our Gold sample, which range from $10^{51}$ to $10^{55}$ erg, with a
typical value $\log (E_{\rm \gamma,iso} \rm /erg)=53.15\pm0.69$
(Fig.\ref{Edistri}(a)).

The isotropic kinetic energy of the afterglow $E_{\rm K,iso}$ can be derived
from the afterglow data. In general, broad-band modeling is needed to
precisely measure $E_{\rm K,iso}$ \citep{panaitescu01,panaitescu02}.
Most GRBs do not have adequate data to perform such an analysis.
More conveniently, one may use the X-ray data only to constrain $E_{\rm K,iso}$,
since the X-ray band is usually above $\nu_c$, so that the X-ray flux does
not depend on the ambient density and only weakly depends on $\epsilon_e$
\citep{kumar99,freedman01,berger03,lloydronning04,zhang07}, see
Appendix for detailed derivations. For the cases with energy injection,
$E_{\rm K,iso}$ is a function of time during the energy injection phase.
Following \cite{zhang07}, we calculate $E_{\rm K,iso}$ at two different
epochs, one at the break time $t_b$, when energy injection is over,
and another at a putative deceleration time $t_{\rm dec} \sim {\rm max}
(60~{\rm s}, T_{90})$. We use the X-ray flux at $t_b$ to derive
$E_{\rm K,end}$, and then derive $E_{\rm K,dec}=E_{\rm K,end}
(t_{\rm dec} / t_b)^{1-q}$. The total injected energy is calculated
as $E_{\rm K,inj}=E_{\rm K,end}-E_{\rm K,dec}$. Based on our constraints
on $\epsilon_B$, we take $\epsilon_B = 10^{-5}$ for all the GRBs in our
calculations. Other parameters are taken as typical values:
$\epsilon_e = 0.1$, $n=1$ or $A_\ast=1$, and $Y=1$.

\begin{figure*}
\includegraphics[angle=0,scale=1.0]{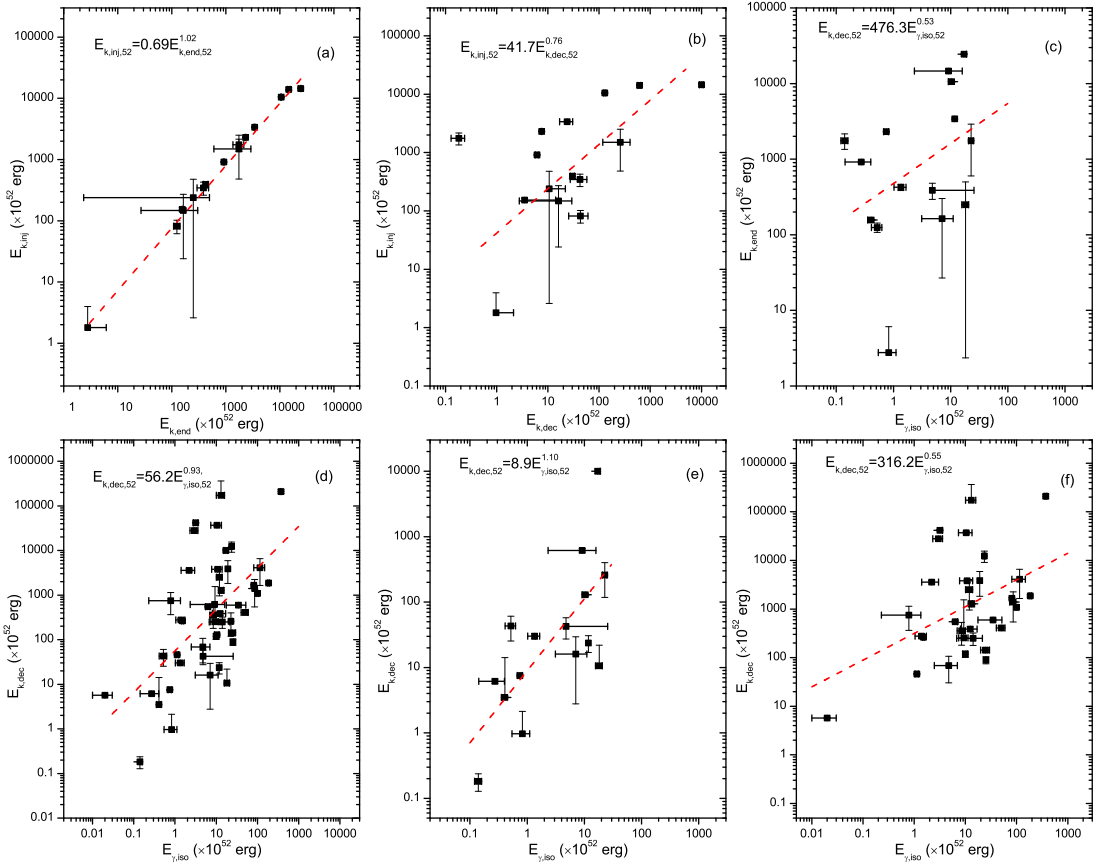}
\caption{The correlations among different energy components: (a) $E_{\rm K,end} - E_{\rm K,inj}$: $E_{\rm K,inj,52}=0.69E^{1.02\pm0.02}_{\rm K,end,52}$; (b) $E_{\rm K,dec} - E_{\rm K,inj}$: $E_{\rm K,inj,52}=41.7E^{0.76\pm0.20}_{\rm K,dec,52}$; (c) $E_{\rm \gamma,iso} - E_{\rm K,end}$: $E_{\rm K,end,52}=476.3E^{0.53\pm0.23}_{\rm \gamma,iso,52}$; (d), (e) and (f):  $E_{\rm \gamma,iso} - E_{\rm K,dec}$ for the entire Gold sample, the energy injection sub-sample, and the sub-sample without energy injection, with $E_{\rm K,dec,52}=56.2E^{0.93\pm0.19}_{\rm \gamma,iso,52}$, $E_{\rm K,dec,52}=8.9E^{1.10\pm0.29}_{\rm \gamma,iso,52}$ and $E_{\rm K,dec,52}=316.2E^{0.55\pm0.21}_{\rm \gamma,iso,52}$, respectively.}\label{Erelation}
\end{figure*}

In Figure \ref{Edistri}, we present several statistical results of the $E_{\rm K,iso}$ calculations. For the GRBs with energy injection, the distributions are log ($E_{\rm K,end} \rm /erg)=54.99\pm0.86$ (Fig.\ref{Edistri}b), log ($E_{\rm K,inj}\rm /erg)=54.95\pm 0.61 $ (Fig.\ref{Edistri}c), and log ($E_{\rm K,dec} \rm /erg)=53.29\pm0.45$ (Fig.\ref{Edistri}d). For the entire Gold sample, one has log ($E_{\rm K,dec}\rm /erg)=54.66\pm1.18$ (Fig.\ref{Edistri}d). It is interesting to see that the energetics of the energy-injection sample reache a similar level as the no-energy-injection sample after the energy injection is over. Clear correlations are found among different energy components: $E_{\rm K,end} - E_{\rm K,inj}$ relation $E_{\rm K,inj,52}=0.69E^{1.02\pm0.02}_{\rm K,end,52}$ (Fig.\ref{Erelation}a), $E_{\rm K,dec} - E_{\rm K,inj}$ relation $E_{\rm K,inj,52}=41.7E^{0.76\pm0.20}_{\rm K,dec,52}$ (Fig.\ref{Erelation}b), $E_{\rm \gamma,iso} - E_{\rm K,end}$ relation $E_{\rm K,end,52}=476.3E^{0.53\pm0.23}_{\rm \gamma,iso,52}$ (Fig.\ref{Erelation}c), $E_{\rm \gamma,iso} - E_{\rm K,dec}$ relations in the entire Gold sample, $E_{\rm K,dec,52}=56.2E^{0.93\pm0.19}_{\rm \gamma,iso,52}$ (Fig.\ref{Erelation}d), in the energy injection sample, $E_{\rm K,dec,52}=8.9E^{1.10\pm0.29}_{\rm \gamma,iso,52}$ (Fig.\ref{Erelation}e), and in the no-energy-injection sample, $E_{\rm K,dec,52}=316.2E^{0.55\pm0.21}_{\rm \gamma,iso,52}$ (Fig.\ref{Erelation}f), respectively. In particular, the$E_{\rm K,inj,52}=41.7E^{0.76\pm0.20}_{\rm K,dec,52}$ correlation suggests a substantial energy injection during the shallow decay phase for most GRBs.

\subsection{Radiative efficiency $\eta_{\gamma}$}
The GRB radiative efficiency, defined as \citep{lloydronning04}
\begin{equation}\label{efficency}
\eta_{\gamma}=\frac{E_{\rm \gamma,iso}}{E_{\rm \gamma,iso}+E_{\rm K,iso}},
\end{equation}
is an essential parameter to probe how efficient a burst converts its global energy to prompt $\gamma$-ray emission.
As mentioned above, if there is continuous energy injection, the kinetic energy of the afterglow $E_{\rm K,iso}$ takes different values if one chooses different epochs. In principle, $\eta_\gamma$ can be defined for two different epochs, $t_{\rm dec}$ and $t_b$, which have different physical meanings (see a discussion in \citealt{zhang07}).

\begin{figure*}
\includegraphics[angle=0,scale=1.0]{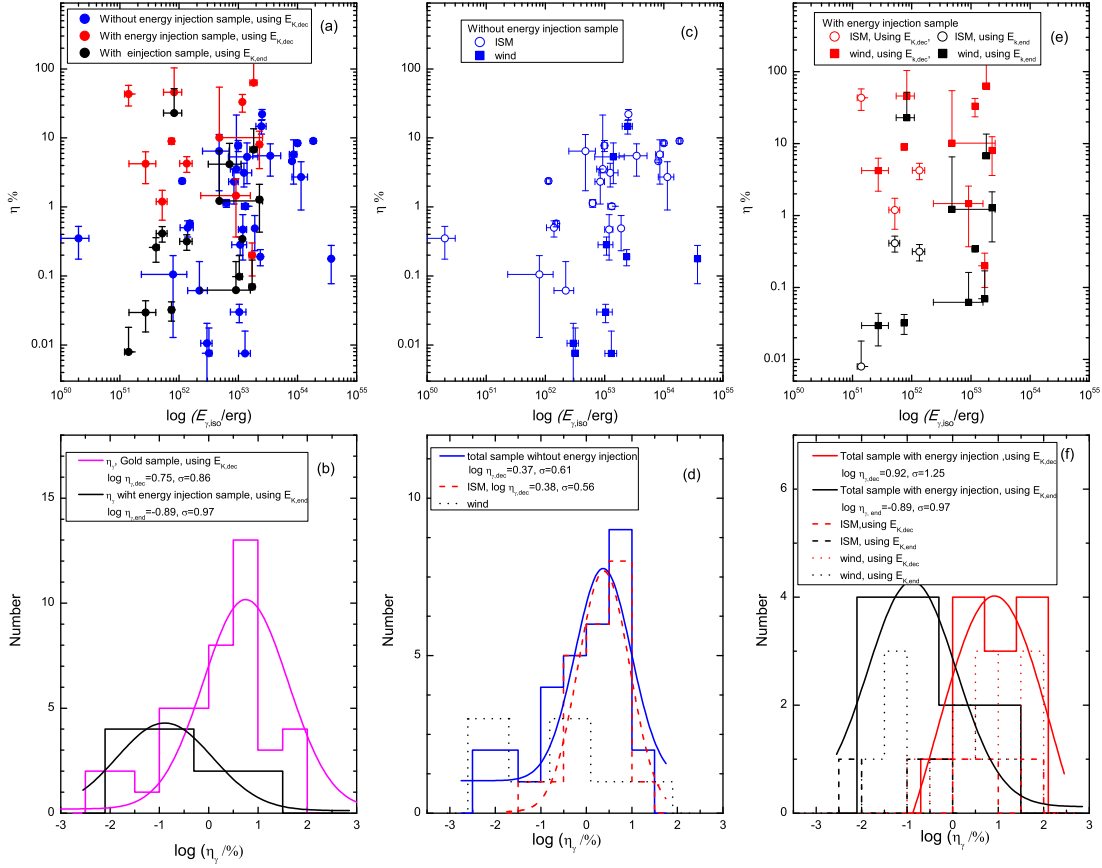}
\caption{The radiative efficiency $\eta_{\rm \gamma}$ of the Gold sample as a function of $E_{\rm \gamma,iso}$ (upper row) and histograms (lower row):  (a) and (b): For the entire Gold sample. Among them, the GRBs in the energy-injection sub-sample are marked in red and black. Two efficiencies are calculated for each GRB: one using $E_{\rm K,dec}$ (red) and the other using $E_{\rm end}$ (black). Those GRBs without energy injection are marked in blue. Log-normal fits to the efficiencies derived from $E_{\rm K,dec}$ and $E_{\rm end}$ give $\rm log (\eta_{\rm \gamma,dec}/\%)=0.75\pm0.86$ and $\rm log (\eta_{\rm \gamma,end}/\%)=-0.89\pm0.97$, respectively; (c) and (d): for the sub-sample of GRBs without energy injection, with best fit $\rm log (\eta_{\rm \gamma,dec}/\%)=0.36\pm0.61$. Open circles denote the ISM medium and solid squares denote the wind medium; (e) and (f): for the sub-sample of GRBs with energy injection, with best fits $\rm log (\eta_{\rm \gamma,dec}/\%)=0.92\pm1.25$ and $\rm log (\eta_{\rm \gamma,end}/\%)=-0.89\pm0.97$, respectively. Again ISM and wind cases are denoted as open circles and solid squares, respectively.}
\label{efficiency}
\end{figure*}

Figure \ref{efficiency} shows the radiative efficiencies calculated at $t_{\rm dec}$ and $t_b$ as a function of $E_{\rm \gamma,iso}$ along with their histograms. No significant correlation between $\eta_\gamma$ and $E_{\rm \gamma,iso}$ is found. The fireball internal shock model predicts a relatively small efficiency of a few per cent \citep[e.g.,][]{kumar99,panaitescu99,maxham09,gaomeszaros15}. Previous constraints on GRB radiative efficiencies give relatively large values, as large as above $90\%$ \citep[e.g.,][]{lloydronning04,zhang07,racusin11} for some GRBs. This challenges the internal shock models, and favors alternative prompt emission models, such as dissipation of magnetic fields \citep{zhangyan11} or photospheric emission \citep{lazzati13}.

The derived efficiencies can be fit with rough log-normal distributions (Fig.\ref{efficiency}b, d, f). For the entire sample, one has $\rm log (\eta_{\rm \gamma,dec}/\%)=0.75\pm0.86$. For the sub-sample GRBs without energy injection, the radiative efficiency is lower, with $\rm log (\eta_{\rm \gamma,dec}/\%)=0.37\pm0.61$. For the sub-sample GRBs with energy injection, the radiative efficiencies read $\rm log (\eta_{\rm \gamma,dec}/\%)=0.92\pm1.25$ for $t_{\rm dec}$, and $\rm log (\eta_{\rm \gamma,end}/\%)=-0.89\pm0.97$ for $t_b$, respectively.

The derived efficiencies are somewhat smaller than the values derived in previous work \citep[e.g.][]{zhang07}. The main reason is the adoption of a smaller value of $\epsilon_B \sim 10^{-5}$, so that the derived $E_{\rm K,iso}$ are systematically larger.  This greatly alleviates the low-efficiency problem of the internal shock models. Nonetheless, some GRBs still have tens of percent efficiency, which demands a contrived setup for the internal shock models   \citep[e.g.,][]{beloborodov00,kobayashisari01}. If $t_{\rm dec}$ is adopted, which is more natural for most prompt emission model to calculate efficiency (see \citealt{zhang07} for a detailed discussion), $\eta_\gamma$ is still typically too large for the internal shock model. This is on the other hand consistent with the suggestion that internal collision-induced magnetic reconnection and turbulence (ICMART) is the dominant process to power GRB prompt emission in the majority of GRBs,  which can typically gives tens of percent radiative efficiency \citep{zhangyan11,deng15}. This conclusion is also consistent with independent studies of modeling the GRB prompt emission spectrum \citep{uhmzhang14} and quasi-thermal photosphere emission component \citep{gaozhang15}.

\subsection{Jet opening angle and geometrically-corrected gamma-ray energy}

In the Gold sample, 8/45 GRBs show a jet break. These include five GRBs of without
energy injection and one more with energy injection. The ambient medium type of
all 6 GRBs is ISM. Five out of these six GRBs have redshift information
(Table \ref{table:jet}).

\begin{figure*}
\includegraphics[angle=0,scale=1.0]{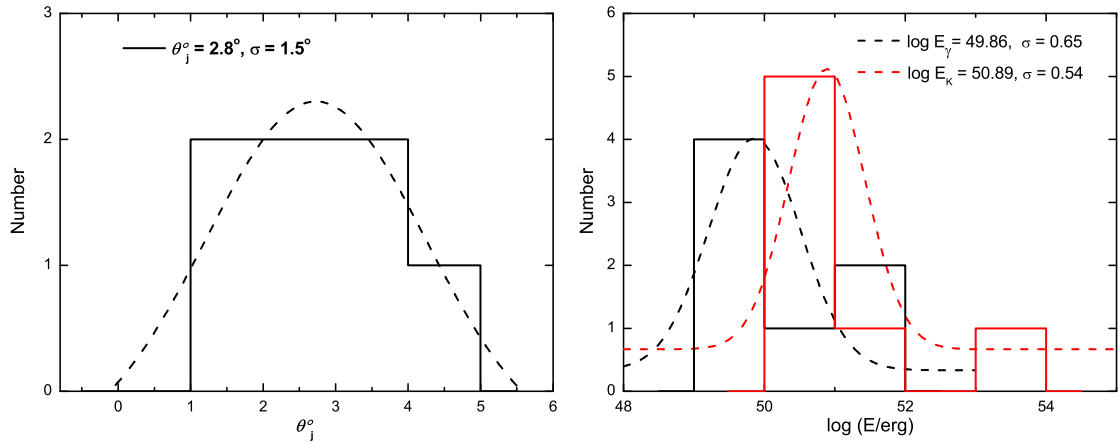}
\caption{The distributions of the jet opening angle ($\theta_{j}$), geometrically corrected $\gamma$-ray energy ($E_{\rm \gamma}$) and kinetic energy ($E_{\rm K}$), respectively, derived from the jet break sub-sample of the Gold sample. The
dashed lines are the best Gaussian fits, with $\theta_{j}=(2.8\pm1.5)^{\rm o}$, $\log (E_{\rm \gamma} \rm /erg)=(49.86\pm0.65)$,
and $\log (E_{\rm K} \rm /erg)=(50.89\pm0.54)$, respectively.}\label{jetbreak}
\end{figure*}

Under the assumption of a conical jet, one can derive the jet opening angle
based on observational data \citep{rhoads99,sari99,frail01}:
\begin{equation}\label{jet}
\theta_{j}=0.070 ~{\rm rad}~\left(\frac{t_{b}}{1\ \rm
day}\right)^{3/8} \left(\frac{1+z}{2}\right)^{-3/8}
\left(\frac{E_{\rm K,iso}}{10^{53}\ \rm ergs}\right)^{-1/8}
\left(\frac{n}{0.1\ \rm cm^{-3}}\right)^{1/8}.
\end{equation}
We then calculate the geometrically corrected $\gamma$-ray energy
\begin{equation}\label{Eiso}
E_{\gamma}=(1-\cos\theta_{j})E_{\rm \gamma,iso},
\end{equation}
and kinetic energy
\begin{equation}\label{Eiso}
E_{\rm K}=(1-\cos\theta_{j})E_{\rm K,iso}.
\end{equation}
Here $E_{\rm K,iso}$ is taken as $E_{\rm K,end}$ for the energy injection sample.
The medium density is taken as $n=1$ cm$^{-3}$.The results are presented in Table \ref{table:jet} and
Figure \ref{jetbreak}. The best fitting results give $\theta_{j} = (2.8 \pm 1.5)^{\rm \circ} $, $\log (E_{\gamma}/\rm erg) = 49.86 \pm 0.65$, and $\log (E_{\rm K}/\rm erg) = 50.89 \pm 0.54$.

\section{Conclusions and Discussion\label{sec:normal}}

The chromatic afterglow behavior observed in some GRBs has raised the concern regarding
whether the external forward shock models are still adequate to interpret
the broad-band afterglows of GRBs and whether alternative ideas, e.g., a long-lasting
engine-driven afterglow, are needed to account for the data.
In order to answer ``how bad/good the external shock models are'',
in this paper, we systematically studied 85 {\em Swift} GRBs up to March 2014, which
all have high-quality X-ray and optical light curves and spectral data to allow us
to study the compliance
of the data to the external forward shock models. The results of this study
can be summarized as the following.

Based on how well the data abide by the external forward shock afterglow models,
we categorized GRBs into five grades and three samples:

\begin{itemize}
  \item A Gold sample (Grade I and II) includes 45/85 GRBs.
These GRBs are fully consistent
with the theoretical predictions of the external shock models, including having
an acceptable achromatic break and fulfilling various closure relations between
the temporal decay indices $\alpha$ and spectral indices $\beta$.
  \item A Silver sample (Grade III and IV) includes 37/85 GRBs. These GRBs
are also consistent with having an acceptable achromatic break, even though one or more
afterglow segments do not comply with the closure relations. These GRBs are
potentially interpretable within the framework of external shock models.
  \item A Bad sample (Grade V) only includes 3/85 GRBs. These GRBs show direct
evidence of chromatic behaviors, which cannot be accounted for within
single-component external shock models.
\end{itemize}

The bottom line of this study is to address how bad/good the external shock models
are. Our results show that external shock models work very well for at least
$\sim$ 53\% of GRBs (our Gold sample). These GRBs can be interpreted within the
simplest afterglow models. If more advanced modeling invoking other factors (e.g.,
structured jet or long-lasting reverse shock) is carried out, up to $\sim 96\%$
of GRBs (including the Silver sample) {\em may be} accounted for within the
external shock models. Only less than 4\% GRBs
truly violate the basic expectations of the external shock models, and demand
another emission component (e.g., central engine afterglow) to account for emission
in at least one band (e.g., the X-ray band).

Several caveats deserve mentioning. First, we only focused on the main afterglow
components (SPL, BPL or TPL) of the X-ray and optical lightcurves. In some
GRBs, there are additional components overlapping with these main components,
such as the X-ray steep decay phase, X-ray flares, and optical re-brightening
features, which are not included in the analysis. These features are usually
chromatic, and demand additional emission components to interpret
the data. The general conclusion that a lot of GRBs have extended central
engine activities \citep{zhang06} remain valid. The true duration of the
GRB central engine activities may be much longer than what is measured by
the GRB duration $T_{90}$ \citep{zhangbb14}.
Second, we adopted a relatively ``loose'' criterion ( $\chi^2_{total} \leq3$) to define ``achromaticity''
by requiring the X-ray and optical lightcurves to have a same break time. Searching for break times independently in the two bands often results in
somewhat different break times, but many GRBs can be made being consistent
with achromatic. The relatively large $\chi^2_{total} \sim 3$ in some GRBs
is mostly caused by the additional features (small flares and fluctuations which we
do not care) in the otherwise (broken) power law lightcurves.
We therefore believe that our approach is appropriate
to address the question of ``how bad the models are''. On the other hand, if in
the future high-quality data indeed show slight chromatic behaviors with high confidence,
one should take cautious to the fraction numbers presented in this paper, and consider
how such slight chromatic behaviors may impact the models.
Finally, we only studied 85 GRBs that have both bright X-ray and bright optical
emission data to allow us to perform the test. There are more GRBs detected
by {\em Swift} ($\sim 900$ with X-ray lightcurves and $\sim 260$ with optical
lightcurves). Due to the complicated sample selection effects,
we do not guarantee that the fractions of Gold, Silver and Bad
samples are reliable numbers for the entire GRB population. In any case,
85 GRBs represent a reasonably large sample, so that our statistics are valid
at least for the ``bright'' sample of GRBs.

With the Gold sample, we further performed a series of statistical analyses
of various observational properties and model parameters. Following interesting
conclusions can be drawn:
\begin{itemize}
  \item Temporal index $\alpha$: The temporal indices $\alpha$ in different bands
and different temporal segments satisfy the afterglow model predictions. On
average, the X-ray lightcurves are steeper than optical. For BPL lightcurves,
the degrees of the break, $\Delta\alpha$, are consistent with the theoretical
predictions of the energy injection models or jet break models;
  \item Spectral index $\beta$: The spectral indices $\beta$
in the optical and X-ray bands are
$\beta_{O}=0.70\pm0.15$, $\beta_{X}=0.98\pm0.15$, respectively.
Some (17/45) have X-ray and optical bands in the same spectral segment,
while most (28/45) have the two bands separated by $\nu_c$ or in the grey zone.
Statistically, $\Delta\beta = 0.37 \pm 0.18$ is consistent with the theoretical
value 0-0.5, a range of $\Delta\beta$, including those expected
in the grey zone \citep{uhm14a}.
  \item Electron spectral index $p$: The typical value $p=2.33\pm0.48$ is very consistent with the theoretical predictions for relativistic shocks. A wide range of $p$ values are observed, which is consistent with previous findings.
  \item Break time $t_b$: The typical break time is found to be $\log (t_b \rm /ks) = 3.8 \pm 0.9$. The break time of energy injection sample ($\log (t_b \rm /ks) = 3.96\pm 1.9$) is statistically earlier than that of the jet break break sample ($\log (t_b \rm /ks) = 3.9 \pm 0.7$).
  \item Energy injection parameter $q$: the central value is  $q=0.22\pm0.11$, and the ISM and wind models are consistent with each other, with $q=0.20\pm0.12$ and $q=0.23\pm0.13$, respectively.

  \item Magnetization parameter $\epsilon_B$: The derived upper limits of $\epsilon_B$ suggests that the typical value of this parameter is low (say, $10^{-5}$), which is consistent with previous work \citep{kumar09,kumar10,santana14,duran14}.

  \item Energetics: The typical isotropic $\gamma$-ray energy is $\log (E_{\rm \gamma,iso}\rm /erg) = 53.15 \pm 0.69$. For the energy injection case, the typical isotropic kinetic energy in the blastwave is log ($E_{\rm K,dec}\rm /erg)=53.29\pm0.45$ at the deceleration time, and log ($E_{\rm K,end}\rm /erg)=54.99\pm0.86$ when energy injection is over. For GRBs without energy injection, the typical blastwave kinetic energy is log ($E_{\rm K,dec}\rm /erg)=54.66\pm1.18$. Clear correlations among various energy components are found.
  \item Radiative efficiency $\eta_\gamma$: With a small $\epsilon_B \sim 10^{-5}$ adopted, the derived radiative efficiency $\eta_\gamma$ is lower than previous studies. For the entire Gold sample, $\rm log (\eta_{\rm \gamma,dec}/\%)=0.75\pm0.86$. Yet, the efficiency is still large for some GRBs, especially the ones with energy injection. For these GRBs, the efficiency measure at the deceleration time has $\rm log (\eta_{\rm \gamma,dec}/\%)=0.92\pm1.25$, which still challenges the internal shock model.

  \item Jet opening angle $\theta_j$: For the jet break sample, we derived the typical jet opening angle as $\theta_{j} = (2.5 \pm 1.5)^{\rm \circ}$. The jet-corrected $\gamma$-ray energy and kinetic energy are $\log (E_{\gamma}\rm /erg) = 49.86 \pm 0.65$ and $\log (E_{\rm K}\rm /erg) = 50.89 \pm 0.54$, respectively.
\end{itemize}

\acknowledgments
We thank an anonymous referee for helpful suggestions, and M. De Pasquale for a comment. This work is supported by the National Basic Research Program (973 Programme) of China (Grant No. 2014CB845800),  the National Natural Science Foundation of China (Grants 11303005, U1331202), the Guangxi Science Foundation (2013GXNSFFA019001), the Strategic Priority Research Program ``The Emergence of Cosmological Structures''(Grant No. XDB09000000) and CSC follow support. BZ acknowledges NASA NNX14AF85G for support. HG acknowledges NASA NNX 13AH50G. DAK acknowledges support by DFG grants Kl 766/16-1 and Kl 766/16-3."

\clearpage

\appendix

\section{Expressions of $E_{\rm K,iso}$ and $\epsilon_{B}$}

In this Appendix, we present expressions of $E_{\rm K,iso}$ and $\epsilon_B$ of the external forward shock afterglow models.

\subsection{The ISM model}

For the $p>2$ case, the forward shock emission can be characterized as \citep{yost03,zhang07}
\begin{eqnarray}
&&\nu_m=3.3\times 10^{12}~{\rm Hz}~\left(\frac{p-2}{p-1}\right)^2(1+z)^{1/2}\epsilon_{B,-2}^{1/2}\epsilon_{e,-1}^2E_{\rm K,iso,52}^{1/2}t_d^{-3/2}, \nonumber\\
&&\nu_c=6.3\times 10^{15}~{\rm Hz}~(1+z)^{-1/2}(1+Y)^{-2}\epsilon_{B,-2}^{-3/2}E_{\rm K,iso,52}^{-1/2}n^{-1}t_d^{-1/2}, \nonumber\\
&&F_{\nu,\rm{max}}=1.6~{\rm mJy}~(1+z)D^{-2}_{28}\epsilon_{B,-2}^{1/2}E_{\rm K,iso,52}n^{1/2},
\label{ISM1}
\end{eqnarray}
where $E_{\rm K,iso,52}$ is the isotropic kinetic energy (in units of $10^{52}$ erg) in the blastwave, $t_d$ is the time since trigger (in units of days), $n$ is the density of the constant ambient medium,  $D_L=10^{28}~{\rm cm}~ D_{28}$ is the luminosity distance, and
\begin{eqnarray}
Y=[-1+(1+4 \eta_1\eta_2 \epsilon_e / \epsilon_B)^{1/2}]/2
\label{Y}
\end{eqnarray}
is the inverse Compton parameter, with $\eta_1 = {\rm min} [1,(\nu_c/
\nu_m)^{(2-p)/2}]$  \citep{sari01},
and $\eta_2 \leq 1$ is a correction factor introduced by the
Klein-Nishina correction.

For $p>2$ and in the $\nu>\rm{max}(\nu_m,\nu_c)$ regime, one has \citep{zhang07}
\begin{eqnarray}
\nu F_{\nu}(\nu=10^{18}~\rm{Hz})&=&F_{\nu, \rm max}\nu_c^{1/2}\nu_m^{(p-1)/2}\nu_X^{(2-p)/2}\nonumber\\
&=&5.2\times10^{-14} ~{\rm erg~s^{-1}~cm^{-2}}~D^{-2}_{28}(1+z)^{(p+2)/4}(1+Y)^{-1}f_{p1}\epsilon_{B,-2}^{(p-2)/4}\epsilon_{e,-1}^{p-1}\nonumber\\
&&\times E_{\rm K,iso,52}^{(p+2)/4} t_d^{(2-3p)/4}\nu_{18}^{(2-p)/2},
\label{ISMufu1}
\end{eqnarray}
where $\nu f_{\nu} (\nu= 10^{18} Hz)$ is the energy flux at $10^{18}$ Hz (in units of ergs s$^{-1}$ cm$^{-1}$), and
\begin{eqnarray}
f_{p1}=6.73\left(\frac{p-2}{p-1}\right)^{p-1}(3.3\times 10^{-6})^{(p-2.3)/2}
\label{fp1}
\end{eqnarray}
is a function of electron spectral index $p$ \citep{zhang07}. One can then derive
\begin{eqnarray}
E_{\rm K,iso,52}&=&\left(\frac{\nu F_{\nu}(\nu=10^{18}~\rm{Hz})}{5.2\times10^{-14}~{\rm erg~s^{-1}~cm^{-2}}}\right)^{4/(p+2)}D^{8/(p+2)}_{28}(1+z)^{-1}(1+Y)^{4/(p+2)}f_{p1}^{-4/(p+2)}\epsilon_{B,-2}^{(2-p)/(p+2)}\nonumber\\
&&\times \epsilon_{e,-1}^{4(1-p)/(p+2)}t_d^{(3p-2)/(p+2)}\nu_{18}^{2(p-2)/(p+2)},
\label{ISMEk1}
\end{eqnarray}
\begin{eqnarray}
\epsilon_{B,-2}&=&\left(\frac{6.3\times 10^{15}~\rm{Hz}}{\nu_c}\right)^{(p+2)/(p+4)}\left(\frac{\nu F_{\nu}(\nu=10^{18}~\rm{Hz})}{5.2\times10^{-14}\rm{erg~s^{-1}~cm^{-2}}}\right)^{-2/(p+4)}D^{-4/(p+4)}_{28}\nonumber\\
&&\times(1+Y)^{-2(p+3)/(p+4)}n^{-(p+2)/(p+4)}f_{p1}^{2/(p+4)}\epsilon_{e,-1}^{2(p-1)/(p+4)}t_d^{-2p/(p+4)}\nu_{18}^{(2-p)/(p+4)}.
\label{ISMeb1}
\end{eqnarray}

For $p>2$ and in the $\nu_m<\nu<\nu_c$ regime, one has
\begin{eqnarray}
\nu F_{\nu}(\nu=10^{18}~{\rm Hz})&=&F_{\nu, \rm max}\nu_m^{(p-1)/2}\nu_X^{(3-p)/2}\nonumber\\
&=&6.5\times10^{-13}~{\rm erg~s^{-1}~cm^{-2}}~D^{-2}_{28}(1+z)^{(p+3)/4}f_{p1}\epsilon_{B,-2}^{(p+1)/4}\epsilon_{e,-1}^{p-1}E_{\rm K,iso,52}^{(p+3)/4}\nonumber\\
&&\times n^{1/2}t_d^{(3-3p)/4}\nu_{18}^{(3-p)/2},
\label{ISMufu1}
\end{eqnarray}
so that
\begin{eqnarray}
E_{\rm K,iso,52}&=&\left(\frac{\nu F_{\nu}(\nu=10^{18}~\rm{Hz})}{6.5\times10^{-13}\rm{erg~s^{-1}~cm^{-2}}}\right)^{4/(p+3)}D^{8/(p+3)}_{28}(1+z)^{-1}f_{p1}^{-4/(p+3)}\epsilon_{B,-2}^{-(p+1)/(p+3)}\nonumber\\
&&\times \epsilon_{e,-1}^{4(1-p)/(p+3)}n^{-2/(p+3)}t_d^{(3p-3)/(p+3)}\nu_{18}^{2(p-3)/(p+3)},
\label{ISMEk2}
\end{eqnarray}
\begin{eqnarray}
\epsilon_{B,-2}&=&\left(\frac{6.3\times 10^{15}~\rm{Hz}}{\nu_c}\right)^{(p+3)/(p+4)}\left(\frac{\nu F_{\nu}(\nu=10^{18}~\rm{Hz})}{6.5\times10^{-13}~{\rm erg~s^{-1}~cm^{-2}}}\right)^{-2/(p+4)}D^{-4/(p+4)}_{28}\nonumber\\
&&\times(1+Y)^{-2(p+3)/(p+4)} n^{-(p+2)/(p+4)}f_{p1}^{2/(p+4)}\epsilon_{e,-1}^{2(p-1)/(p+4)}t_d^{-2p/(p+4)}\nu_{18}^{(3-p)/(p+4)}
\label{ISMeb2}
\end{eqnarray}

Following \cite{gao13}, below we derive the expressions in the $p<2$ case.

In the $\nu>\rm{max}(\nu_m,\nu_c)$ regime, one has
\begin{eqnarray}
E_{\rm K,iso}&=&(\nu F_{\nu}(\nu=10^{18}~{\rm Hz}))^{16/(p+14)}D^{32/(p+14)}f_{p2}^{-16/(p+14)}n^{(p-2)/(p+14)}t^{(3p+10)/(p+14)}(1+Y)^{16/(p+14)}\nonumber\\
&&\times (1+z)^{-12/(p+14)}\epsilon_{e}^{-16/(p+14)}\nu_{18}^{8(p-2)/(p+14)},
\label{ISMEk3}
\end{eqnarray}
\begin{eqnarray}
f_{p2}=0.00529\times e^{0.767p}\left(\frac{2-p}{p-1}\right),
\label{fp2}
\end{eqnarray}
\begin{eqnarray}
\epsilon_{B}&=&9.44\times10^{27}\nu_c^{-2/3}(\nu F_{\nu}(\nu=10^{18}~\rm{Hz}))^{-16/3(p+14)}D^{-32/3(p+14)}f_{p2}^{16/3(p+14)}n^{-(3p+26)/(3p+42)}\nonumber\\
&&\times t^{-4(p+6)/3(p+14)}(1+Y)^{-4(p+18)/3(p+14)}(1+z)^{-(p+2)/3(p+14)}\epsilon_{e}^{16/3(p+14)}\nu_{18}^{-8(p-2)/3(p+14)}. \nonumber \\
\label{ISMeb3}
\end{eqnarray}

In the $\nu_m<\nu<\nu_c$ regime, one has
\begin{eqnarray}
E_{\rm K,iso}&=&(\nu F_{\nu}(\nu=10^{18}~{\rm Hz}))^{16/(p+18)}D^{32/(p+18)}f_{p3}^{-16/(p+18)}n^{(p-10)/(p+18)}t^{3(p+2)/(p+18)}\nonumber\\
&&\times (1+z)^{-16/(p+18)}\epsilon_{e}^{-16/(p+18)}\epsilon_{B}^{-12/(p+18)}\nu_{18}^{8(p-3)/(p+18)},
\label{ISMEk4}
\end{eqnarray}
\begin{eqnarray}
\epsilon_{B}&=&e^{(1159.5+64.4p)/(p+14)}\nu_c^{-2(p+18)/3(p+14)}(\nu F_{\nu}(\nu=10^{18}~\rm{Hz}))^{-16/3(p+14)}D^{-32/3(p+14)}f_{p3}^{16/3(p+14)}\nonumber\\
&&\times n^{-(3p+26)/(3p+42)}t^{-4(p+6)/3(p+14)}(1+Y)^{-4(p+18)/3(p+14)}(1+z)^{-(p+2)/3(p+14)}\nonumber\\
&&\times \epsilon_{e}^{16/3(p+14)}\nu_{18}^{-8(p-3)/3(p+14)},
\label{ISMeb4}
\end{eqnarray}
and
\begin{eqnarray}
f_{p3}=5.53\times 10^{-15}e^{0.767p}\left(\frac{2-p}{p-1}\right).
\label{fp3}
\end{eqnarray}

\subsection{The wind model}

The following derivations follow \cite{gao13,chevalier00,lv14}.

For the $p>2$ case, the forward shock emission can be characterized as
\begin{eqnarray}
&&\nu_m=5.2\times 10^{11}~{\rm Hz}~\left(\frac{p-2}{p-1}\right)^2(1+z)^{1/2}\epsilon_{B,-2}^{1/2}\epsilon_{e,-1}^2E_{\rm K,iso,52}^{1/2}t_d^{-3/2}, \nonumber\\
&&\nu_c=1.7\times 10^{18}~{\rm Hz}~(1+z)^{-3/2}(1+Y)^{-2}\epsilon_{B,-2}^{-3/2}E_{\rm K,iso,52}^{1/2}A_{*,-1}^{-2}t_d^{1/2}, \nonumber\\
&&F_{\nu,\rm{max}}=1.6~{\rm mJy}~(1+z)^{3/2}D^{-2}_{28}\epsilon_{B,-2}^{1/2}E_{\rm K,iso,52}^{1/2}A_{*,-1}t_{d}^{-1/2},
\label{wind1}
\end{eqnarray}
where $A_\ast$ is the density parameter of the stellar wind medium.

For $p>2$ and in the $\nu>\rm{max}(\nu_m,\nu_c)$ regime, one has
\begin{eqnarray}
\nu F_{\nu}(\nu=10^{18}~\rm{Hz})&=&F_{\nu, \rm max}\nu_c^{1/2}\nu_m^{(p-1)/2}\nu_X^{(2-p)/2}\nonumber\\
&=&2.6\times10^{-13}~{\rm erg~s^{-1}~cm^{-2}}~D^{-2}_{28}(1+z)^{(p+2)/4}(1+Y)^{-1}f_{p4}\epsilon_{B,-2}^{(p-2)/4}\epsilon_{e,-1}^{p-1}\nonumber\\
&&\times E_{\rm K,iso,52}^{(p+2)/4}t_d^{(2-3p)/4}\nu_{18}^{(2-p)/2},
\label{windufu1}
\end{eqnarray}
\begin{eqnarray}
f_{p4}=6.73\left(\frac{p-2}{p-1}\right)^{p-1}(5.2\times 10^{-7})^{(p-2.3)/2},
\label{fp4}
\end{eqnarray}
\begin{eqnarray}
E_{\rm K,iso,52}&=&\left(\frac{\nu F_{\nu}(\nu=10^{18}~\rm{Hz})}{2.6\times10^{-13}~{\rm erg~s^{-1}~cm^{-2}}}\right)^{4/(p+2)}D^{8/(p+2)}_{28}(1+z)^{-1}(1+Y)^{4/(p+2)}f_{p4}^{-4/(p+2)}\epsilon_{B,-2}^{(2-p)/(p+2)}\nonumber\\
&&\times \epsilon_{e,-1}^{4(1-p)/(p+2)}t_d^{(3p-2)/(p+2)}\nu_{18}^{2(p-2)/(p+2)},
\label{windEk1}
\end{eqnarray}
\begin{eqnarray}
\epsilon_{B,-2}&=&\left(\frac{1.7\times 10^{18}~\rm{Hz}}{\nu_c}\right)^{(p+2)/(2p+2)}\left(\frac{\nu F_{\nu}(\nu=10^{18}~\rm{Hz})}{2.6\times10^{-13}~{\rm erg~s^{-1}~cm^{-2}}}\right)^{1/(p+1)}D^{2/(p+1)}_{28}(1+z)^{-(p+2)/(p+1)}\nonumber\\
&&\times (1+Y)^{-1}f_{p4}^{-1/(p+1)}\epsilon_{e,-1}^{(1-p)/(p+1)}A_{*,-1}^{-(p+2)/(p+1)}t_d^{p/(p+1)}\nu_{18}^{(p-2)/2(p+1)}
\label{windeb1}
\end{eqnarray}

For $p>2$ and in the $\nu_m<\nu<\nu_c$ regime, one has
\begin{eqnarray}
\nu F_{\nu}(\nu=10^{18}~{\rm Hz})&=&F_{\nu, \rm max}\nu_m^{(p-1)/2}\nu_X^{(3-p)/2}\nonumber\\
&=&2.0\times10^{-13}~{\rm erg~s^{-1}~cm^{-2}}D^{-2}_{28}(1+z)^{(p+5)/4}f_{p4}\epsilon_{B,-2}^{(p+1)/4}\epsilon_{e,-1}^{p-1}E_{\rm K,iso,52}^{(p+1)/4}\nonumber\\
&&\times A_{*,-1}t_d^{(1-3p)/4}\nu_{18}^{(3-p)/2},
\label{windufu2}
\end{eqnarray}
\begin{eqnarray}
E_{\rm K,iso,52}&=&\left(\frac{\nu F_{\nu}(\nu=10^{18}~{\rm Hz})}{2.0\times10^{-13}~{\rm erg~s^{-1}~cm^{-2}}}\right)^{4/(p+1)}D^{8/(p+1)}_{28}(1+z)^{-(p+5)/(p+1)}f_{p4}^{-4/(p+1)}\epsilon_{B,-2}^{-1}\nonumber\\
&&\times \epsilon_{e,-1}^{4(1-p)/(p+1)}A_{*,-1}^{-4/(p+1)}t_d^{(3p-1)/(p+1)}\nu_{18}^{2(p-3)/(p+1)},
\label{windEk2}
\end{eqnarray}
\begin{eqnarray}
\epsilon_{B,-2}&=&\left(\frac{1.7\times 10^{18}~\rm{Hz}}{\nu_c}\right)^{1/2}\left(\frac{\nu F_{\nu}(\nu=10^{18}~{\rm Hz})}{2.0\times10^{-13}~{\rm erg~s^{-1}~cm^{-2}}}\right)^{1/(p+1)}D^{2/(p+1)}_{28}(1+z)^{-(p+2)/(p+1)}\nonumber\\
&& \times  (1+Y)^{-1} A_{*,-1}^{-(p+2)/(p+1)} f_{p4}^{-1/(p+1)}\epsilon_{e,-1}^{(1-p)/(p+1)}t_d^{p/(p+1)}\nu_{18}^{(p-3)/2(p+1)}.
\label{windeb2}
\end{eqnarray}

For $p<2$ and in the $\nu>\rm{max}(\nu_m,\nu_c)$ regime, one has
\begin{eqnarray}
E_{\rm K,iso}&=&(\nu F_{\nu}(\nu=10^{18}~{\rm Hz}))^{8/(p+6)}D^{16/(p+6)}f_{p5}^{-8/(p+6)}A^{(p-2)/(p+6)}t\nonumber\\
&& \times (1+Y)^{8/(p+6)}\times (1+z)^{-6/(p+6)}\epsilon_{e}^{-8/(p+6)}\nu_{18}^{4(p-2)/(p+6)},
\label{windEk3}
\end{eqnarray}
\begin{eqnarray}
f_{p5}=405854\times3^{3p/8}5^{-p/4}e^{-7p}\pi^{(4-3p)/8}\left(\frac{2-p}{p-1}\right),
\label{fp5}
\end{eqnarray}
\begin{eqnarray}
\epsilon_{B}&=&6.87\times 10^{-11}\nu_c^{-2/3}(\nu F_{\nu}(\nu=10^{18}~{\rm Hz}))^{8/3(p+6)}D^{16/3(p+6)}f_{p5}^{-8/3(p+6)}A^{-(3p+26)/(3p+18)}t^{2/3}\nonumber\\
&&\times (1+Y)^{-4(p+4)/3(p+6)}(1+z)^{-(p+8)/(p+6)}\epsilon_{e}^{-8/3(p+6)}\nu_{18}^{4(p-2)/3(p+6)}.
\label{windeb3}
\end{eqnarray}

For $p<2$ and in the $\nu_m<\nu<\nu_c$ regime, one has
\begin{eqnarray}
E_{\rm K,iso}&=&(\nu F_{\nu}(\nu=10^{18}~{\rm Hz}))^{8/(p+4)}D^{16/(p+4)}f_{p6}^{-8/(p+4)}A^{(p-10)/(p+4)}t^{(p+8)/(p+4)}\nonumber\\
&&\times (1+z)^{-12/(p+4)}\epsilon_{e}^{-8/(p+4)}\epsilon_{B}^{-6/(p+4)}\nu_{18}^{4(p-3)/(p+4)},
\label{windEk3}
\end{eqnarray}
\begin{eqnarray}
f_{p6}=1.72\times 10^{22}\times2^{-17p/2}3^{3p/8}5^{-37p/4}e^{13.38p}\pi^{(4-3p)/8}\left(\frac{2-p}{p-1}\right),
\label{fp6}
\end{eqnarray}
and
\begin{eqnarray}
\epsilon_{B}&=&(5.69\times 10^{-16})^{2(p+4)/3(p+6)}\nu_c^{-2(p+4)/3(p+6)}(\nu F_{\nu}(\nu=10^{18}~{\rm Hz}))^{8/3(p+6)}D^{16/3(p+6)}f_{p6}^{-8/3(p+6)}\nonumber\\
&&\times A^{-(3p+26)/(3p+18)}t^{2/3}(1+Y)^{-4(p+4)/3(p+6)}(1+z)^{-(p+8)/(p+6)}\epsilon_{e}^{-8/3(p+6)}\nu_{18}^{4(p-3)/3(p+6)}. \nonumber \\
\label{windeb4}
\end{eqnarray}

\clearpage



\begin{deluxetable}{ccccccccccccccccccccccccc}

\tabletypesize{\tiny}
\tablecaption{The temporal and spectral parameters of 85 GRBs. }
\tablewidth{0pt}
\tabcolsep=2.5pt

\tablehead{ \colhead{GRB}& \colhead{$\beta_{O}$}&
\colhead{$\beta_{X}$}&
\colhead{$\alpha_{O,1}$\tablenotemark{a}}& \colhead{$\alpha_{O,2}$}&
\colhead{$\omega$}& \colhead{Function}&
\colhead{$\alpha_{X,1}$\tablenotemark{a}}& \colhead{$\alpha_{X,2}$}&
\colhead{$\omega$}& \colhead{Function}&
\colhead{$\bigtriangleup \alpha_{\rm X,O}$\tablenotemark{b}}&
\colhead{$\bigtriangleup \beta_{\rm X,O}$\tablenotemark{c}}&
\colhead{$t_{b}$\tablenotemark{d}}}

\startdata

Grade I																																												 \\
\hline																																												 \\
050408	&	0.28 	$\pm$	0.33 	&	1.14 	$\pm$	0.14 	&	0.49 	$\pm$	0.01 	&	 1.29 	 $\pm$	0.11 	&	3 	&	 BPL	&	0.73 	$\pm$	0.15 	&	1.17 	$\pm$	 0.22 	 &	3	 &	 BPL	&	-0.12 	$\pm$	0.33 	&	0.86 	$\pm$	 0.47 	&	40.7 		 \\
050801	&	0.69 	$\pm$	0.34 	&	0.92 	$\pm$	0.17 	&	0.07 	$\pm$	0.01 	&	 1.20 	 $\pm$	0.01 	&	3 	&	 BPL	&	0.24 	$\pm$	0.11 	&	1.18 	$\pm$	 0.03 	 &	3	 &	 BPL	&	-0.02 	$\pm$	0.04 	&	0.23 	$\pm$	 0.51 	&	0.2 		 \\
050820A	&	0.72 	$\pm$	0.03 	&	0.89 	$\pm$	0.05 	&	0.91 	$\pm$	0.02 	&	 1.67 	 $\pm$	0.09 	&	3 	&	 BPL	&	1.12 	$\pm$	0.08 	&	1.89 	$\pm$	 0.11 	 &	3	 &	 BPL	&	0.22 	$\pm$	0.20 	&	0.17 	$\pm$	 0.08 	&	2379.0 		 \\
050922C	&	0.51 	$\pm$	0.05 	&	1.06 	$\pm$	0.11 	&	0.82 	$\pm$	0.11 	&	 1.53 	 $\pm$	0.09 	&	3 	&	 BPL	&	1.04 	$\pm$	0.12 	&	1.71 	$\pm$	 0.19 	 &	3	 &	 BPL	&	0.18 	$\pm$	0.28 	&	0.55 	$\pm$	 0.16 	&	8.0 		 \\
051028	&	0.60 	$\pm$	0.00 	&	0.95 	$\pm$	0.15 	&	0.99 	$\pm$	0.06 	&				 &		&	SPL	&	 1.16 	$\pm$	0.08 	&				&		&	SPL	&	0.17 	$\pm$	 0.14 	 &	0.35 	$\pm$	0.15 	&			\\
051109A	&	0.70 	$\pm$	0.05 	&	0.98 	$\pm$	0.08 	&	0.64 	$\pm$	0.08 	&	 1.07 	 $\pm$	0.12 	&	3 	&	 BPL	&	0.24 	$\pm$	0.04 	&	1.22 	$\pm$	 0.11 	 &	3	 &	 BPL	&	0.15 	$\pm$	0.23 	&	0.28 	$\pm$	 0.13 	&	3.5 		 \\
060111B	&	0.70 	$\pm$	0.10 	&	0.95 	$\pm$	0.18 	&	0.80 	$\pm$	0.07 	&	 1.55 	 $\pm$	0.08 	&	3 	&	 BPL	&	0.90 	$\pm$	0.15 	&	1.59 	$\pm$	 0.12 	 &	3	 &	 BPL	&	0.04 	$\pm$	0.20 	&	0.25 	$\pm$	 0.28 	&	7.2 		 \\
060206	&	0.73 	$\pm$	0.05 	&	1.20 	$\pm$	0.31 	&	0.42 	$\pm$	0.09 	&	 1.43 	 $\pm$	0.10 	&	3 	&	 BPL	&	0.40 	$\pm$	0.09 	&	1.50 	$\pm$	 0.06 	 &	3	 &	 BPL	&	0.07 	$\pm$	0.16 	&	0.47 	$\pm$	 0.36 	&	12.5 		 \\
060418	&	0.78 	$\pm$	0.09 	&	1.10 	$\pm$	0.10 	&	1.23 	$\pm$	0.07 	&		   ... 		&	3 	&	BPL	&	 1.33 	$\pm$	0.06 	&		   ... 		&	3	&	BPL	&	 0.10 	 $\pm$	0.13 	&	0.32 	$\pm$	0.19 	&			 \\
060512	&	0.68 	$\pm$	0.05 	&	1.04 	$\pm$	0.10 	&	0.81 	$\pm$	0.05 	&		   ... 		&		&	SPL	&	 1.20 	$\pm$	0.07 	&		   ... 		&		&	SPL	&	 0.39 	 $\pm$	0.12 	&	0.36 	$\pm$	0.15 	&			 \\
060714	&	0.44 	$\pm$	0.04 	&	1.10 	$\pm$	0.19 	&	0.15 	$\pm$	0.06 	&	 1.04 	 $\pm$	0.15 	&	3 	&	 BPL	&	0.48 	$\pm$	0.09 	&	1.34 	$\pm$	 0.11 	 &	3	 &	 BPL	&	0.30 	$\pm$	0.26 	&	0.66 	$\pm$	 0.23 	&	5.9 		 \\
060729	&	0.78 	$\pm$	0.03 	&	1.02 	$\pm$	0.04 	&	0.10 	$\pm$	0.05 	&	 1.40 	 $\pm$	0.15 	&	3 	&	 BPL	&	0.05 	$\pm$	0.01 	&	1.45 	$\pm$	 0.11 	 &	3	 &	 BPL	&	0.05 	$\pm$	0.26 	&	0.24 	$\pm$	 0.07 	&	53.0 		 \\
060904B	&	1.11 	$\pm$	0.10 	&	1.19 	$\pm$	0.15 	&	1.10 	$\pm$	0.05 	&		   ... 		&	3 	&	BPL	&	 1.41 	$\pm$	0.18 	&		   ... 		&	3	&	BPL	&	 0.31 	 $\pm$	0.23 	&	0.08 	$\pm$	0.25 	&	2.4 		 \\
060912A	&	0.60 	$\pm$	0.15 	&	0.62 	$\pm$	0.20 	&	0.94 	$\pm$	0.03 	&		   ... 		&		&	SPL	&	 1.07 	$\pm$	0.02 	&		   ... 		&		&	SPL	&	 0.13 	 $\pm$	0.05 	&	0.02 	$\pm$	0.35 	&			 \\
060927	&	0.61 	$\pm$	0.05 	&	0.77 	$\pm$	0.20 	&	1.30 	$\pm$	0.10 	&		   ... 		&	3 	&	BPL	&	 1.30 	$\pm$	0.07 	&		   ... 		&	3	&	BPL	&	 0.00 	 $\pm$	0.17 	&	0.16 	$\pm$	0.25 	&	0.9 		 \\
061007	&	1.02 	$\pm$	0.05 	&	1.00 	$\pm$	0.10 	&	1.62 	$\pm$	0.08 	&		   ... 		&	3 	&	BPL	&	 1.66 	$\pm$	0.07 	&		   ... 		&		&	SPL	&	 0.04 	 $\pm$	0.15 	&	-0.02 	$\pm$	0.15 	&			 \\
061126	&	0.82 	$\pm$	0.09 	&	0.85 	$\pm$	0.17 	&	1.29 	$\pm$	0.04 	&		   ... 		&	3 	&	BPL	&	 1.34 	$\pm$	0.05 	&		   ... 		&		&	SPL	&	 0.05 	 $\pm$	0.09 	&	0.03 	$\pm$	0.26 	&	6.0 	 $^{O}$	\\
070318	&	0.78 	$\pm$	0.10 	&	0.97 	$\pm$	0.11 	&	1.02 	$\pm$	0.10 	&		   ... 		&	3 	&	BPL	&	 1.03 	$\pm$	0.02 	&		   ... 		&		&	SPL	&	 0.01 	 $\pm$	0.12 	&	0.19 	$\pm$	0.21 	&			 \\
070411	&	0.75 			&	1.24 	$\pm$	0.22 	&	0.50 	$\pm$	0.08 	&	1.50 	 $\pm$	 0.11 	&	3 	&	BPL	&	 1.10 	$\pm$	0.06 	&	1.40 	$\pm$	0.09 	&	3	 &	 BPL	 &	 -0.10 	$\pm$	0.20 	&	0.49 	$\pm$	0.22 	&	 65.0 		\\
070518	&	0.80 			&	1.20 	$\pm$	0.34 	&	0.70 	$\pm$	0.07 	&	1.80 	 $\pm$	 0.11 	&	3 	&	BPL	&	 0.41 	$\pm$	0.06 	&	1.51 	$\pm$	0.09 	&	3	 &	 BPL	 &	 -0.29 	$\pm$	0.20 	&	0.40 	$\pm$	0.34 	&	 40.1 		\\
071025	&	0.96 	$\pm$	0.14 	&	1.08 	$\pm$	0.11 	&	1.43 	$\pm$	0.06 	&		   ... 		&	3 	&	BPL	&	 1.52 	$\pm$	0.08 	&		   ... 		&		&	SPL	&	 0.09 	 $\pm$	0.14 	&	0.12 	$\pm$	0.25 	&			 \\
071031	&	0.64 	$\pm$	0.05 	&	0.71 	$\pm$	0.14 	&	0.79 	$\pm$	0.05 	&		   ... 		&	3 	&	BPL	&	 0.82 	$\pm$	0.05 	&		   ... 		&		&	SPL	&	 0.03 	 $\pm$	0.10 	&	0.07 	$\pm$	0.19 	&			 \\
080319C	&	0.98 	$\pm$	0.42 	&	0.61 	$\pm$	0.10 	&	1.12 	$\pm$	0.13 	&		   ... 		&	3 	&	BPL	&	 1.33 	$\pm$	0.08 	&		   ... 		&		&	SPL	&	 0.21 	 $\pm$	0.21 	&	-0.37 	$\pm$	0.52 	&			 \\
080413A	&	0.52 	$\pm$	0.37 	&	1.15 	$\pm$	0.24 	&	1.54 	$\pm$	0.05 	&		   ... 		&	3 	&	BPL	&	 1.68 	$\pm$	0.09 	&		   ... 		&	3	&	BPL	&	 0.14 	 $\pm$	0.14 	&	0.63 	$\pm$	0.61 	&	0.3 		 \\
080603A	&	0.98 	$\pm$	0.04 	&	1.01 	$\pm$	0.10 	&	0.95 	$\pm$	0.03 	&		   ... 		&	3 	&	BPL	&	 0.96 	$\pm$	0.05 	&		   ... 		&		&	SPL	&	 0.01 	 $\pm$	0.08 	&	0.03 	$\pm$	0.14 	&			 \\
080710	&	0.80 	$\pm$	0.09 	&	1.00 	$\pm$	0.11 	&	0.39 	$\pm$	0.05 	&	 1.32 	 $\pm$	0.11 	&	3 	&	 BPL	&	0.34 	$\pm$	0.04 	&	1.57 	$\pm$	 0.14 	 &	3	 &	 BPL	&	0.25 	$\pm$	0.25 	&	0.20 	$\pm$	 0.20 	&	6.8 		 \\
080804	&	0.43 			&	0.82 	$\pm$	0.10 	&	0.87 	$\pm$	0.01 	&		   ... 		 &		&	SPL	&	1.11 	 $\pm$	0.01 	&		   ... 		&		&	SPL	&	 0.24 	$\pm$	 0.02 	&	0.39 	$\pm$	0.10 	&			\\
080913	&	0.79 	$\pm$	0.03 	&	1.01 	$\pm$	0.23 	&	0.98 	$\pm$	0.02 	&		   ... 		&		&	SPL	&	 1.32 	$\pm$	0.15 	&		   ... 		&		&	SPL	&	 0.34 	 $\pm$	0.17 	&	0.22 	$\pm$	0.26 	&			 \\
080928	&	1.32 	$\pm$	0.22 	&	1.14 	$\pm$	0.10 	&	2.02 	$\pm$	0.12 	&		   ... 		&	3 	&	BPL	&	 1.81 	$\pm$	0.11 	&		   ... 		&	3	&	BPL	&	 -0.21 	 $\pm$	0.23 	&	-0.18 	$\pm$	0.32 	&	7.1 		 \\
081008	&	0.40 	$\pm$	0.23 	&	0.98 	$\pm$	0.11 	&	0.64 	$\pm$	0.06 	&	 1.60 	 $\pm$	0.09 	&	3 	&	 BPL	&	0.87 	$\pm$	0.15 	&	1.68 	$\pm$	 0.08 	 &	3	 &	 BPL	&	0.08 	$\pm$	0.17 	&	0.58 	$\pm$	 0.34 	&	9.5 		 \\
081203A	&	0.60 			&	1.04 	$\pm$	0.10 	&	1.15 	$\pm$	0.07 	&	1.87 	 $\pm$	 0.13 	&	3 	&	BPL	&	 1.04 	$\pm$	0.09 	&	1.89 	$\pm$	0.11 	&	3	 &	 BPL	 &	 0.02 	$\pm$	0.24 	&	0.44 	$\pm$	0.10 	&	 7.1 		\\
090102	&	0.74 	$\pm$	0.22 	&	0.79 	$\pm$	0.11 	&	0.20 	$\pm$	0.04 	&	 1.16 	 $\pm$	0.09 	&	3 	&	 BPL	&	0.31 	$\pm$	0.05 	&	1.41 	$\pm$	 0.09 	 &	3	 &	 BPL	&	0.25 	$\pm$	0.18 	&	0.05 	$\pm$	 0.33 	&	1.0 		 \\
090323	&	0.74 	$\pm$	0.15 	&	0.87 	$\pm$	0.22 	&	1.55 	$\pm$	0.05 	&		   ... 		&		&	SPL	&	 1.62 	$\pm$	0.09 	&		   ... 		&		&	SPL	&	 0.07 	 $\pm$	0.14 	&	0.13 	$\pm$	0.37 	&			 \\
090328	&	1.19 	$\pm$	0.21 	&	0.90 	$\pm$	0.30 	&	1.84 	$\pm$	0.08 	&		   ... 		&		&	SPL	&	 1.67 	$\pm$	0.11 	&		   ... 		&		&	SPL	&	 -0.17 	 $\pm$	0.19 	&	-0.29 	$\pm$	0.51 	&			 \\
090426	&	0.76 	$\pm$	0.14 	&	1.03 	$\pm$	0.15 	&	0.14 	$\pm$	0.09 	&	 1.25 	 $\pm$	0.04 	&	3 	&	 BPL	&	0.13 	$\pm$	0.02 	&	1.04 	$\pm$	 0.05 	 &	3	 &	 BPL	&	-0.21 	$\pm$	0.09 	&	0.27 	$\pm$	 0.29 	&	0.2 		 \\
090618	&	0.50 	$\pm$	0.05 	&	0.92 	$\pm$	0.05 	&	0.76 	$\pm$	0.11 	&	 1.53 	 $\pm$	0.11 	&	3 	&	 BPL	&	0.93 	$\pm$	0.09 	&	1.74 	$\pm$	 0.10 	 &	3	 &	 BPL	&	0.21 	$\pm$	0.21 	&	0.42 	$\pm$	 0.10 	&	45.1 		 \\
090926A	&	0.72 	$\pm$	0.17 	&	0.98 	$\pm$	0.15 	&	1.34 	$\pm$	0.05 	&		   ... 		&		&	SPL	&	 1.41 	$\pm$	0.03 	&		   ... 		&		&	SPL	&	 0.07 	 $\pm$	0.08 	&	0.26 	$\pm$	0.32 	&			 \\
091127	&	0.18 			&	0.68 	$\pm$	0.11 	&	0.55 	$\pm$	0.11 	&	1.50 	 $\pm$	 0.11 	&	3 	&	BPL	&	 0.96 	$\pm$	0.05 	&	1.59 	$\pm$	0.12 	&	3	 &	 BPL	 &	 0.09 	$\pm$	0.23 	&	0.50 	$\pm$	0.11 	&	 35.3 		\\
100418A	&	0.98 	$\pm$	0.09 	&	1.04 	$\pm$	0.29 	&	0.11 	$\pm$	0.01 	&	 1.60 	 $\pm$	0.10 	&	3 	&	 BPL	&	-0.12 	$\pm$	0.03 	&	1.57 	$\pm$	 0.11 	 &	3	 &	 BPL	&	-0.03 	$\pm$	0.21 	&	0.06 	$\pm$	 0.38 	&	90.1 		 \\
100901A	&	0.52 	$\pm$	0.10 	&	1.00 	$\pm$	0.30 	&	1.42 	$\pm$	0.02 	&		   ... 		&	3 	&	BPL	&	 1.41 	$\pm$	0.02 	&		   ... 		&	3	&	BPL	&	 -0.01 	 $\pm$	0.04 	&	0.48 	$\pm$	0.40 	&	29.8 		 \\
101024A	&	0.70 	$\pm$	0.40 	&	0.82 	$\pm$	0.13 	&	0.01 	$\pm$	0.05 	&	 1.07 	 $\pm$	0.08 	&	3 	&	 BPL	&	-0.09 	$\pm$	0.07 	&	1.37 	$\pm$	 0.10 	 &	3	 &	 BPL	&	0.30 	$\pm$	0.18 	&	0.12 	$\pm$	 0.53 	&	1.0 		 \\
120326A	&	0.75 	$\pm$	0.08 	&	0.77 	$\pm$	0.06 	&	1.52 	$\pm$	0.10 	&		   ... 		&	3 	&	BPL	&	 1.69 	$\pm$	0.09 	&		   ... 		&	3	&	BPL	&	 0.17 	 $\pm$	0.19 	&	0.02 	$\pm$	0.14 	&	35.50 		 \\
130427A	&	0.69 	$\pm$	0.01 	&	0.68 	$\pm$	0.16 	&	1.02 	$\pm$	0.09 	&	 1.84 	 $\pm$	0.11 	&	3 	&	 BPL	&	1.09 	$\pm$	0.07 	&	1.63 	$\pm$	 0.09 	 &	3	 &	 BPL	&	-0.21 	$\pm$	0.20 	&	-0.01 	$\pm$	 0.17 	&	127.5 		 \\
\hline	&				&				&	&			&				&								 &				&		&						 &				&			\\
Grade II	&				&				&	&			&				&								 &				&		 &						&				&			\\
\hline	&				&				&	&			&				&								 &				&		&						 &				&			\\
051111	&	0.78 	$\pm$	0.07 	&	1.24 	$\pm$	0.17 	&	1.56 	$\pm$	0.11 	&		   ... 		&	3 	&	BPL	&	 1.60 	$\pm$	0.12 	&		   ... 		&		&	SPL	&	 0.04 	 $\pm$	0.23 	&	0.46 	$\pm$	0.24 	&	3.0 	 $^{O}$	\\
090313	&	0.74 	$\pm$	0.40 	&	1.08 	$\pm$	0.17 	&	1.55 	$\pm$	0.12 	&		   ... 		&	3 	&	BPL	&	 1.67 	$\pm$	0.10 	&		   ... 		&		&	SPL	&	 0.12 	 $\pm$	0.22 	&	0.34 	$\pm$	0.57 	&	20.5 	 $^{O}$	\\
\hline	&				&				&	&			&				&								 &				&		&						 &				&			\\
Grade III	&				&				&	&			&				&								 &				&		 &						&				&			\\
\hline	&				&				&	&			&				&								 &				&		&						 &				&			\\
050319	&	0.74 	$\pm$	0.42 	&	1.01 	$\pm$	0.07 	&	0.39 	$\pm$	0.06 	&	 1.02 	 $\pm$	0.04 	&	3 	&	 BPL	&	0.58 	$\pm$	0.07 	&	1.72 	$\pm$	 0.11 	 &	3	 &	 BPL	&	0.70 	$\pm$	0.15 	&	0.27 	$\pm$	 0.49 	&	55.0 		 \\
050401	&	0.50 	$\pm$	0.20 	&	0.79 	$\pm$	0.13 	&	0.50 	$\pm$	0.08 	&	 0.89 	 $\pm$	0.08 	&	3 	&	 SPL	&	0.76 	$\pm$	0.05 	&	1.67 	$\pm$	 0.11 	 &	3	 &	 BPL	&	0.78 	$\pm$	0.19 	&	0.29 	$\pm$	 0.33 	&	4.3 	 $^{X}$	\\
050416A	&	1.30 			&	1.07 	$\pm$	0.11 	&	0.26 	$\pm$	0.07 	&	1.12 	 $\pm$	 0.12 	&	3 	&	BPL	&	 0.66 	$\pm$	0.12 	&	0.99 	$\pm$	0.02 	&	3	 &	 BPL	 &	 -0.13 	$\pm$	0.14 	&	-0.23 	$\pm$	0.11 	&	 11.0 		\\
050525A	&	0.52 	$\pm$	0.08 	&	1.09 	$\pm$	0.16 	&	1.46 	$\pm$	0.07 	&		   ... 		&	3 	&	BPL	&	 1.57 	$\pm$	0.04 	&		   ... 		&	3	&	BPL	&	 0.11 	 $\pm$	0.11 	&	0.57 	$\pm$	0.24 	&	4.2 	 $^{O}$	\\
050603	&	0.20 	$\pm$	0.10 	&	1.02 	$\pm$	0.13 	&	1.70 	$\pm$	0.13 	&		   ... 		&		&	SPL	&	 1.71 	$\pm$	0.05 	&		   ... 		&		&	SPL	&	 0.01 	 $\pm$	0.18 	&	0.82 	$\pm$	0.23 	&			 \\
050721	&	1.16 	$\pm$	0.35 	&	0.85 	$\pm$	0.22 	&	0.60 	$\pm$	0.03 	&		   ... 		&		&	SPL	&	 1.01 	$\pm$	0.08 	&		   ... 		&		&	SPL	&	 0.41 	 $\pm$	0.11 	&	-0.31 	$\pm$	0.57 	&			 \\
050730	&	0.52 	$\pm$	0.05 	&	1.62 	$\pm$	0.04 	&	0.48 	$\pm$	0.05 	&	 1.47 	 $\pm$	0.06 	&	3 	&	 BPL	&	0.45 	$\pm$	0.13 	&	2.64 	$\pm$	 0.20 	 &	3	 &	 BPL	&	1.17 	$\pm$	0.26 	&	1.10 	$\pm$	 0.09 	&	90.1 		 \\
051221A	&	0.64 	$\pm$	0.05 	&	1.06 	$\pm$	0.14 	&	0.34 	$\pm$	0.07 	&	 1.24 	 $\pm$	0.04 	&	3 	&	 BPL	&	0.35 	$\pm$	0.08 	&	1.34 	$\pm$	 0.04 	 &	3	 &	 BPL	&	0.10 	$\pm$	0.08 	&	0.42 	$\pm$	 0.19 	&	25.1 		 \\
060210	&	0.37 	$\pm$	0.08 	&	1.08 	$\pm$	0.08 	&	0.53 	$\pm$	0.05 	&	 1.77 	 $\pm$	0.14 	&	3 	&	 BPL	&	0.53 	$\pm$	0.06 	&	1.30 	$\pm$	 0.12 	 &	3	 &	 BPL	&	-0.47 	$\pm$	0.26 	&	0.71 	$\pm$	 0.16 	&	5.0 		 \\
060526	&	0.51 	$\pm$	0.32 	&	0.90 	$\pm$	0.11 	&	0.56 	$\pm$	0.10 	&	 1.93 	 $\pm$	0.10 	&	1 	&	 BPL	&	0.67 	$\pm$	0.08 	&	2.06 	$\pm$	 0.15 	 &	3	 &	 BPL	&	0.13 	$\pm$	0.25 	&	0.39 	$\pm$	 0.43 	&	50.1 		 \\
060605	&	1.06 			&	1.02 	$\pm$	0.09 	&	0.13 	$\pm$	0.09 	&	2.64 	 $\pm$	 0.15 	&	3 	&	BPL	&	 0.55 	$\pm$	0.08 	&	2.82 	$\pm$	0.15 	&	3	 &	 BPL	 &	 0.18 	$\pm$	0.30 	&	-0.04 	$\pm$	0.09 	&	 15.0 		\\
060614	&	0.47 	$\pm$	0.04 	&	0.90 	$\pm$	0.09 	&	-0.35 	$\pm$	0.04 	&	 1.90 	 $\pm$	0.12 	&	3 	&	 BPL	&	0.11 	$\pm$	0.10 	&	1.97 	$\pm$	 0.15 	 &	3	 &	 BPL	&	0.07 	$\pm$	0.27 	&	0.43 	$\pm$	 0.13 	&	44.0 		 \\
060906	&	0.56 	$\pm$	0.02 	&	1.08 	$\pm$	0.17 	&	1.45 	$\pm$	0.21 	&		   ... 		&	3 	&	BPL	&	 1.25 	$\pm$	0.11 	&		   ... 		&	3	&	BPL	&	 -0.20 	 $\pm$	0.32 	&	0.52 	$\pm$	0.19 	&	1.3 		 \\
060906	&	0.56 	$\pm$	0.02 	&	1.08 	$\pm$	0.17 	&	1.44 	$\pm$	0.20 	&		   ... 		&	3 	&	BPL	&	 1.90 	$\pm$	0.11 	&		   ... 		&	3	&	BPL	&	 0.46 	 $\pm$	0.31 	&	0.52 	$\pm$	0.19 	&	10.5 		 \\
060908	&	0.24 	$\pm$	0.20 	&	1.13 	$\pm$	0.19 	&	0.71 	$\pm$	0.04 	&	 1.11 	 $\pm$	0.09 	&	3 	&	 BPL	&	0.54 	$\pm$	0.05 	&	1.66 	$\pm$	 0.05 	 &	3	 &	 BPL	&	0.55 	$\pm$	0.14 	&	0.89 	$\pm$	 0.39 	&	1.1 		 \\
070110	&	0.55 	$\pm$	0.04 	&	1.08 	$\pm$	0.11 	&	0.16 	$\pm$	0.06 	&	 1.67 	 $\pm$	0.12 	&	3 	&	 BPL	&	0.30 	$\pm$	0.10 	&	5.10 	$\pm$	 0.30 	 &	3	 &	 BPL	&	3.43 	$\pm$	0.42 	&	0.53 	$\pm$	 0.15 	&	20.3 		 \\
070125	&	0.59 	$\pm$	0.10 	&	1.03 	$\pm$	0.20 	&	2.96 	$\pm$	0.05 	&		   ... 		&		&	BPL	&	 2.12 	$\pm$	0.04 	&		   ... 		&	3	&	BPL	&	 -0.84 	 $\pm$	0.09 	&	0.44 	$\pm$	0.30 	&	101.1 		 \\
070306	&	0.70 	$\pm$	0.10 	&	0.95 	$\pm$	0.09 	&	1.22 	$\pm$	0.11 	&		   ... 		&	3 	&	BPL	&	 2.03 	$\pm$	0.04 	&		   ... 		&	3	&	BPL	&	 0.81 	 $\pm$	0.15 	&	0.25 	$\pm$	0.19 	&	36.9 		 \\
070311	&	1.00 	$\pm$	0.20 	&	1.00 	$\pm$	0.24 	&	0.73 	$\pm$	0.02 	&		   ... 		&		&	SPL	&	 1.09 	$\pm$	0.06 	&		   ... 		&		&	SPL	&				 &	0.00 	$\pm$	0.44 	&			\\
070419A	&	0.48 	$\pm$	0.48 	&	1.20 	$\pm$	0.30 	&	1.28 	$\pm$	0.04 	&		   ... 		&	3 	&	BPL	&	 0.60 	$\pm$	0.02 	&		   ... 		&	3	&	BPL	&	 -0.68 	 $\pm$	0.06 	&	0.72 	$\pm$	0.78 	&			 \\
071010A	&	0.61 	$\pm$	0.12 	&	1.30 	$\pm$	0.40 	&	2.19 	$\pm$	0.08 	&		   ... 		&	3 	&	BPL	&	 1.89 	$\pm$	0.07 	&		   ... 		&	3	&	BPL	&	 -0.30 	 $\pm$	0.15 	&	0.69 	$\pm$	0.52 	&	70.1 		 \\
071112C	&	0.63 	$\pm$	0.29 	&	0.67 	$\pm$	0.13 	&	0.30 	$\pm$	0.07 	&	 0.95 	 $\pm$	0.11 	&	3 	&	 BPL	&	0.50 	$\pm$	0.08 	&	1.49 	$\pm$	 0.11 	 &	3	 &	 BPL	&	0.54 	$\pm$	0.22 	&	0.04 	$\pm$	 0.42 	&	1.5 		 \\
080310	&	0.42 	$\pm$	0.12 	&	0.95 	$\pm$	0.18 	&	0.11 	$\pm$	0.01 	&	 1.24 	 $\pm$	0.02 	&	3 	&	 BPL	&	0.03 	$\pm$	0.06 	&	1.24 	$\pm$	 0.08 	 &	3	 &	 BPL	&	0.00 	$\pm$	0.10 	&	0.53 	$\pm$	 0.30 	&	5.1 		 \\
080319A	&	0.77 	$\pm$	0.02 	&	0.89 	$\pm$	0.10 	&	0.65 	$\pm$	0.07 	&		   ... 		&	3 	&	BPL	&	 0.94 	$\pm$	0.05 	&		   ... 		&		&		&	 0.29 	 $\pm$	0.12 	&	0.12 	$\pm$	0.12 	&			 \\
080319B	&	0.51 	$\pm$	0.26 	&	0.81 	$\pm$	0.07 	&	0.93 	$\pm$	0.11 	&	 1.60 	 $\pm$	0.12 	&	3 	&	 BPL	&	0.73 	$\pm$	0.05 	&	2.73 	$\pm$	 0.16 	 &	3	 &	 BPL	&	1.13 	$\pm$	0.28 	&	0.30 	$\pm$	 0.33 	&	3.0 		 \\
080319B	&	0.51 	$\pm$	0.26 	&	0.81 	$\pm$	0.07 	&	1.03 	$\pm$	0.10 	&	 1.69 	 $\pm$	0.09 	&	3 	&	 BPL	&	1.43 	$\pm$	0.09 	&	2.19 	$\pm$	 0.11 	 &	3	 &	 BPL	&	0.50 	$\pm$	0.20 	&	0.30 	$\pm$	 0.33 	&	690.7 		 \\
080413B	&	0.25 	$\pm$	0.07 	&	0.94 	$\pm$	0.07 	&	0.31 	$\pm$	0.15 	&	 1.89 	 $\pm$	0.22 	&	3 	&	 BPL	&	0.92 	$\pm$	0.16 	&	1.91 	$\pm$	 0.23 	 &	3	 &	 BPL	&	0.02 	$\pm$	0.45 	&	0.69 	$\pm$	 0.14 	&	148.5 		 \\
080721	&	0.68 	$\pm$	0.02 	&	0.94 	$\pm$	0.06 	&	1.17 	$\pm$	0.03 	&	 1.31 	 $\pm$	0.05 	&	3 	&	 BPL	&	0.81 	$\pm$	0.01 	&	1.65 	$\pm$	 0.07 	 &	3	 &	 BPL	&	0.34 	$\pm$	0.12 	&	0.26 	$\pm$	 0.08 	&	3.1 		 \\
090510	&	0.85 	$\pm$	0.05 	&	0.75 	$\pm$	0.12 	&	0.84 	$\pm$	0.05 	&		   ... 		&	3 	&	BPL	&	 2.27 	$\pm$	0.06 	&		   ... 		&	3	&	BPL	&	 1.43 	 $\pm$	0.11 	&	-0.10 	$\pm$	0.17 	&	1.5 		 \\
090812	&	0.36 			&	0.89 	$\pm$	0.14 	&	1.27 	$\pm$	0.05 	&		   ... 		 &	3 	&	BPL	&	1.22 	 $\pm$	0.09 	&		   ... 		&		&	SPL	&	 -0.05 	$\pm$	 0.14 	&	0.53 	$\pm$	0.14 	&			\\
091029	&	0.49 	$\pm$	0.12 	&	1.12 	$\pm$	0.08 	&	0.48 	$\pm$	0.06 	&	 1.34 	 $\pm$	0.09 	&	3 	&	 BPL	&	0.32 	$\pm$	0.09 	&	1.35 	$\pm$	 0.08 	 &	3	 &	 BPL	&	0.01 	$\pm$	0.17 	&	0.63 	$\pm$	 0.20 	&	20.8 		 \\
100219A	&	0.56 			&	0.69 	$\pm$	0.23 	&	0.74 	$\pm$	0.08 	&	1.91 	 $\pm$	 0.12 	&	3 	&	BPL	&	 0.54 	$\pm$	0.07 	&	1.65 	$\pm$	0.15 	&	3	 &	 BPL	 &	 -0.26 	$\pm$	0.27 	&	0.13 	$\pm$	0.23 	&	 1.8 		\\
100219A	&	0.56 			&	0.69 	$\pm$	0.23 	&	2.21 	$\pm$	0.13 	&		   ... 		 &	3 	&	BPL	&	2.51 	 $\pm$	0.16 	&		   ... 		&	3	&	BPL	&	 0.30 	$\pm$	 0.29 	&	0.13 	$\pm$	0.23 	&	20.5 		\\
110205A	&	0.49 	$\pm$	0.08 	&	0.78 	$\pm$	0.06 	&	1.51 	$\pm$	0.08 	&		   ... 		&	3 	&	SPL	&	 1.59 	$\pm$	0.02 	&		   ... 		&		&	SPL	&	 0.08 	 $\pm$	0.10 	&	0.29 	$\pm$	0.14 	&			 \\
110918A	&	0.42 	$\pm$	0.18 	&	0.89 	$\pm$	0.30 	&	1.65 	$\pm$	0.07 	&		   ... 		&		&	SPL	&	 1.61 	$\pm$	0.12 	&		   ... 		&		&	SPL	&	 -0.04 	 $\pm$	0.19 	&	0.47 	$\pm$	0.48 	&			 \\
120729A	&	1.00 	$\pm$	0.10 	&	0.80 	$\pm$	0.17 	&	0.94 	$\pm$	0.05 	&	 2.27 	 $\pm$	0.09 	&	3 	&	 BPL	&	1.09 	$\pm$	0.06 	&	2.40 	$\pm$	 0.16 	 &	3	 &	 BPL	&	0.13 	$\pm$	0.25 	&	-0.20 	$\pm$	 0.27 	&	6.61		 \\
120815A	&	0.78 	$\pm$	0.01 	&	0.72 	$\pm$	0.11 	&	0.63 	$\pm$	0.04 	&		   ... 		&	3 	&	BPL	&	 0.86 	$\pm$	0.06 	&		   ... 		&		&	SPL	&	 0.23 	 $\pm$	0.10 	&	-0.06 	$\pm$	0.12 	&			 \\
\hline	&				&				&				&				&								 &				&		&						 &				&			\\
Grade IV	&				&				&				&				&								 &				&		 &						&				&			\\
\hline	&				&				&				&				&								 &				&		&						 &				&			\\
070611	&	0.73 	$\pm$	0.00 	&	0.83 	$\pm$	0.31 	&	0.58 	$\pm$	0.04 	&		   ... 		&	3 	&	BPL	&	 1.34 	$\pm$	0.12 	&		   ... 		&	3	&	BPL	&	 0.76 	 $\pm$	0.16 	&	0.10 	$\pm$	0.31 	&	33.7 	 $^{X}$	\\
071003	&	0.35 	$\pm$	0.11 	&	0.91 	$\pm$	0.12 	&	1.62 	$\pm$	0.11 	&		   ... 		&	3 	&	BPL	&	 1.63 	$\pm$	0.02 	&		   ... 		&		&	SPL	&	 0.01 	 $\pm$	0.13 	&	0.56 	$\pm$	0.23 	&			 \\
120711A	&	0.52 	$\pm$	0.02 	&	0.81 	$\pm$	0.10 	&	0.96 	$\pm$	0.04 	&		   ... 		&	3 	&	BPL	&	 1.64 	$\pm$	0.05 	&		   ... 		&		&	SPL	&	 0.68 	 $\pm$	0.09 	&	0.29 	$\pm$	0.12 	&			 \\
\hline	&				&				&				&				&								 &				&		&						 &				&			\\
Grade V	&				&				&				&				&								 &				&		&						 &				&			\\
\hline	&				&				&				&				&								 &				&		&						 &				&			\\
060607A	&	0.72 	$\pm$	0.27 	&	0.62 	$\pm$	0.06 	&	-0.93 			&	4.60 			 &	3 	&	BPL	&	0.36 	 $\pm$	0.03 	&	3.10 	$\pm$	0.12 	&	3 	&	BPL	&				 &	-0.10 	$\pm$	0.33 	&	9.5 	$^{X}$	\\
070208	&	0.68 			&	1.20 	$\pm$	0.20 	&	0.49 	$\pm$	0.06 	&		   ... 		 &	3 	&	BPL	&	0.43 	 $\pm$	0.03 	&	1.78 	$\pm$	0.13 	&	3 	&	 BPL	&				 &	0.52 	$\pm$	0.20 	&	9.0 	$^{X}$	\\
070420	&	   ... 			&	   ... 			&	-1.43 	$\pm$	0.04 	&	0.90 	$\pm$	 0.08 	 &	3	&	BPL	&	0.12 	 $\pm$	0.01 	&	1.46 	$\pm$	0.05 	&	3	&	BPL	 &				 &				&	3.0 	$^{X}$	\\

\enddata
\tablenotetext{a}{For single power-law (SPL) decay lightcurves (as described in
section 3), the decay indices are also denoted as $\alpha_1$; }
\tablenotetext{b}{$\bigtriangleup \alpha_{\rm X,O}=\alpha_{\rm X,2}-\alpha_{\rm O,2}$;}
\tablenotetext{c}{$\bigtriangleup \beta_{\rm X,O}=\beta_{X}-\beta_{O}$;}
\tablenotetext{d}{In units of ks. The symbols ``X'' and ``O'' denote
the X-ray and optical bands, respectively.}

\label{table:sample}
\end{deluxetable}

\clearpage \thispagestyle{empty} \setlength{\voffset}{-18mm}
\begin{deluxetable}{ccccccccccccccccccccccccc}
\tabletypesize{\scriptsize}

\tablecaption{The temporal decay index $\alpha$ and spectral index
$\beta$ in different afterglow models.}
\tablehead{ \colhead{CMB}& \colhead{Spectral regime}&
\colhead{$\beta(p)$}& \colhead{$\alpha(p)/\alpha(p,q)$}&
\colhead{$\alpha(\beta)/\alpha(\beta,q)$}&
\colhead{$\alpha(p)/\alpha(p,q)$}&
\colhead{$\alpha(\beta)/\alpha(\beta,q)$}}
 \startdata
& & & \multicolumn{2}{c}{$p>2$}& \multicolumn{2}{c}{$1<p<2$} \\
\hline
     & \multicolumn{4}{c}{  Adiabatic deceleration without energy injection}                    \\
\hline ISM & $\nu_m<\nu<\nu_c$   &  ${{p-1 \over 2}}$  &
${3(p-1)\over 4}$
&  $\alpha={3\beta \over 2}$ &  ${3(p+2)\over 16}$  & $\alpha={6\beta+9 \over 16}$ \\
 &$\nu>\nu_c$   &  ${{p\over 2}}$   &   ${3p-2 \over 4}$  & $\alpha={3\beta-1 \over 2}$ &${3p+10 \over 16}$  & $\alpha={3\beta+5 \over 8}$\\

Wind & $\nu_m<\nu<\nu_c$   &  ${p-1\over 2}$  &   ${3p-1\over 4}$ &
$\alpha={3\beta+1 \over 2}$ &   ${p+8\over 8}$    &   $\alpha={2\beta+9 \over 8}$  \\
     & $\nu>\nu_c$   &  ${p\over 2}$   &   ${3p-2\over 4}$    &  $\alpha={3\beta-1
     \over 2}$   &   ${p+6\over 8}$    &  $\alpha={2\beta+6 \over 8}$  \\
\hline
     & \multicolumn{4}{c}{  Adiabatic deceleration with energy injection}                   \\
\hline
ISM & $\nu_m<\nu<\nu_c$   &  ${{p-1 \over 2}}$ & ${(2p-6)+(p+3)q \over 4}$   & $\alpha=(q-1)+\frac{(2+q)\beta}{2}$& $-{12-18q-p(q+2) \over 16}$   & $\alpha=\frac{19q-10}{16}+\frac{(2+q)\beta}{8}$\\
 &$\nu>\nu_c$   &  ${{p\over 2}}$   & ${(2p-4)+(p+2)q\over 4}$   & $\alpha=\frac{q-2}{2}+\frac{(2+q)\beta}{2}$ & ${14q+p(q+2)-4\over 16}$   & $\alpha=\frac{7q-2}{8}+\frac{(2+q)\beta}{8}$\\

Wind & $\nu_m<\nu<\nu_c$   &  ${p-1\over 2}$ & ${(2p-2)+(p+1)q \over 4}$ & $\alpha=\frac{q}{2}+\frac{(2+q)\beta}{2}$ & ${4+(p+4)q \over 8}$ & $\alpha=\frac{5q+4}{8}+\frac{\beta q}{4}$\\
   & $\nu>\nu_c$   & ${p\over 2}$  &   ${(2p-4)+(p+2)q\over 4}$  &  $\alpha=\frac{q-2}{2}+\frac{(2+q)\beta}{2}$&   ${(6+p)q\over 8}$  &  $\alpha=\frac{(\beta+3)q}{4}$\\

\hline
     & \multicolumn{4}{c}{   Post jet break phase without energy injection}                   \\
\hline ISM & $\nu_m<\nu<\nu_c$   &  ${{p-1 \over 2}}$  &  ${3p \over
4}$
   &  $\alpha={6\beta+3 \over 4}$ & ${3(p+6) \over 16}$   & $\alpha={3(2\beta+7) \over 16}$\\
& $\nu>\nu_c$   &  ${{p\over 2}}$   &   ${3p+1 \over 4}$  & $\alpha={6\beta+1 \over 4}$ & ${3p+22 \over 16}$   & $\alpha={3\beta+11 \over 8}$ \\
Wind &$\nu_m<\nu<\nu_c$   &  ${p-1\over 2}$   &   ${3p+1\over 4}$    &   $\alpha={3\beta+2 \over 2}$ & ${p+12 \over 8}$ & $\alpha={2\beta+13 \over 8}$\\
&$\nu>\nu_c$   &  ${p\over 2}$   &   ${3p\over 4}$    &  $\alpha={3\beta \over 2}$   &   ${p+10 \over 8}$  &  $\alpha={\beta+5 \over 4}$\\
\hline
     & \multicolumn{4}{c}{   Post jet break phase with energy injection}                   \\
\hline

ISM & $\nu_m<\nu<\nu_c$   &  ${{p-1 \over 2}}$  &  ${p(q+2)-4(1-q)
\over 4}$
   &  $\alpha=\frac{5q-2}{4}+\frac{(2+q)\beta}{2}$ & ${22q-4+p(q+2) \over 16}$   & $\alpha=\frac{11q-2}{8}+\frac{(2+q)\beta}{8}$\\

& $\nu>\nu_c$   &  ${{p\over 2}}$   &   ${3q-2+p(q+2) \over 4}$  & $\alpha={3q-2+2\beta(q+2) \over 4}$ & ${18q+4+p(q+2) \over 16}$   & $\alpha={9q+2+\beta(q+2) \over 8}$ \\

Wind & $\nu_m<\nu<\nu_c$   &  ${p-1\over 2}$   &   ${3q-2+p(q+2)\over 4}$    &   $\alpha=q+\frac{(2+q)\beta}{2}$ & ${pq+8q+4 \over 8}$ & $\alpha=\frac{1}{2}+\frac{(2\beta+9)q}{8}$\\
& $\nu>\nu_c$   &  ${p\over 2}$   &   ${p(q+2)-4(1-q)\over 4}$    &  $\alpha={\beta(q+2)-2(1-q)\over 2}$   &   ${(p+10)q \over 8}$  &  $\alpha={(\beta+5)q \over 4}$\\

\hline

\enddata

 \label{Tab:alpha-beta}
\end{deluxetable}

\clearpage
\begin{deluxetable}{ccccccccccccccccccccccccc}
\tabletypesize{\scriptsize}
\tablecaption{The $\Delta \alpha_{X,O}$ and $\Delta \beta_{X,O}$ values in different afterglow models, $p>2$.}

\tablewidth{0pt}
\tabletypesize{\tiny}
\tablehead{ \colhead{}&
\colhead{$$}&
\colhead{$p>2$}&
\colhead{$$}&
\colhead{$$}}
 \startdata
Same regime$^{a}$& \multicolumn{2}{c}{Different regimes$^{b}$} & \multicolumn{2}{c}{Grey zone$^{c}$} \\
\hline
ISM,wind & ISM & wind  & ISM & wind \\
\hline
   && \multicolumn{2}{c}{SPL}                    \\
\hline
$\Delta \beta_{X,O}$=0 & $\Delta \beta_{X,O}$=$\frac{1}{2}$ & $\Delta \beta_{X,O}$=$\frac{1}{2}$ & $\Delta \beta_{X,O}$=(0,$\frac{1}{2}$) & $\Delta \beta_{X,O}$=(0,$\frac{1}{2}$) \\
$\Delta \alpha_{X,O}$=0 & $\Delta \alpha_{X,O}$=$\frac{1}{4}$ & $\Delta \alpha_{X,O}$=-$\frac{1}{4}$  & $\Delta \alpha_{X,O}$=(0,$\frac{1}{4}$) & $\Delta \alpha_{X,O}$=(-$\frac{1}{4}$,0) \\
\hline
   && \multicolumn{2}{c}{ Energy injection break} \\
\hline
$\Delta \beta_{X,O}$=0 & $\Delta \beta_{X,O}$=$\frac{1}{2}$ & $\Delta \beta_{X,O}$=$\frac{1}{2}$ & $\Delta \beta_{X,O}$=(0,$\frac{1}{2}$) & $\Delta \beta_{X,O}$=(0,$\frac{1}{2}$) \\
$\Delta \alpha_{1,X,O}$=0 & $\Delta \alpha_{1,X,O}=\frac{2-q}{4}$ & $\Delta \alpha_{1,X,O}=\frac{-2+q}{4}$  & $\Delta \alpha_{1,X,O}=(0,\frac{2-q}{4})$ & $\Delta \alpha_{1,X,O}=(\frac{-2+q}{4},0)$ \\
$\Delta \alpha_{2,X,O}$=0 & $\Delta \alpha_{2,X,O}$=$\frac{1}{4}$ & $\Delta \alpha_{2,X,O}$=-$\frac{1}{4}$ & $\Delta \alpha_{2,X,O}$=(0,$\frac{1}{4}$) & $\Delta \alpha_{2,X,O}$=(-$\frac{1}{4}$,0) \\
\hline
 && \multicolumn{2}{c}{ Jet break}                   \\
\hline
$\Delta \beta_{X,O}$=0 & $\Delta \beta_{X,O}$=$\frac{1}{2}$ & $\Delta \beta_{X,O}$=$\frac{1}{2}$ & $\Delta \beta_{X,O}$=(0,$\frac{1}{2}$) & $\Delta \beta_{X,O}$=(0,$\frac{1}{2}$) \\
$\Delta \alpha_{1,X,O}$=0 & $\Delta \alpha_{1,X,O}$=$\frac{1}{4}$ & $\Delta \alpha_{1,X,O}$=-$\frac{1}{4}$  & $\Delta \alpha_{1,X,O}$=(0,$\frac{1}{4}$) & $\Delta \alpha_{1,X,O}$=(-$\frac{1}{4}$,0) \\
$\Delta \alpha_{2,X,O}$=0 & $\Delta \alpha_{2,X,O}=\frac{1}{4}$ & $\Delta \alpha_{2,X,O}$=-$\frac{1}{4}$  & $\Delta \alpha_{2,X,O}$=(0,$\frac{1}{4}$) & $\Delta \alpha_{2,X,O}$=(-$\frac{1}{4}$,0) \\
\hline
  &&\multicolumn{2}{c}{  Energy injection with jet break}                    \\
\hline
$\Delta \beta_{X,O}$=0 & $\Delta \beta_{X,O}$=$\frac{1}{2}$ & $\Delta \beta_{X,O}$=$\frac{1}{2}$ & $\Delta \beta_{X,O}$=(0,$\frac{1}{2}$) & $\Delta \beta_{X,O}$=(0,$\frac{1}{2}$) \\
$\Delta \alpha_{1,X,O}$=0 & $\Delta \alpha_{1,X,O}=\frac{2-q}{4}$ & $\Delta \alpha_{1,X,O}=\frac{-2+q}{4}$ & $\Delta \alpha_{1,X,O}=(0,\frac{2-q}{4})$ & $\Delta \alpha_{1,X,O}=(0,\frac{2-q}{4})$ \\
$\Delta \alpha_{2,X,O}$=0 & $\Delta \alpha_{2,X,O}=\frac{2-q}{4}$ & $\Delta \alpha_{2,X,O}=\frac{-2+q}{4}$ & $\Delta \alpha_{2,X,O}=(0,\frac{2-q}{4})$ & $\Delta \alpha_{2,X,O}=(0,\frac{2-q}{4})$ \\
\enddata
\tablenotetext{a}{Same regime: X-ray and optical bands are in the same spectral regime, regime I ($\nu>\nu_{c}$) or regime II ($\nu_{m}<\nu<\nu_{c}$). In this table, $\bigtriangleup \beta_{\rm X,O}=\beta_{X}-\beta_{O}$, and $\bigtriangleup \alpha_{\rm X,O}=\alpha_{\rm X}-\alpha_{\rm O}$. The subsripts ``1'' and ``2''
denote the pre- and post-break segments for the BPL lightcurves, respectively.}
\tablenotetext{b}{Different regimes: X-ray band in regime I ($\nu>\nu_{c}$), and optical band in regime II ($\nu_{m}<\nu<\nu_{c}$); }
\tablenotetext{c}{Grey zone: One band or two bands in the grey zone  regime I$-$ II. }

\label{Tab:delta-alpha-betahigh2}
\end{deluxetable}

\clearpage
\begin{deluxetable}{ccccccccccccccccccccccccc}
\tablecaption{The $\Delta \alpha_{X,O}$ and $\Delta \beta_{X,O}$ values in different afterglow models, $1<p<2$.}

\tablewidth{0pt}
\tabletypesize{\tiny}
\tablehead{ \colhead{}&
\colhead{$$}&
\colhead{$1<p<2$}&
\colhead{$$}&
\colhead{$$}}
 \startdata
Same regime$^{a}$& \multicolumn{2}{c}{Different regimes$^{b}$} & \multicolumn{2}{c}{Grey zone$^{c}$} \\
\hline
ISM,wind & ISM & wind  & ISM & wind \\
\hline
   && \multicolumn{2}{c}{SPL}                    \\
\hline
 $\Delta \beta_{X,O}$=0 & $\Delta \beta_{X,O}$=$\frac{1}{2}$ & $\Delta \beta_{X,O}$=$\frac{1}{2}$ & $\Delta \beta_{X,O}$=(0,$\frac{1}{2}$) & $\Delta \beta_{X,O}$=(0,$\frac{1}{2}$) \\
 $\Delta \alpha_{X,O}$=0 & $\Delta \alpha_{X,O}$=$\frac{1}{2}$ & $\Delta \alpha_{X,O}$=$\frac{1}{4}$ & $\Delta \alpha_{X,O}$=(0,$\frac{1}{2}$) & $\Delta \alpha_{X,O}$=(0,$\frac{1}{4}$) \\
\hline
   && \multicolumn{2}{c}{ Energy injection break} \\
\hline
 $\Delta \beta_{X,O}$=0 & $\Delta \beta_{X,O}$=$\frac{1}{2}$ & $\Delta \beta_{X,O}$=$\frac{1}{2}$ & $\Delta \beta_{X,O}$=(0,$\frac{1}{2}$) & $\Delta \beta_{X,O}$=(0,$\frac{1}{2}$) \\
 $\Delta \alpha_{1,X,O}$=0 & $\Delta \alpha_{1,X,O}=\frac{2-q}{4}$ & $\Delta \alpha_{1,X,O}=\frac{-2+q}{4}$ & $\Delta \alpha_{1,X,O}=(0,\frac{2-q}{4})$ & $\Delta \alpha_{1,X,O}=(\frac{-2+q}{4},0)$ \\
 $\Delta \alpha_{2,X,O}$=0 & $\Delta \alpha_{2,X,O}$=$\frac{1}{2}$ & $\Delta \alpha_{2,X,O}$=$\frac{1}{4}$ & $\Delta \alpha_{2,X,O}$=(0,$\frac{1}{2}$) & $\Delta \alpha_{2,X,O}$=(0,$\frac{1}{4}$) \\
\hline
 && \multicolumn{2}{c}{ Jet break}                   \\
\hline
 $\Delta \beta_{X,O}$=0 & $\Delta \beta_{X,O}$=$\frac{1}{2}$ & $\Delta \beta_{X,O}$=$\frac{1}{2}$ & $\Delta \beta_{X,O}$=(0,$\frac{1}{2}$) & $\Delta \beta_{X,O}$=(0,$\frac{1}{2}$) \\
$\Delta \alpha_{1,X,O}$=0 & $\Delta \alpha_{1,X,O}$=$\frac{1}{2}$ & $\Delta \alpha_{1,X,O}$=$\frac{1}{4}$ & $\Delta \alpha_{1,X,O}$=(0,$\frac{1}{2}$) & $\Delta \alpha_{1,X,O}$=(0,$\frac{1}{4}$) \\
 $\Delta \alpha_{2,X,O}$=0 & $\Delta \alpha_{2,X,O}$=$\frac{1}{4}$ & $\Delta \alpha_{2,X,O}$=-$\frac{1}{4}$ & $\Delta \alpha_{2,X,O}$=(0,$\frac{1}{4}$) & $\Delta \alpha_{2,X,O}$=(-$\frac{1}{4}$,0) \\
\hline
  &&\multicolumn{2}{c}{  Energy injection with jet break}                    \\
\hline
 $\Delta \beta_{X,O}$=0 & $\Delta \beta_{X,O}$=$\frac{1}{2}$ & $\Delta \beta_{X,O}$=$\frac{1}{2}$ & $\Delta \beta_{X,O}$=(0,$\frac{1}{2}$) & $\Delta \beta_{X,O}$=(0,$\frac{1}{2}$) \\
 $\Delta \alpha_{1,X,O}$=0 & $\Delta \alpha_{1,X,O}=\frac{2-q}{4}$ & $\Delta \alpha_{1,X,O}=\frac{-2+q}{4}$ & $\Delta \alpha_{1,X,O}=(0,\frac{2-q}{4})$ & $\Delta \alpha_{1,X,O}=(\frac{-2+q}{4},0)$ \\
$\Delta \alpha_{2,X,O}$=0 & $\Delta \alpha_{2,X,O}=\frac{2-q}{4}$ & $\Delta \alpha_{2,X,O}=\frac{-2+q}{4}$ & $\Delta \alpha_{2,X,O}=(0,\frac{2-q}{4})$ & $\Delta \alpha_{2,X,O}=(\frac{-2+q}{4},0)$ \\
\enddata
\tablenotetext{a}{Same regime: X-ray and optical bands are in the same spectral regime, regime I ($\nu>\nu_{c}$) or regime II ($\nu_{m}<\nu<\nu_{c}$). In this table, $\bigtriangleup \beta_{\rm X,O}=\beta_{X}-\beta_{O}$, and $\bigtriangleup \alpha_{\rm X,O}=\alpha_{\rm X}-\alpha_{\rm O}$. The subsripts ``1'' and ``2''
denote the pre- and post-break segments for the BPL lightcurves, respectively.}
\tablenotetext{b}{Different regimes: X-ray band in regime I ($\nu>\nu_{c}$), and optical band in regime II ($\nu_{m}<\nu<\nu_{c}$); }
\tablenotetext{c}{Grey zone: One band or two bands in the grey zone  regime I$-$ II. }

\label{Tab:delta-alpha-betalow2}
\end{deluxetable}

\clearpage
\begin{deluxetable}{ccccccccccccccccccccccccccc}
\tabletypesize{\scriptsize} \tablecaption{The criteria for GRB
grades} \tablewidth{0pt}
\tablehead{ \colhead{Grades}& \colhead{Light curve behavior}&
\colhead{Closure-relation}& \colhead{Fraction}} \startdata
Grade I &   Achromatic break or SPL decay in both bands  &   Yes &   43/85   \\
Grade II    &   Break missing in one band, consistent with being achromatic &   Yes &   2/85   \\
Grade III   &   Achromatic break or SPL decay in both bands  &   No  &   34/85   \\
Grade IV    &   Break missing in one band, consistent with being achromatic     &   No  &   3/85   \\
Grade VI    &   Chromatic   &   No  &   3/85    \\
\enddata
\label{table:grades}
\end{deluxetable}

\begin{deluxetable}{cccccccccccccccccccccccccc}
\tablecaption{The Gold sample GRBs and their derived parameters.}
\rotate
\tabletypesize{\tiny}
\tabcolsep=2.5pt
\tablewidth{0pt}
\tablehead{ \colhead{GRB}&
\colhead{$z$}&
\colhead{$Optical$\tablenotemark{a}}&
\colhead{$X ray$\tablenotemark{a}}&
\colhead{$p$}&
\colhead{$q$}&
\colhead{$\epsilon_{B}$\tablenotemark{b}}&
\colhead{$ E_{\rm \gamma,iso}$\tablenotemark{c}}&
\colhead{$ E_{\rm K,end}$\tablenotemark{c}}&
\colhead{$ E_{\rm K,in}$\tablenotemark{c}}&
\colhead{$ E_{\rm K,dec}$\tablenotemark{c}}&
\colhead{$ \eta_{\rm \gamma,dec}$\tablenotemark{d}}&
\colhead{$ \eta_{\rm \gamma,end}$\tablenotemark{d}}&
\colhead{type}\tablenotemark{e}}

\startdata

\hline																																											 \\
050408	&	1.24 	&	windII	&	windI-windII	&	2.12	$\pm$	0.12	&	0.25	 $\pm$	 0.17	&	   ... 	&	11.76			 &	3410.39 	$\pm$	76.12 	&	3386.60 	 $\pm$	 75.06 	 &	23.80 	$\pm$	6.84 	&	33 	$\pm$	10 	&	0.3 	$\pm$	 0.0 	&	 1	\\
050801	&	1.56 	&	ISMII	&	ISMII	&	2.78	$\pm$	0.28	&	0.22	$\pm$	 0.15	 &	 2.7E-05	&	0.52 	$\pm$	 0.11 	&	124.89 	$\pm$	17.39 	&	81.83 	$\pm$	 20.10 	&	 43.06 	 $\pm$	17.65 	&	1 	$\pm$	1 	&	0.4 	$\pm$	 0.1 	&	1	\\
051109A	&	2.35 	&	windI-windII	&	windI-windII	&	2.4	$\pm$	0.1	&	0.22	 $\pm$	 0.06	&	   ... 	&	9.15 	 $\pm$	6.83 	&	14675.86 	$\pm$	763.28 	&	 14060.42 	 $\pm$	 730.61 	&	615.44 	$\pm$	38.42 	&	1 	$\pm$	1 	 &	0.1 	 $\pm$	0.1 	&	 1	\\
060206	&	4.05 	&	windI-windII	&	windI	&	2.46	$\pm$	0.1	&	0.41	$\pm$	 0.06	 &	   ... 	&	4.78 	 $\pm$	20.79 	&	386.76 	$\pm$	93.02 	&	344.28 	$\pm$	 83.18 	&	 42.47 	$\pm$	15.11 	&	10 	$\pm$	44 	&	1 	 $\pm$	5 	&	1	\\
060714	&	2.71 	&	windI-windII	&	windI-windII	&	1.89	$\pm$	0.07	&	 0.19	 $\pm$	0.05	&	   ... 	&	 18.22 	$\pm$	2.53 	&	250.46 	$\pm$	248.11 	&	 239.83 	 $\pm$	 237.24 	&	10.62 	$\pm$	11.25 	&	63 	$\pm$	67 	 &	7 	$\pm$	7 	 &	1	\\
060729	&	0.54 	&	windI-windII	&	windI-windII	&	2.04	$\pm$	0.08	&	 0.06	 $\pm$	0.01	&	   ... 	&	 0.75 	$\pm$	0.07 	&	2312.14 	$\pm$	33.54 	 &	2304.60 	 $\pm$	33.43 	&	7.54 	$\pm$	0.12 	&	9 	$\pm$	 1 	&	0.03 	 $\pm$	0.01 	&	1	 \\
070411	&	2.95 	&	windI-windII	&	windI-windII	&	2.48	$\pm$	0.02	&	 0.86	 $\pm$	0.09	&	   ... 	&	 17.06 	$\pm$	1.84 	&	24456.57 	$\pm$	256.42 	 &	14460.59 	 $\pm$	794.78 	&	9995.98 	$\pm$	782.91 	&	0.2 	 $\pm$	0.1 	&	 0.1 	$\pm$	0.1 	 &	1	\\
070518	&	1.16 	&	windI-windII	&	windI-windII	&	2.4	$\pm$	0.2	&	0.24	 $\pm$	 0.08	&	   ... 	&	0.27 	 $\pm$	0.13 	&	917.47 	$\pm$	22.99 	&	911.31 	 $\pm$	 22.81 	 &	6.16 	$\pm$	0.52 	&	4 	$\pm$	2 	&	0.03 	 $\pm$	0.01 	&	 1	\\
080710	&	0.85 	&	ISMII	&	ISMII	&	2.78			&	0.32			&	9.6E-06	 &	 1.34 	$\pm$	0.32 	&	 424.26 	$\pm$	40.68 	&	393.99 	$\pm$	37.78 	&	 30.27 	 $\pm$	 2.90 	&	4 	$\pm$	1 	&	0.3 	$\pm$	0.1 	&	 1	\\
090102	&	1.55 	&	windI-windII	&	windI-windII	&	1.58	$\pm$	0.22	&	 0.33	 $\pm$	0.05	&	   ... 	&	 22.74 	$\pm$	2.12 	&	1750.25 	$\pm$	1149.59 	 &	 1490.68 	 $\pm$	1009.33 	&	259.57 	$\pm$	141.45 	&	8 	 $\pm$	4 	&	1 	 $\pm$	1 	 &	1	\\
090426	&	2.61 	&	windI-windII	&	windI	&	2.06	$\pm$	0.3	&	0.13	$\pm$	 0.02	 &	   ... 	&	0.83 	 $\pm$	0.28 	&	2.77 	$\pm$	3.33 	&	1.80 	$\pm$	 2.17 	&	 0.97 	$\pm$	1.17 	&	46 	$\pm$	58 	&	23 	 $\pm$	29 	&	1	\\
100418A	&	0.62 	&	ISMII	&	ISMII	&	2.96	$\pm$	0.18	&	0.05	$\pm$	 0.02	 &	 2.9E-06	&	0.14 	$\pm$	 0.02 	&	1754.24 	$\pm$	403.22 	&	1754.06 	 $\pm$	 403.18 	 &	0.18 	$\pm$	0.05 	&	43 	$\pm$	14 	&	0.01 	 $\pm$	0.01 	&	 1	\\
101024A	&	   ... 	&	ISMII	&	ISMII	&	2.64	$\pm$	0.26	&	0.06	$\pm$	 0.11	 &	   ... 	&		   ... 		 &		   ... 		&		   ... 		&		   ... 		 &		   ... 		 &		   ... 		&	1	\\
050820A	&	2.61 	&	ISMI-ISMII	&	ISMI-ISMII	&	1.78	$\pm$	0.1	&		   ... 		 &	   ... 	&	114.67 	$\pm$	 33.59 	&		   ... 		&		   ... 		&	4117.25 	 $\pm$	 2462.50 	&	3 	$\pm$	2 	&		   ... 		&	2	 \\
050922C	&	2.20 	&	ISMII	&	ISMI	&	2.06	$\pm$	0.05	&		   ... 		&	   ... 	 &	9.93 	$\pm$	1.06 	 &		   ... 		&		   ... 		&	118.25 	$\pm$	 17.29 	&	 8 	 $\pm$	1 	&		   ... 		&	2	\\
060111B	&	   ... 	&	ISMI	&	ISMI	&	1.57	$\pm$	0.03	&		   ... 		&	   ... 	 &	11.00 	$\pm$	5.00 	 &		   ... 		&		   ... 		&		   ... 		 &		   ... 		 &		   ... 		&	2	\\
081008	&	1.97 	&	ISMII	&	ISMI	&	1.96	$\pm$	0.18	&		   ... 		&	   ... 	 &	9.30 	$\pm$	1.91 	 &		   ... 		&		   ... 		&	253.98 	$\pm$	 93.28 	&	 4 	 $\pm$	3 	&		   ... 		&	2	\\
081203A	&	2.10 	&	ISMI-ISMII	&	ISMI-ISMII	&	2.2	$\pm$		&		   ... 		&	   ... 	 &	34.71 	$\pm$	17.12 	 &		   ... 		&		   ... 		&	594.07 	$\pm$	 14.99 	&	 6 	 $\pm$	3 	&		   ... 		&	2	\\
090618	&	0.54 	&	ISMII	&	ISMI	&	1.92	$\pm$	0.02	&		   ... 		&	   ... 	 &	25.30 	$\pm$	0.00 	 &		   ... 		&		   ... 		&	89.10 	$\pm$	 14.81 	&	 22 	 $\pm$	4 	&		   ... 		&	2	\\
091127	&	0.49 	&	ISMII	&	ISMI	&	1.36	$\pm$		&		   ... 		&	   ... 	 &	 1.52 	$\pm$	0.08 	&		   ... 		&		   ... 		&	263.04 	$\pm$	 22.94 	&	 1 	 $\pm$	0.1 	&		   ... 		&	2	\\
130427A	&	0.34 	&	ISMII	&	ISMII	&	2.38	$\pm$	0.02	&		   ... 		&	 9.1E-06	 &	81.00 			&		   ... 		&		   ... 		&	1665.96 	$\pm$	 62.14 	 &	 5 	 $\pm$	0.2 	&		   ... 		&	2	\\
051028	&	3.70 	&	ISMI-SIMII	&	ISMI-ISMII	&	1.9	$\pm$	0.1	&		   ... 		&	   ... 	 &	11.94 	$\pm$	1.87 	 &		   ... 		&		   ... 		&	2515.02 	 $\pm$	 1557.52 	 &	0.5 	$\pm$	0.3 	&		   ... 		&	4	 \\
060418	&	1.49 	&	ISMII	&	ISMI-ISMII	&	2.58	$\pm$	0.18	&		   ... 		 &	   ... 	&	12.45 	$\pm$	 4.44 	&		   ... 		&		   ... 		&	386.85 	 $\pm$	 42.95 	 &	3 	$\pm$	1 	&		   ... 		&	4	\\
060512	&	0.44 	&	ISMI-SIMII	&	ISMI-ISMII	&	1.94	$\pm$	0.06	&		   ... 		 &	   ... 	&	0.02 	$\pm$	 0.01 	&		   ... 		&		   ... 		&	5.70 	 $\pm$	 0.12 	 &	0.3 	$\pm$	0.2 	&		   ... 		&	4	 \\
060904B	&	0.70 	&	ISMI-SIMII	&	ISMI-ISMII	&	3.25	$\pm$	0.17	&		   ... 		 &	   ... 	&	0.79 	$\pm$	 0.56 	&		   ... 		&		   ... 		&	750.73 	 $\pm$	 388.34 	 &	0.1 	$\pm$	0.1 	&		   ... 		&	 4	\\
060912A	&	0.94 	&	ISMII	&	ISMII	&	2.2	$\pm$	0.3	&		   ... 		&	8.7E-05	 &	 1.12 	$\pm$	0.08 	&		   ... 		&		   ... 		&	46.23 	$\pm$	1.16 	 &	2 	 $\pm$	0.2 	&		   ... 		&	4	\\
060927	&	5.46 	&	windII	&	windI-windII	&	2.22	$\pm$	0.1	&		   ... 		 &	   ... 	&	10.37 	$\pm$	 3.03 	&		   ... 		&		   ... 		&	37039.49 	 $\pm$	 2373.94 	&	0.03 	$\pm$	0.01 	&		   ... 		 &	4	\\
061007	&	1.26 	&	ISMII	&	ISMII	&	2.04	$\pm$	0.1	&		   ... 		&	 7.3E-05	 &	99.81 	$\pm$	7.22 	&		   ... 		&		   ... 		&	1085.73 	 $\pm$	18.15 	 &	 8 	$\pm$	1 	&		   ... 		&	4	\\
061126	&	1.16 	&	ISMII	&	ISMII	&	2.64	$\pm$	0.18	&		   ... 		&	 1.0E-05	 &	13.26 	$\pm$	1.90 	 &		   ... 		&		   ... 		&	1282.64 	 $\pm$	 177.08 	 &	1 	$\pm$	0.1 	&		   ... 		&	4	\\
070318	&	0.84 	&	ISMII-ISMII	&	ISMI-ISMII	&	2.74	$\pm$	0.02	&		   ... 		 &	   ... 	&	1.39 	$\pm$	 0.35 	&		   ... 		&		   ... 		&	277.28 	 $\pm$	 22.56 	 &	0.5 	$\pm$	0.1 	&		   ... 		&	4	 \\
071025	&	1.55 	&	ISMII	&	ISMII	&	3.07	$\pm$	0.13	&		   ... 		&	 9.2E-06	 &	85.42 	$\pm$	12.42 	 &		   ... 		&		   ... 		&	1394.60 	 $\pm$	 851.47 	 &	6 	$\pm$	4 	&		   ... 		&	4	\\
071031	&	2.69 	&	ISMI	&	ISMI	&	2.01	$\pm$	0.27	&		   ... 		&	   ... 	 &	4.73 	$\pm$	9.26 	 &		   ... 		&		   ... 		&	68.53 	$\pm$	 38.27 	&	 6 	 $\pm$	5 	&		   ... 		&	4	\\
080319C	&	1.95 	&	windI-II	&	windI-windII	&	2.22	$\pm$	0.2	&		   ... 		 &	   ... 	&	23.54 	$\pm$	 1.47 	&		   ... 		&		   ... 		&	12314.17 	 $\pm$	 3183.23 	&	0.2 	$\pm$	0.1 	&		   ... 		 &	4	\\
080413A	&	2.43 	&	windII	&	windI-windII	&	2.3	$\pm$	0.48	&		   ... 		 &	   ... 	&	14.09 	$\pm$	 7.30 	&		   ... 		&		   ... 		&	249.98 	 $\pm$	 71.42 	 &	5 	$\pm$	3 	&		   ... 		&	4	\\
080603A	&	1.68 	&	ISMI	&	ISMI	&	2.96	$\pm$	0.08	&		   ... 		&	   ... 	 &	2.20 	$\pm$	0.80 	 &		   ... 		&		   ... 		&	3577.17 	 $\pm$	 272.70 	 &	0.1 	$\pm$	0.1 	&		   ... 		&	4	 \\
080804	&	2.20 	&	windI-windII	&	windI-windII	&	1.86			&		   ... 		 &	   ... 	&	24.61 	$\pm$	 4.79 	&		   ... 		&		   ... 		&	143.09 	 $\pm$	 16.57 	 &	15 	$\pm$	3 	&		   ... 		&	4	\\
080913	&	6.70 	&	ISMI-SIMII	&	ISMI-ISMII	&	2.79	$\pm$	0.27	&		   ... 		 &	   ... 	&	8.44 	$\pm$	 1.55 	&		   ... 		&		   ... 		&	354.60 	 $\pm$	 174.38 	 &	2 	$\pm$	1 	&		   ... 		&	4	 \\
080928	&	1.69 	&	ISMII	&	ISMII	&	2.44	$\pm$	0.04	&		   ... 		&	 5.7E-05	 &	6.30 	$\pm$	0.75 	 &		   ... 		&		   ... 		&	548.00 	$\pm$	 26.66 	 &	 1 	 $\pm$	0.1 	&		   ... 		&	4	\\
090323	&	3.57 	&	windII	&	windII	&	2.65	$\pm$	0.13	&		   ... 		&	 9.5E-04	 &	372.38 	$\pm$	16.86 	 &		   ... 		&		   ... 		&	209526.73 	 $\pm$	 18842.90 	 &	0.2 	$\pm$	0.1 	&		   ... 		&	4	 \\
090328	&	0.74 	&	ISMII	&	ISMII	&	3.19	$\pm$	0.21	&		   ... 		&	 3.4E-06	 &	19.03 			&		   ... 		&		   ... 		&	3861.76 	$\pm$	 2039.23 	 &	 0.5 	 $\pm$	0.3 	&		   ... 		&	4	\\
090926A	&	2.11 	&	ISMII	&	ISMII	&	2.29	$\pm$	0.19	&		   ... 		&	   ... 	 &	185.13 	$\pm$	9.10 	 &		   ... 		&		   ... 		&	1859.10 	 $\pm$	 111.85 	 &	9 	$\pm$	1 	&		   ... 		&	4	\\
100901A	&	1.41 	&	windII	&	windI-windII	&	2.1	$\pm$	0.1	&		   ... 		&	   ... 	 &	2.95 	$\pm$	0.63 	 &		   ... 		&		   ... 		&	27847.74 	 $\pm$	 871.45 	 &	0.01 	$\pm$	0.01 	&		   ... 		&	4	 \\
120326A	&	1.80 	&	windII	&	windII	&	2.51	$\pm$	0.09	&		   ... 		&	 5.2E-04	 &	3.18 	$\pm$	0.40 	 &		   ... 		&		   ... 		&	41757.87 	 $\pm$	 1069.46 	 &	0.01 	$\pm$	0.01 	&		   ... 		&	4	 \\
\hline			&		&		&				&				&		&				&																 &				&		\\
Grade II			&		&		&				&				&		&				&																 &				&		\\
\hline			&		&		&				&				&		&				&																 &				&		\\
051111	&	1.55 	&	windII	&	windI-windII	&	2.52	$\pm$	0.14	&		   ... 		 &	   ... 	&	10.79 	$\pm$	 3.07 	&		   ... 		&		   ... 		&	3793.13 	 $\pm$	 273.16 	&	0.3 	$\pm$	0.1 	&		   ... 		 &	4	\\
090313	&	3.38 	&	windII	&	windI-windII	&	2.54	$\pm$	0.72	&		   ... 		 &	 5.0E-06	&	13.02 	$\pm$	 2.94 	&		   ... 		&		   ... 		&	173028.58 	 $\pm$	 189747.00 	&	0.01 	$\pm$	0.01 	&		   ... 		 &	4	\\

\enddata
\tablenotetext{a}{ISMI: the ISM model in the spectral regime I ($\nu>\nu_{c}$);
ISMII: the ISM model in the spectral regime II ($\nu_{m}<\nu<\nu_{c}$);
windI: the wind model in the spectral regime I;
windII: the wind model in the spectral regime II:
ISMI-ISMII: the ISM model in the grey zone (between $\nu> \nu_{c}$ and $\nu_{m}<\nu<\nu_{c}$);
windI-windII: the wind model in the grey zone.}

\tablenotetext{b}{Upper limit, calculate with the X-ray data in the spectral regime II ($\nu_{m}<\nu<\nu_{c}$);}

\tablenotetext{c}{In units of $10^{52} erg$. $E_{\rm K,end}$ is the kinetic energy at the end of energy injection ($t_{\rm end}$); $E_{\rm K,in}$ is the injected kinetic energy during the energy injection phase; and $E_{\rm K,dec}$ is the kinetic energy at the fireball deceleration time ($t_{\rm dec}$);}

\tablenotetext{d}{In units of $\%$; $\eta_{\rm \gamma,dec}$ is the radiative efficiency calculated using $E_{\rm K,dec}$; and $\eta_{\rm \gamma,dec}$ is the radiative efficiency calculated using $E_{\rm K,end}$}

\tablenotetext{e}{ 1: energy injection break; 2: jet break; 3: jet break with energy injection; 4: SPL decay.}

\label{table:parameter}
\end{deluxetable}

\begin{deluxetable}{ccccccccccccccccccccccccc}
\tabletypesize{\scriptsize}

\tablecaption{Parameters of the jet break sample.}

\tablewidth{0pt}

\tablehead{ \colhead{GRB}&
\colhead{$\theta_{j}^{o}$}&
\colhead{$\log (E_{\gamma}/{\rm erg})$} &
\colhead{$\log (E_{K}/{\rm erg})$}}

\startdata

050820A	&	4.5 	$\pm$	3.0 	&	51.54 	$^{+	0.24 		}	_{	-0.57 	}$	&	53.05 	 $^{+	0.28 	}	_{	-0.99 	}$	\\
050922C	&	1.8 	$\pm$	0.3 	&	49.69 	$^{+	0.08 		}	_{	-0.10 	}$	&	50.13 	 $^{+	0.09 	}	_{	-0.12 	}$	\\
060111B$^{a}$	&		   ... 		&					   ... 				&				   ... 				 \\
081008	&	1.3 	$\pm$	0.4 	&	49.41 	$^{+	0.14 		}	_{	-0.21 	}$	&	50.54 	 $^{+	0.17 	}	_{	-0.29 	}$	\\
081203A	&	1.0 	$\pm$	0.6 	&	49.76 	$^{+	0.24 		}	_{	-0.56 	}$	&	50.72 	 $^{+	0.19 	}	_{	-0.33 	}$	\\
090618	&	3.5 	$\pm$	1.3 	&	50.68 	$^{+	0.14 		}	_{	-0.21 	}$	&	50.84 	 $^{+	0.15 	}	_{	-0.23 	}$	\\
091127	&	2.7 	$\pm$	0.4 	&	49.22 	$^{+	0.07 		}	_{	-0.08 	}$	&	50.73 	 $^{+	0.07 	}	_{	-0.09 	}$	\\
130427A	&	3.8 	$\pm$	0.3 	&	51.25 	$^{+	0.04 		}	_{	-0.04 	}$	&	51.54 	 $^{+	0.04 	}	_{	-0.04 	}$	\\

\label{table:jet}
\enddata

\tablenotetext{a}{no redshift $z$ available.}

\end{deluxetable}

\clearpage
\setlength{\voffset}{-18mm}

\clearpage
\setlength{\voffset}{-18mm}
\begin{figure*}
\includegraphics[angle=0,scale=0.35,width=0.325\textwidth,height=0.30\textheight]{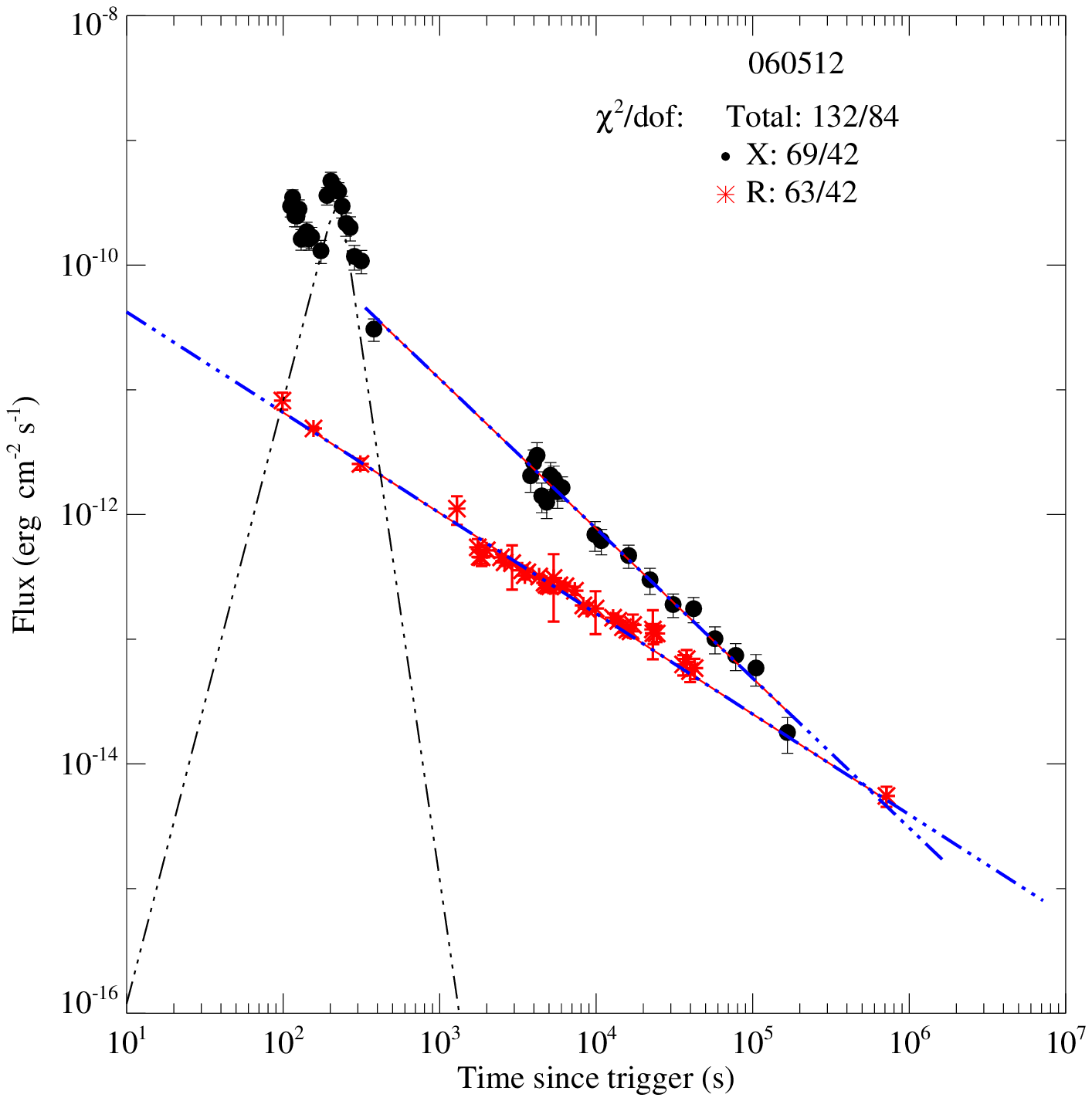}
\includegraphics[angle=0,scale=0.35,width=0.325\textwidth,height=0.30\textheight]{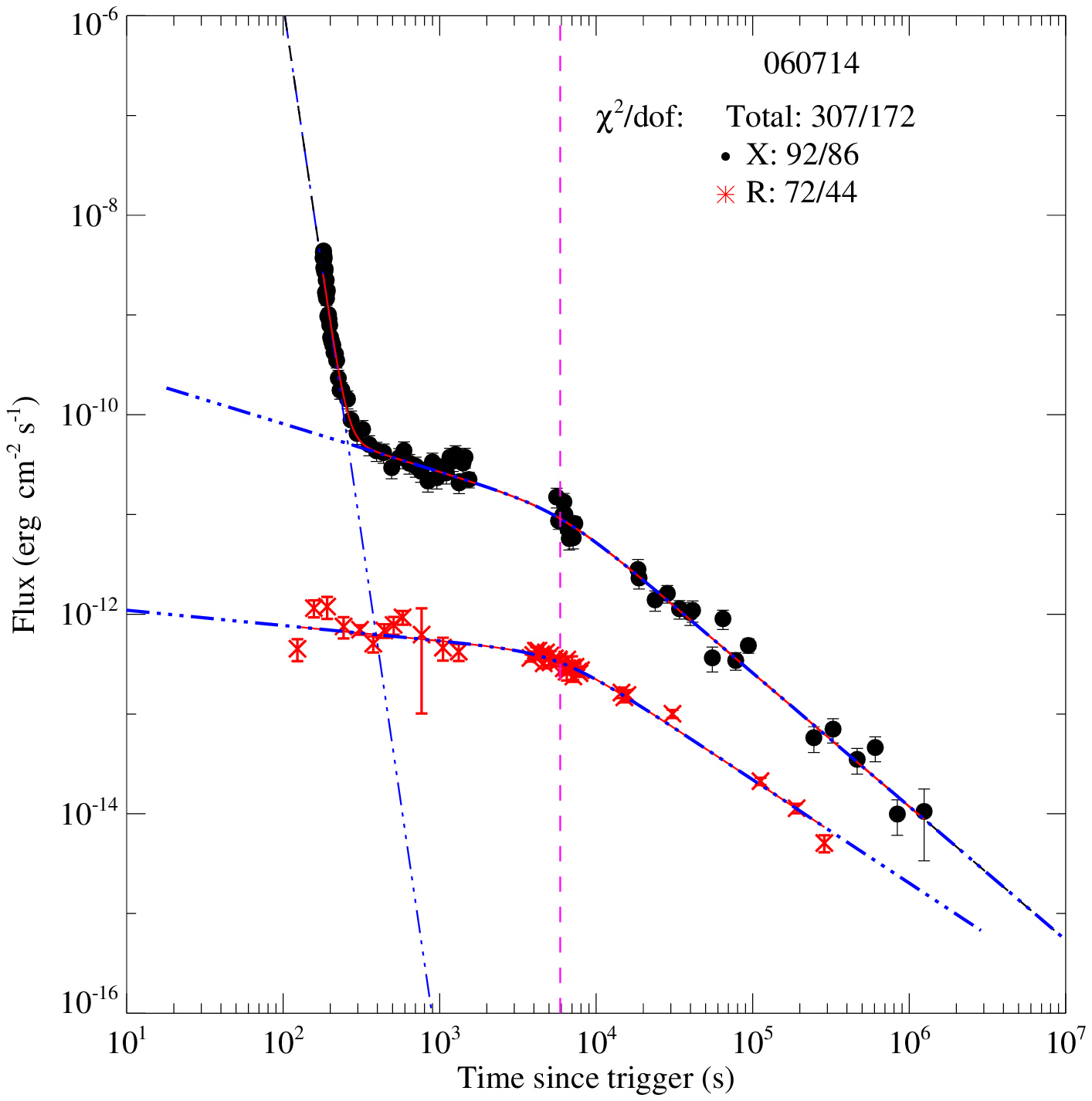}
\includegraphics[angle=0,scale=0.35,width=0.325\textwidth,height=0.30\textheight]{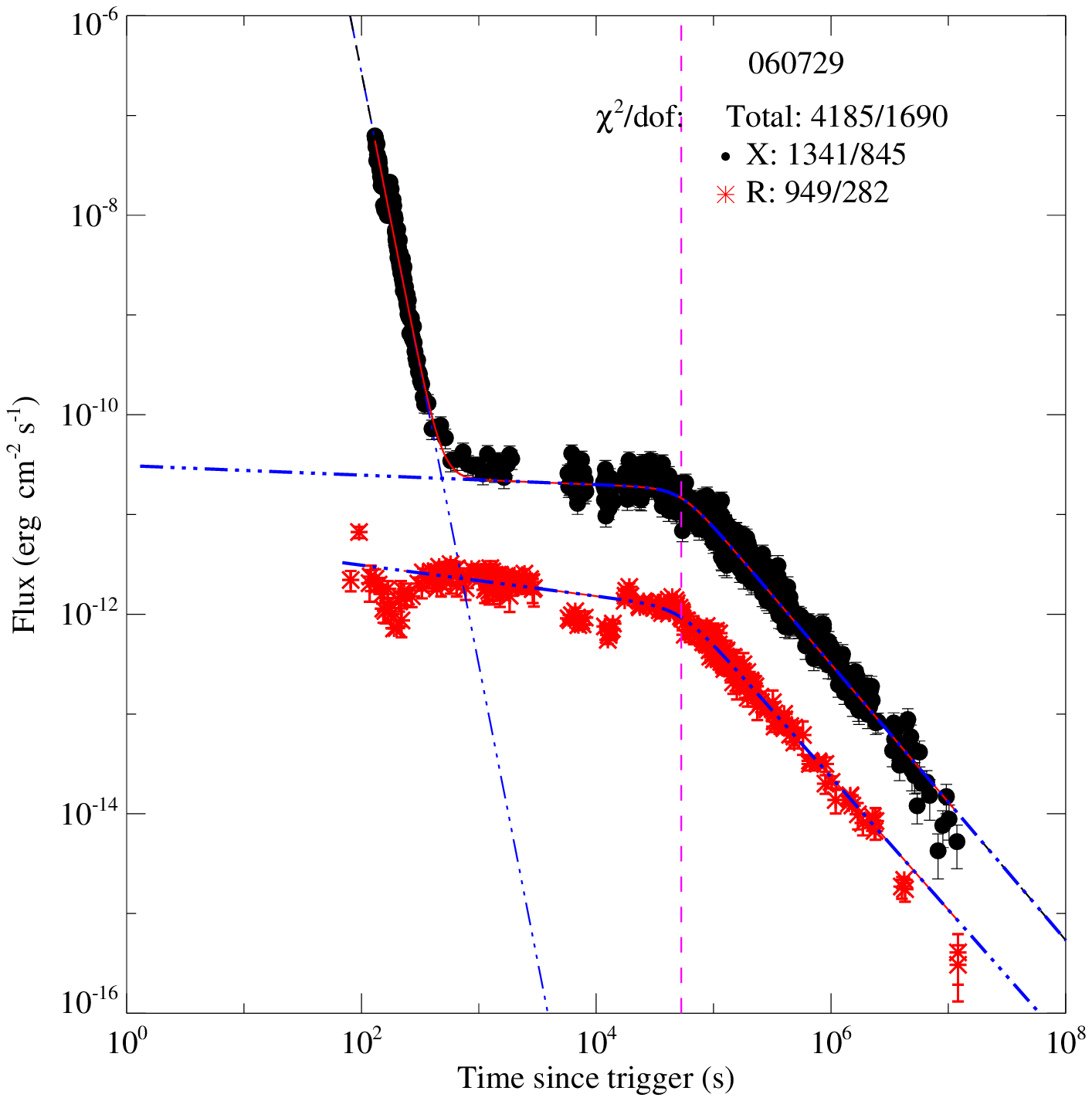}
\includegraphics[angle=0,scale=0.35,width=0.325\textwidth,height=0.30\textheight]{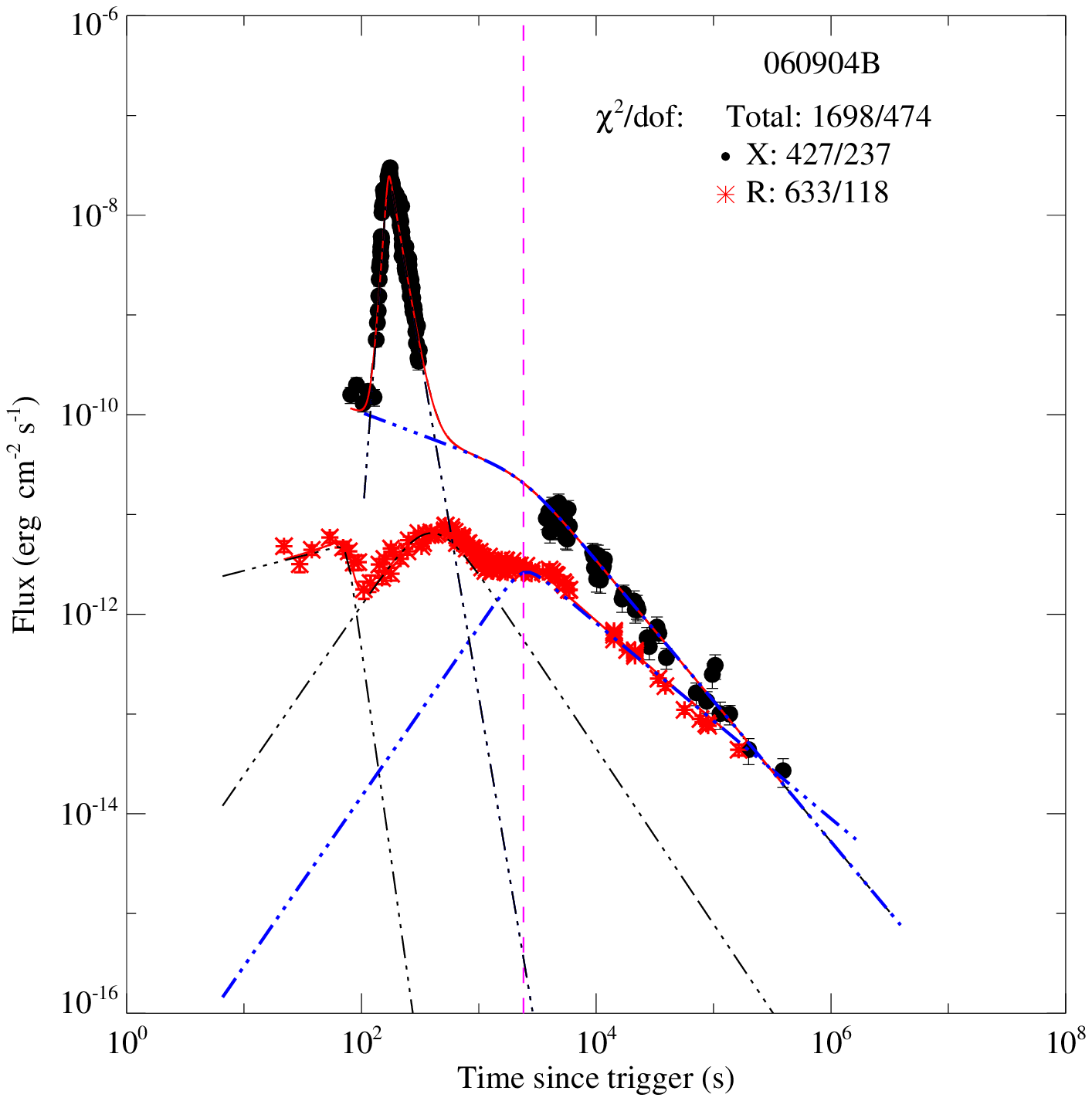}
\includegraphics[angle=0,scale=0.35,width=0.325\textwidth,height=0.30\textheight]{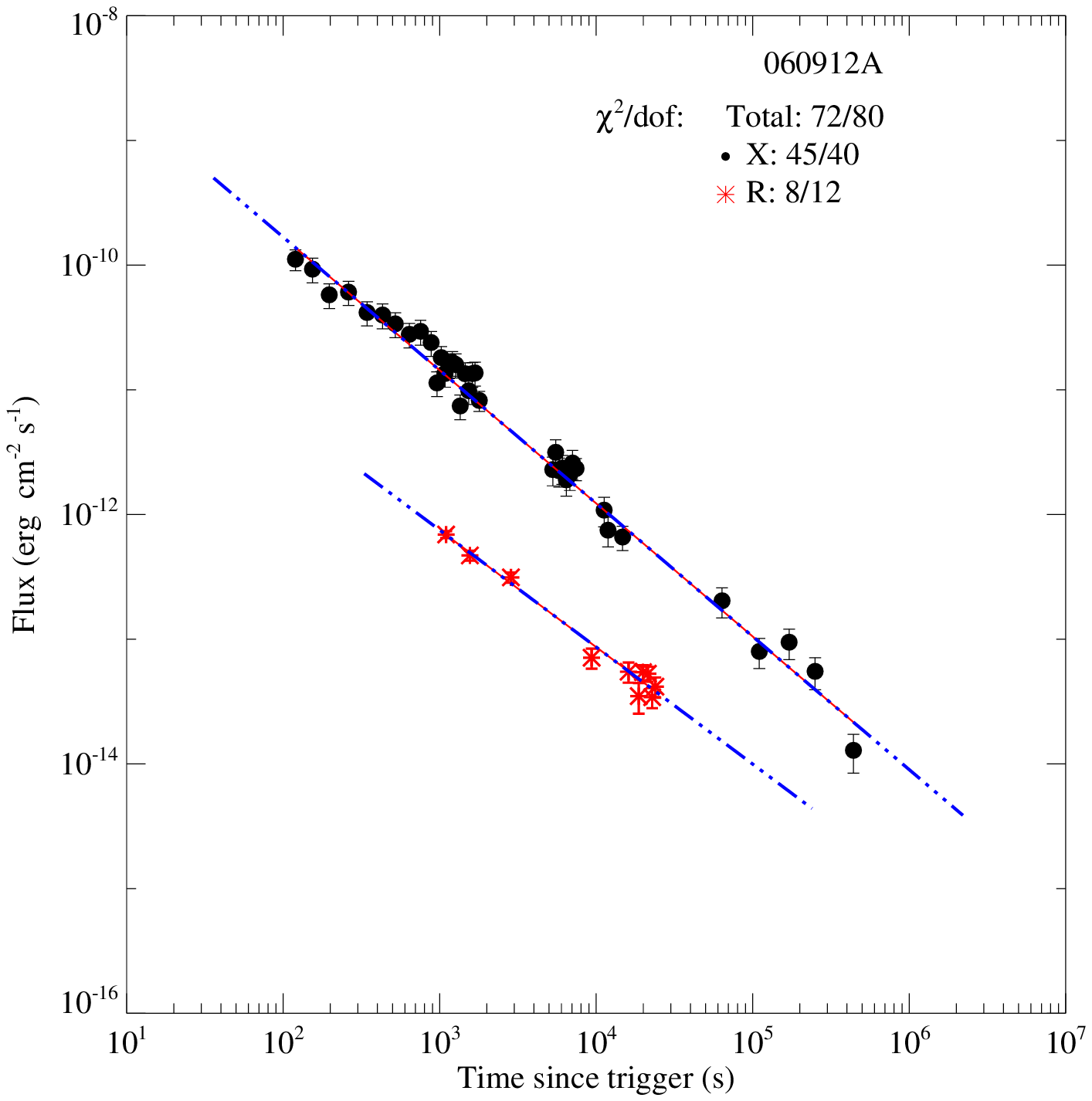}
\includegraphics[angle=0,scale=0.35,width=0.325\textwidth,height=0.30\textheight]{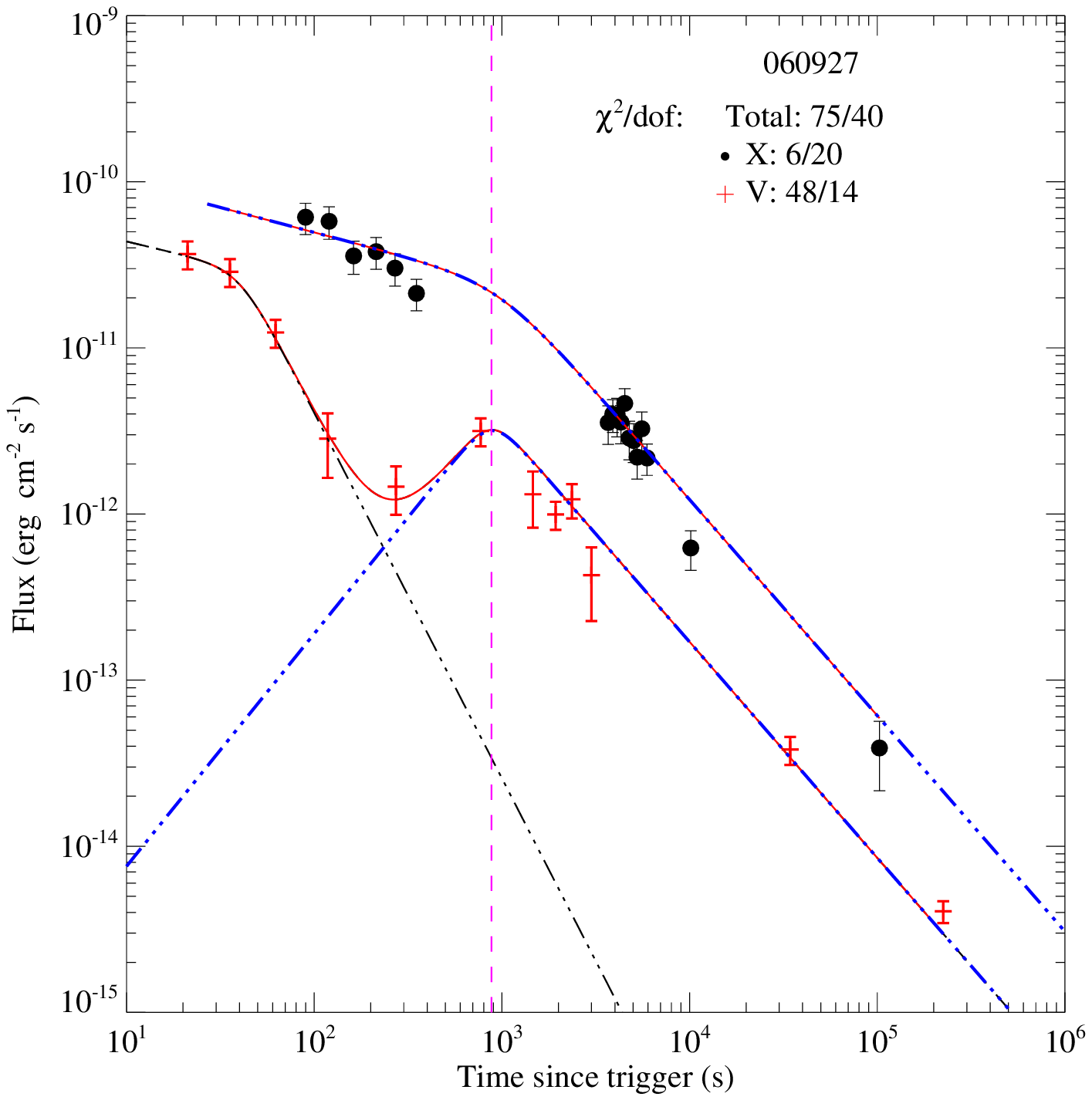}
\includegraphics[angle=0,scale=0.35,width=0.325\textwidth,height=0.30\textheight]{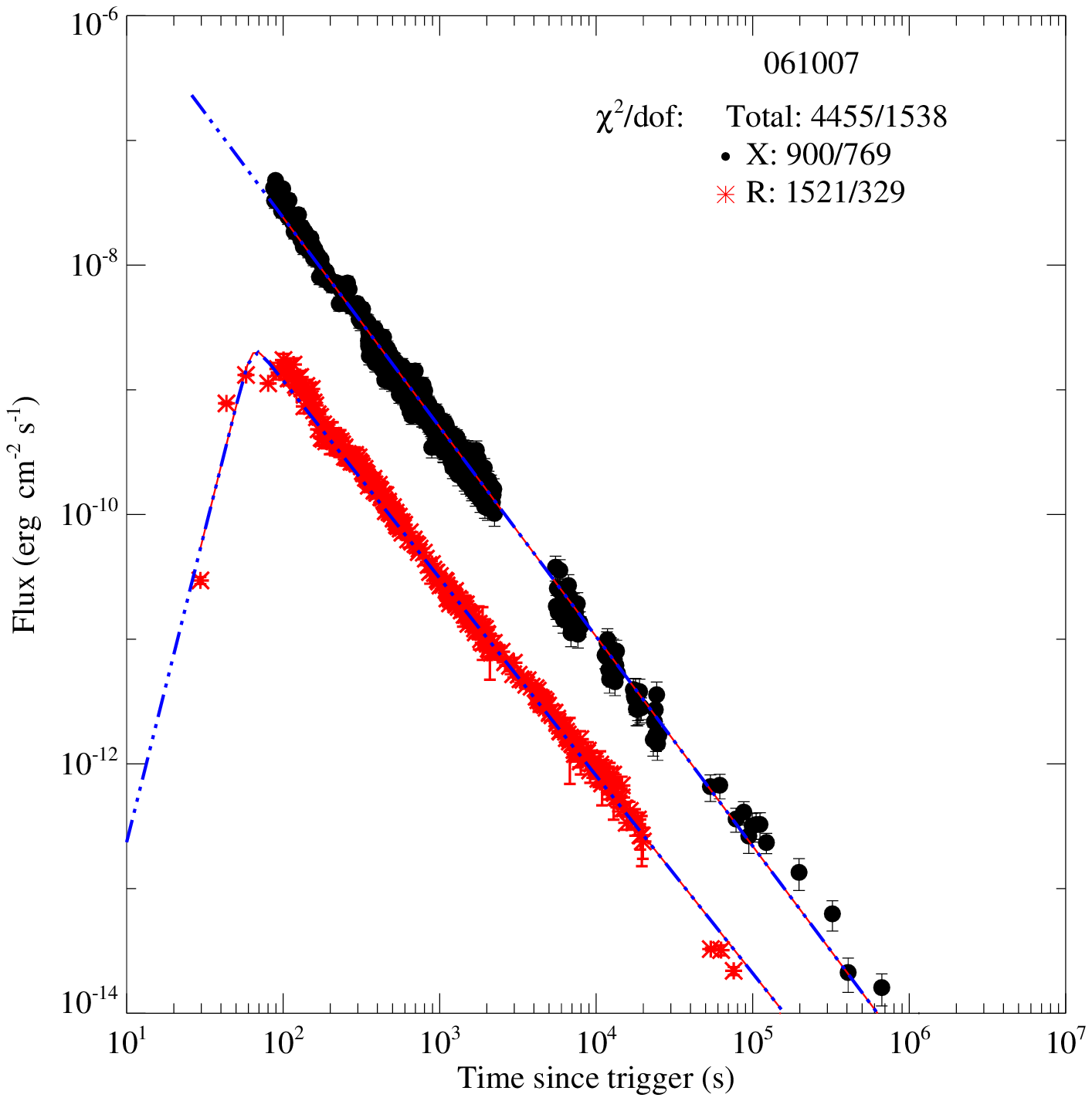}
\includegraphics[angle=0,scale=0.35,width=0.325\textwidth,height=0.30\textheight]{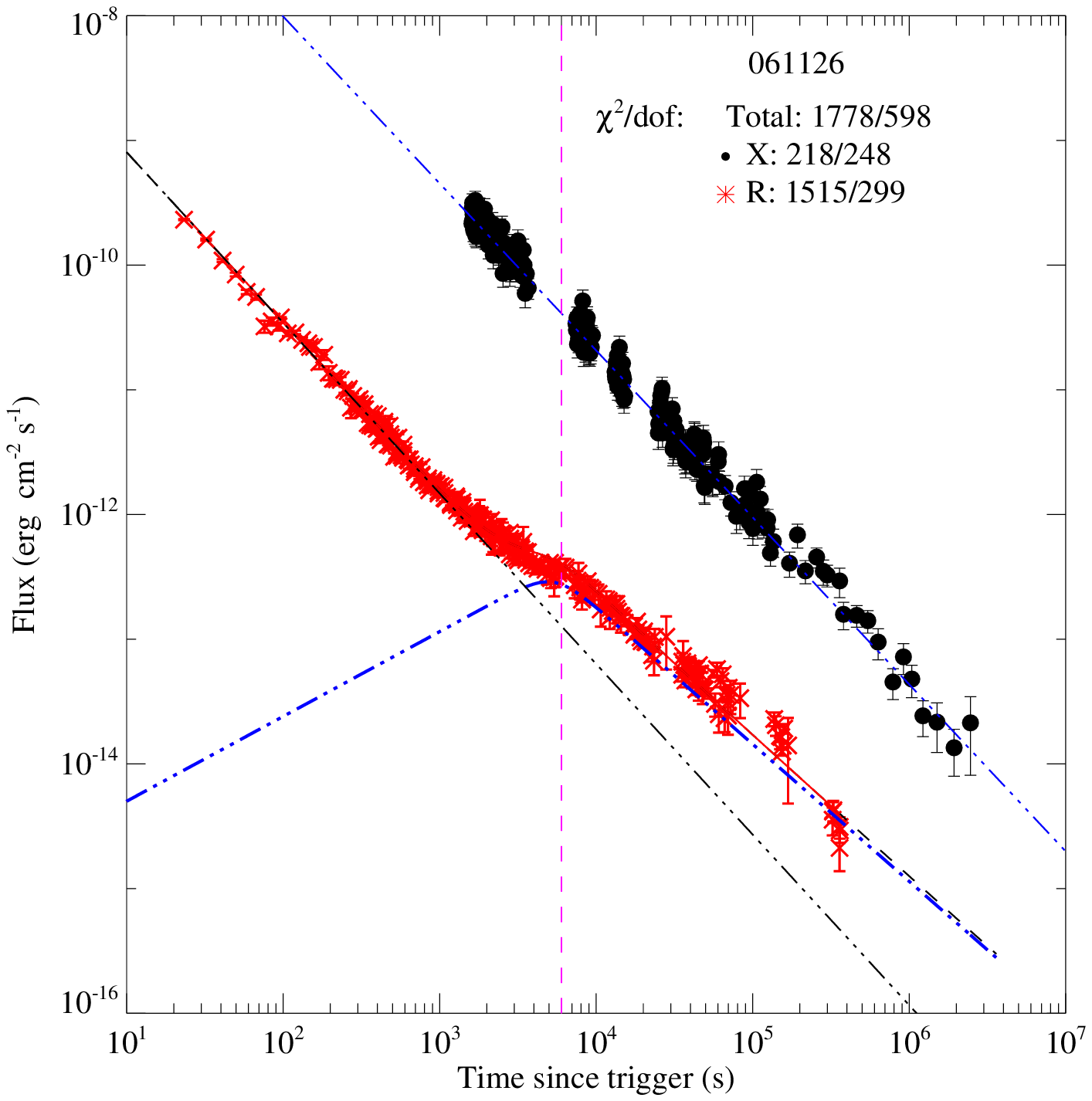}
\includegraphics[angle=0,scale=0.35,width=0.325\textwidth,height=0.30\textheight]{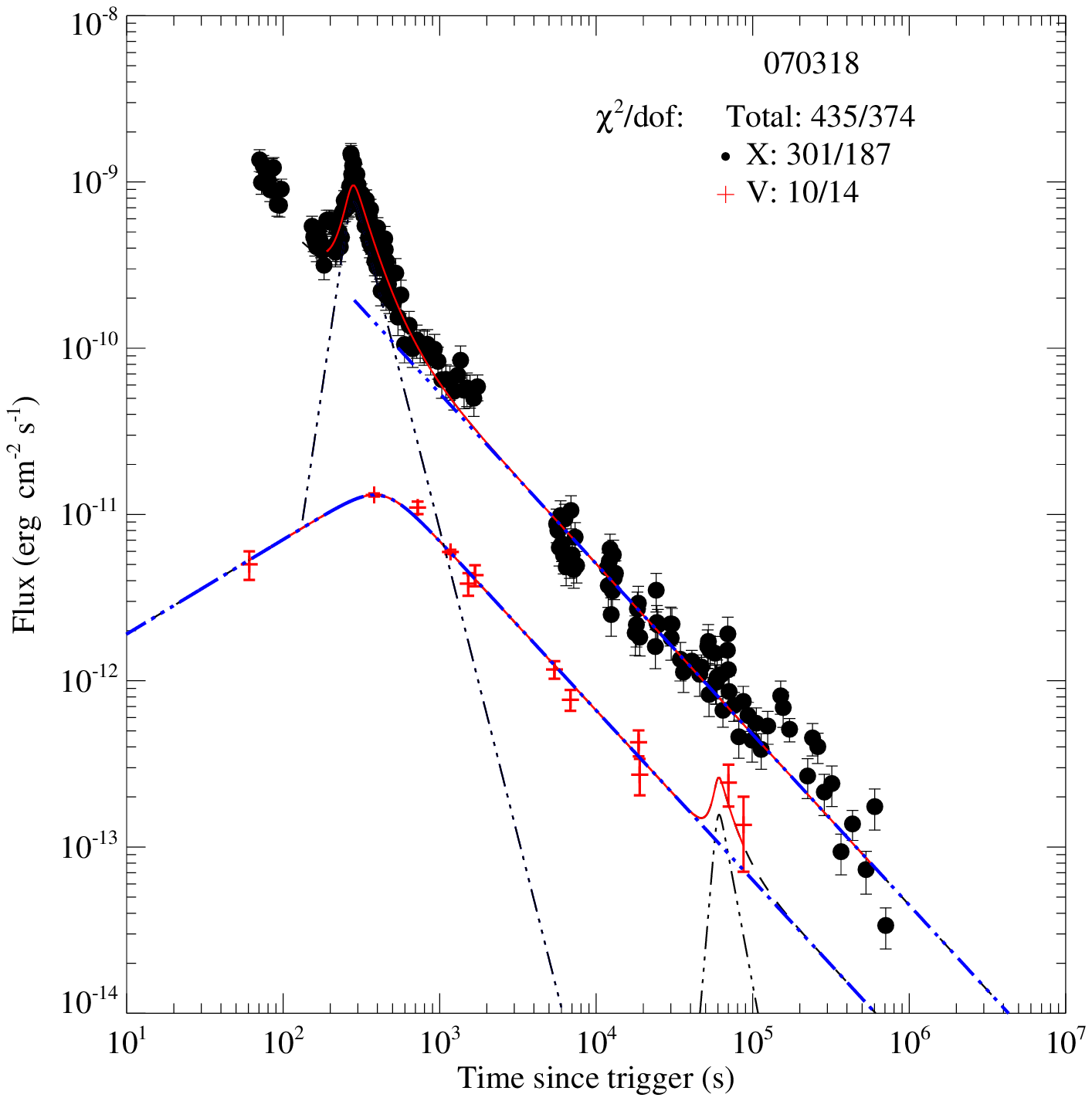}

\center{Fig. \ref{gradeI}---Continued}
\end{figure*}

\clearpage
\setlength{\voffset}{-18mm}
\begin{figure*}

\includegraphics[angle=0,scale=0.35,width=0.325\textwidth,height=0.30\textheight]{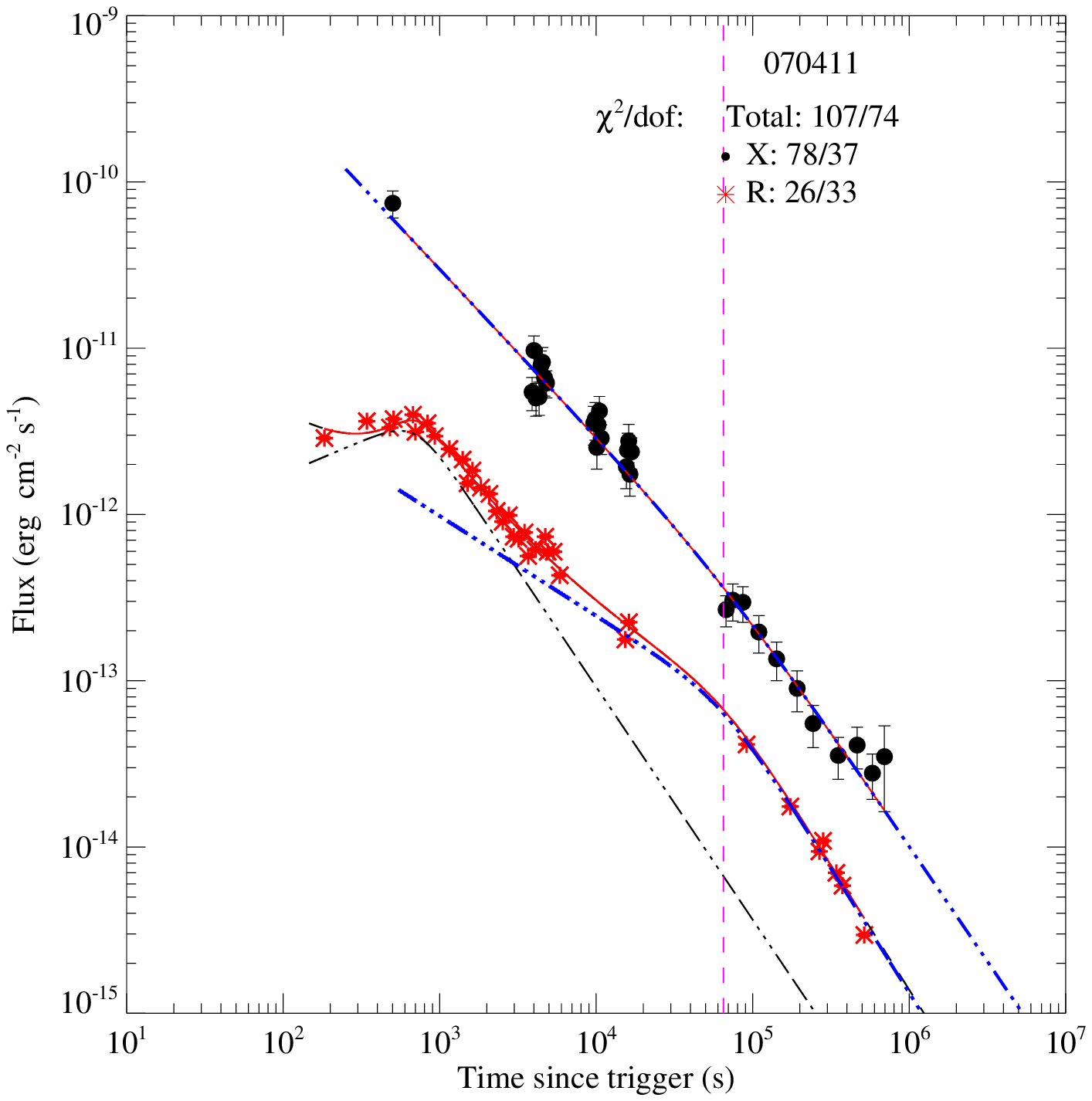}
\includegraphics[angle=0,scale=0.35,width=0.325\textwidth,height=0.30\textheight]{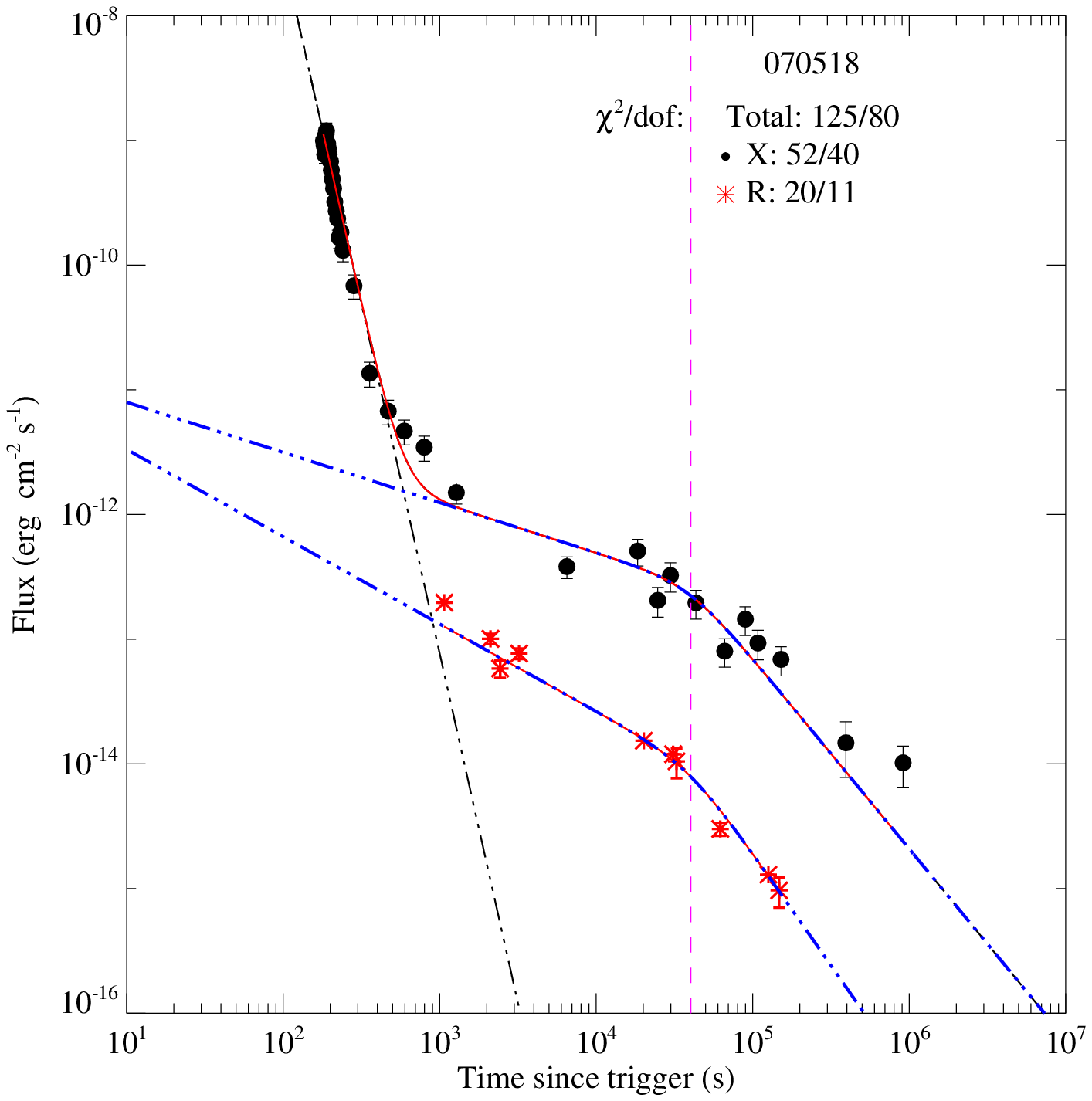}
\includegraphics[angle=0,scale=0.35,width=0.325\textwidth,height=0.30\textheight]{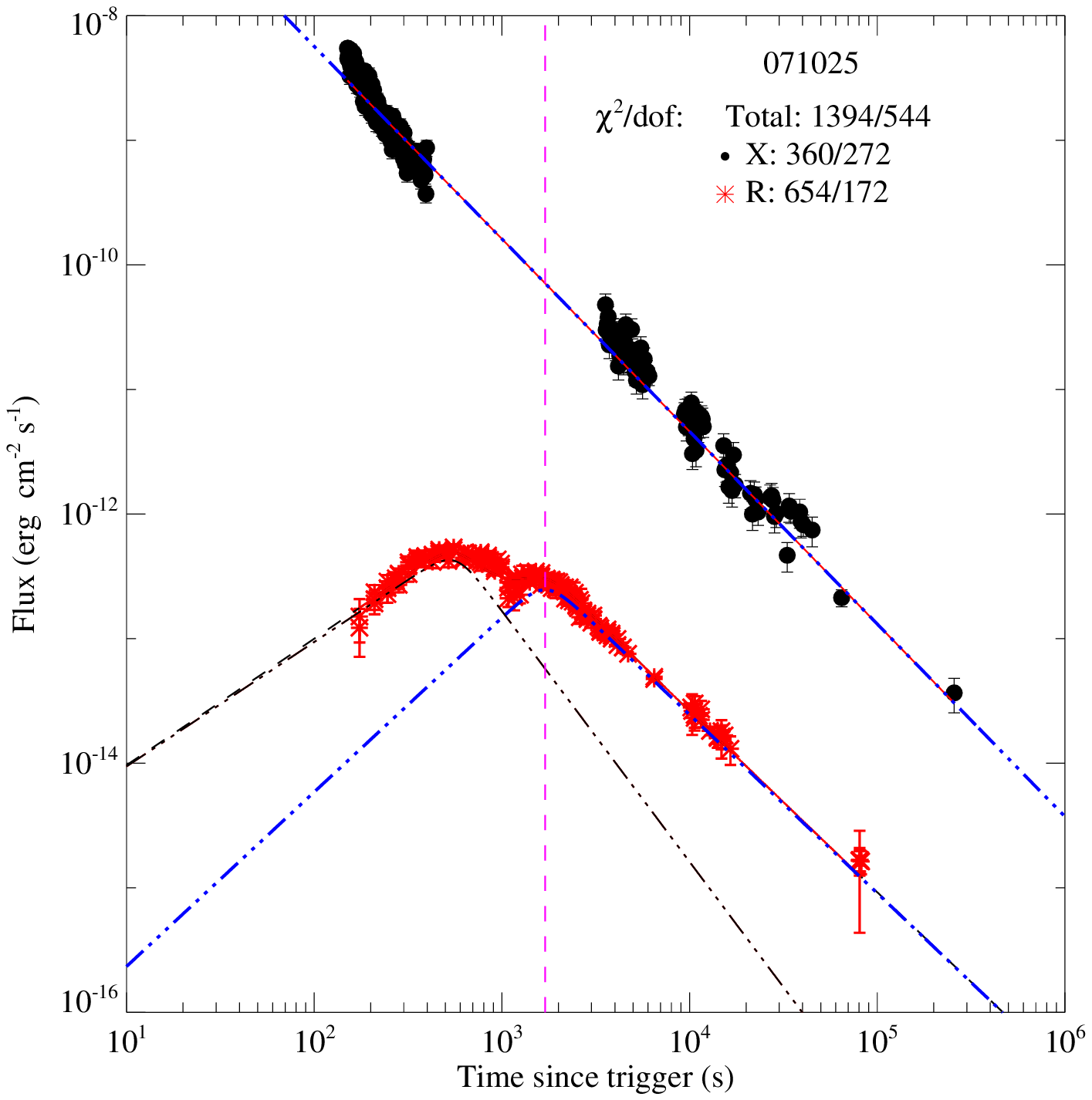}
\includegraphics[angle=0,scale=0.35,width=0.325\textwidth,height=0.30\textheight]{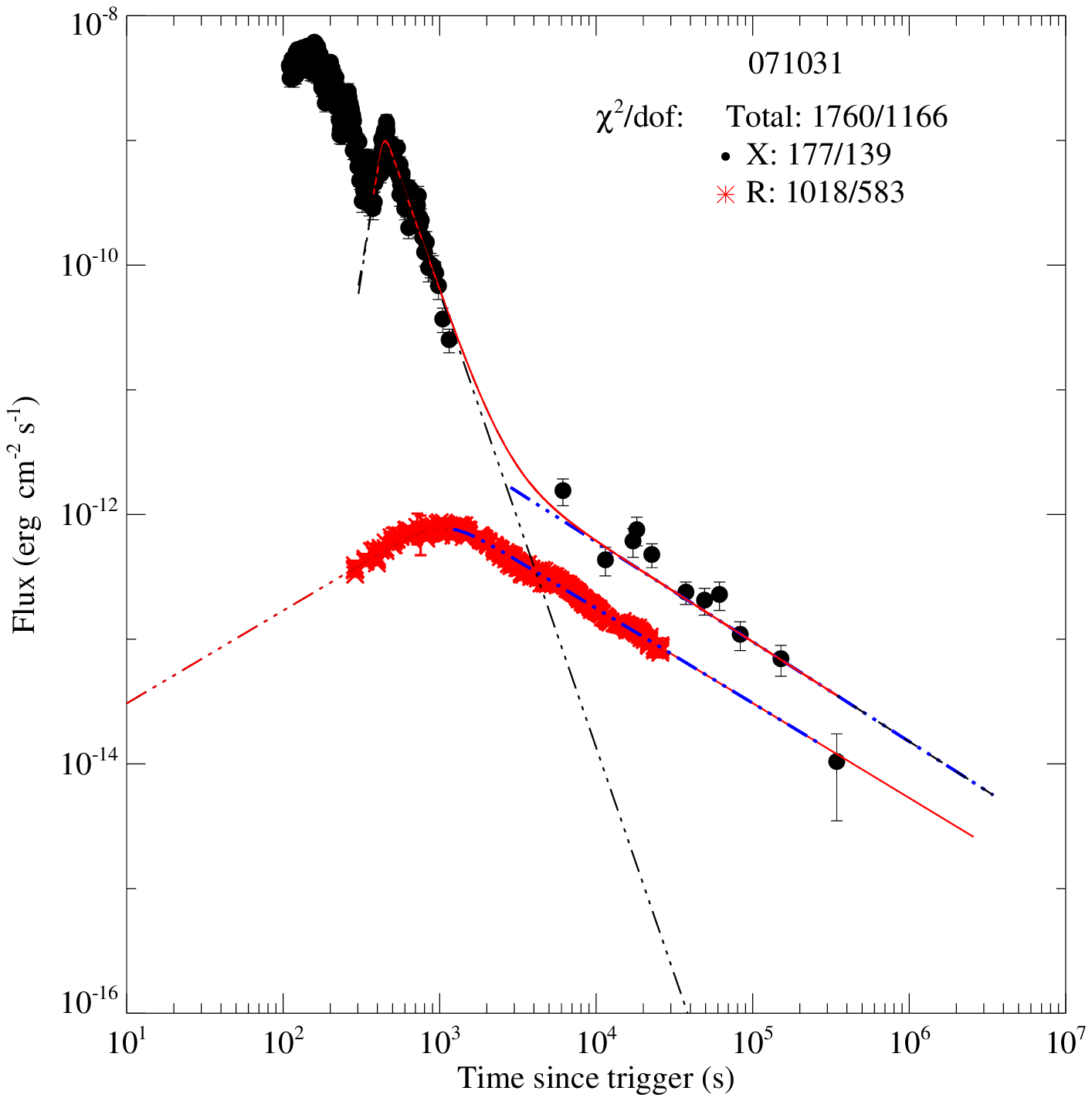}
\includegraphics[angle=0,scale=0.35,width=0.325\textwidth,height=0.30\textheight]{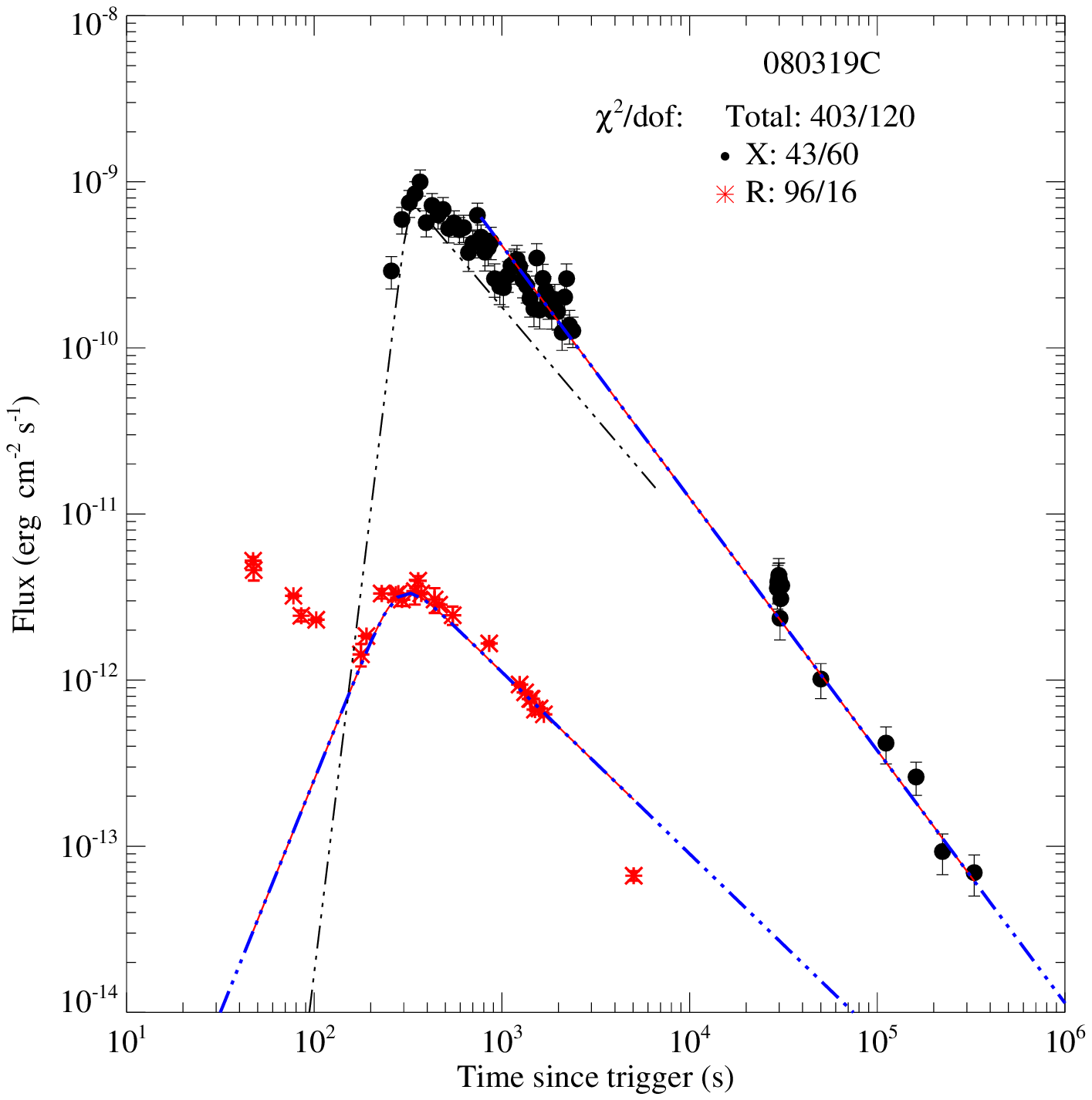}
\includegraphics[angle=0,scale=0.35,width=0.325\textwidth,height=0.30\textheight]{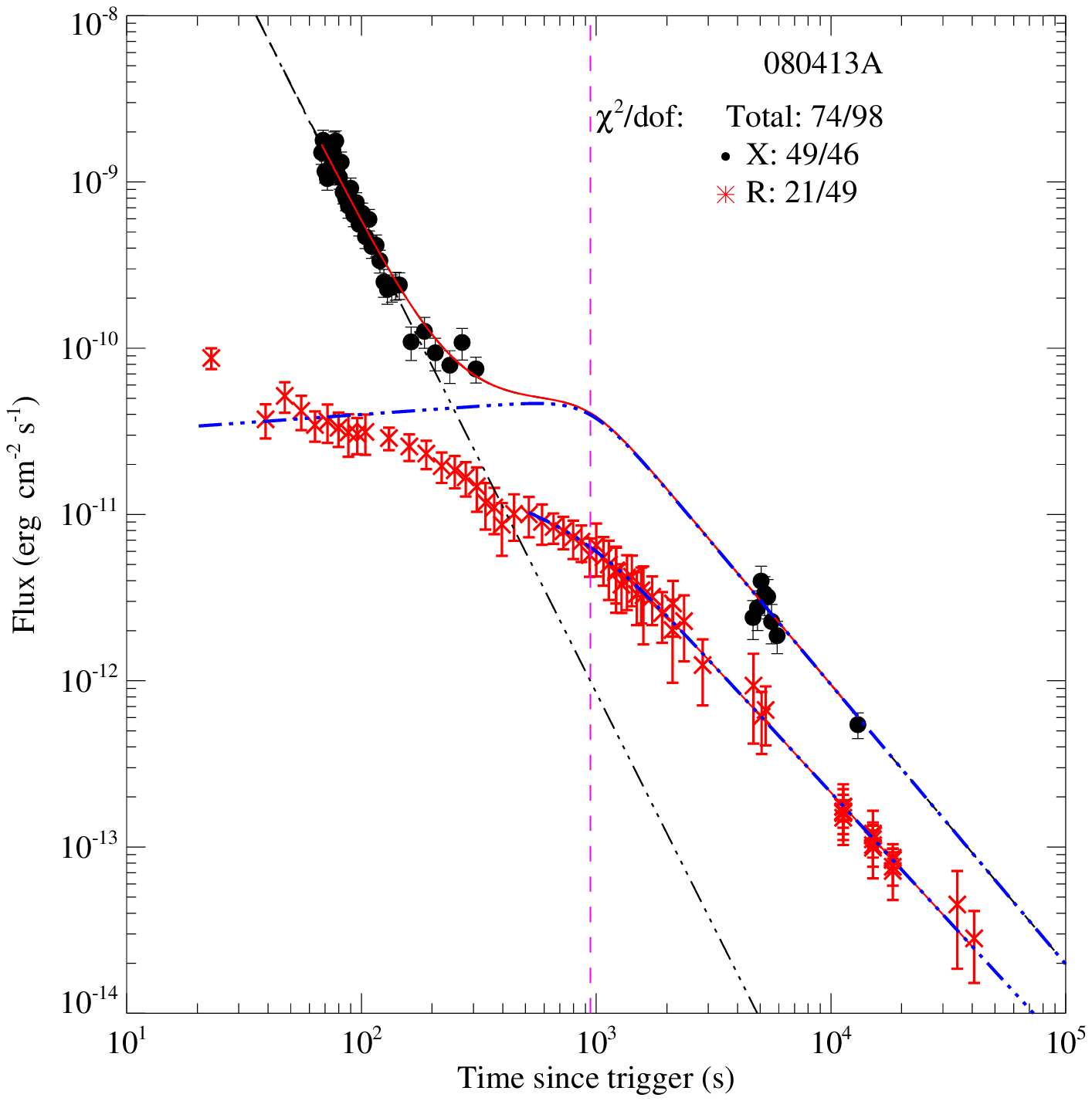}
\includegraphics[angle=0,scale=0.35,width=0.325\textwidth,height=0.30\textheight]{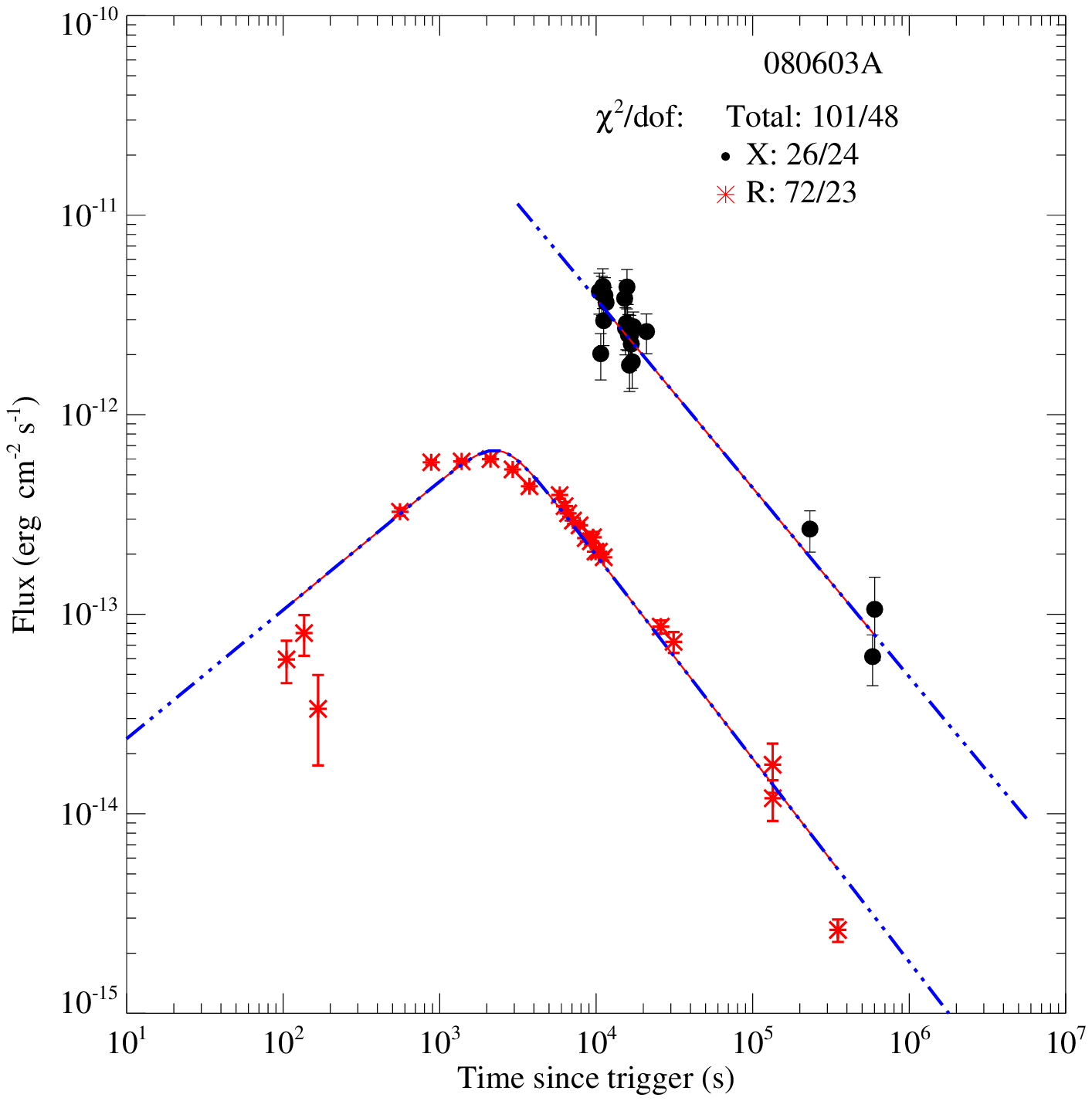}
\includegraphics[angle=0,scale=0.35,width=0.325\textwidth,height=0.30\textheight]{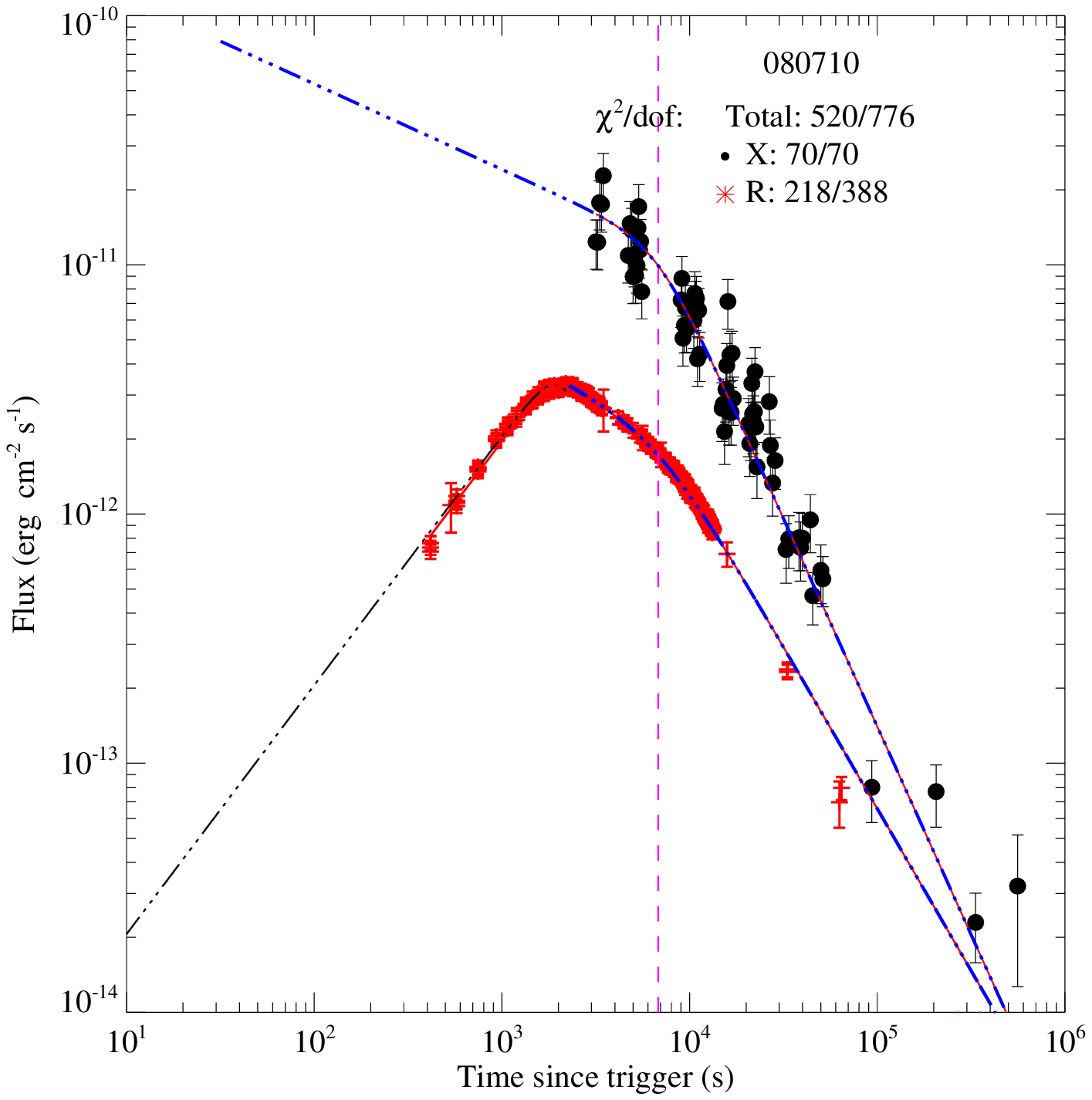}
\includegraphics[angle=0,scale=0.35,width=0.325\textwidth,height=0.30\textheight]{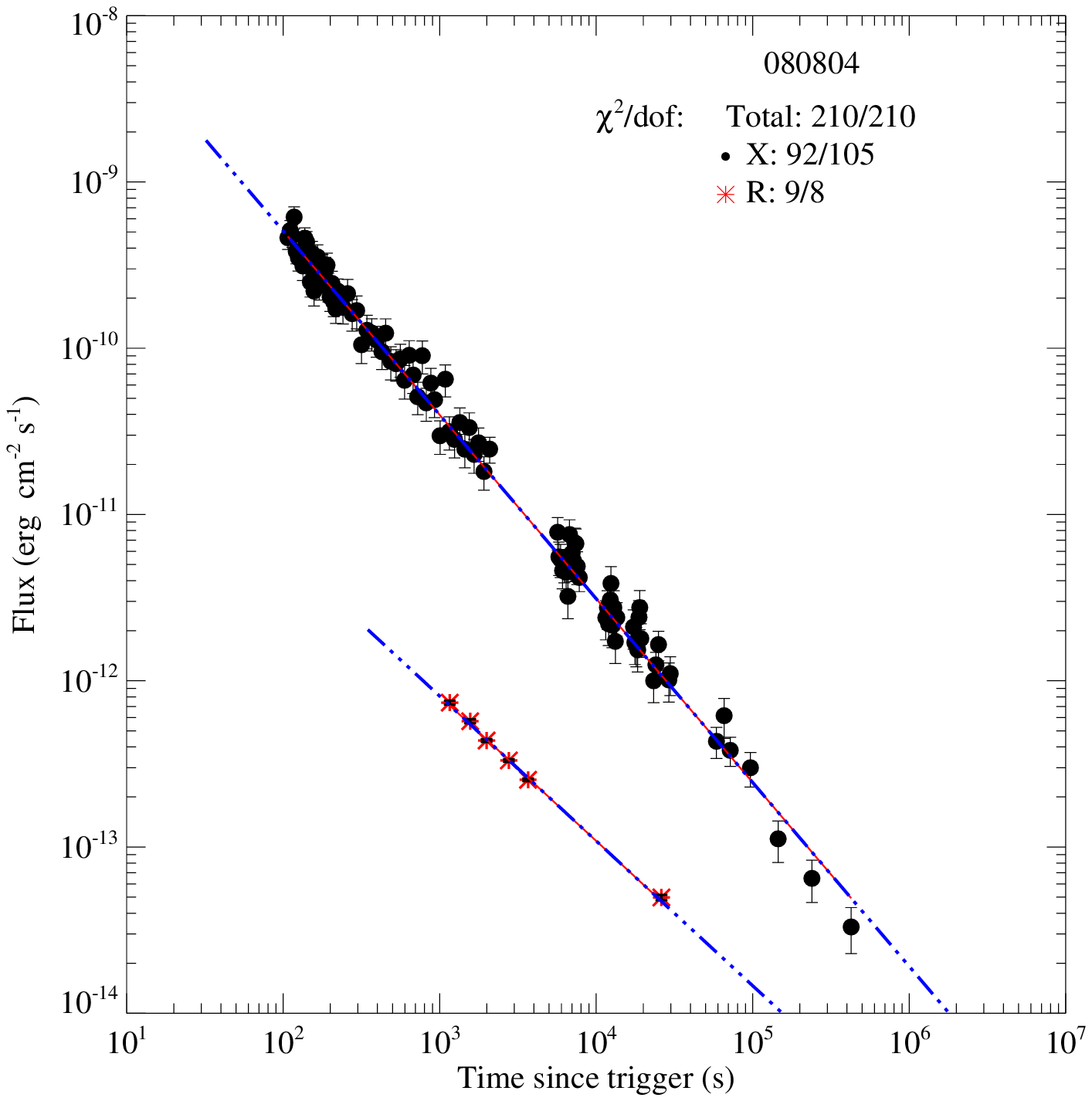}

\center{Fig. \ref{gradeI}---Continued}
\end{figure*}

\clearpage
\setlength{\voffset}{-18mm}
\begin{figure*}

\includegraphics[angle=0,scale=0.35,width=0.325\textwidth,height=0.30\textheight]{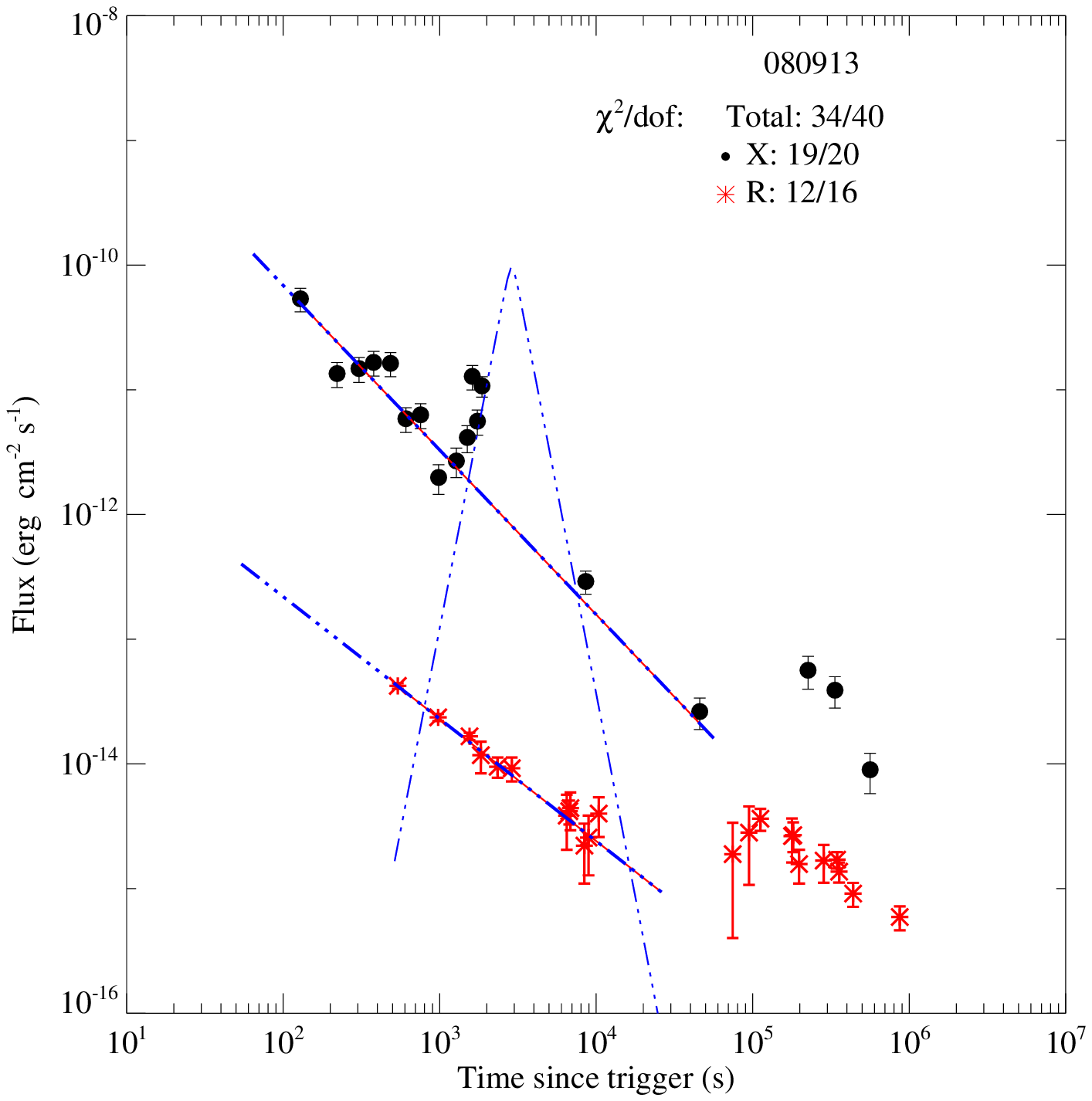}
\includegraphics[angle=0,scale=0.35,width=0.325\textwidth,height=0.30\textheight]{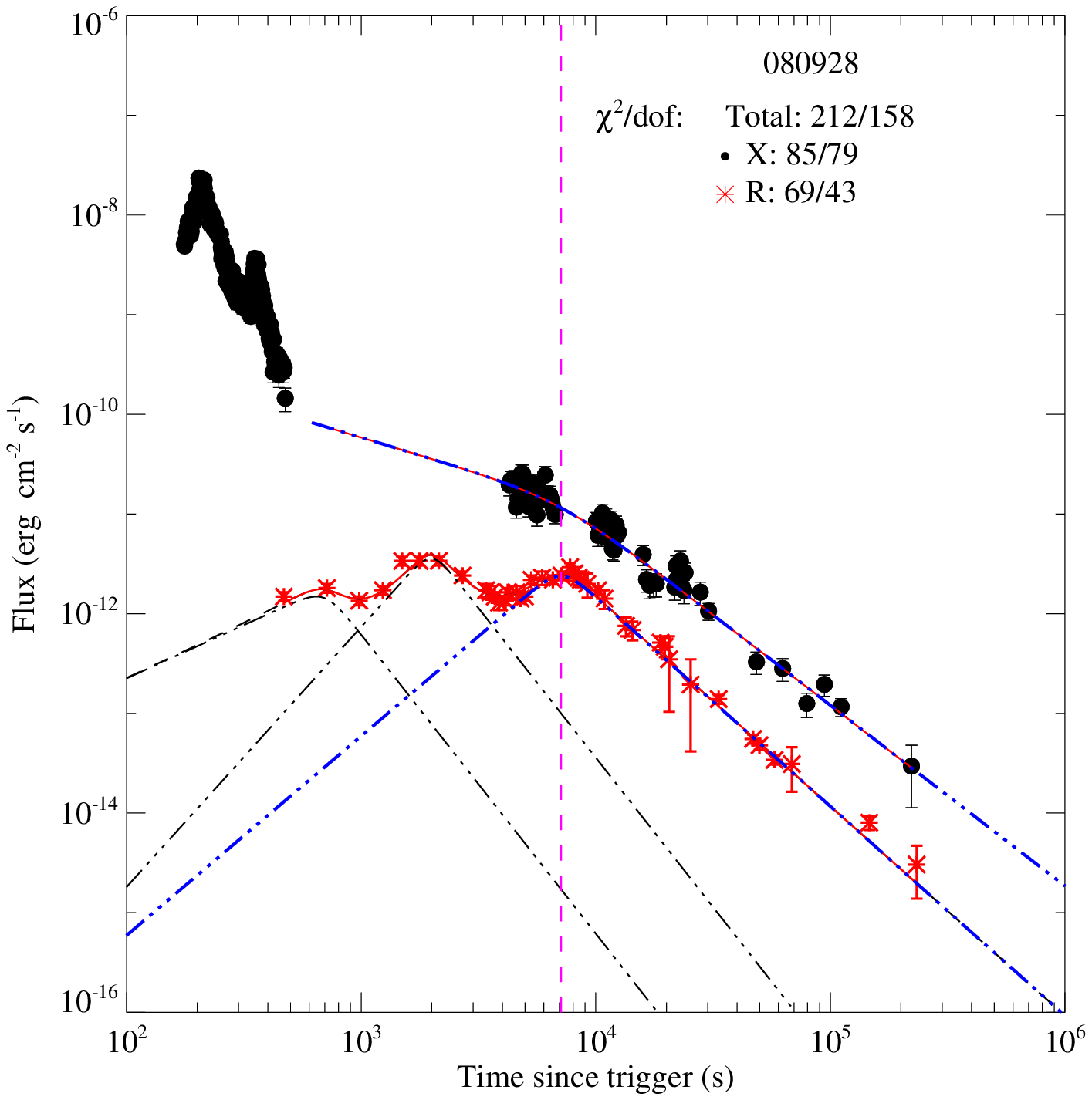}
\includegraphics[angle=0,scale=0.35,width=0.325\textwidth,height=0.30\textheight]{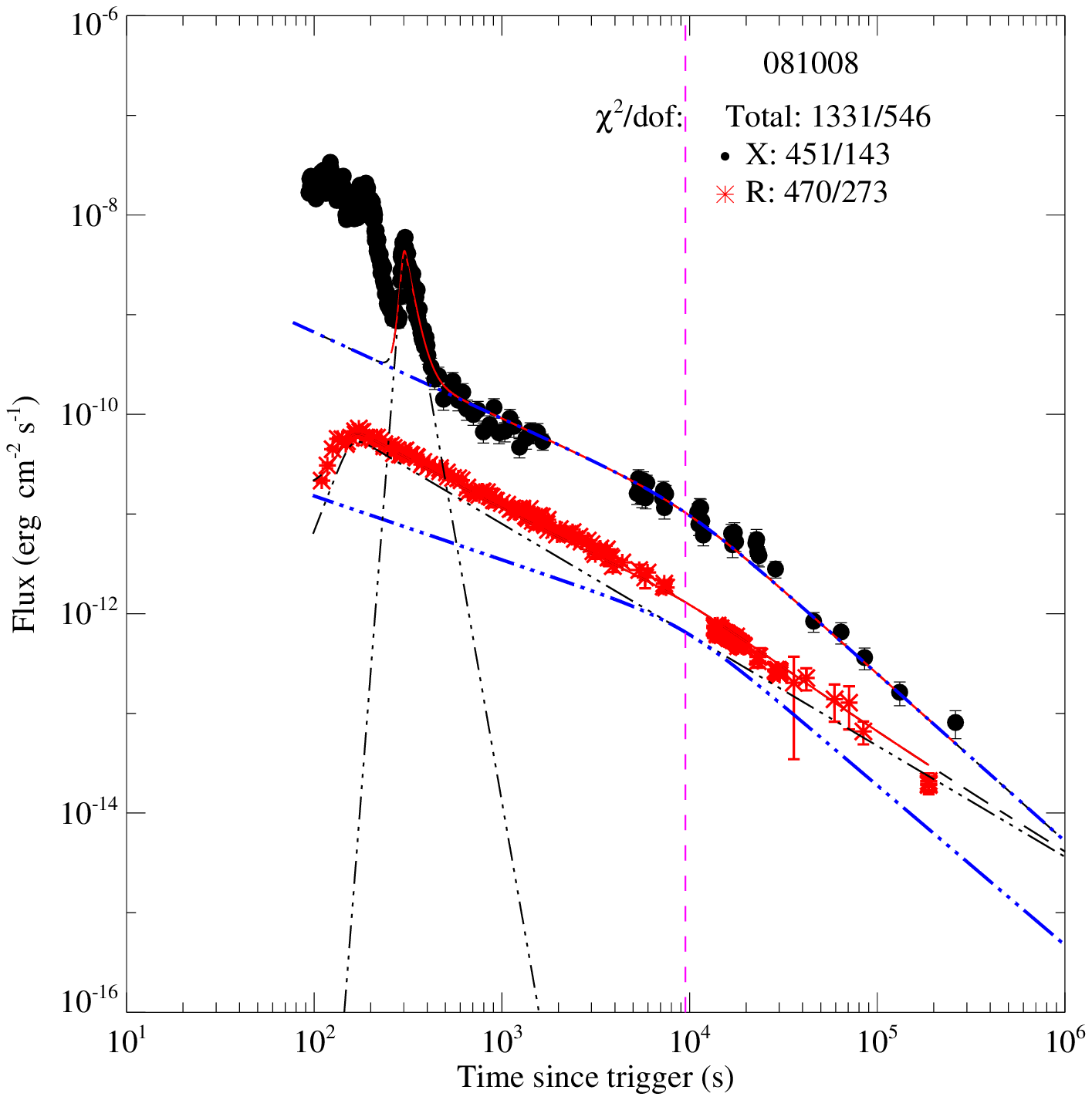}
\includegraphics[angle=0,scale=0.35,width=0.325\textwidth,height=0.30\textheight]{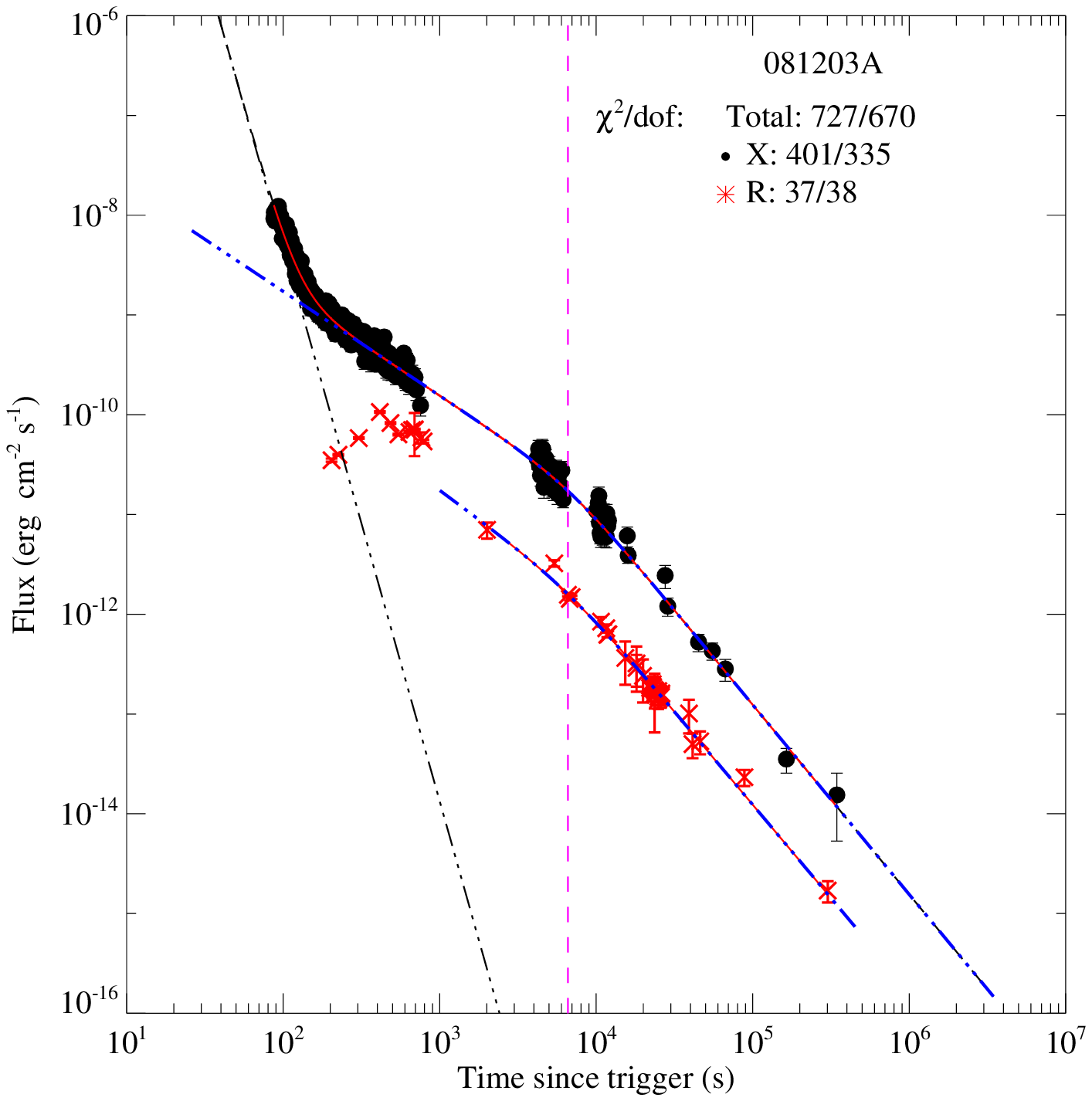}
\includegraphics[angle=0,scale=0.35,width=0.325\textwidth,height=0.30\textheight]{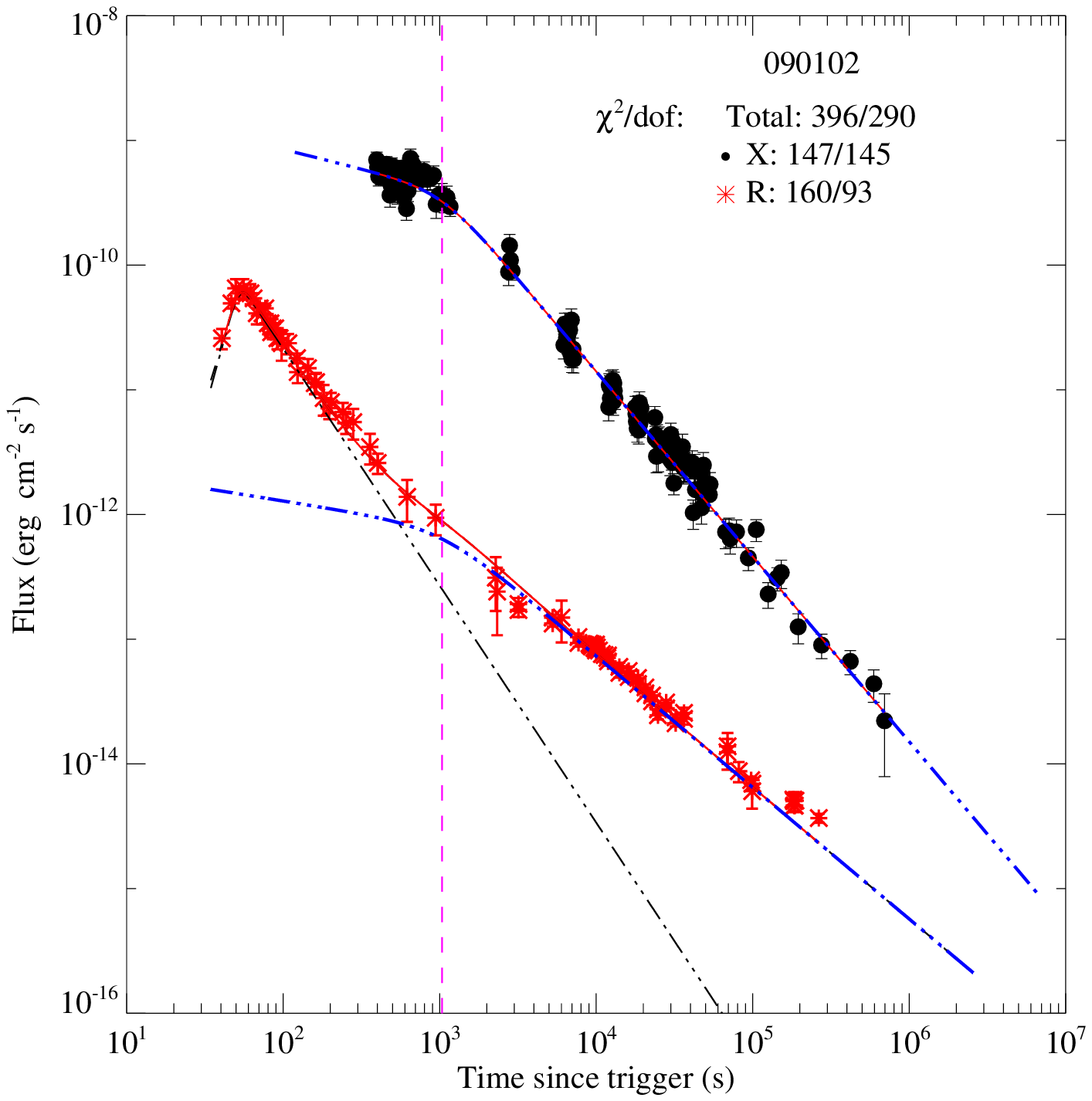}
\includegraphics[angle=0,scale=0.35,width=0.325\textwidth,height=0.30\textheight]{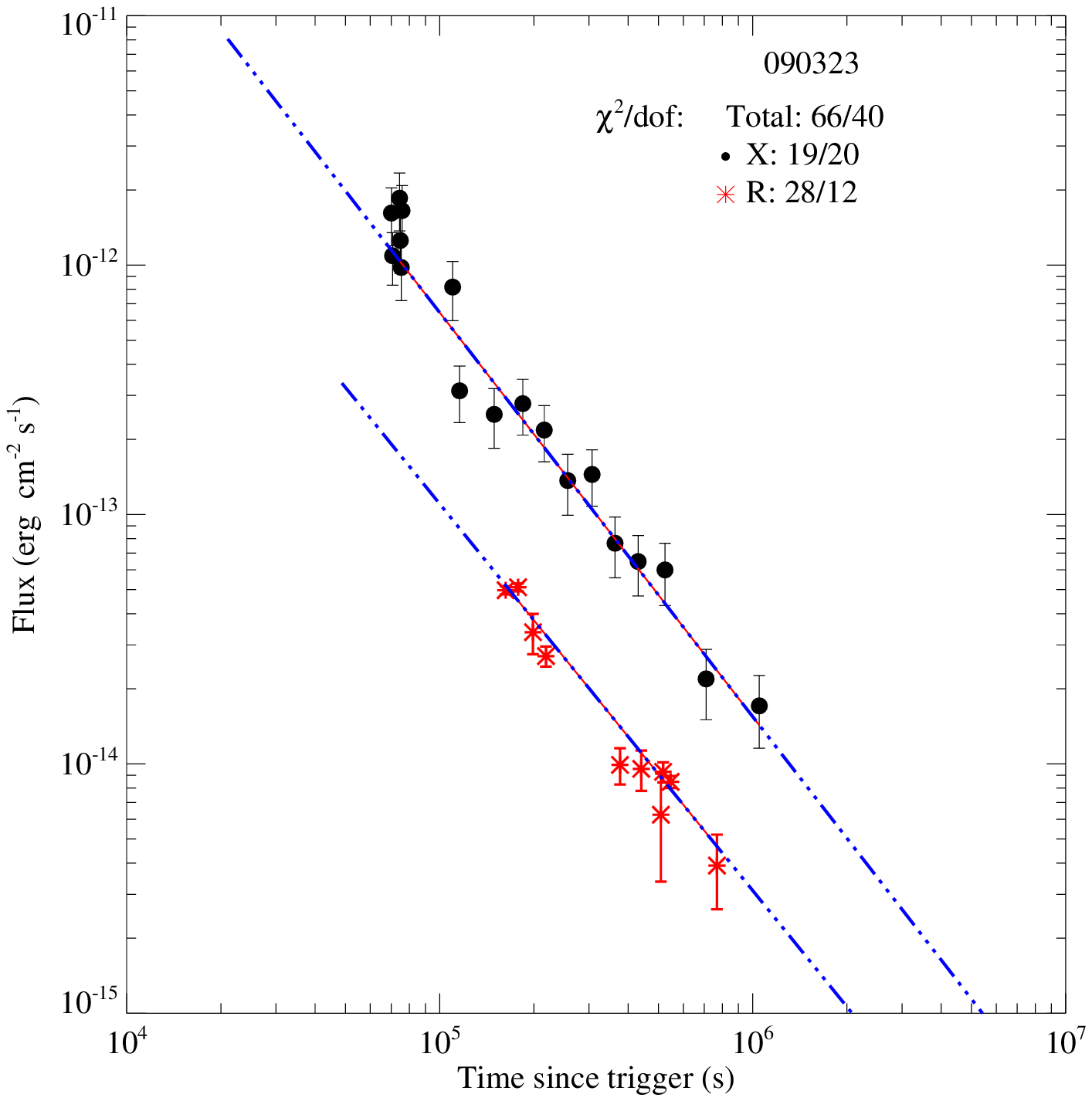}
\includegraphics[angle=0,scale=0.35,width=0.325\textwidth,height=0.30\textheight]{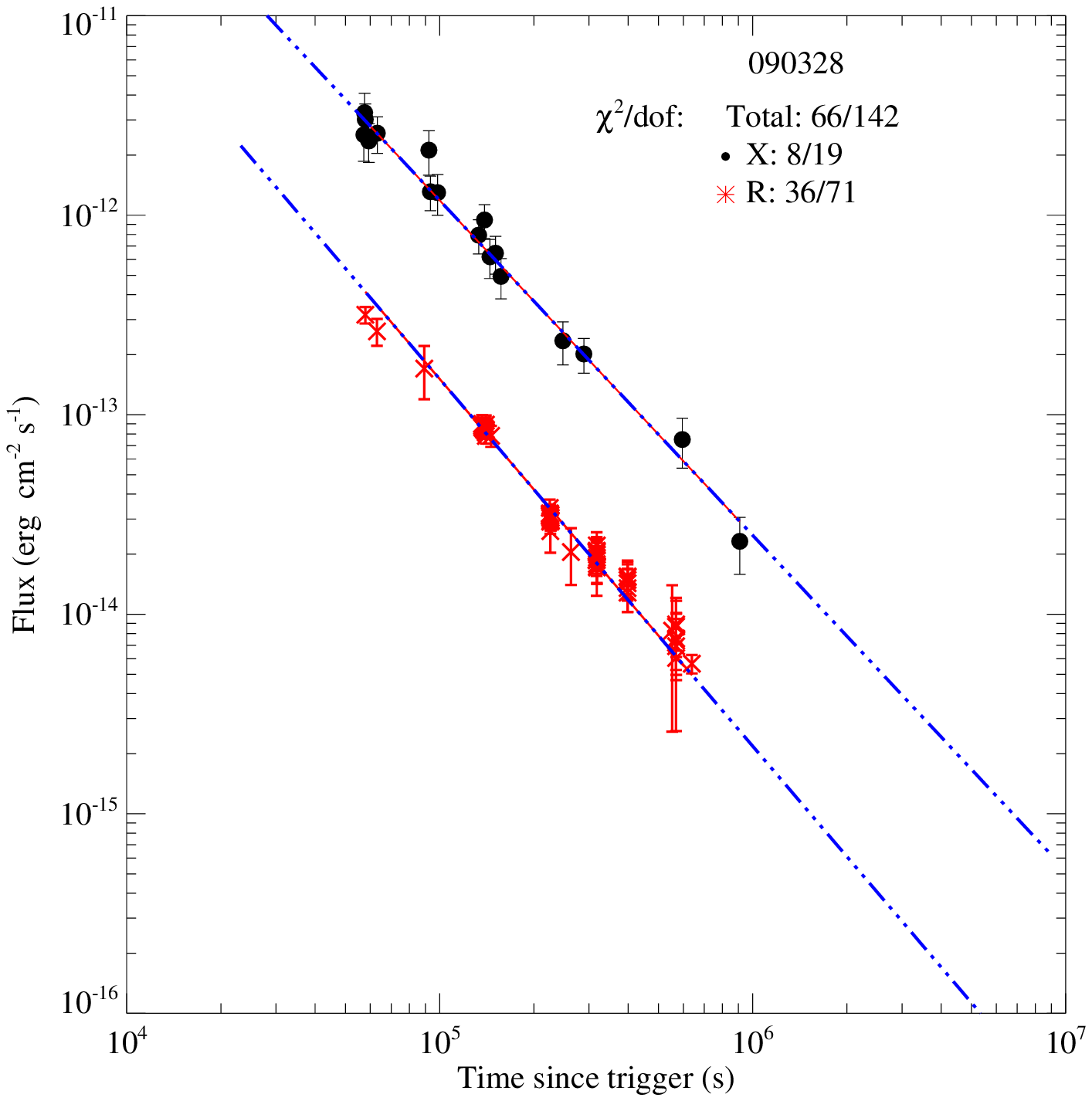}
\includegraphics[angle=0,scale=0.35,width=0.325\textwidth,height=0.30\textheight]{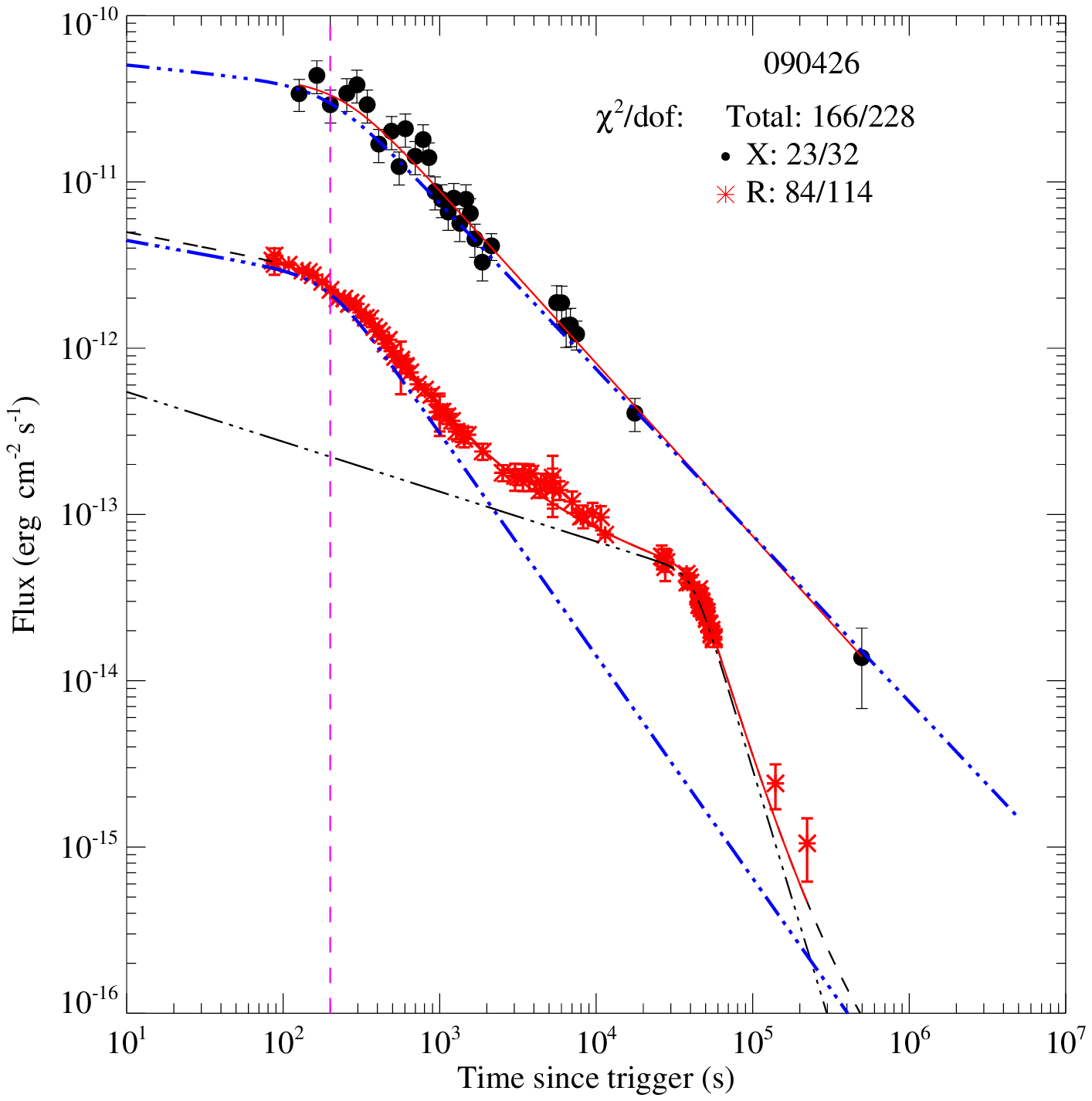}
\includegraphics[angle=0,scale=0.35,width=0.325\textwidth,height=0.30\textheight]{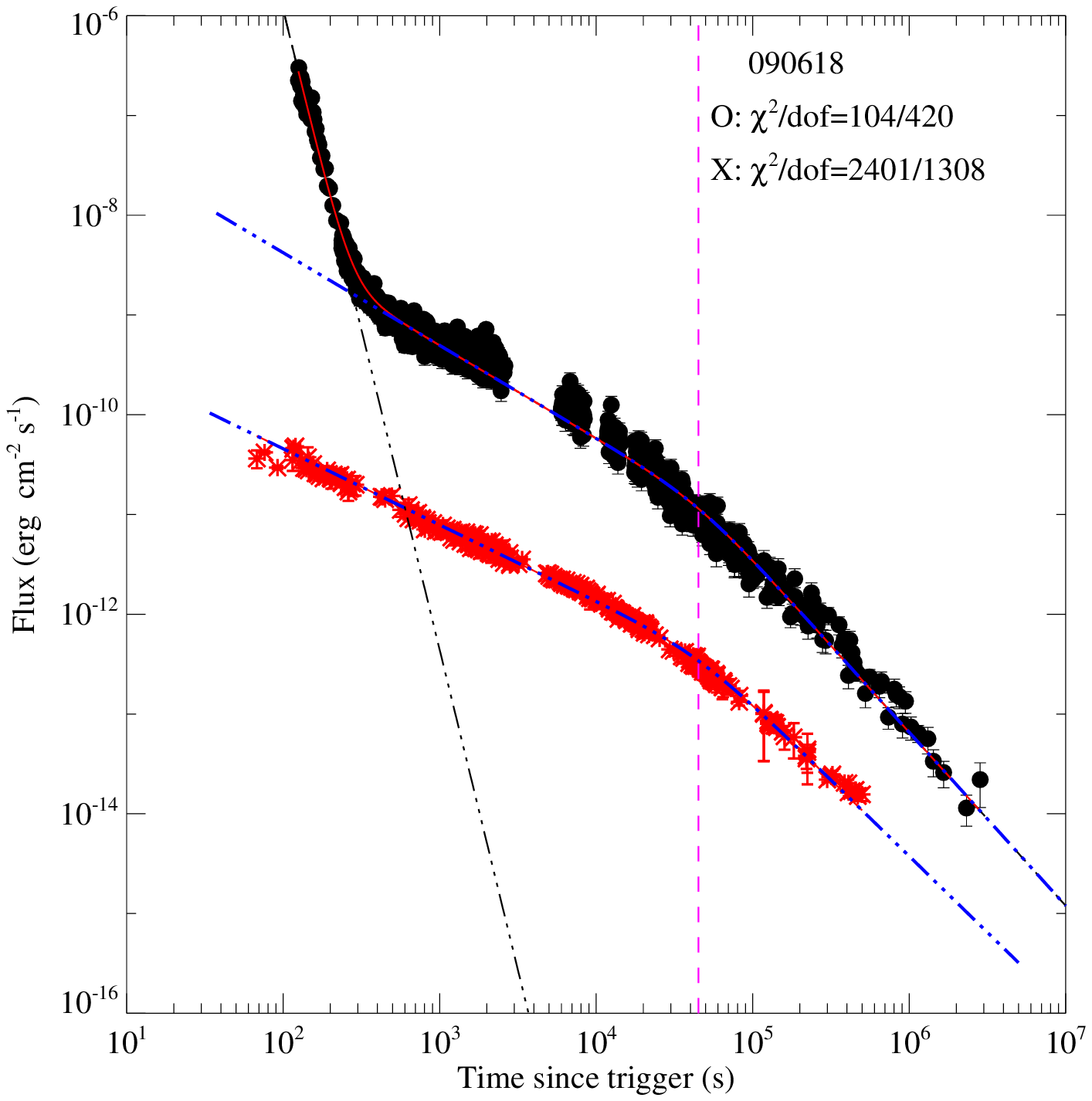}

\center{Fig. \ref{gradeI}---Continued}
\end{figure*}

\clearpage
\setlength{\voffset}{-18mm}
\begin{figure*}
\includegraphics[angle=0,scale=0.35,width=0.325\textwidth,height=0.30\textheight]{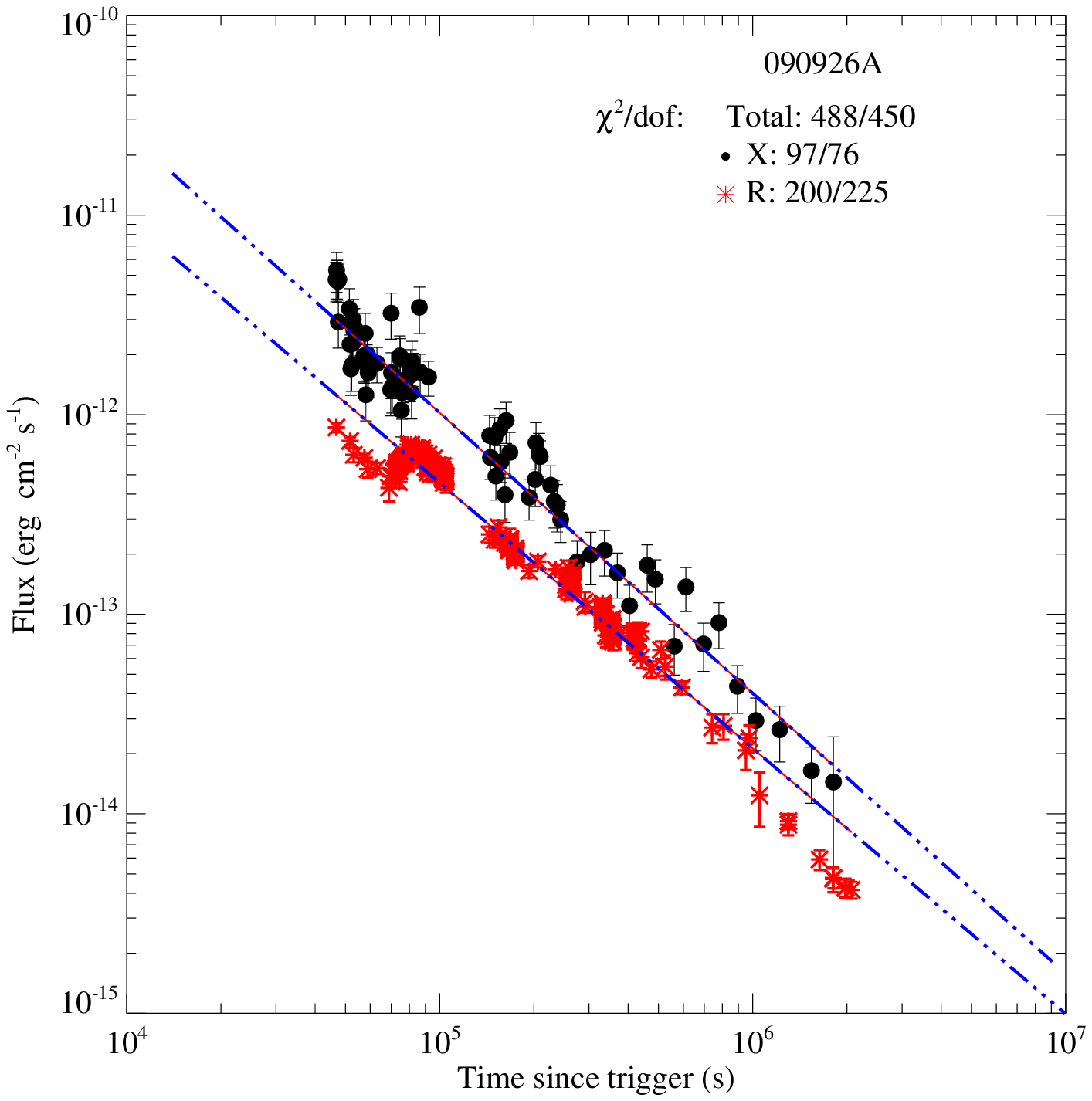}
\includegraphics[angle=0,scale=0.35,width=0.325\textwidth,height=0.30\textheight]{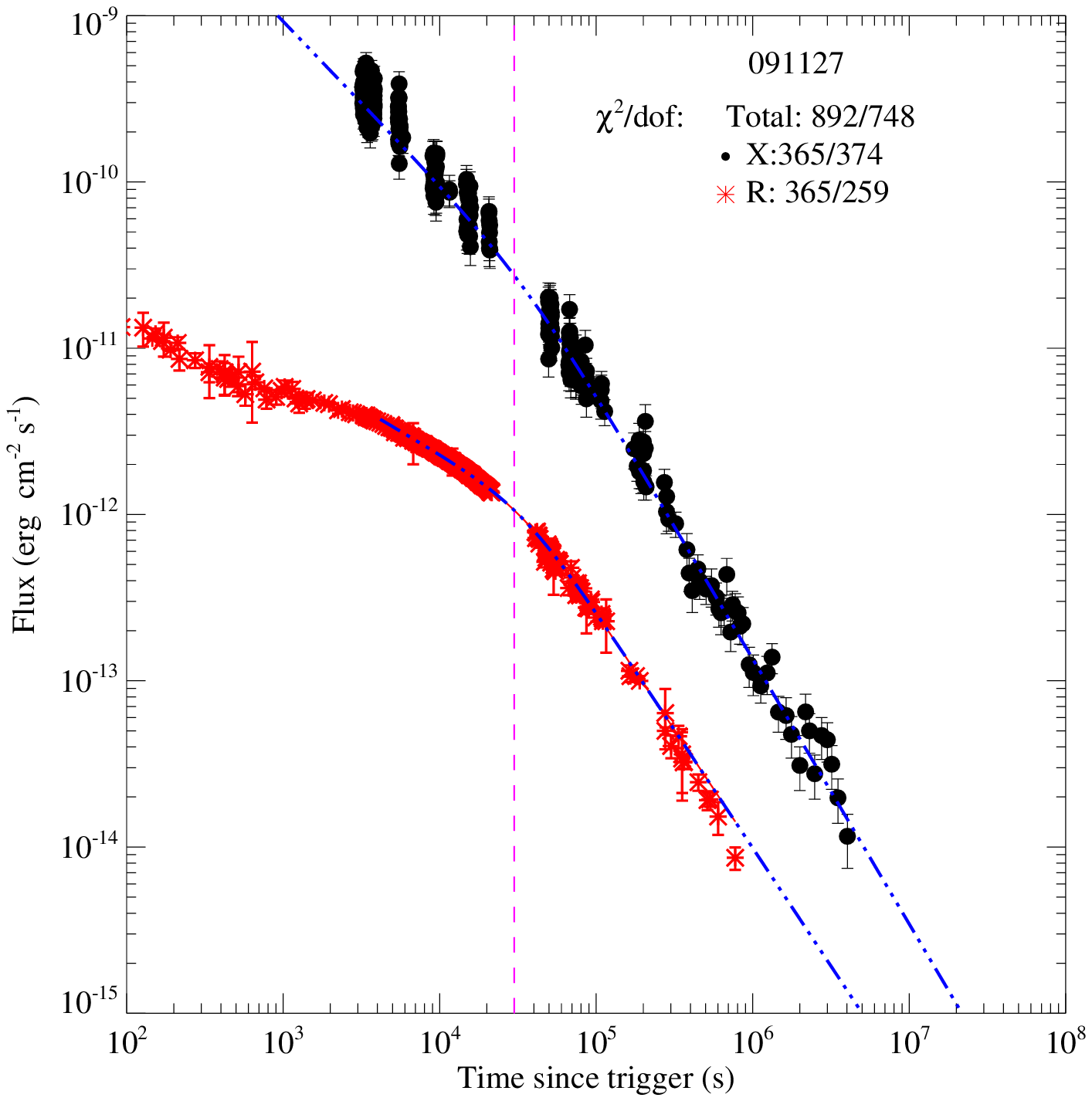}
\includegraphics[angle=0,scale=0.35,width=0.325\textwidth,height=0.30\textheight]{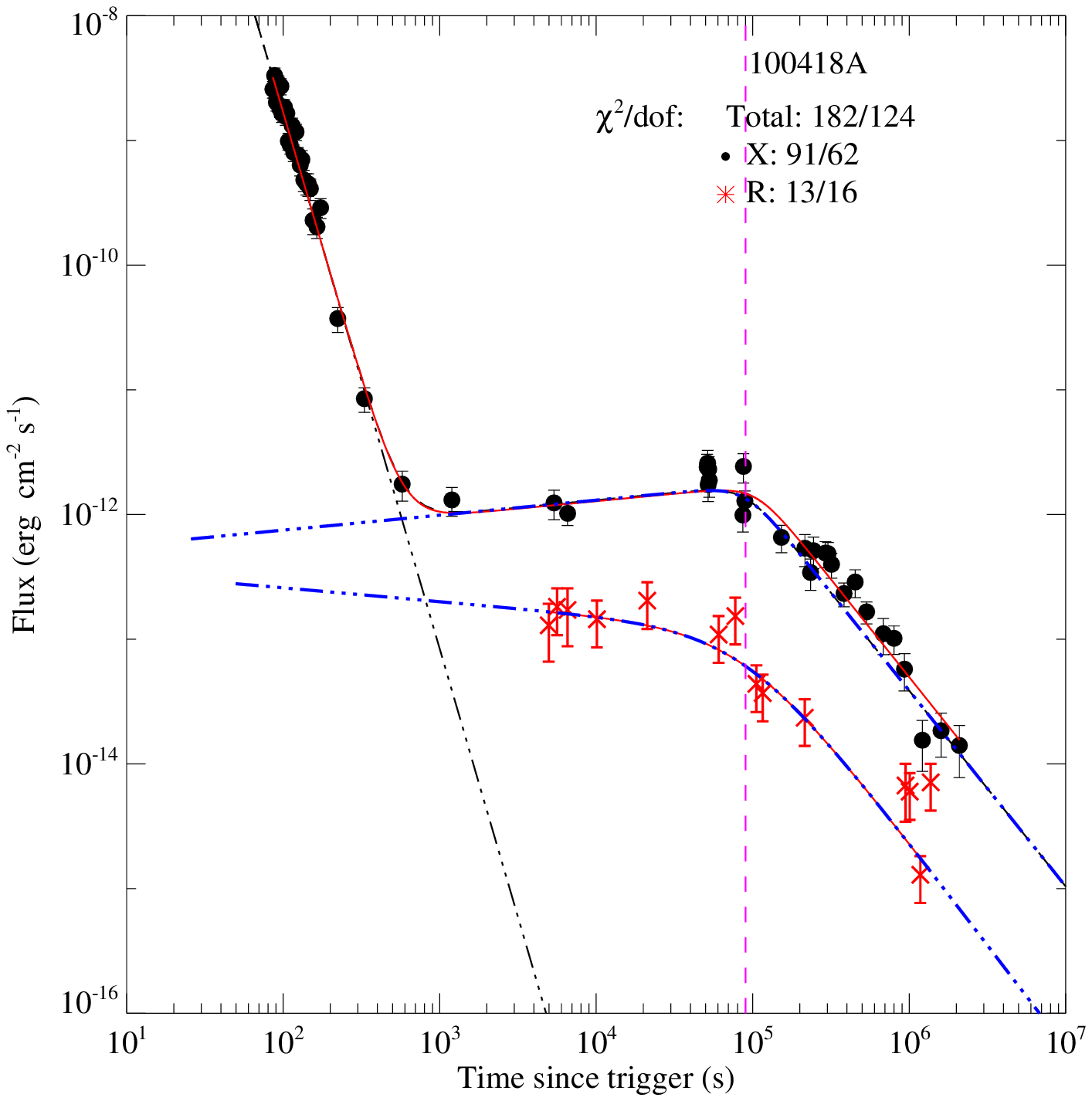}
\includegraphics[angle=0,scale=0.35,width=0.325\textwidth,height=0.30\textheight]{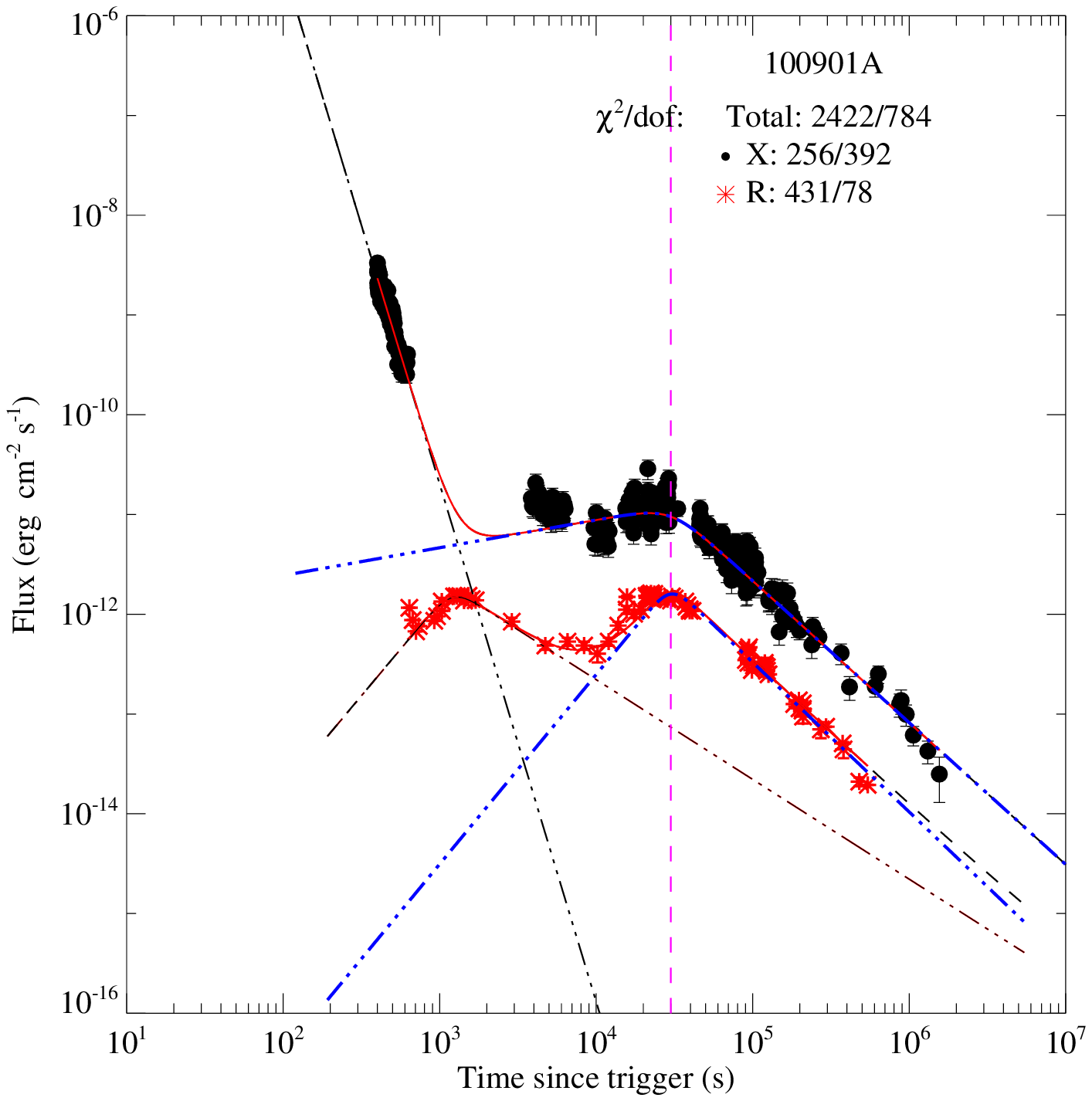}
\includegraphics[angle=0,scale=0.35,width=0.325\textwidth,height=0.30\textheight]{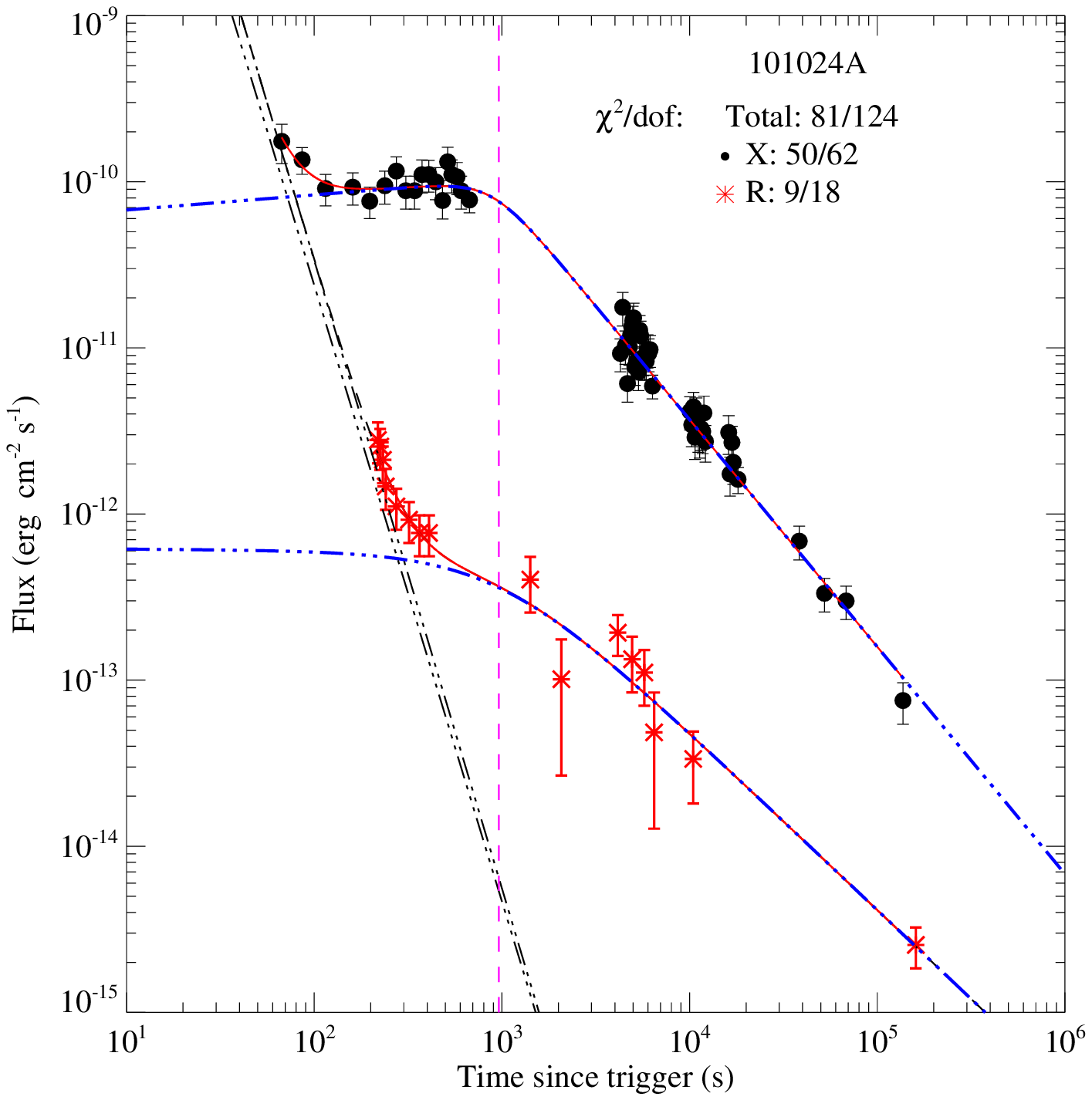}
\includegraphics[angle=0,scale=0.35,width=0.325\textwidth,height=0.30\textheight]{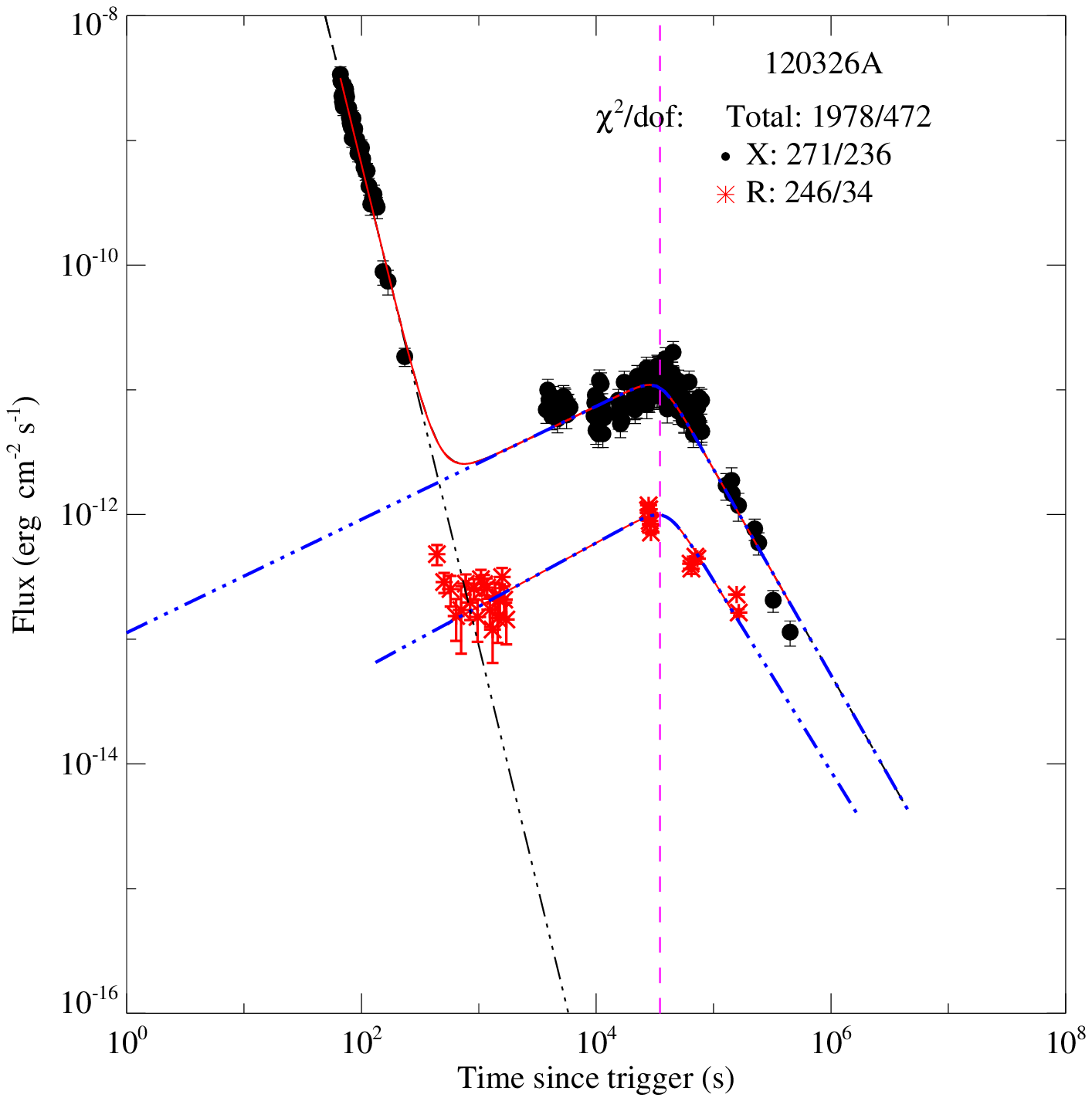}
\includegraphics[angle=0,scale=0.35,width=0.325\textwidth,height=0.30\textheight]{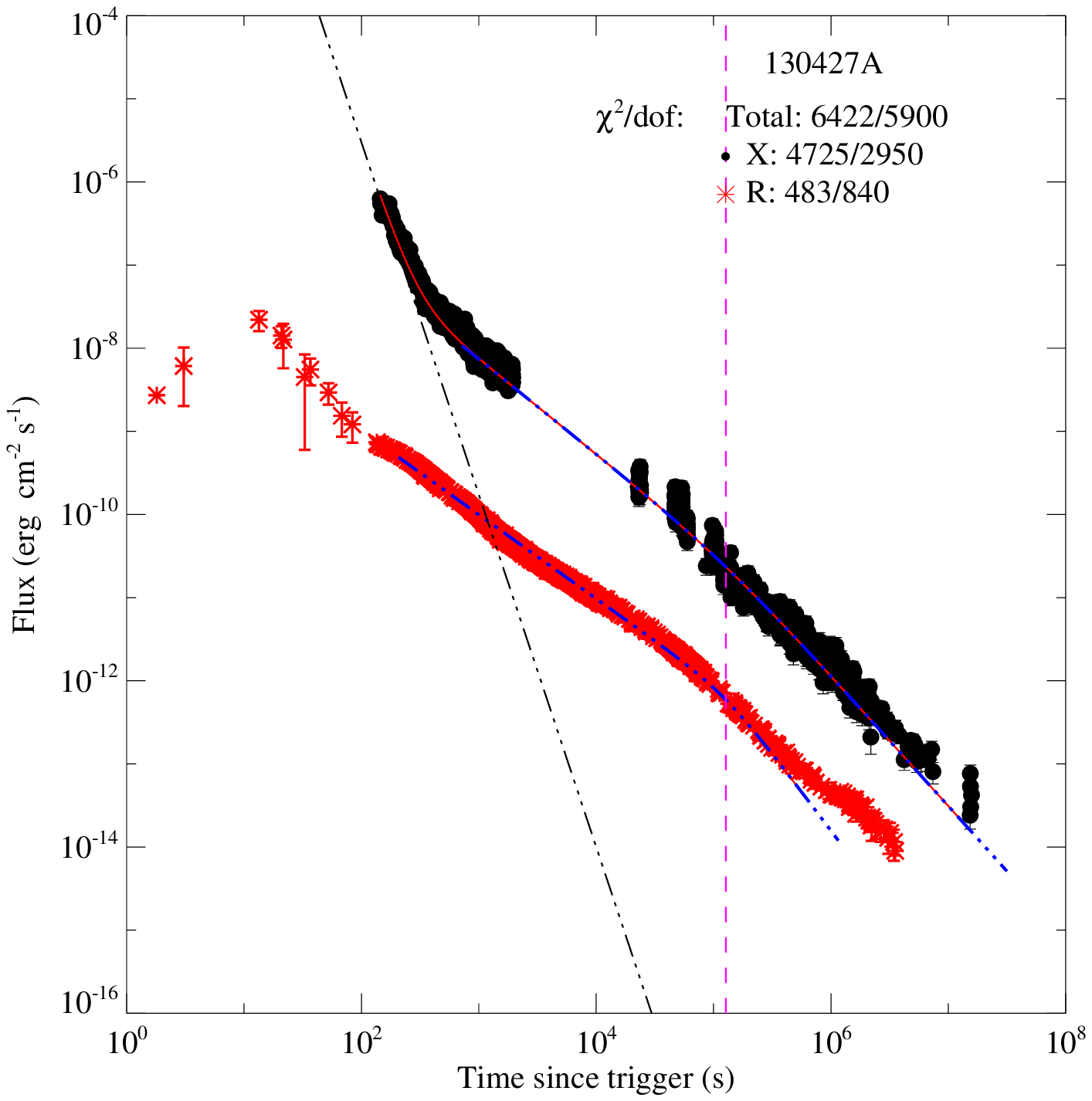}
\center{Fig. \ref{gradeI}---Continued}
\end{figure*}

\clearpage
\setlength{\voffset}{-18mm}

\clearpage
\setlength{\voffset}{-18mm}

\clearpage
\setlength{\voffset}{-18mm}
\begin{figure*}
\includegraphics[angle=0,scale=0.35,width=0.325\textwidth,height=0.30\textheight]{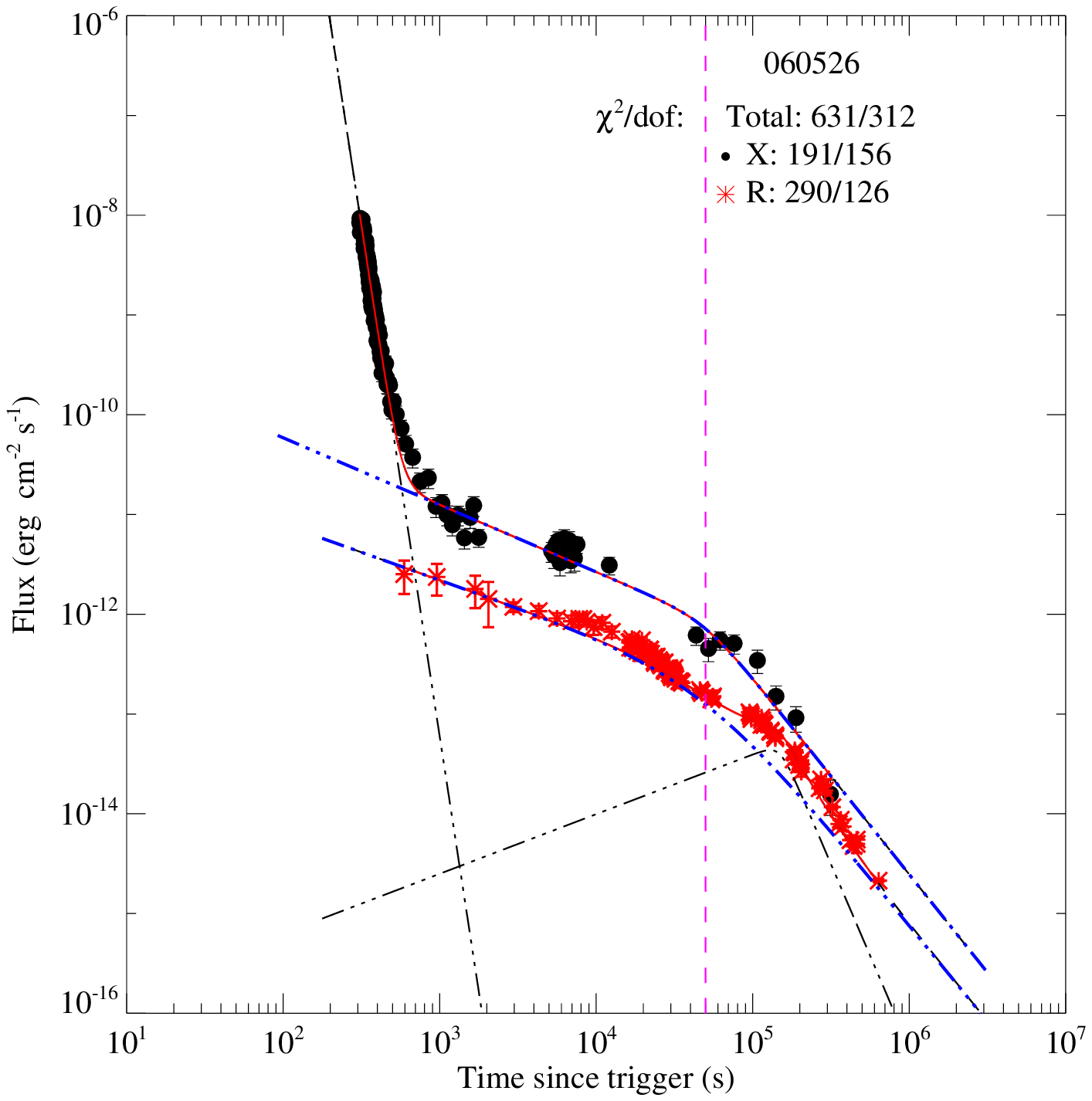}
\includegraphics[angle=0,scale=0.35,width=0.325\textwidth,height=0.30\textheight]{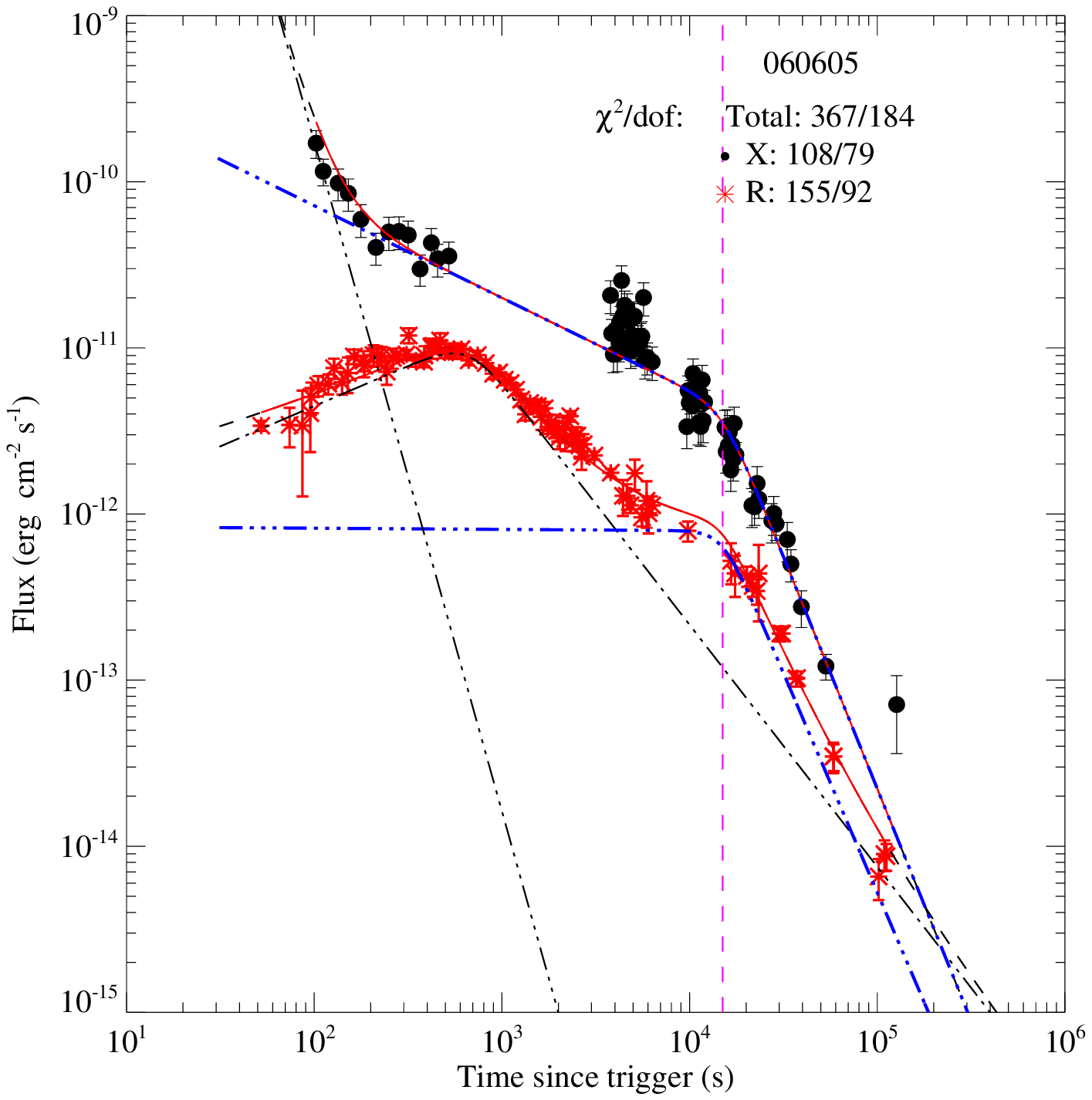}
\includegraphics[angle=0,scale=0.35,width=0.325\textwidth,height=0.30\textheight]{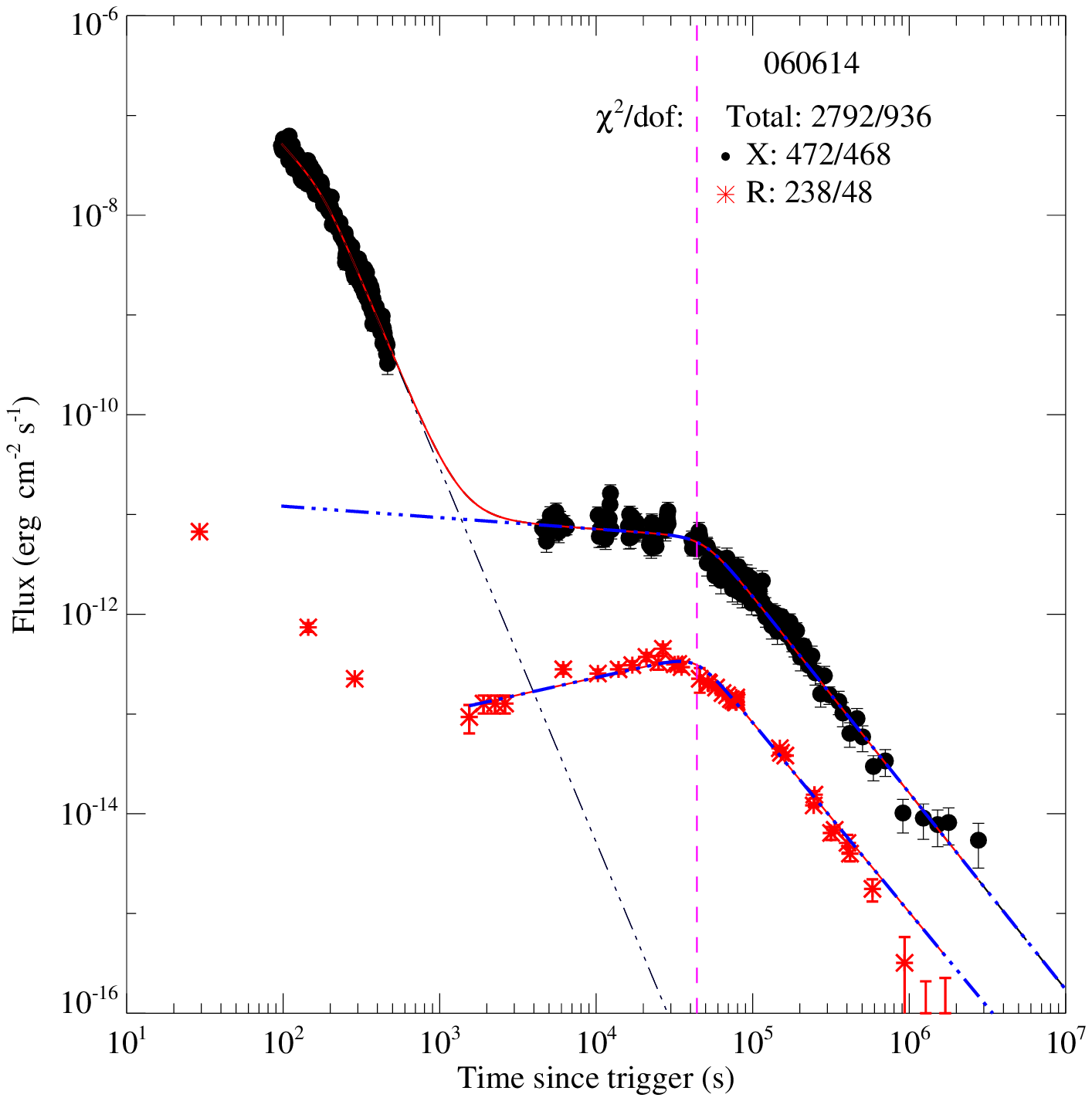}
\includegraphics[angle=0,scale=0.35,width=0.325\textwidth,height=0.30\textheight]{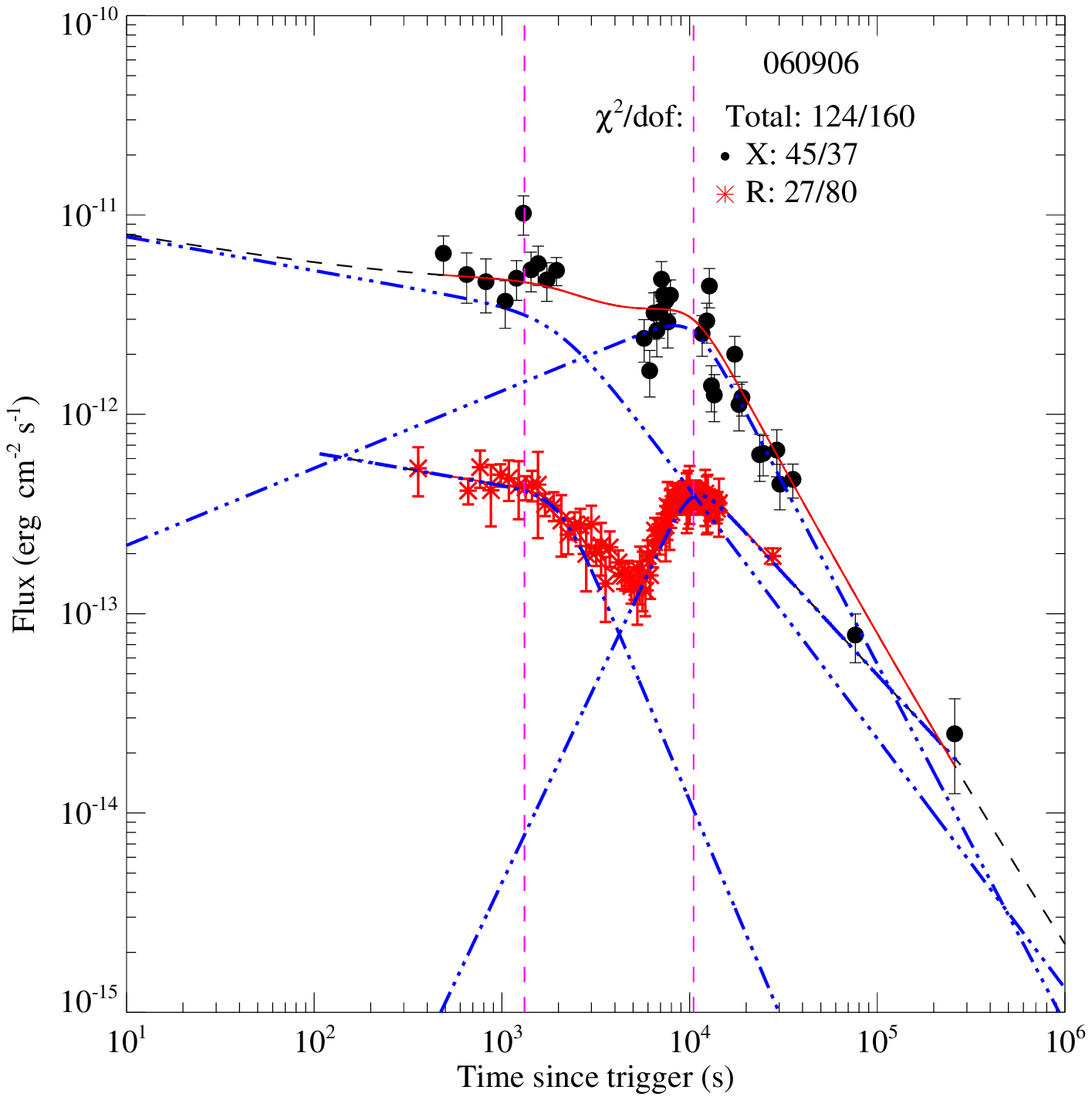}
\includegraphics[angle=0,scale=0.35,width=0.325\textwidth,height=0.30\textheight]{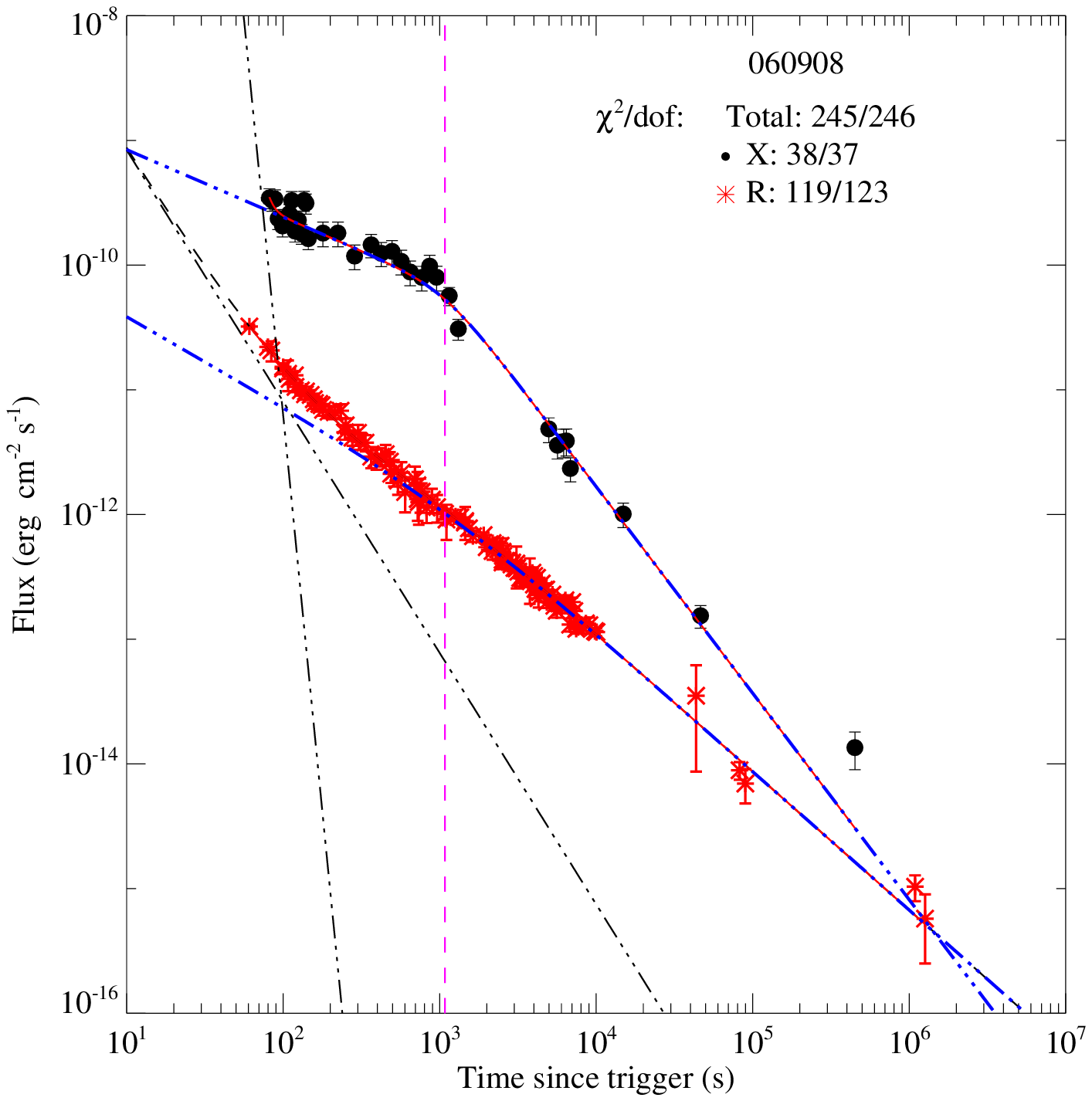}
\includegraphics[angle=0,scale=0.35,width=0.325\textwidth,height=0.30\textheight]{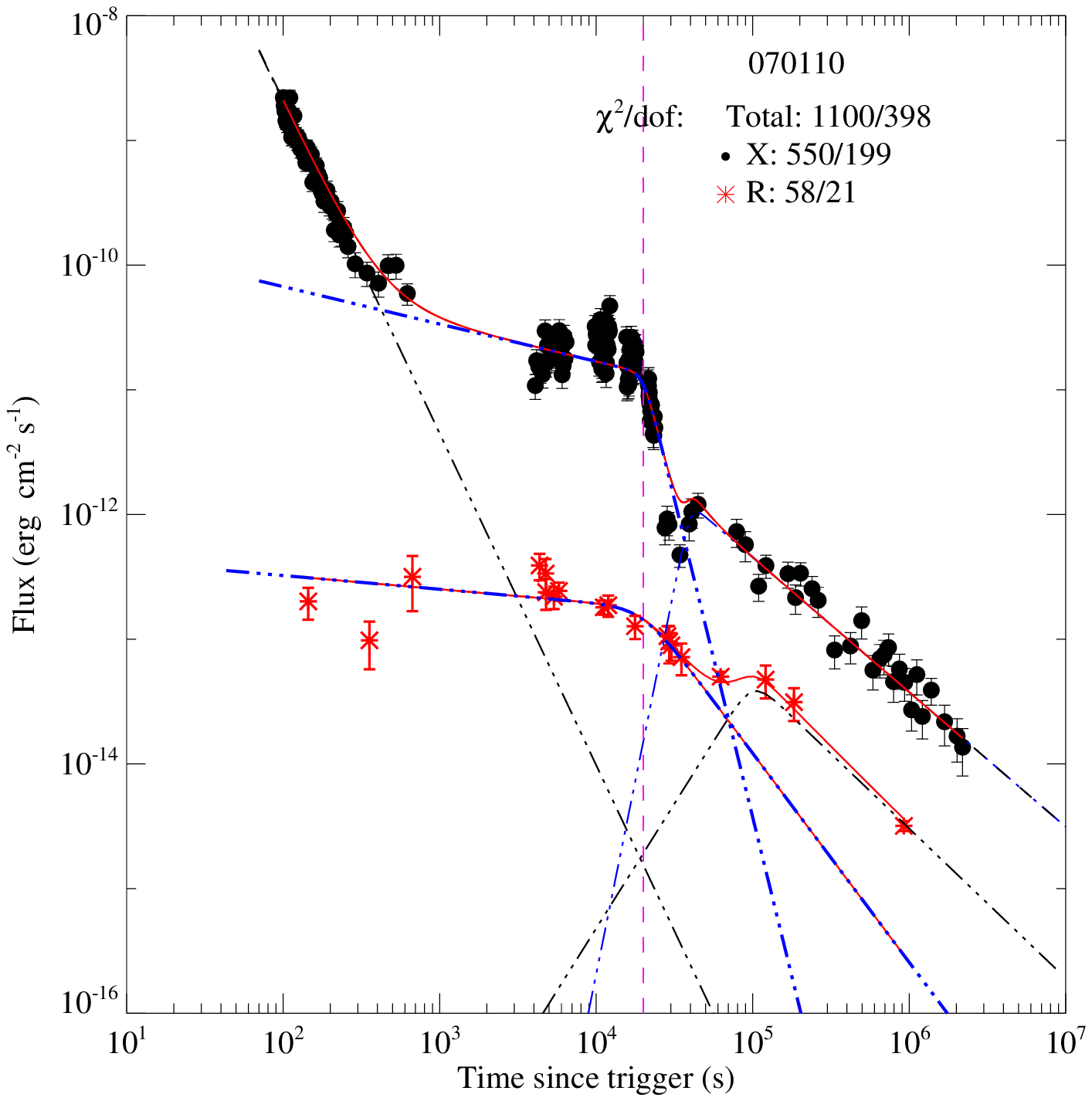}
\includegraphics[angle=0,scale=0.35,width=0.325\textwidth,height=0.30\textheight]{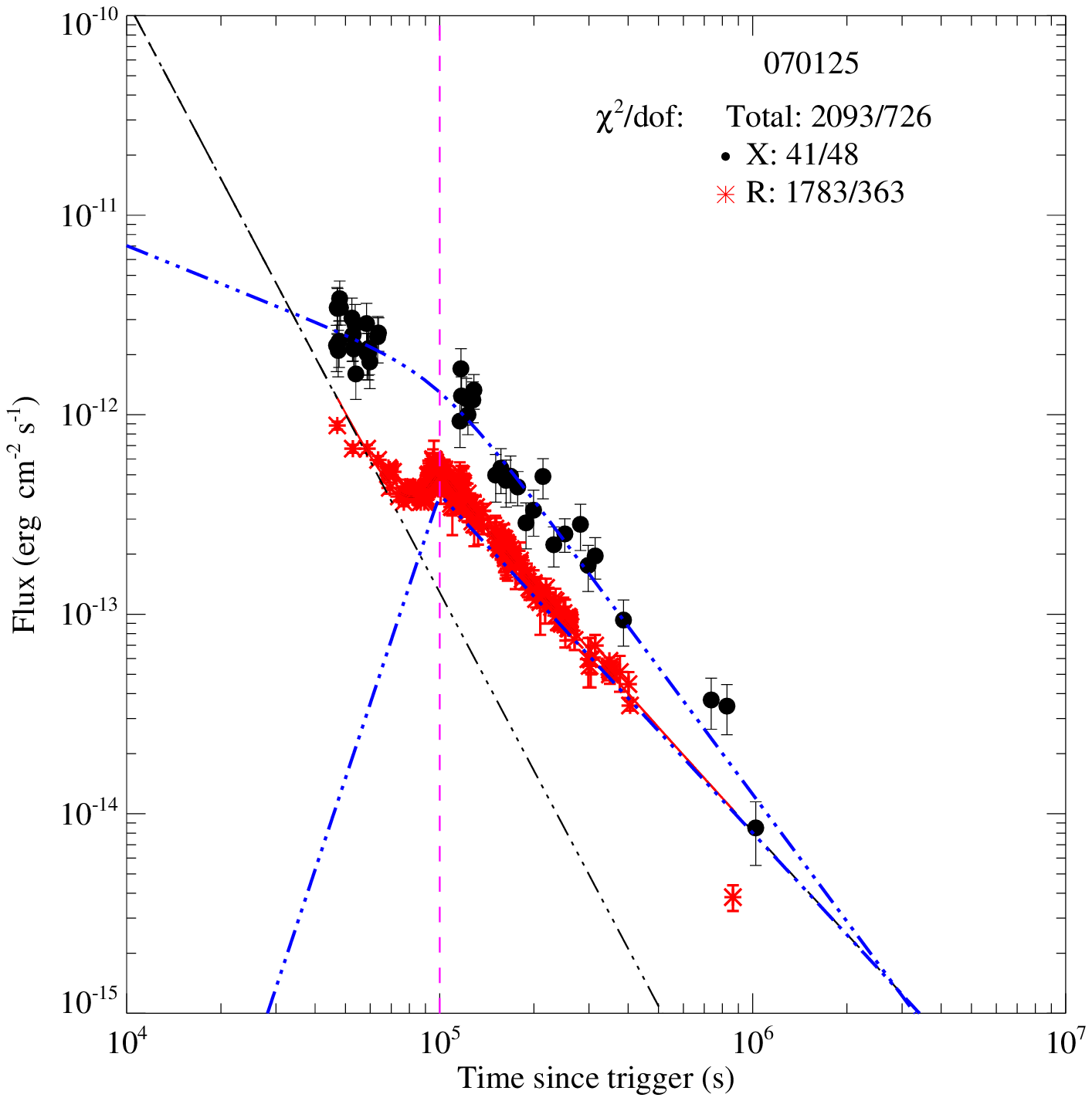}
\includegraphics[angle=0,scale=0.35,width=0.325\textwidth,height=0.30\textheight]{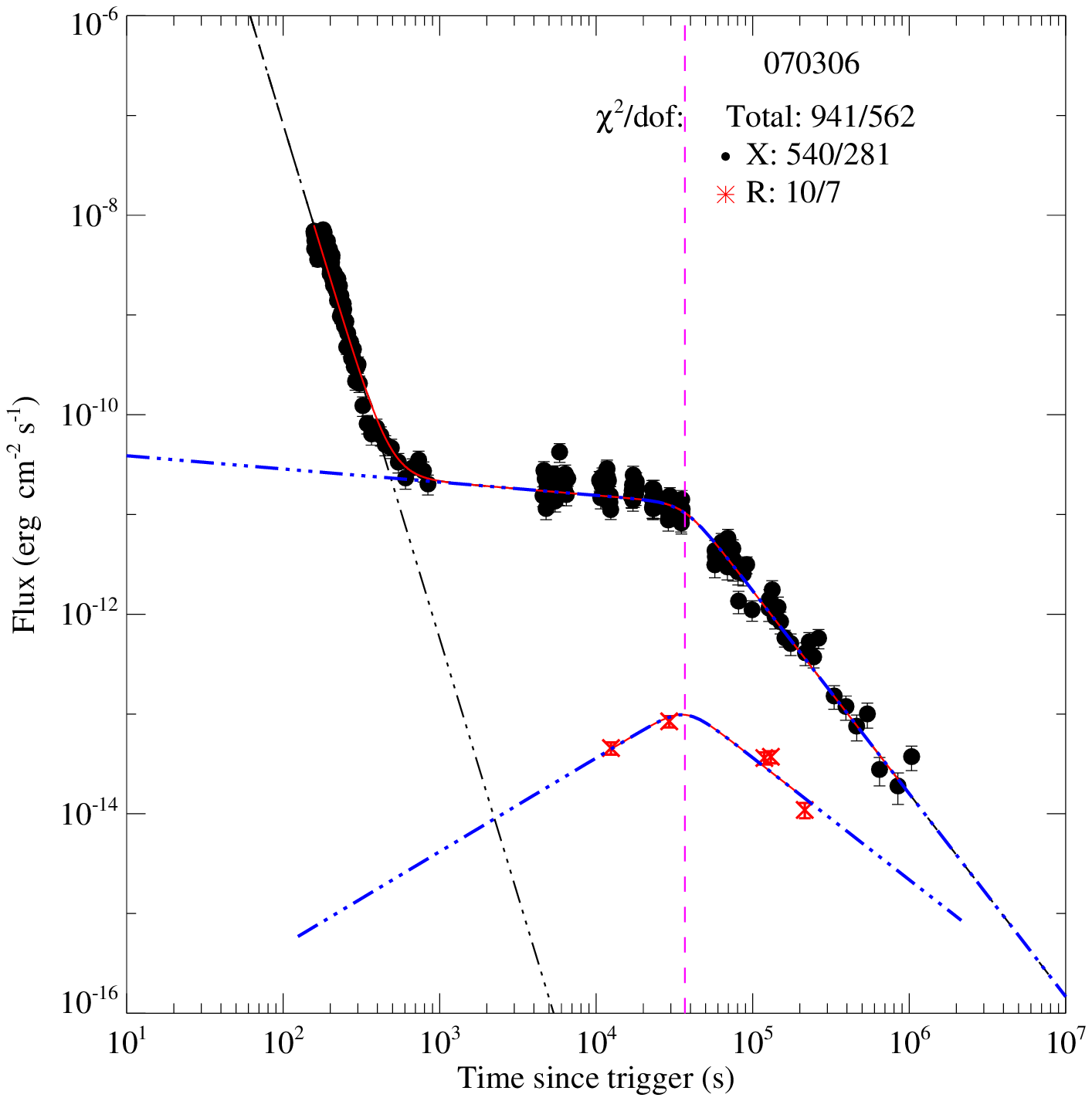}
\includegraphics[angle=0,scale=0.35,width=0.325\textwidth,height=0.30\textheight]{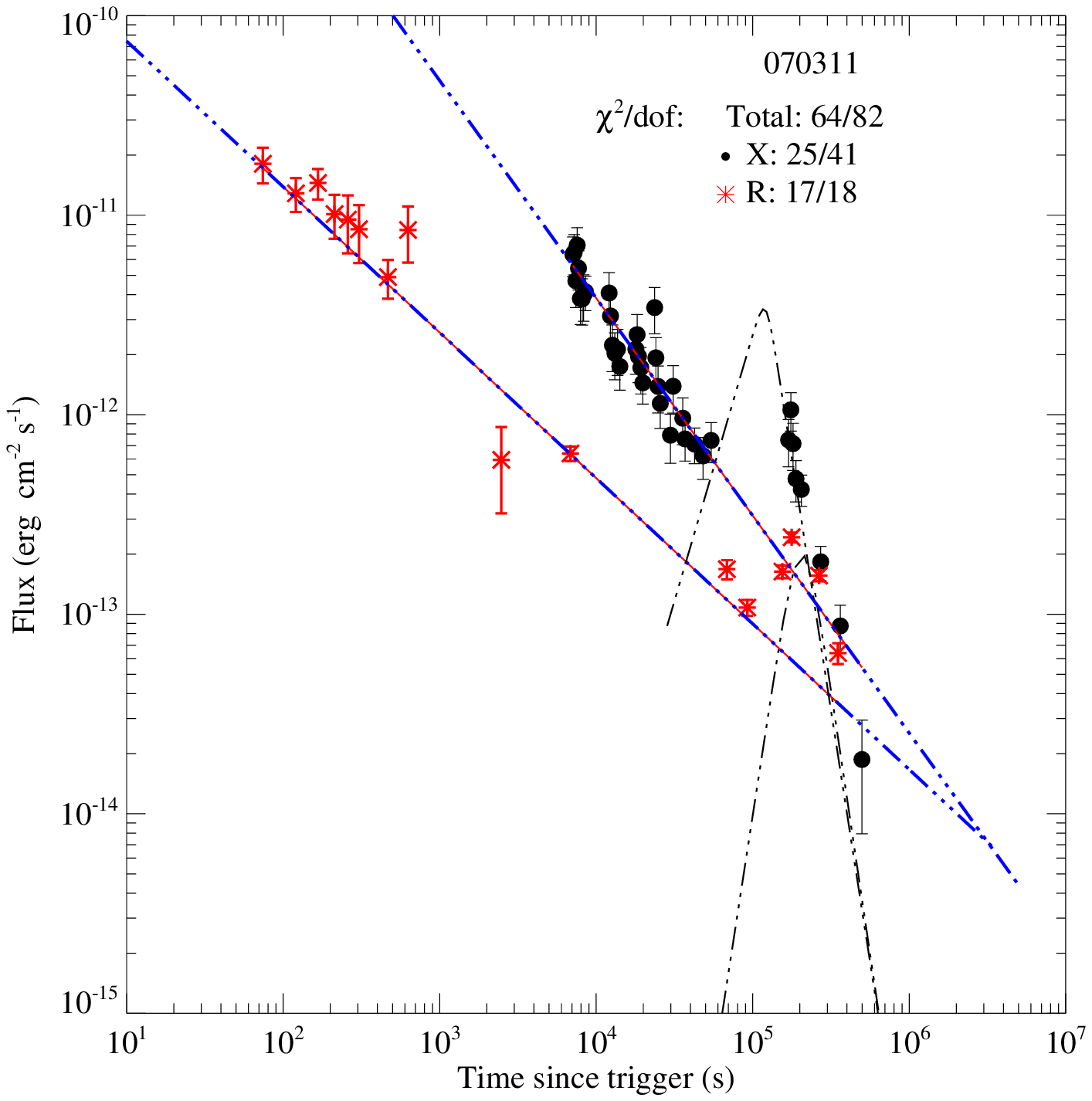}

\center{Fig. \ref{gradeIII}---Continued}
\end{figure*}

\clearpage
\setlength{\voffset}{-18mm}
\begin{figure*}

\includegraphics[angle=0,scale=0.35,width=0.325\textwidth,height=0.30\textheight]{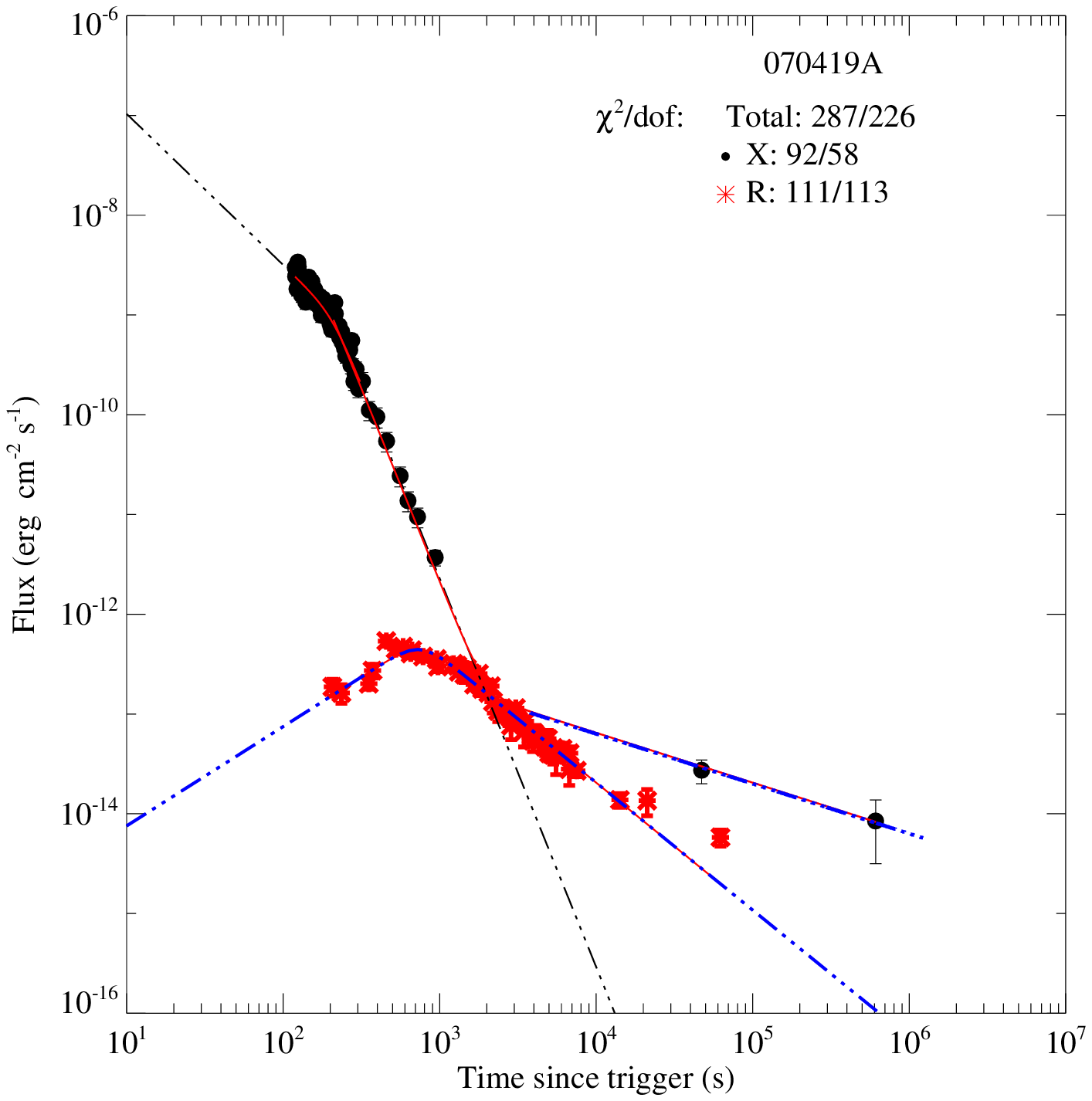}
\includegraphics[angle=0,scale=0.35,width=0.325\textwidth,height=0.30\textheight]{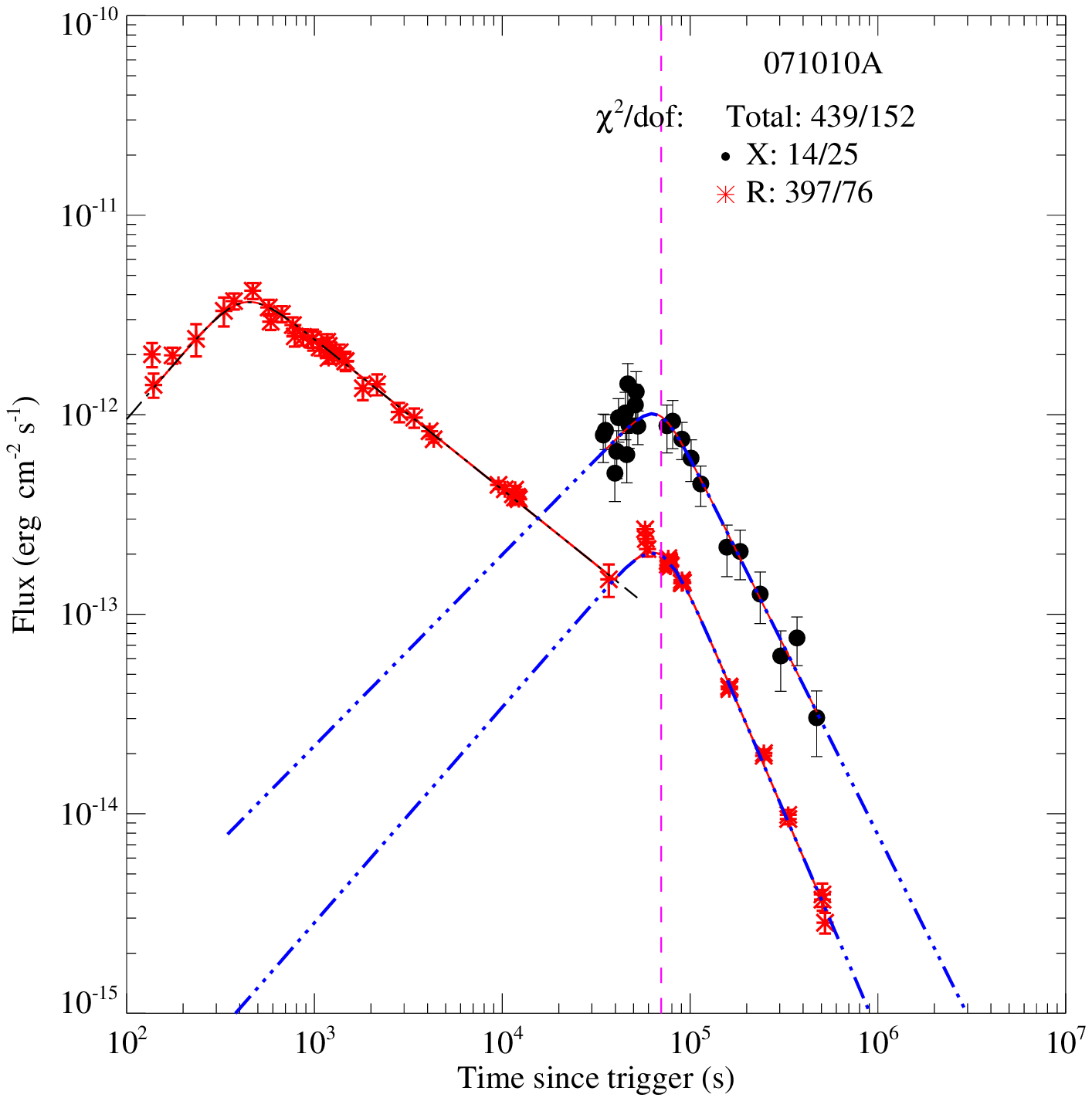}
\includegraphics[angle=0,scale=0.35,width=0.325\textwidth,height=0.30\textheight]{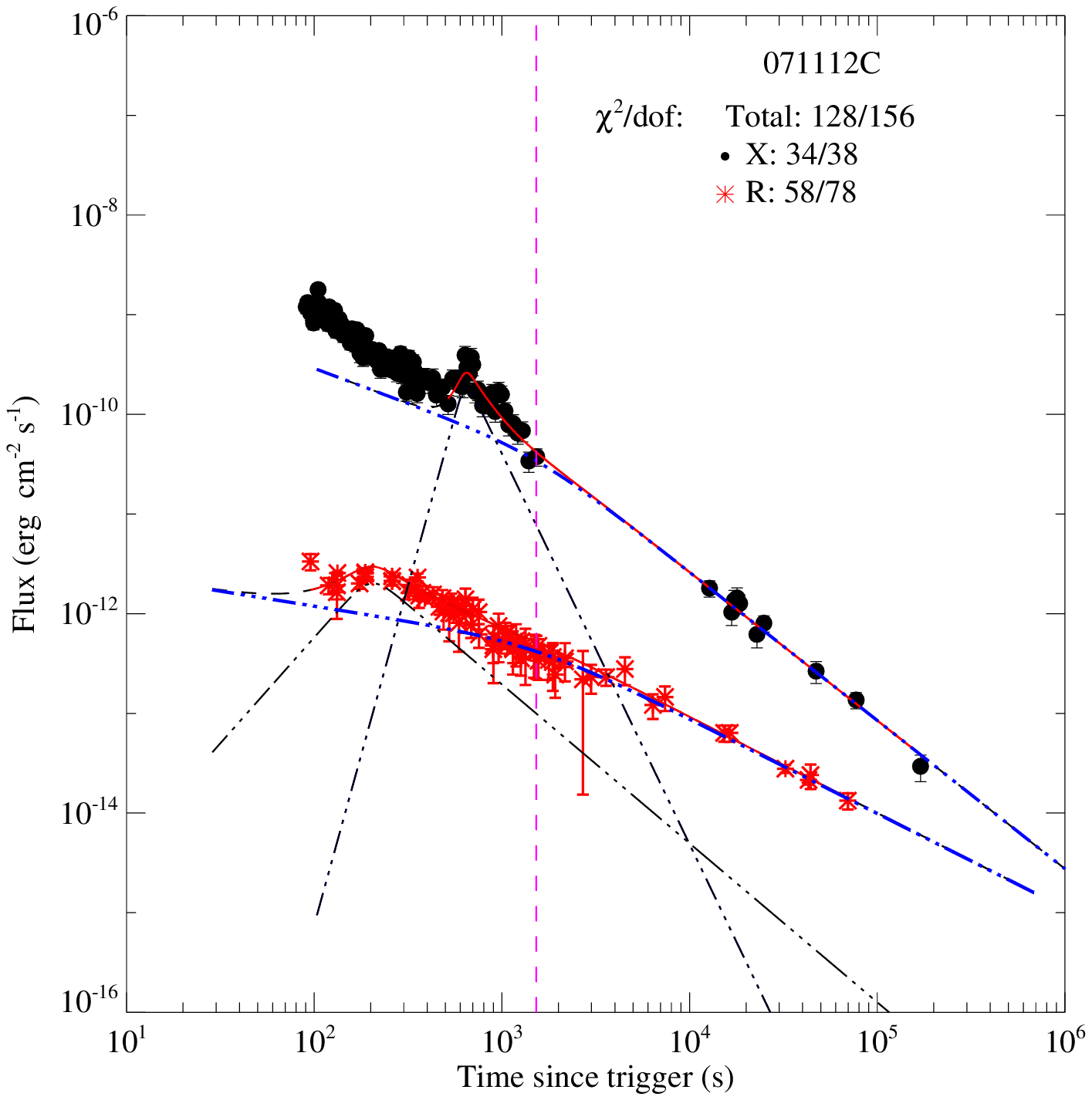}
\includegraphics[angle=0,scale=0.35,width=0.325\textwidth,height=0.30\textheight]{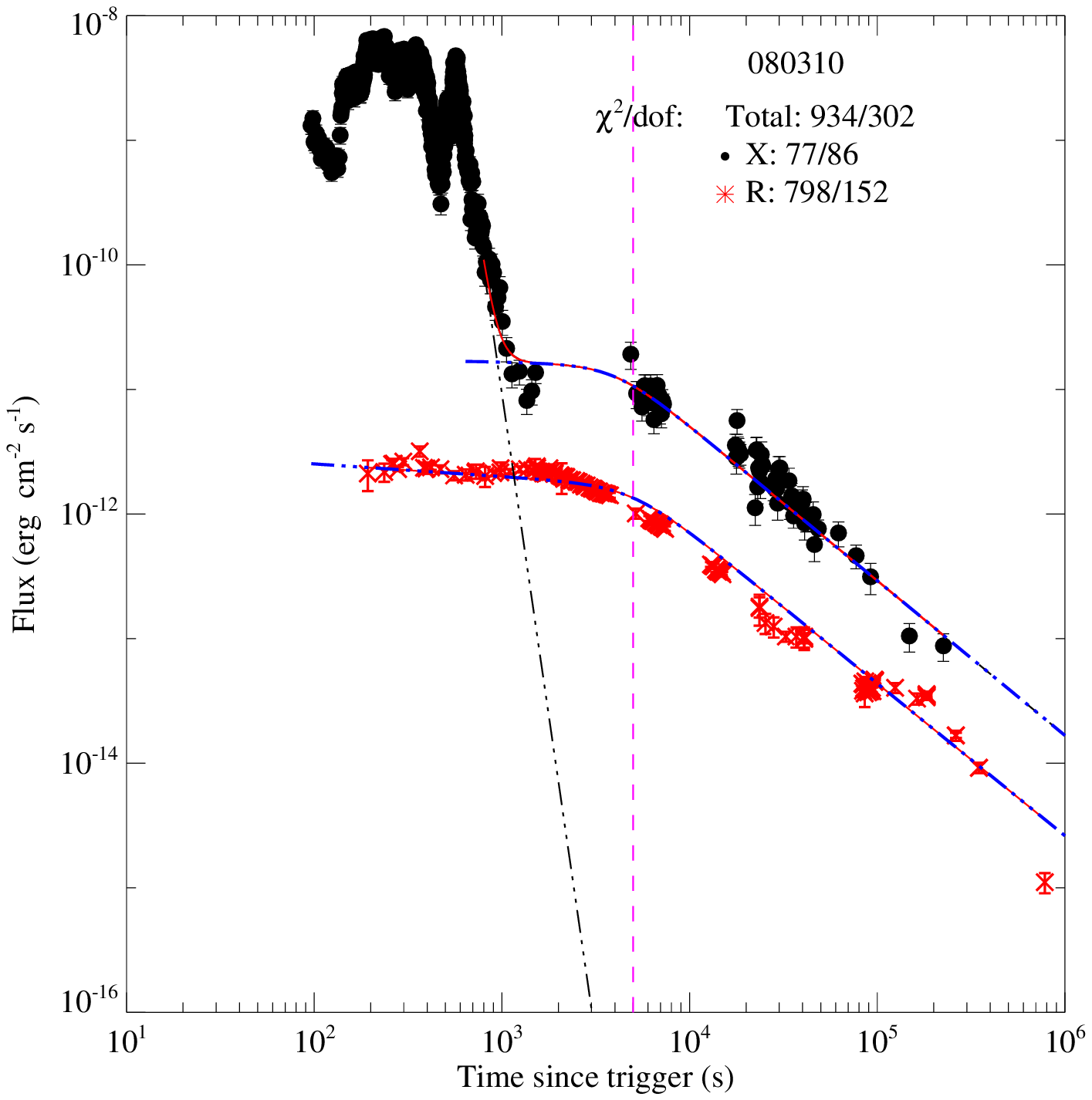}
\includegraphics[angle=0,scale=0.35,width=0.325\textwidth,height=0.30\textheight]{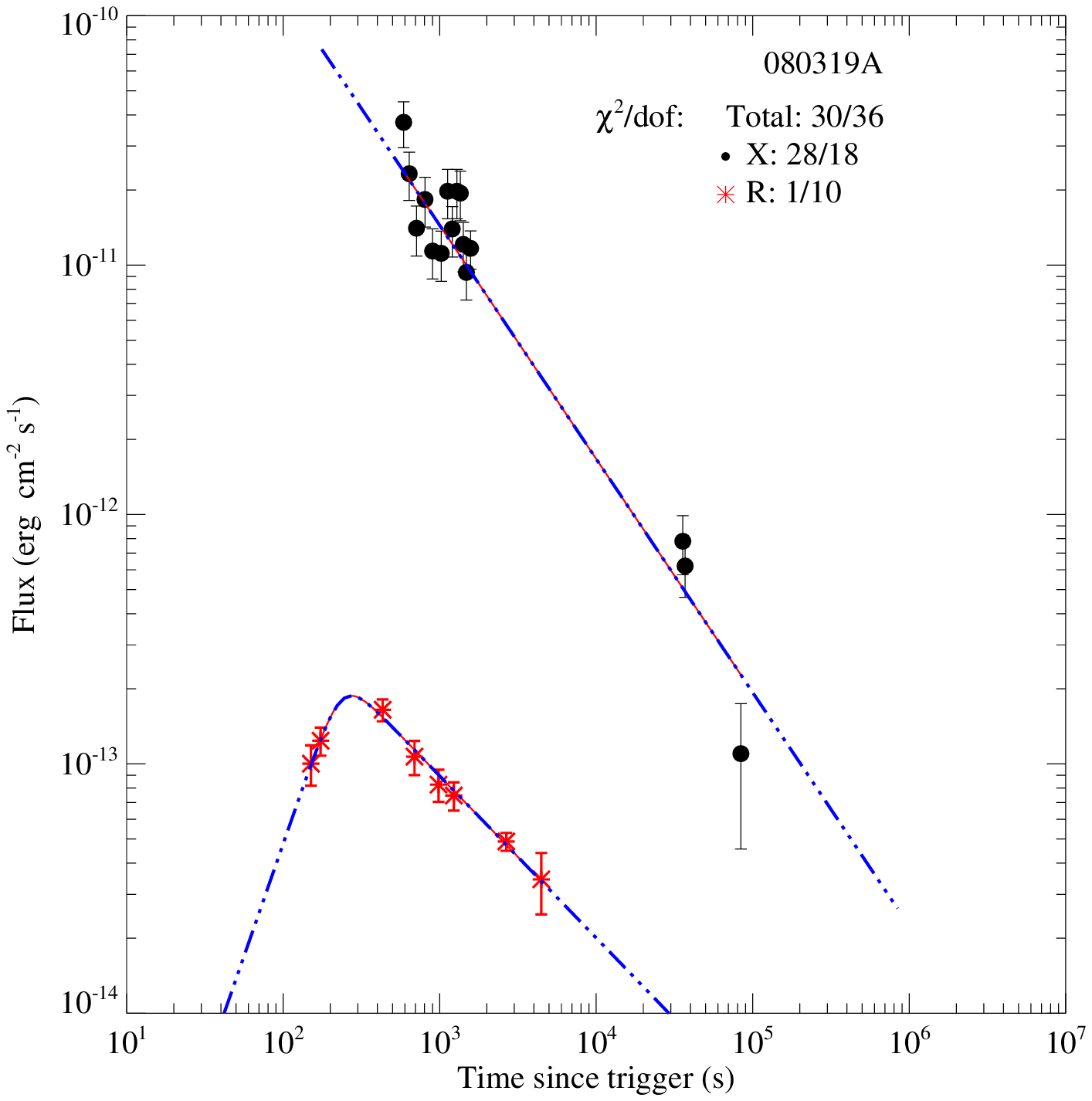}
\includegraphics[angle=0,scale=0.35,width=0.325\textwidth,height=0.30\textheight]{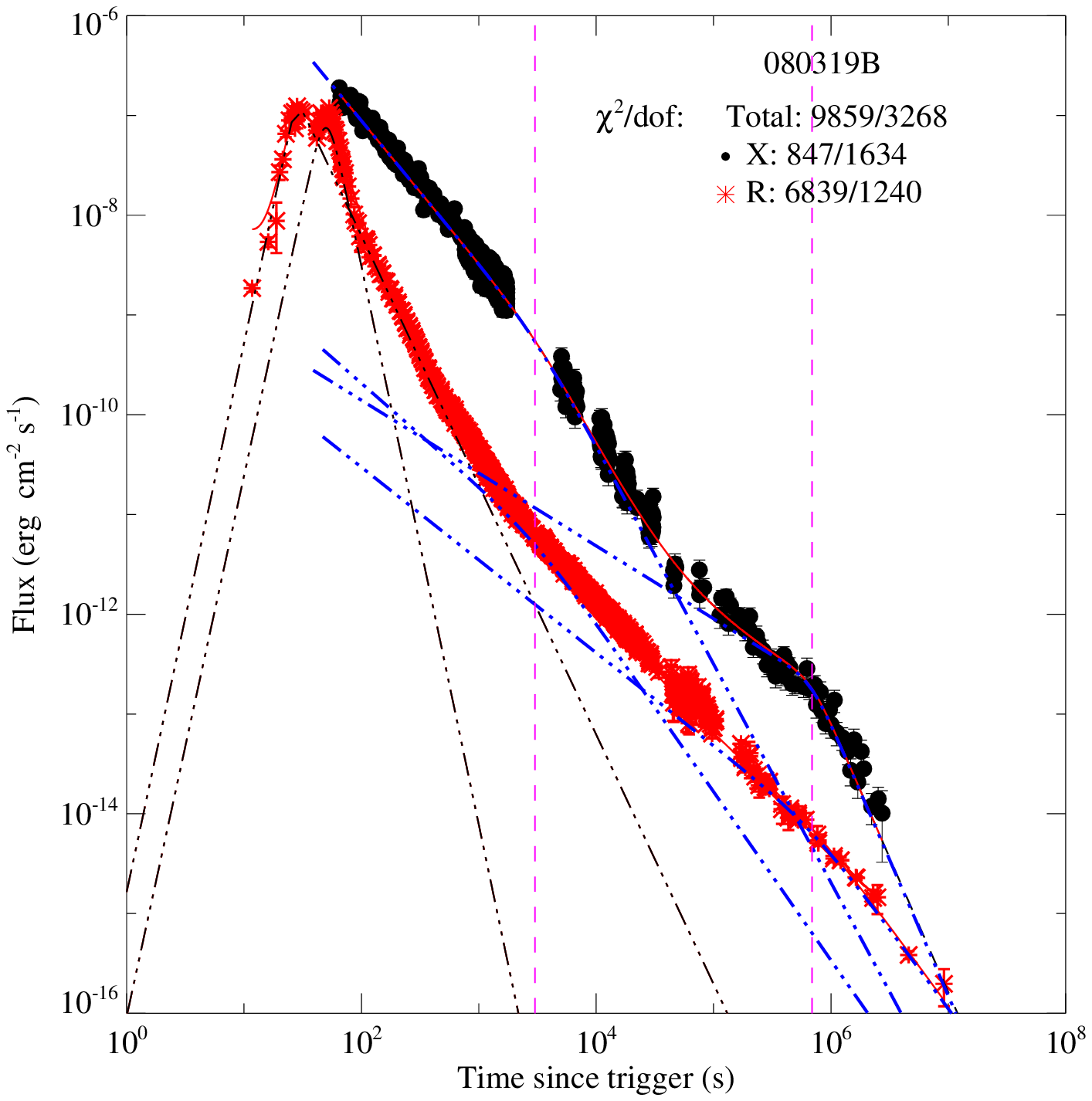}
\includegraphics[angle=0,scale=0.35,width=0.325\textwidth,height=0.30\textheight]{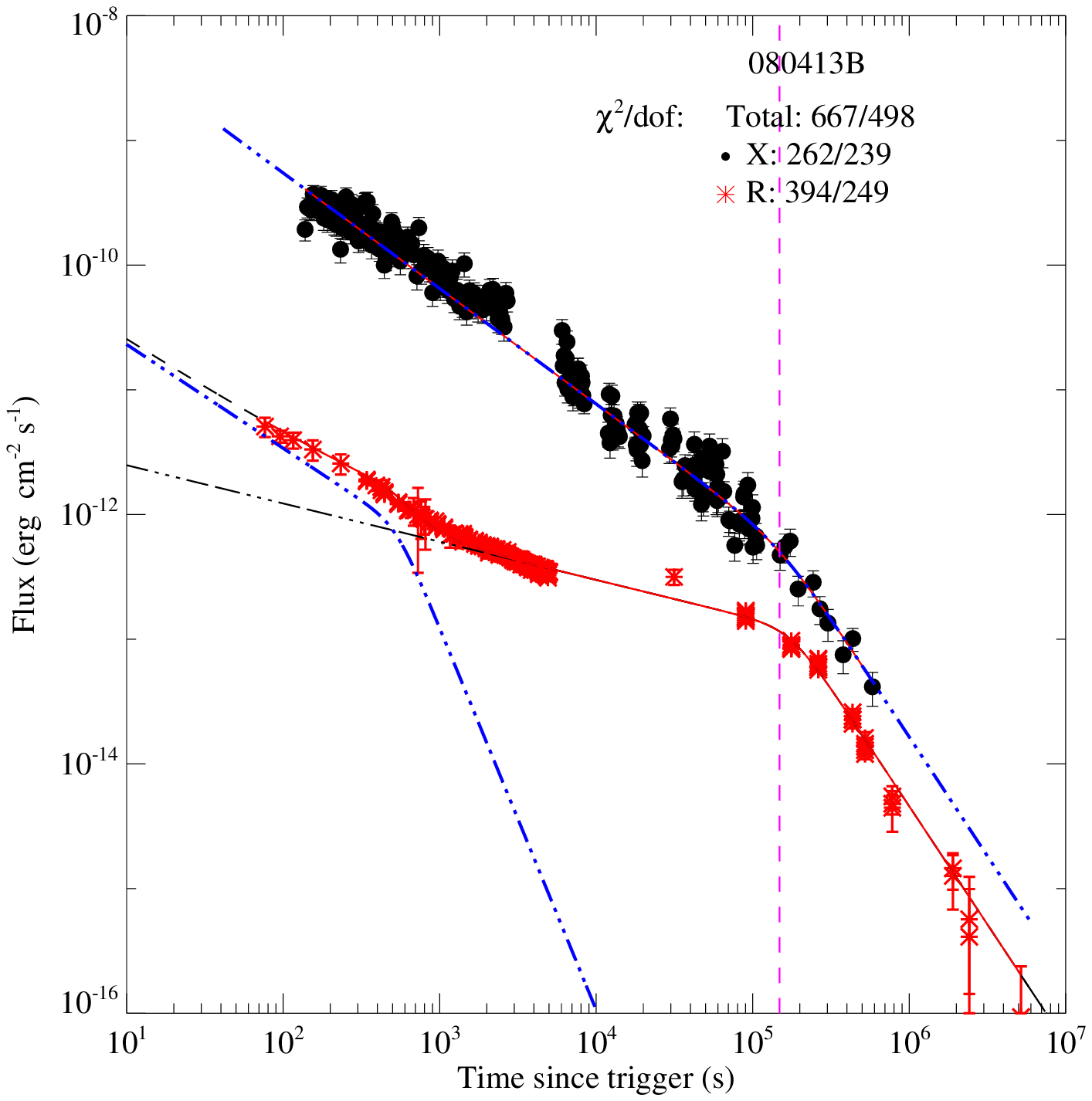}
\includegraphics[angle=0,scale=0.35,width=0.325\textwidth,height=0.30\textheight]{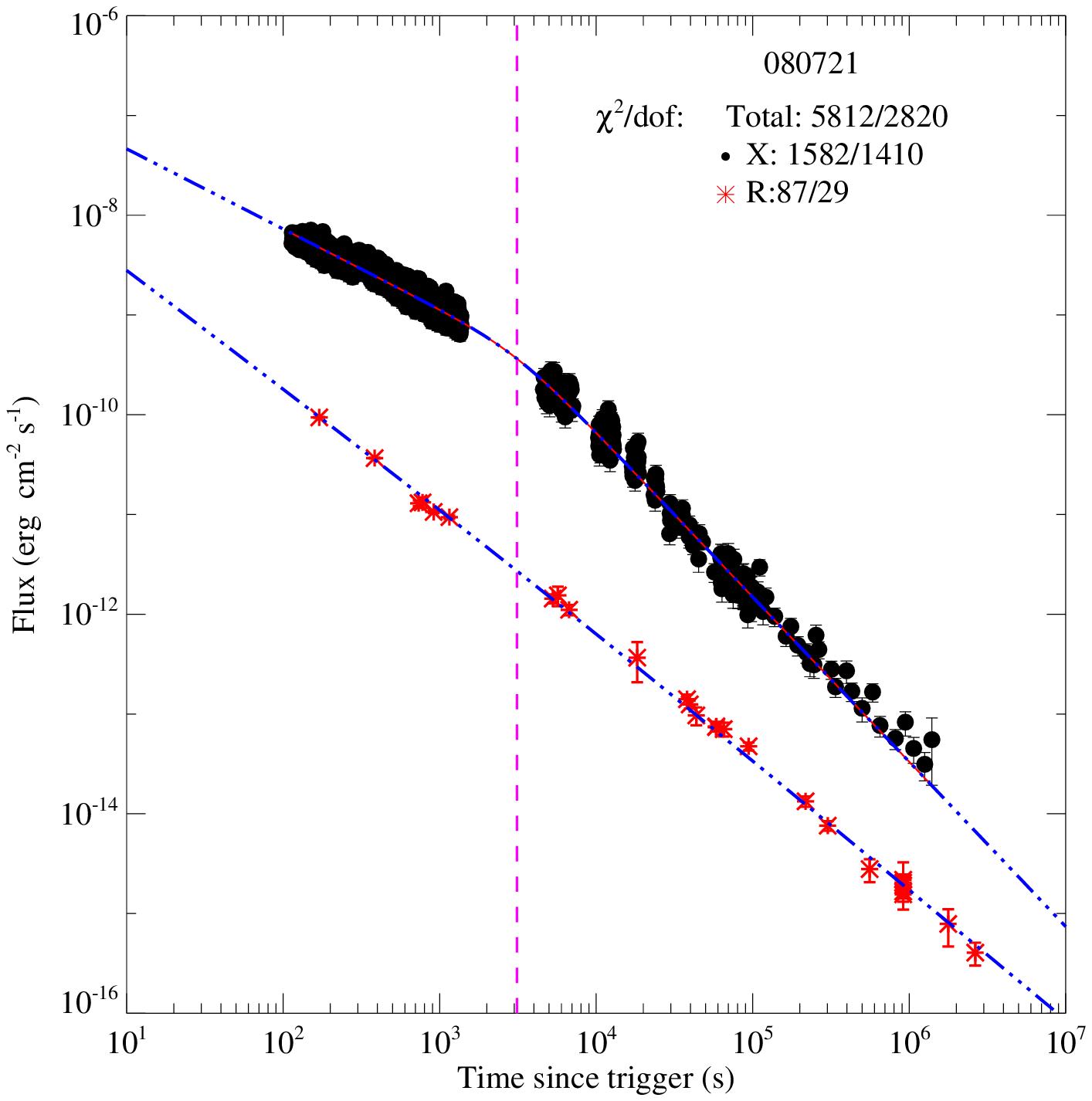}
\includegraphics[angle=0,scale=0.35,width=0.325\textwidth,height=0.30\textheight]{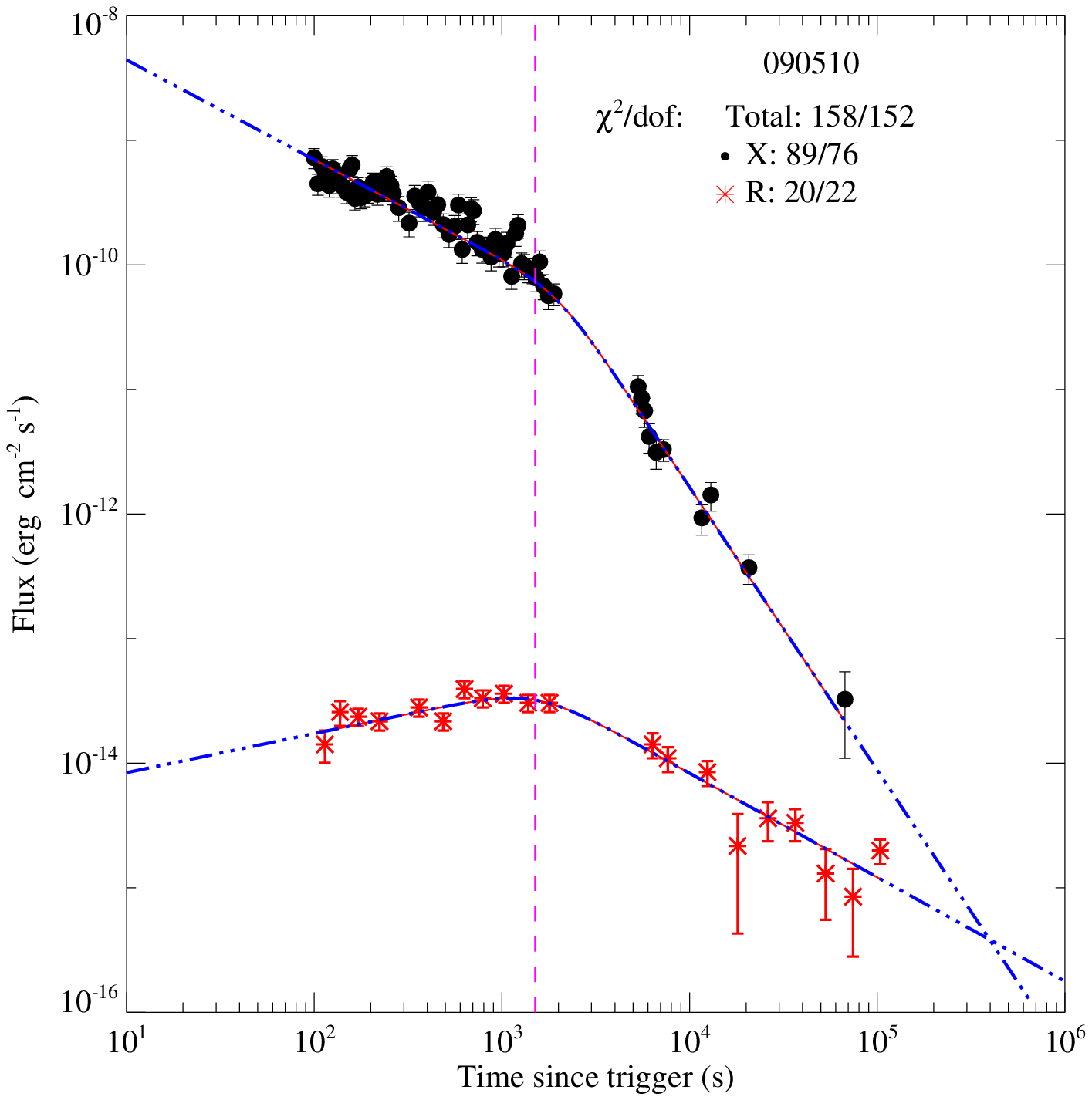}

\center{Fig. \ref{gradeIII}---Continued}
\end{figure*}

\clearpage
\setlength{\voffset}{-18mm}
\begin{figure*}
\includegraphics[angle=0,scale=0.35,width=0.325\textwidth,height=0.30\textheight]{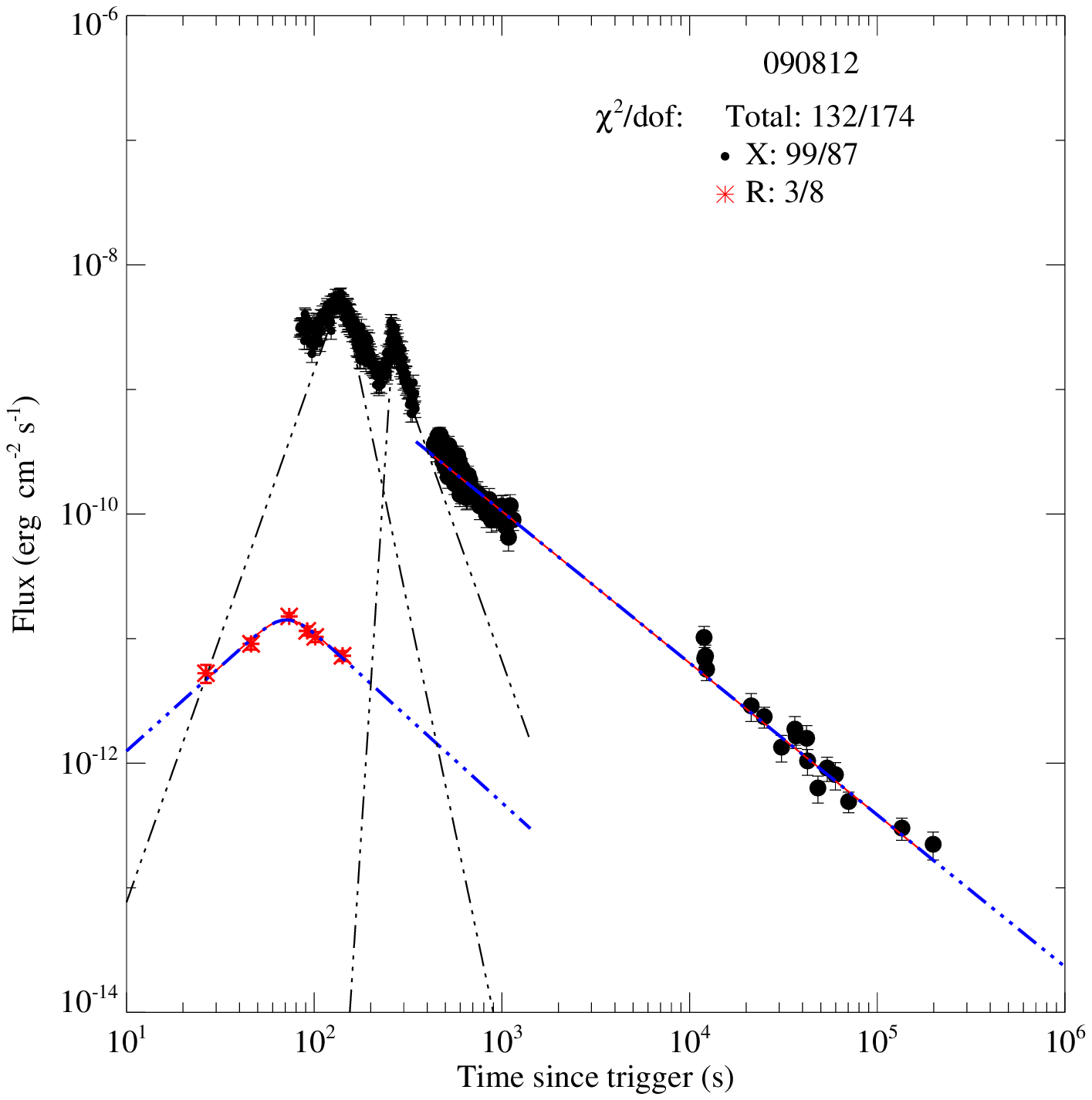}
\includegraphics[angle=0,scale=0.35,width=0.325\textwidth,height=0.30\textheight]{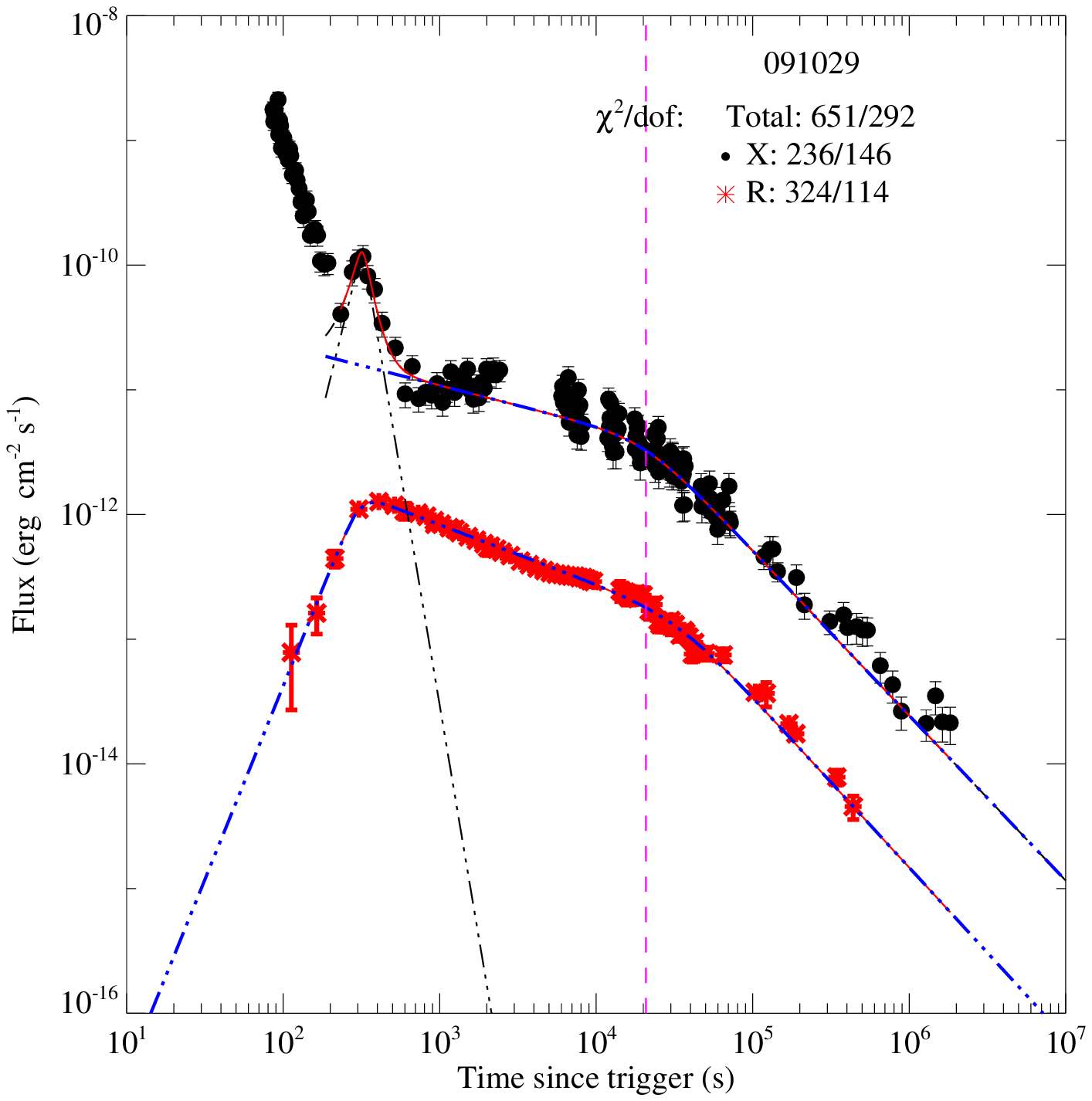}
\includegraphics[angle=0,scale=0.35,width=0.325\textwidth,height=0.30\textheight]{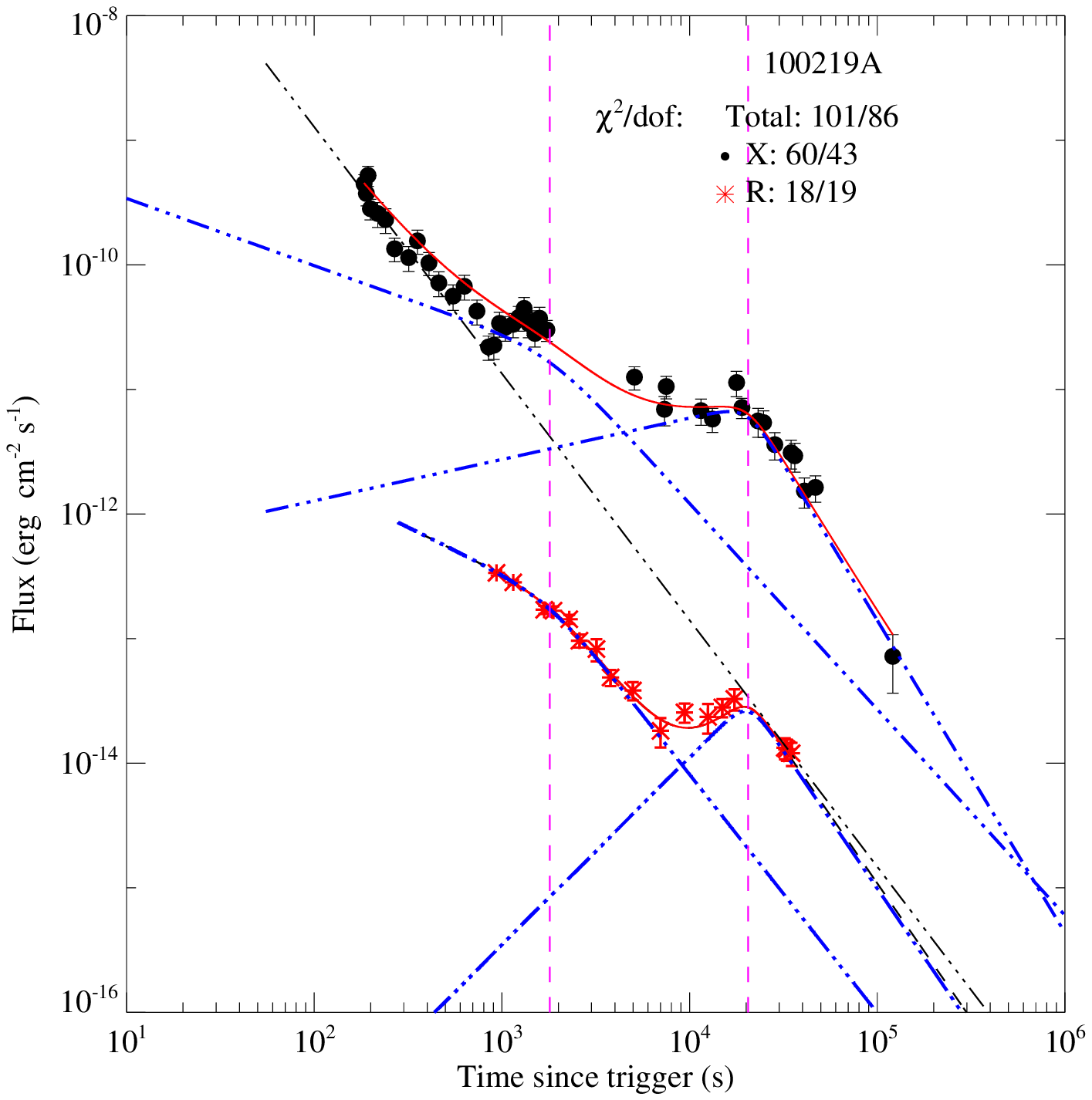}
\includegraphics[angle=0,scale=0.35,width=0.325\textwidth,height=0.30\textheight]{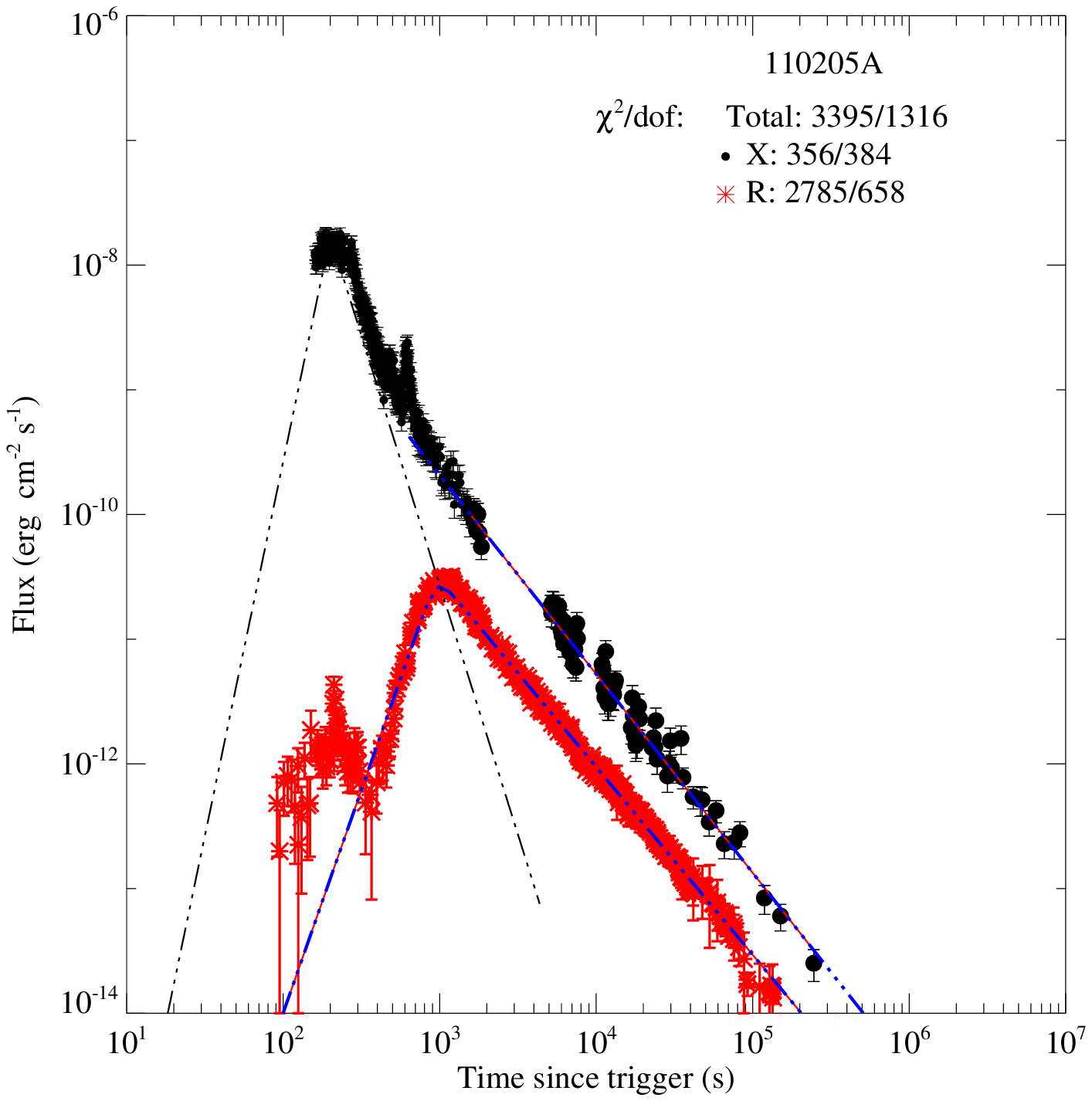}
\includegraphics[angle=0,scale=0.35,width=0.325\textwidth,height=0.30\textheight]{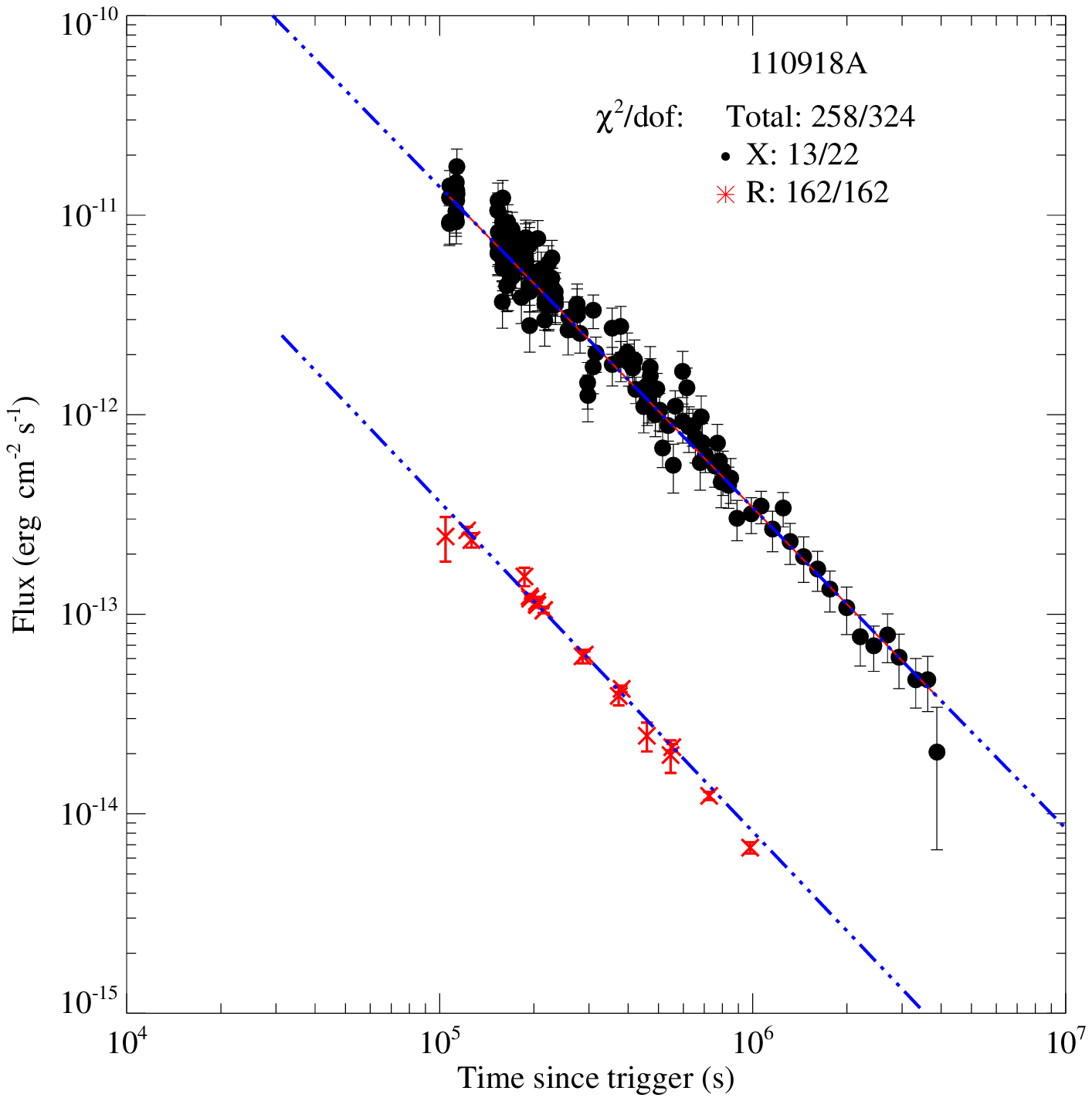}
\includegraphics[angle=0,scale=0.35,width=0.325\textwidth,height=0.30\textheight]{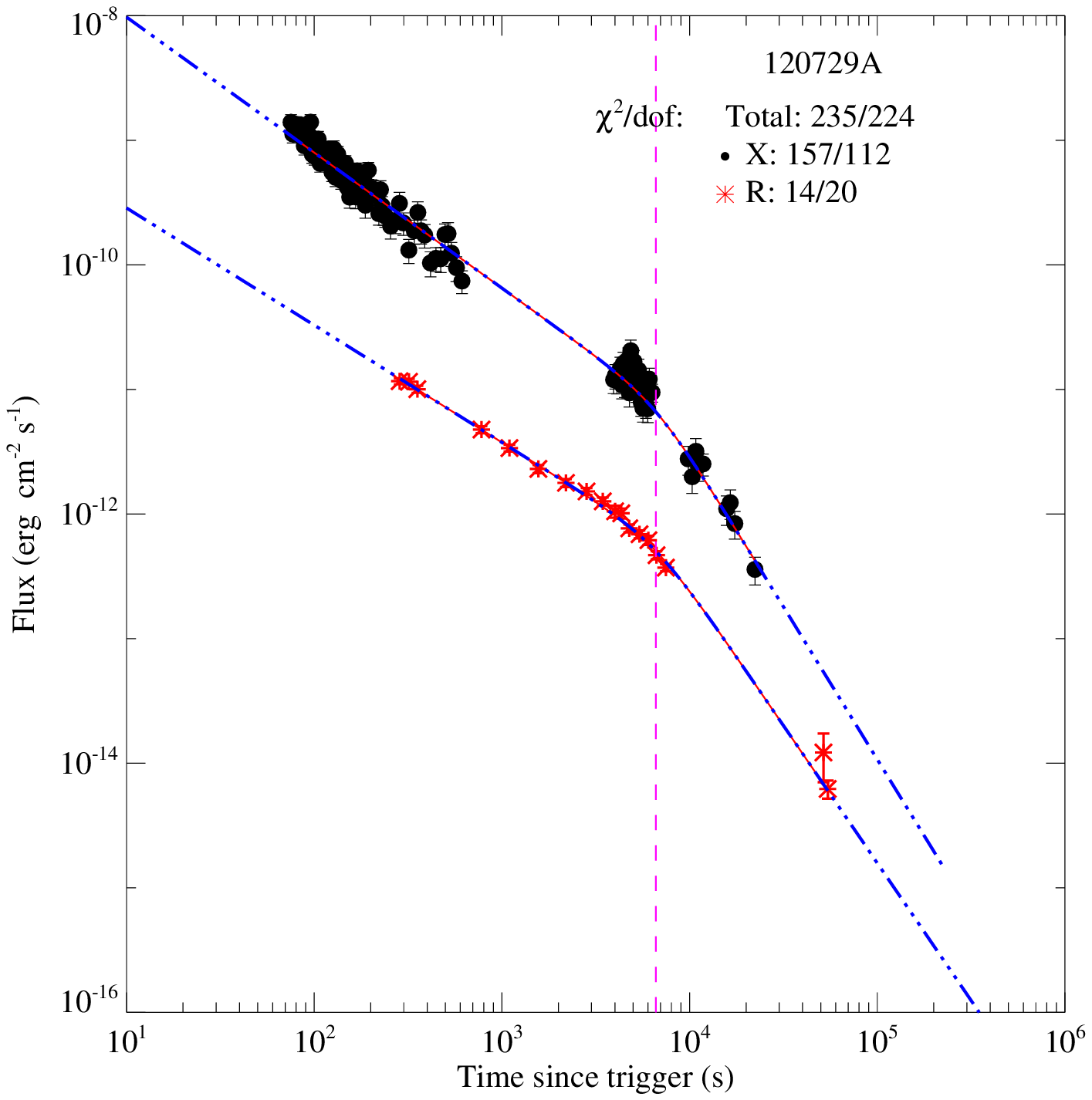}
\includegraphics[angle=0,scale=0.35,width=0.325\textwidth,height=0.30\textheight]{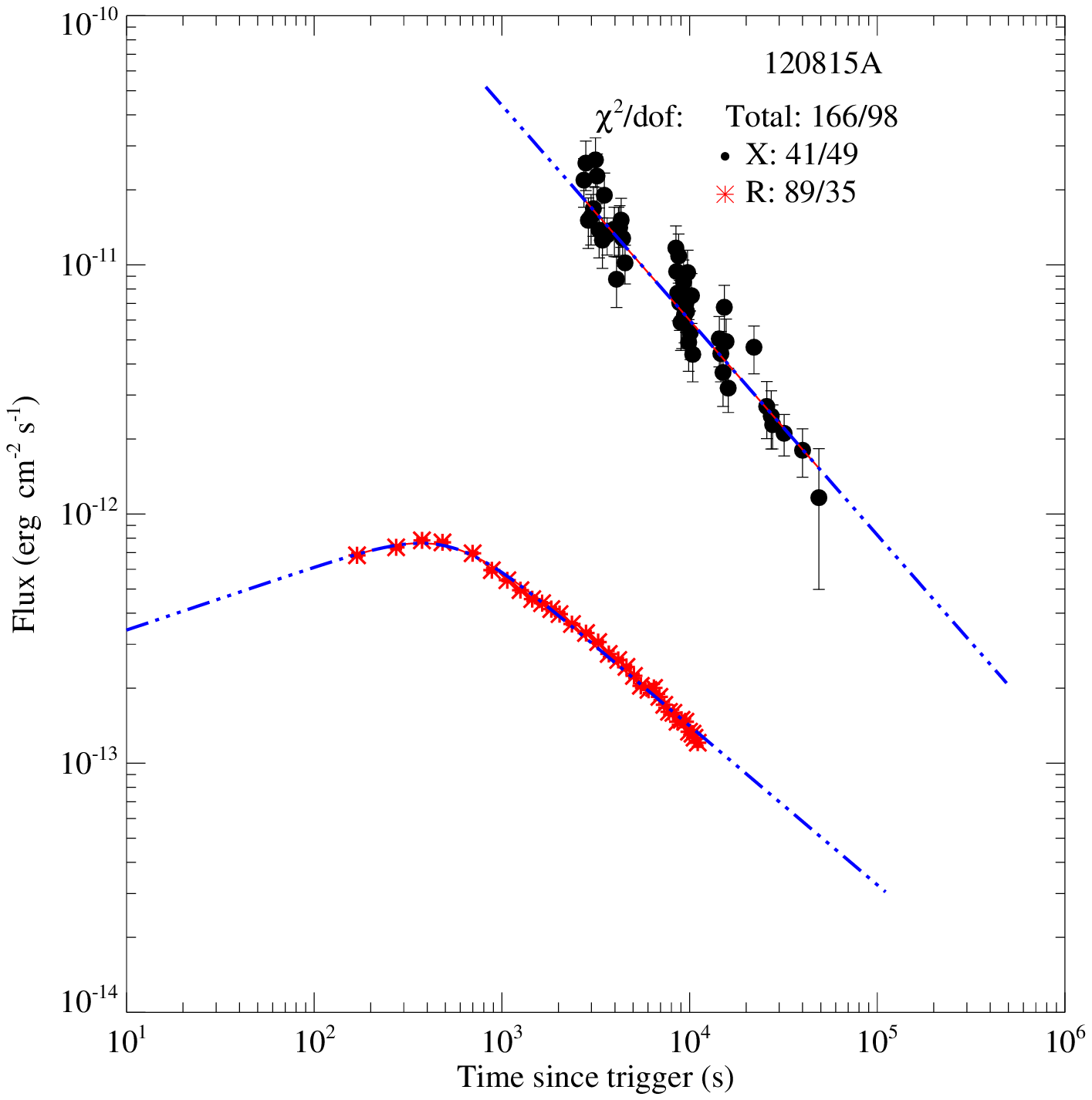}
\center{Fig. \ref{gradeIII}---Continued}
\end{figure*}

\clearpage
\setlength{\voffset}{-18mm}

\clearpage
\setlength{\voffset}{-18mm}

\clearpage 

\clearpage 

\clearpage 

\clearpage 

\clearpage 

\clearpage 

\clearpage 

\clearpage 

\clearpage 

\clearpage 

\clearpage 

\clearpage 

\clearpage 

\clearpage 

\end{document}